\documentclass[a4paper,useAMS,usenatbib]{mn2e}

\usepackage{graphicx}
\usepackage[T1]{fontenc}
\usepackage{amsmath}
\usepackage{aas_macros}
\usepackage{mwe}    % loads ÈblindtextÇ and ÈgraphicxÇ
\usepackage{times}
\usepackage{ amssymb }
\usepackage{color}
\usepackage[total={17.8cm,24.0cm},centering]{geometry}

\usepackage[a4paper]{lscape}

\newcommand{\kms}{km\,s$^{-1}$}

\newcommand{\reff}{R$_\mathrm{eff}$}
\newcommand{\Qb}{Q$_{\rm_b}$}

\title[The BaLROG project - II. Bar influence on stellar populations]{The BaLROG project - II. Quantifying the influence of bars on the stellar populations of nearby galaxies}

\author[Seidel et al.]{M. K. Seidel$^{1,2,3}$\thanks{E-mail: mseidel@carnegiescience.edu},
J. Falc\'{o}n-Barroso$^{1,2}$,
I. Mart\'inez-Valpuesta$^{1,2}$,
P. S\'anchez-Bl\'azquez$^{4}$ 
\newauthor
I.~P\'erez$^{5,6}$,
R.Peletier$^{7}$,
A. Vazdekis$^{1,2}$  \\
$^{1}$Instituto de Astrof\'isica de Canarias, E-38200 La Laguna, Tenerife, Spain\\
$^{2}$Departamento de Astrof\'isica, Universidad de La Laguna, E-38205 La Laguna, Tenerife, Spain\\
$^{3}$Observatories of the Carnegie Institution of Washington, Pasadena, CA 91101, USA\\
$^{4}$Dpto. de F\'isica Te\'orica, Universidad Aut\'onoma de Madrid, E-28049
Cantoblanco, Spain\\
$^{5}$Dpto. de F\'isica Te\'orica y del Cosmos, Universidad de Granada,
Facultad de Ciencias (Edificio Mecenas), 18071 Granada, Spain\\
$^{6}$Instituto Universitario Carlos I de F\'isica Te\'orica y Computacional,
Universidad de Granada, 18071 Granada, Spain\\
$^{7}$Kapteyn Astronomical Institute, University of Groningen, Postbus 800, 9700 AV Groningen, the Netherlands}

\begin{document}

\date{ }

\pagerange{\pageref{firstpage}--\pageref{lastpage}} \pubyear{2016}

\maketitle
\label{firstpage}
%=============================================================================
\begin{abstract}
We continue the exploration of the \textit{BaLROG} (Bars in Low Redshift Optical Galaxies) sample: 16 large mosaics of barred galaxies observed with the integral field unit SAURON. We quantify the influence of bars on the composition of the stellar component. We derive line-strength indices of H$\beta$, Fe5015 and Mg$b$. Based on single stellar population (SSP) models, we calculate ages, metallicities and [Mg/Fe] abundances and their gradients along the bar major and minor axes. The high spatial resolution of our data allows us to identify breaks among index and SSP profiles, commonly at 0.13$\pm$0.06 bar length, consistent with kinematic features. Inner gradients are about ten times steeper than outer gradients and become larger when there is a central rotating component, implying that the gradients are not independent of dynamics and orbits. Central ages appear to be younger for stronger bars. Yet, the bar regions are usually old. We find a flattening of the iron (Fe5015) and magnesium (Mg$b$) outer gradients along the bar major axis, translating into a flattening of the metallicity gradient. This gradient is found to be 0.03$\pm$0.07~dex/kpc along the bar major axis while the mean value of the bar minor axis compares well with that of an unbarred control sample and is significantly steeper, namely $-$0.20$\pm$0.04~dex/kpc. These results confirm recent simulations and discern the important $localized$ influence of bars. The elevated [Mg/Fe] abundances of bars and bulges compared to the lower values of discs suggest an early formation, in particular for early type galaxies.

\end{abstract}
%=============================================================================


%###############################################################################
%=============================================================================
\section{Introduction} \label{sec:intro}

Barred galaxies are ubiquitous in the local universe \citep[e.g.,][]{2000Eskridge, 2000Knapen, 2002Whyte, 2007Marinova, 2007MenendezD,
2008Barazza, 2009Aguerri, 2010MendezA, 2011Masters, 2014MendezA, 2014Cisternas} and can even be found at higher redshifts ($z$) \citep[e.g.,][]{1996Abraham,
2004Elmegreen, 2004Jogee, 2014Simmons}, but their fraction at higher $z$ is still under debate and also might depend on the galaxy's mass \citep[e.g.,][]{2008Sheth, 2010NairA,2014Melvin}. Their presence at different redshifts along with their prominence at lower redshifts has motivated many theoretical as well as observational studies on their formation and their influence on the evolution of their host galaxy. 

Bars are proposed to act as a major driver of the so-called \textit{secular evolution} \citep[e.g.,][]{korm_rev,2013seg..book..305A,2014Sellwood}. They are predicted to have a crucial influence on their host galaxies due to their ability to redistribute angular momentum as well as gas, which can lead to a central mass concentration or build-up of bulges \citep[e.g.,][]{1999Bureau, 1999Sakamoto, 1995Knapen, 2004Chung,2003Fathi,2013Cheung}. They thus seem to influence the inner regions of galaxies. And yet, there is no consensus on stellar population parameters among studies of central regions and bulges in barred and unbarred galaxies. Simulations and observations mostly agree on the influence on the gas phase, such as an increase of the gaseous metallicity in the bulge \citep[e.g.,][]{1994Friedli, 1995FriedliB, 2013Martel, 2011Ellison}, although some recent works disagree \citep[e.g.,][]{2014Cacho}. Concerning the stellar metallicity however, authors find different results. Some find that it remains unchanged in the central parts \citep[e.g.,][]{1994Friedli,2011CoelhoG,2012Williams,2014Cacho}, also in agreement with \citet{2015Cheung} who find that the central stellar population in general are not affected by bars, while others find a mild increase \citep[e.g.,][]{2006MoorthyHoltzmann,Perez2011}, which is however marginal. Simulations furthermore predict that in the absence of star formation, both stellar and gas metallicities of bulges in barred galaxies are supposed to decrease because lower metallicity stars and gas are funneled towards the center and thus dilute the initial negative metallicity gradient \citep[e.g.,][]{2013DiMatteo}. In the presence of star formation however, the central metallicity can rise. The disagreement also exists for the bulge ages, where \citet{2011CoelhoG} found a difference between barred and unbarred galaxies, the former showing younger ages than the latter (although only for the most massive spirals), while others did not \citep{Perez2011, 2012Williams, 2014Cacho}. Furthermore, numerous recent studies suggest that bars do not seem to alter AGN activity or be related to its presence \citep[e.g.,][]{2014Cisternas,2015Cheung}, in contrast to early predictions suggesting their influence \citep[e.g.,][]{1989Shlosman} and also other works that claim that there is a slight connection between bar presence and AGN activity \citep[e.g.,][]{2000Knapen,2002Laine, 2015Galloway}.

Apart from their influence on the central parts, there is evidence for bars being responsible for a redistribution of the stellar component \citep[e.g.,][]{2001Gadotti}. Due to their non-axisymmetric structure, bars have been predicted to increase radial motions and thus to act as agents reshuffling the stellar content resulting in a flattening of radial gradients \citep[e.g.,][]{2010Minchev,2013Kubryk, 2013DiMatteo}. In particular, large-scale bars are proposed as a major driver of radial mixing for a long time \citep[e.g.,][]{1993FriedliB}, not only in the inner regions but especially throughout the disc. In combination with spiral arms \citep[e.g.,][]{2002SellwoodB}, they can create a resonance overlap \citep[e.g.,][]{1997MassetT, 2010Minchev,2011Shevchenko, 2012Minchev} which can lead to a further increase of radial mixing. This type of mixing is especially dominant at larger radii, even outside of the bar corotation radius, such that inner gradients can differ. Former studies yet again find different results, similarly to the central values. 
\citet{2009PerezSBZ, 2007Perez} find a variety of age and metallicity gradients within the bar region and no systematic difference between barred and unbarred galaxies. A deeper analysis of 2 of them by \citet{2011MNRAS.415..709S} however shows flatter age and metallicity gradients \textit{along the bar} compared to those along the disk. This is confirmed by flatter gradients in edge-on boxy/peanut-shaped bulges (indicative for the presence of a bar \citet{2005Atha}) compared to unbarred early-type galaxies \citep{2012Williams}. Simulations of the Milky Way (boxy) bulge confirm this trend \citep{2013MVG}. Recent studies using large statistics by means of the CALIFA \citep{2012Sanchez} and Sloan Digital Sky Survey (SDSS, \citealt{2002Strauss,2009Abazajian}) however find no difference between the stellar population gradients of barred and unbarred galaxies, both along the entire disc as well as in their bulges (\citealt{2014SB2} and \citealt{2015bCheung} respectively). The reason for the differences obtained in the gradients are related to the different areas examined and might als result from the resolution and methodology of those latter studies which are significantly different to the former ones (see Sec.~\ref{sec:discuss2}).

As outlined above, major efforts have already been undertaken to identify the influence of bars on the chemical evolution of galaxies, but no consensus has been reached. Thanks to technical developments within the
last decade, stellar and gas contributions in the spectra 
could be separated \citep[e.g.,][]{sarzi2006} and improved instrumentation enabled the distinction of
fainter (sub-)structures  \citep[e.g.,][]{macarthur2009,
Perez2011, 2011MNRAS.415..709S, 2015Seidela}. Hence on the one hand, this has pushed the analysis in detail 
forward. On the other hand, recent surveys have improved the statistics of stellar population results \citep[e.g.,][]{2014SB2,2015McDermid}, but without the capability to resolve a high level of detailed substructures within the bulge and inner regions. The BaLROG sample tries to reconcile the two extremes by providing a small and yet representative sample of barred galaxies in the local Universe covering different bar morphologies and strengths (see Paper I), while sampling them in unprecedented spatial detail, necessary to detect and quantify the effect of bars.

This is the second paper of a series using the BaLROG (Bars in Low Redshift Optical Galaxies) sample (see Paper I, \citealt{2015Seidelb}). We presented the sample, measured bar strengths and studied the influence of the bars on the stellar and gas kinematics in Paper I. Our analysis, in particular of the stellar kinematics, revealed tentative evidence for an increase of radial mixing with bar strength, comparing stellar velocity dispersion gradients along bar major and minor axes. Most importantly however, we observed an important influence of the bar on central features. We found a relation between the magnitude of inner kinematic features with bar strength. Also, stellar angular momentum profiles present a dip commonly at around 0.2$\pm$0.1 bar lengths. Within this region, about 50\% of our sample exhibits an anticorrelation of h$_3$ and stellar velocity over velocity dispersion. In this work, we will focus on the stellar content and quantify the effect of bars on stellar population gradients. Also, we will try to connect our kinematic results with this stellar population analysis. We briefly summarize the data and data reduction, as well as necessary corrections in Sec.~\ref{sec:data}. In Sec. ~\ref{sec:methods}, we explain the method of obtaining line strength index measurements and single stellar populations. In Sec.~\ref{sec:results1} we will show the results of the line strength analysis and in Sec.~\ref{sec:results2} the corresponding population parameters. We discuss our results in Sec.~\ref{sec:discuss} and summarize them briefly in Sec.~\ref{sec:summ}.

%###############################################################################
\section{Data and Reduction}
\label{sec:data}

\subsection{Observations}

We have obtained large-scale two-dimensional spectroscopic data for a sample of 16 barred galaxies, using the integral field unit (IFU) SAURON (Spectrographic Areal Unit for Research on Optical Nebulae; \citealt{2001Bacon}) at the William Herschel Telescope (WHT) in La Palma at the Observatorio del Roque de los Muchachos. The sample is drawn from the Spitzer Survey of Stellar Structure in Galaxies (S$^4$G; \citealt{2010Sheth}), restricting in declination, inclination, magnitude and redshift (for details, please refer to Paper I). We sample both early- and late-type barred galaxies, covering various Hubble types and bar strengths representative for a sample of local barred galaxies.  We sample galaxies with morphological types from SB0 to SBbc and thus have a slight bias towards early types. Throughout the paper we will call early-types to the SBa while we will consider the rest (SBb and later) as late type galaxies. We list our sample and its basic characteristics, including stellar population properties obtained in this work, in Table~\ref{tab:sample}. While our sample is limited in numbers compared to ongoing integral field surveys such SAMI \citep{2012Croom}, CALIFA \citep{2012Sanchez} or MaNGA \citep{2015Bundy}, it provides us with the necessary spatial detail. We sample at typically 100~pc, even maintained with our Voronoi-binning (within the bar region), which is in most cases at least a factor 10 better than the larger surveys. We achieve the combination of high spatial resolution and large field-of view by creating mosaics using up to 7 SAURON IFU pointings.

The observations were conducted between March 2012 and January 2014 in 4 consecutive runs. We used the low-resolution mode of SAURON which leads to a field of view of 33\arcsec\,$\times$\,41\arcsec, spatial sampling of 0\farcs94\,$\times$\,0\farcs94 per lenslet (1431 in total) and a
spectral resolution of full width at half maximum (FWHM) of 3.9\,\AA. The wavelength coverage ranges from 4760\,\AA\ to 5300\,\AA.

Apart from our own data, the BaLROG dataset, we use subsets of other datasets for this work. Up to date, unfortunately, there is no sample of unbarred galaxies available that perfectly matches the BaLROG dataset in spatial resolution and spatial extensions and it surpassed our means to additionally observe a matching unbarred sample. Since our aim is to best restrict bar driven influences (comparing to unbarred galaxies), we profit from available data and build a subsample of unbarred galaxies from major surveys that used the same instrument, the SAURON IFU. We hence chose unbarred galaxies from several surveys, always matching our inclination and redshift constraints. For early-types, we took galaxies from the ATLAS3D survey \citep{2011Cappellari}. For early-type spirals (Sa), we chose galaxies from the SAURON \citep{FalconBarroso2006} studies, taking data from \citet{peletier2007}. To include later types, we also took galaxies from the late-type extension of the SAURON survey, using data from \citet{ganda2007}. As some of our investigations focus on correlations with the bar strength parameter developed in Paper I, we do not use this comparison for our entire work. Instead, we try to reach out for it, whenever a comparison is possible and helps us to better understand the influence of bars and in particular trying to distinguish it from a Hubble type driven effect.

%-----------------------------------------------------------------------------
\begin{table*}
\centering
\caption{{\textsc Galaxy Sample}.~-~(1) Galaxy name, (2) Hubble type, (3) inclination, (4) systemic velocity, (5) bar strength, (6) bar length, (7) effective radius (\reff) of the galaxy. Objects forming part of the Virgo cluster are marked with a small $v$ next to their name. {\textsc Notes.~-~}All
morphological classifications and V$_{\rm sys}$ are from the
\textit{NASA Extragalactic Database (NED)}. Bar strengths are taken from Paper I,  effective radii of the galaxy
(\reff) are from the S$^4$G P4 \citep{2015Salo} while bar lengths are
determined by Herrera-Endoqui et al. (2015, submitted). }
\begin{tabular}{lcccccccccc}
\hline
Galaxy          &  Hubble       &  Inclination    & V$_\mathrm{sys}$ &  \Qb         & Bar Length & \reff             \\
~               &  Type         &  (deg)          & (\kms)           &  ~           & (arcsec)   & (arcsec)                          \\
(1)             &  (2)          & (3)             & (4)              &  (5)         &  (6)       & (7)                                \\
\hline                                  
NGC\,1015      &  SBa            &  30.5    & 2628                &    0.264    & 21.5       &  19.73             \\
NGC\,2543      &  SB(s)b         &  59.9    & 2471                &    0.357    & 14.9       &  26.56             \\
NGC\,2712      &  SBb            &  60.5    & 1815                &    0.283    & 20.5       &  25.09             \\
NGC\,2859      &  (R)SB0$^+$(r)  &  37.2    & 1687                &    0.166    & 34.4       &  22.41             \\
NGC\,2893      &  SB0-a          &  17.4    & 1703                &    0.164    & 12.0       &  \phantom{0}4.58   \\
NGC\,2962      &  SB0-a          &  49.0    & 1960                &    0.136    & 30.7       &  20.50             \\
NGC\,3485      &  SBb            &  20.4    & 1436                &    0.383    & 21.0       &  26.38             \\
NGC\,3504      &  SBab           &  12.8    & 1539                &    0.256    & 37.1       &  11.13             \\
NGC\,4245      &  SB0/a(r)       &  33.3    & \phantom{0}886      &    0.184    & 36.3       &  23.52             \\
NGC\,4262$^v$  &  SB0$^-$(s)     &  24.5    & 1359                &    0.069    & 13.4       &  \phantom{0}5.99   \\
NGC\,4267$^v$  &  SB0$^-$        &  11.9    & \phantom{0}983      &    0.038    & 16.9       &  21.07              \\
NGC\,4394$^v$  &  (R)SB(r)b      &  30.4    & \phantom{0}922      &    0.234    & 41.4       &  36.79              \\
NGC\,4643      &  SB0/a(rs)      &  36.8    & 1330                &    0.276    & 49.9       &  24.22              \\
NGC\,5350      &  SBbc           &  50.3    & 2321                &    0.440    & 15.2       &  28.06              \\
NGC\,5375      &  SBab           &  29.8    & 2386                &    0.231    & 27.2       &  24.35               \\
NGC\,5701      &  (R)SB0/a(rs)   &  15.2    & 1505                &    0.179    & 39.0       &  25.97              \\ 
\hline
\end{tabular}
\label{tab:sample}
\end{table*}
%-----------------------------------------------------------------------------

\subsection{Data Reduction and Emission line corrections} 

The data were reduced using the SAURON pipeline XSauron described in detail in \citet{2001Bacon}. In Paper I we give a more detailed description, here we only provide a brief summary. The reduction included overscan and bias subtraction, flat fielding with a combination of twilight and continuum exposures and wavelength calibration using arc (neon) lamp exposures. Cosmic rays were removed and a flux calibration applied using spectrophotometric standards. Finally, we merged the mosaics also using the XSAURON software comparing integrated intensity contours with those of a $g$-band SDSS image. 

Similar to our analysis in Paper I we again adopted the Voronoi binning scheme of \citet{2003Cappellari} for our work here. To ensure high-quality spectra for our stellar population analysis we chose in this case a minimum S/N of $\approx$80 per pixel for all galaxies, which is double than that for the kinematic analysis. The central spectra remained unbinned nevertheless in all cases and exceeded this S/N level (e.g., S/N$>$100). The reason to elevate the S/N for this work is the delicacy of the extraction of absorption line strengths (connected with the emission line removal).  Before the binning, we also excluded spaxels with a S/N below 3 in order to avoid contamination by poor quality measurements. The extensions of the maps are therefore not only given by the mosaic, but also by this imposed S/N minimum threshold. This threshold remains the same as in Paper I.

\begin{figure}
\includegraphics[width=\linewidth]{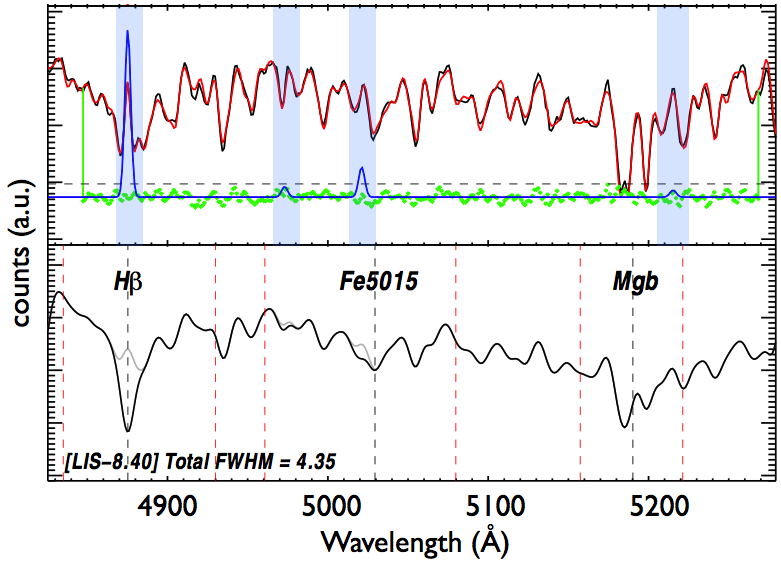} 
\caption{Figure 3.1: A spectra of the disc region of NGC 4394 is shown to illustrate the cleaning of emission lines. The top panel shows the spectrum of the galaxy (black), a model fit (red), residuals of the areas fitted (green) and indications of areas where emission lines are present (blue transparent columns) and the emission lines themselves are shown as additional gaussians in blue. The line-strength measurements are computed from the emission-line-cleaned spectra. This is the same spectrum, but brought to 8.4\,\AA\,(as indicated in the lower left corner, along with the maximum FWHM detected for this spectrum before raising it to 8.4\,\AA). The central bandpasses for H$\beta$, Fe5015 and Mg{\it b} are shown as gray dashed lines and their continuum areas as red dashed lines. The parts where corrections for emission lines have been made are shown as gray lines deviating from the galaxy spectrum (black).}
\label{fig:fits}
\end{figure}

In Paper I, we already used the pPXF -- penalized pixel fitting -- code developed by \cite{capems_2004} in order to extract the stellar kinematics. Further, we already used the Gas AND Absorption Line Fitting ({\tt GANDALF}) package by \citet{sarzi2006} and \citet{FalconBarroso2006} to determine the ionised-gas distribution and kinematics. Several nebular emission lines are found within the SAURON spectral range, namely H$\beta\lambda\lambda$4861 and the doublets [O{\sc iii}]$\lambda\lambda$4959, 5007and [N{\sc i}]$\lambda\lambda$5200, 5202\,\AA. The first, H$\beta$, is present in emission and absorption and it is crucial for line-strength analysis. The [O{\sc iii}] doublet is important to be removed correctly as it can strongly affect the Fe5015 index. The [N{\sc i}]  doublet falls within the Mg{\it b} index and hence needs to also be corrected for. In Fig.~\ref{fig:fits} we show an example spectrum of one of our galaxies indicating the emission lines and their correction in the top panel and the cleaned spectrum at 8.4\,\AA\, in the bottom panel. 

%-----------------------------------------------------------------------------
%-----------------------------------------------------------------------------

The optimal correction was achieved tying spectral lines kinematically to the [O{\sc iii}] doublet to lower the number of free parameters given to {\tt GANDALF}. To
constrain the freedom of the doublet lines during the fitting process even further, known relations, namely $F([$O{\sc iii}$]_{4959})=0.350\cdot F([$O{\sc iii}$]_{5007})$, were imposed. The obtained gas kinematics were presented in Paper I.

%=============================================================================
\section{Stellar population analysis}
\label{sec:methods}

In this paper we focus our analysis on the classical approach of using absorption line-strength 
indices measured on observed spectra.
These indices can be compared with those computed via single stellar population (SSP) models 
in order to derive stellar population parameters from integrated
spectra \citep[e.g.,][]{1973Faber, 1993Davies, 1994Worthey, 1999Vaz, 2005Thomas,
2006SBa, kuntschner2006, macarthur2009, 2015McDermid}. 
%-----------------------------------------------------------------------------

The most commonly used method to probe
the luminosity-weighted age, metallicity and $\alpha$ abundance ratios of specific
elements is the measurement of Lick/IDS indices in the Lick system. This system suffers, however, from intrinsic uncertainties due to a multi-object and multi-instrument approach of the spectra leading to an inhomogeneous spectral resolution \citep{2010Vaz}. To avoid those and profit from flux-calibrated spectra with a constant resolution as a function of wavelength, we chose to measure the absorption line strengths in the Line Index System at Full Width Half Maximum (FWHM) = 8.4 \AA \, (LIS-8.4\AA) \citep{2010Vaz}.
For several galaxies however (namely NGC~2859, NGC~2962, NGC~4262, NGC~4267 and NGC~4643), a few central spectra reach up to
$\approx$9\,\AA\, which exceeds the 8.4 \AA\, value. Testing the resulting effect, we found that this would lead to a shift of $\approx$0.02\,\AA\,in H$\beta$ and
$\approx$0.05\,\AA\,in Mg$b$. This corresponds to a difference of 0.2\,Gyr and up to 1.5\,Gyr (for ages older than 10\,Gyr) in age from
our measured value which is well within our uncertainties for old populations. Given this small
effect, and for simplicity, we chose to not convolve the data further and use the
models at 8.4\,\AA. We made sure that this is not affecting any of our conclusions.

Within the wavelength range of SAURON we can detect the following lines and measure their strengths: H$\beta$ as an age indicator,
Mg$b$ and Fe5015 as proxies for metallicity. We also combine the Mg$b$ and Fe5015 indices to obtain the [MgFe50]' index following \citet{kuntschner2010}:
\begin{equation}
 \rm{[MgFe50]'} = \frac{0.69 \times Mg{\it b} + Fe5015}{2}
\end{equation}
This combined iron-magnesium index is almost insensitive to the [Mg/Fe] overabundance \citep[e.g.,][]{kuntschner2010}. For our analysis, we relate the index measurements to MILES model predictions \citep{2006SB, 2010Vaz, miles}. 

We obtain the mean luminosity-weighted age and
metallicity by employing the {\tt rmodel}\footnote{http://www.ucm.es/info/Astrof/software/rmodel/rmodel.html} code
\citep{cardiel2003}. Here, we use the H$\beta$ index as a proxy for age and we implemented the combined iron-magnesium index, as given above, as a proxy for metallicity. We represent our results in index-index diagrams with the MILES SSP models for different ages and metallicities overlaid, compute gradients and average values for different regions in the galaxy. Throughout this work we assume a Kroupa initial mass function \citep[IMF,][]{2001Kroupa}. Furthermore, we use the difference of the metallicities of magnesium and iron (denoted as [Z$_{\rm Mg}$-Z$_{\rm Fe}$] in our work) as a proxy to determine the [Mg/Fe] abundance. Even though a linear relation between [Mg/Fe] and [Z$_{\rm Mgb}$-Z$_{\rm Fe5015}$] exists \citep{peletier2007, 2010Vaz}, the absolute value of our measurements cannot be compared directly with literature values because most other authors take combined iron indices into account. Due to the limited SAURON spectral range, we are however restricted to the above proxy (which has been successfully used in former SAURON and ATLAS3D studies, e.g., \citet{kuntschner2006}).

We are conscious about the simplification and hence introduced bias when representing the 
stellar populations by an SSP \citep[e.g.,][]{serra2007}. Therefore, this method 
has been commonly used in galaxies where variations in the locally averaged ages and metallicities 
are expected to be minimal, but has also been performed on systems with extended star formation histories \citep[e.g.,][]{peletier2007, ganda2007}.
In particular, this classical approach provides SSP-weighted population parameters which are similar to luminosity-weighted values and very valuable for our study: bars are prominent structures seen visually, as an additional luminous component. Therefore,
we restrict our analysis in this work to the index measurements and SSP parameters
derived from the former.

%###############################################################################
%=============================================================================
\section{Line strength indices}
\label{sec:results1}

This section summarizes the results of the line-strength measurements of the BaLROG sample. We start by investigating global properties comparing with trends found in the literature for both, barred and unbarred galaxies. Later, we will focus on the details revealed by the two-dimensional maps, looking at specific regions and radial trends, as well as gradients. We will discuss our findings in the context of former works and their implications in Sec.~\ref{sec:discuss}. In several parts, we only show a few maps of one galaxy as an example, but maps of our entire sample can be found in the Appendix~\ref{sec:allmaps}. We will use the bar length R$_{\rm bar}$  and fraction of it as reference points, as well as the effective radius R$_{\rm eff}$ (values taken from Paper I).

\subsection{Aperture absorption line index measurements}

%-----------------------------------------------------------------------------
\begin{figure}
\includegraphics[width=1\linewidth]{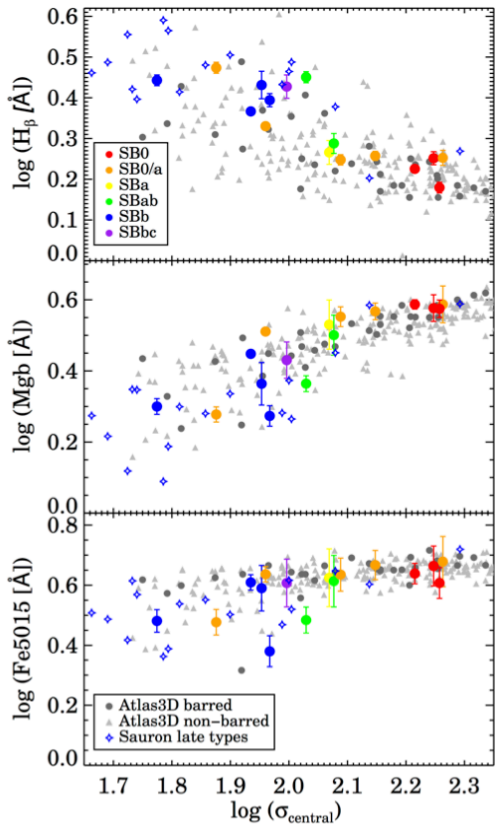} 
\caption{Index measurements within one effective radius ($R_{\rm eff}$) versus central velocity dispersion - all in logarithmic scale. Different Hubble types are colour-coded for our sample, while we show the ATLAS3D sample divided in barred and unbarred galaxies and measurements from late-type SAURON galaxies (also both barred and unbarred, but without distinction in the figure). }
\label{fig:sig-index}
\end{figure}
%-----------------------------------------------------------------------------
We start with a comparison of the global values obtained from the index measurements in order to situate our sample amongst statistically larger sample and detect possible offsets or outliers. We determined line strengths within different circular apertures. Figure~\ref{fig:sig-index} displays these measurements for an aperture size of one effective radius R$_{\rm eff}$ of the galaxy. We also computed relations for half the effective radius and for a central aperture (R$_{\rm eff}$/8) which show a similar good agreement (not displayed here) between barred and unbarred galaxies following the same global relations. This choice of apertures has been applied in various literature works, including long-slit studies \citep[e.g.,][]{Trager2000, 2005Thomas} and allows a direct comparison with the values from the ATLAS3D sample \citep{2015McDermid} as well as spiral galaxies taken from the SAURON study of \citet{peletier2007} and \citet{ganda2007}, recalculating the values from their data to match our resolutions and methodology. These values are shown in the figure, where the ATLAS3D galaxies are shown in gray - dots for barred and triangles for unbarred galaxies and the SAURON sample is shown by open diamonds. Our own sample is colour-coded according to morphological type, where early types lie to the right exhibiting a higher central velocity dispersion. The kinematic values are determined from the BaLROG data (see our kinematic paper, \citealt{2015Seidelb}). 

%-----------------------------------------------------------------------------
\begin{figure*}
\includegraphics[width=0.2\linewidth]{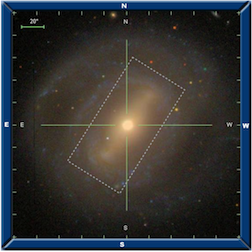}
\includegraphics[width=0.8\linewidth]{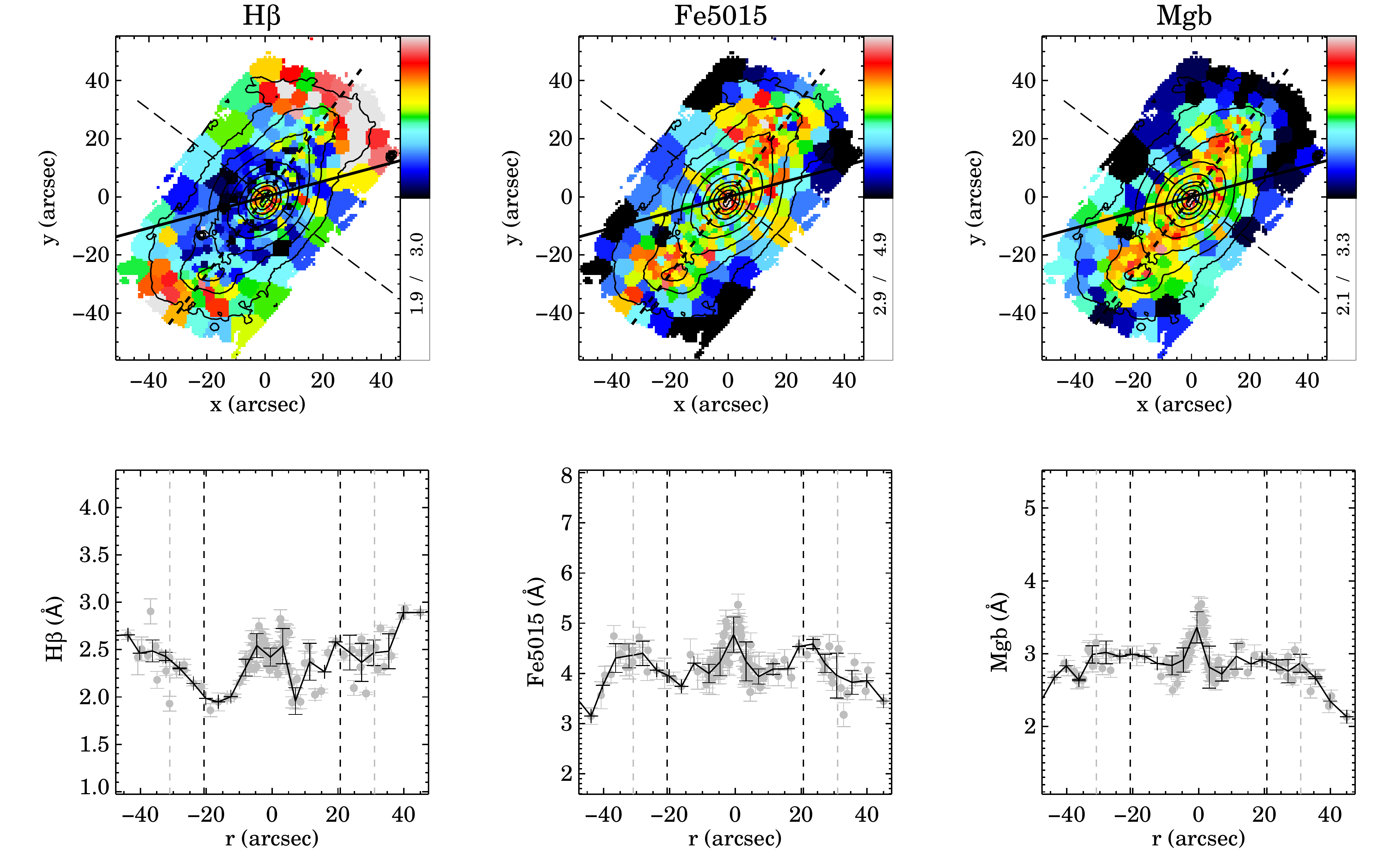}
\caption{The top panel shows a colour SDSS image of NGC~4394 with the IFU mosaic indicated in white dashed lines. The central stellar bar is well distinguished and the colour already reveals different populations: redder (older) insider the bar and bluer in the disk and spiral arms around it. The scale of 20 arcsec is given in the left top corner of the image. Below, in the middle row, we present the index maps of our SAURON mosaic of this galaxy (in this case a composite of 6 pointings). The lines explored in our wavelength range are H$_\beta$, Fe5015 and Mg$b$. The major axis of the galaxy is indicated by a straight black line, the bar major axis by a thick dashed black line and the bar minor axis by a slightly thinner dashed line. The isophotes also clearly show the bar's extensions. The bottom panels show the profiles along the bar major axis directly extracted from the map with a slit of width of 3 arcsec. We show individual measurements in gray and averaged values overplotted in black. Half and three quarters of the bar length are indicated by dashed black and gray lines respectively.}
\label{fig:ind-profiles}
\end{figure*}
%----------------------------------------------------------------------------

The trends found with central velocity dispersion as well as with Hubble type are consistent with what is found in larger surveys, such as the ATLAS3D survey (\citealt{2015McDermid}). There is no difference between barred and unbarred galaxies and our sample does not reveal any outliers. Bars do not seem to have an influence on these global relations, as partly expected since barred and unbarred galaxies share the same Tully-Fisher relations \citep{2003Courteau}. The range of values covers a similar magnitude for early- and late-type galaxies and follows the correlations found by e.g., \citet{Trager2000}: the age-sensitive H$_\beta$ index reveals a negative correlation with velocity dispersion while the metal-sentive indices show positive trends with this parameter. Even the smaller scatter for early types and larger scatter for later types (lower velocity dispersion) are reproduced by our small dataset. The correlations are clearly tighter for the high velocity dispersion end while the lower end exhibits a larger range of values. The stellar population parameters in these systems therefore might be more complex as they display a larger variety of index values at the same central velocity dispersion. This is found generally for later types, usually lower mass galaxies, than for the more massive, more early type systems.

Overall, bars do not seem to affect the global trends, as they share the same aperture line-strength-velocity-dispersion-relations as unbarred galaxies. Locally however, we do detect their influence as outlined in the following sections.

\subsection{Line-strength maps and profiles}
\label{sec:profind}

In this section, we explore the details of each index map and analyse their profiles along different axes to better understand the local influence of bars on their host galaxy. 

Figure~\ref{fig:ind-profiles} shows NGC~4394 as an example of one of the galaxies in our sample. We show a colour ($ugri$) SDSS image\footnote{SDSS images are from \\ http://skyserver.sdss.org/dr7/en/tools/chart/list.asp} in comparison with the line strength maps that we obtain from our SAURON data. Comparing the index measurements with the colour image, several similarities can be seen: the bar itself seems to be dominated by older populations as H$_\beta$ is only starting to increase towards the beginning of the disc where the spiral arms start. Nonetheless, in the very center, we can appreciate higher H$_\beta$ values in the line-strength maps, unlike in the photometric image where no blue star forming region can be seen. The iron and magnesium indices also show high values in the center and along the bar. In particular Fe5015 seems to exhibit higher values towards the edges of the bar. These enhancements can be distinguished as areas of elevated (red) values.  We note, that the enhancement can be an effect of contrast between the bulge and the bar components. Considering the high steep gradient of the inner structure and the light contribution of the bulge, compared with the flatter gradient of the bar and its light, the combination of both can make the gradient in the inner bar to be less steep, flat or even positive. In that sense, the `bumps' could equally be seen as `dips' interior to them, due to the presence of the overlapping structures.
 In the Mg$b$ map, the region seen as the photometric bar is dominated by constant elevated values, while the disc around it exhibits lower values and no bumps are seen at the edges of the bar.  

Below the line-strength maps, we show profiles along the major axis of the bar, in the bottom row of Fig.~\ref{fig:ind-profiles}. We extracted the values directly from the Voronoi-binned maps taking a slit of $\sim$3 arcsec width (in the center slightly less and enlarging its width in a cone-like way towards the outskirts for better sampling). We show individual (gray) and averaged (black) measurements, indicating on each side half and three quarters of the bar length (vertical dashed lines, black and gray respectively). 

The H$_\beta$ profile nicely shows the elevated values around the center, but with a slight central dip. After the centre, we observe a steep decrease followed by increasing values towards the outside of the bar into the disc, where the spiral arms start dominating. The behaviour of the H$_\beta$ profiles depends very much on the galaxy and is not systematic among our sample as it is reflecting the most recent bursts.

%-----------------------------------------------------------------------------
\begin{figure*}
\includegraphics[width=0.8\linewidth]{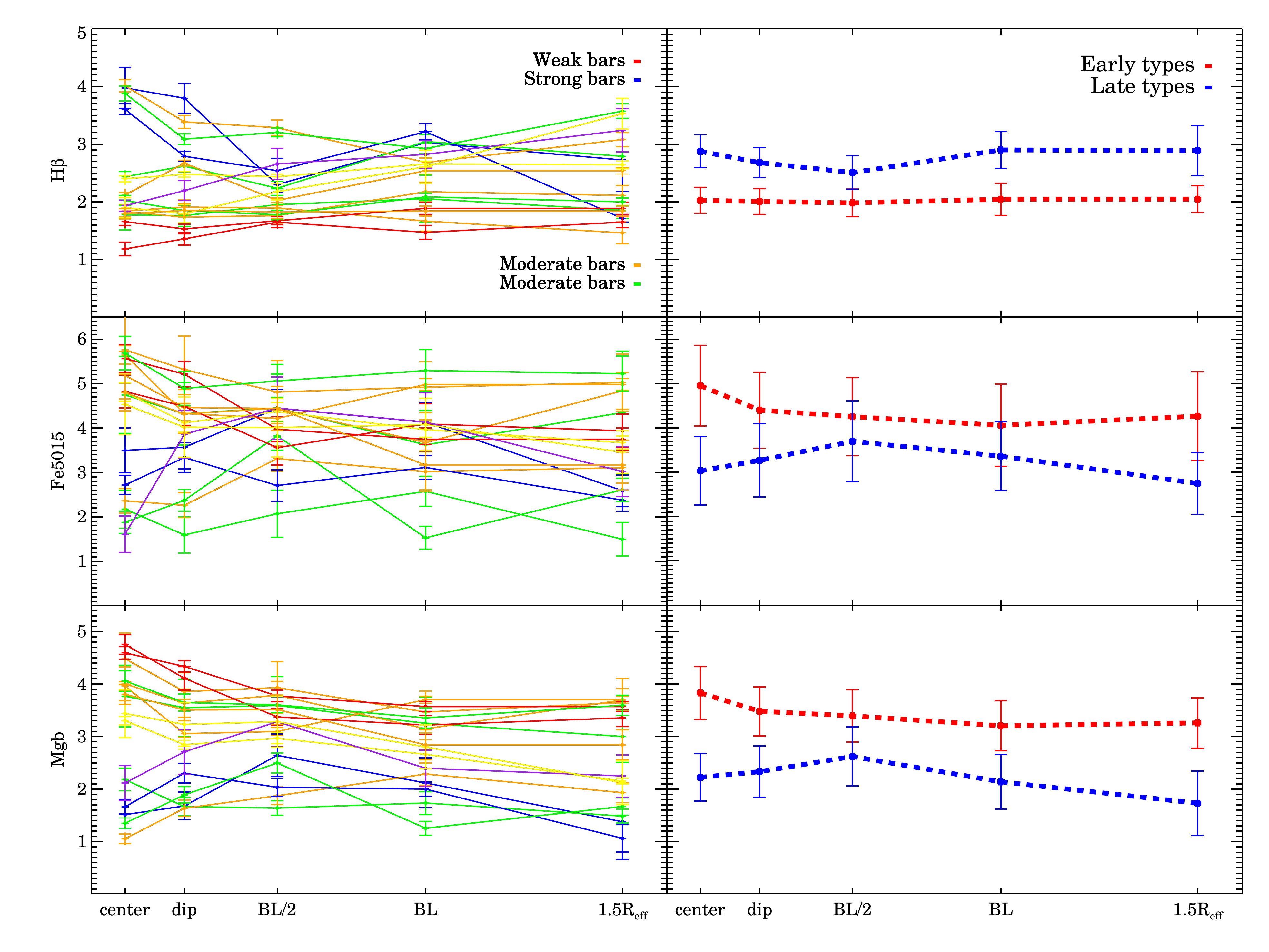}
\caption{Index measurements within different regions along the bar major axis of the galaxies. We sample the center (within 0.01 bar length), the potential spot of bar-induced influence at around 0.13 bar length and denoted as the 'dip' because we found dips in e.g. the angular momentum profiles (see PaperI); then at half the bar length, full bar length and 1.5 effective radii R$_{\rm eff}$ (2.5 scale lengths of the disc). The individual profiles are colour-coded according to increasing bar strength, from red (weakest), over orange, yellow, green, blue up to purple (strongest) in the left panel. The right panel shows averaged profiles according to the Hubble type with dashed lines: red for early types and blue for late types among our sample. }
\label{fig:comp-profiles}
\end{figure*}
%-----------------------------------------------------------------------------

The Fe5015 profile is clearly different from the H$_\beta$ index profile. Instead of a central dip, we find a central peak, dropping off steeply to either side. Following this valley, the profile reveals the aforementioned particular feature: we find enhanced Fe5015 values at around 0.5-0.75 bar length.  Inspecting the profiles among all galaxies in our sample, we can detect this feature in 7 galaxies (NGC~1015, NGC~2893, NGC~4262, NGC~4394, NGC~4643, NGC~5375 and NGC~5701). In some galaxies, it is not seen very clear and only 4 galaxies of our sample do not exhibit this feature at all. In some cases, it is more apparent in the maps while others reveal it better in the profiles. It is however not always found at the same radius - even in this galaxy, the bumps seem to be located at different distances from the center. Correcting for possible deprojection effects we could not account for the offsets.

This feature is very mild - although appreciable - but based on a qualitative examination of the profiles and maps. Due to the varying location, no clear ratio of the higher areas to the lower ones could be established. With even better data in the future, e.g. MUSE, this could be possibly better explored. We could however establish, that this feature is detected more strongly among the early types in our sample than in the later types. It does not depend on entering spiral arms (examined visually).

Mg$b$, similar to Fe5015, displays a central peak value, quickly dropping off to a stable level. No clear enhancements are seen in this index at certain regions along the bar length. Instead, similar to Fe5015, after 0.75-1.0 bar length, the values start to decrease, when entering the disc/spiral arm regime. Overall, the Mg$b$ index seems to best trace the entire bar extensions, i.e. it exhibits elevated values within the entire bar region constrained by the isophotes.

\subsection{Analysis of distinct components}
\label{sec:comp}

Apart from the radial trends, we also compared values of distinct regions in order to check for a common behaviour in potential resonance positions. Based on our previous analysis we obtained radial measurements along the bar major axis in five regions, mainly related to the bar length. We use the parameter of bar length R$_{\rm bar}$  and fraction of it as reference points as it is closely related to the corotation radius R$_{\rm CR}$ of the bar as shown by recent results by \citet{2015Aguerri}. Using the \citet{1984TremaineW} method, they determined that the mean value of the ratio R$_{\rm CR}$/R$_{\rm bar}$ is around 1. Therefore, we use the bar length found from an analysis of $Spitzer$ images \citep{2015HerraraE} as an approximation. We sampled the following regions: the center (here denoted as 0.01 bar length), the potential spot of bar-induced influence at 0.13 bar length, half the bar length, full bar length and 1.5 effective radii R$_{\rm eff}$ (2.5 scale lengths of the disc). For some of the galaxies in our sample we do not reach this last point due to the limited extensions of the mosaic. In those cases we take the value that is the furthest into the disc. 

Figure~\ref{fig:comp-profiles} depicts the result of our analysis. Regardless of the individual color-code, we distinguish several known trends independent of barred or unbarred galaxies, in particular for the early types (red dashed lines). As already seen in \citet{kuntschner2006} for the original SAURON sample of early-type galaxies, we find generally negative gradients (with increasing radius) of the metal line strength maps. Similarly, we confirm flat or slightly positive trends for the H$_{\beta}$ measurements. We also find central or circum-central elevated H$_{\beta}$ values likely connected to recent star formation activity. 

Overall, we observe a higher scatter for the central and circum-central values, than for the barred region (half or full bar length). The spread increases again at 1.5 R$_{\rm eff}$. The fact that the central values as well as disc values exhibit larger scatter, implies likely a more complex star formation history with recent star formation activity in the center, given the elevated H$_{\beta}$ strengths. Qualitatively, we do not find significant difference between major and minor axes, apart from a tentative higher scatter for the minor axis. This is however only based on a few galaxies and the limitations of our sample size are obvious. 

The colour-code indicates the bar strength, starting with weak bars (red), over intermediate (from orange, yellow, green to blue) to strong bars (purple). The bar strengths are also closely coupled with Hubble type for our sample. Hence, the observed trends might be also linked to that latter parameter. As already seen in Fig.~\ref{fig:sig-index}, the early types galaxies show lower H$_\beta$ and higher Mg$b$ values overall than the later types. Considering that bars drive fuel towards the central parts to aliment or even initiate star formation \citep[e.g.,][]{1994Heller, 1995Knapen}, we do find higher values and steeper drop-off slopes in the centre and circumcentral parts for the stronger barred galaxies, seen in the top panel. Several observational studies have already related star-forming nuclear rings to bars \citep[e.g.,][]{Allard2006,2010Knapen, 2013vdLaan} and our finding of higher H$_\beta$ values in those regions for stronger barred galaxies supports this result.

From the indices alone, we could already confirm certain trends and find new evidence for a local bar-driven influence on the host galaxies in our sample, namely along the bar major axis. Global values however, such as aperture absorption line index measurements seem to be unaffected by the presence of a bar. In the following Sec. we will now relate the indices to SSP parameters.  %I prefer to leave this sentence hereÉ. 
%-----------------------------------------------------------------------------
\begin{figure}
\includegraphics[width=\linewidth]{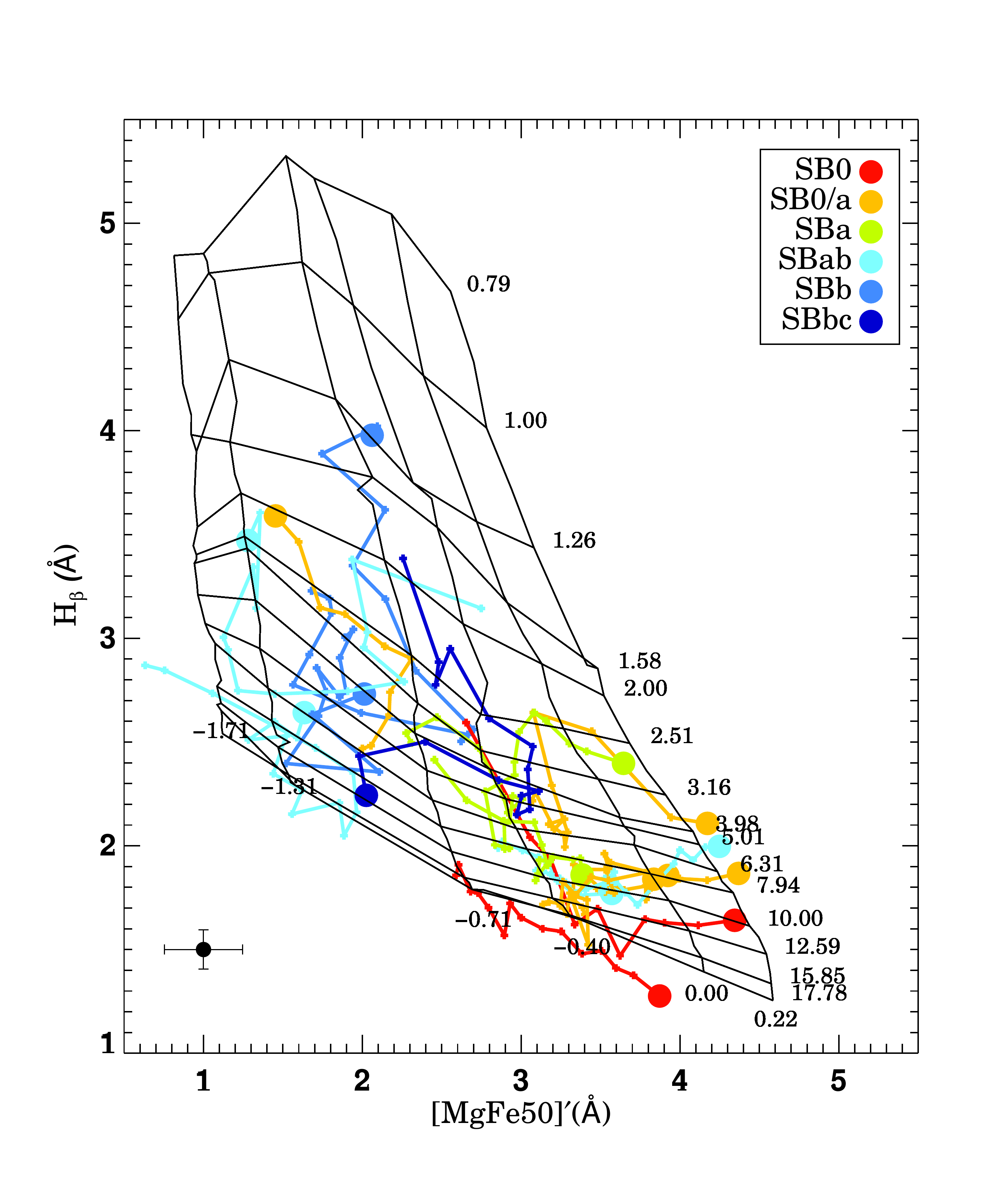}
\caption{Index-index diagram displaying isophotal radial profiles for our entire sample, colour-coded by Hubble type. The values are the average of the index measurements along a given isophote. The centers are always indicated by a large circle. H${\beta}$ is used as an age discriminator versus the combined index of [MgFe50]', employed as a proxy for metallicity. Overplotted
is a model grid of single stellar populations, indicating the ages (in Gyr) and
metallicities (dex) (shown on the side of this grid). In the left lower corner we
indicate a typical uncertainty (weighted mean of individual errors) of the individual points with representative error bars - this error can vary depending on the galaxies and individual radial (as well as complete) index-index diagrams can be found in the appendix for each galaxy.}
\label{fig:grid}
\end{figure}
%-----------------------------------------------------------------------------

%###############################################################################
%=============================================================================
%\section{Star formation histories}
%\label{sec:results}
\section{Single stellar populations}
\label{sec:results2}

In this section we will relate the measured line strength values to SSP parameters deducing the SSP equivalent age, metallicity and [Mg/Fe] abundances for our sample. Since bars are structures seen prominently in the photometry, light-weighted quantities are of great interest. Similar to the index analysis, we first present global trends and then analyse the details of the maps and radial profiles. A quantitive analysis of the profiles' gradients will be discussed in Sec.~\ref{sec:discuss}.

\subsection{Index-Index diagrams}

%-----------------------------------------------------------------------------
\begin{figure}
\includegraphics[width=1\linewidth]{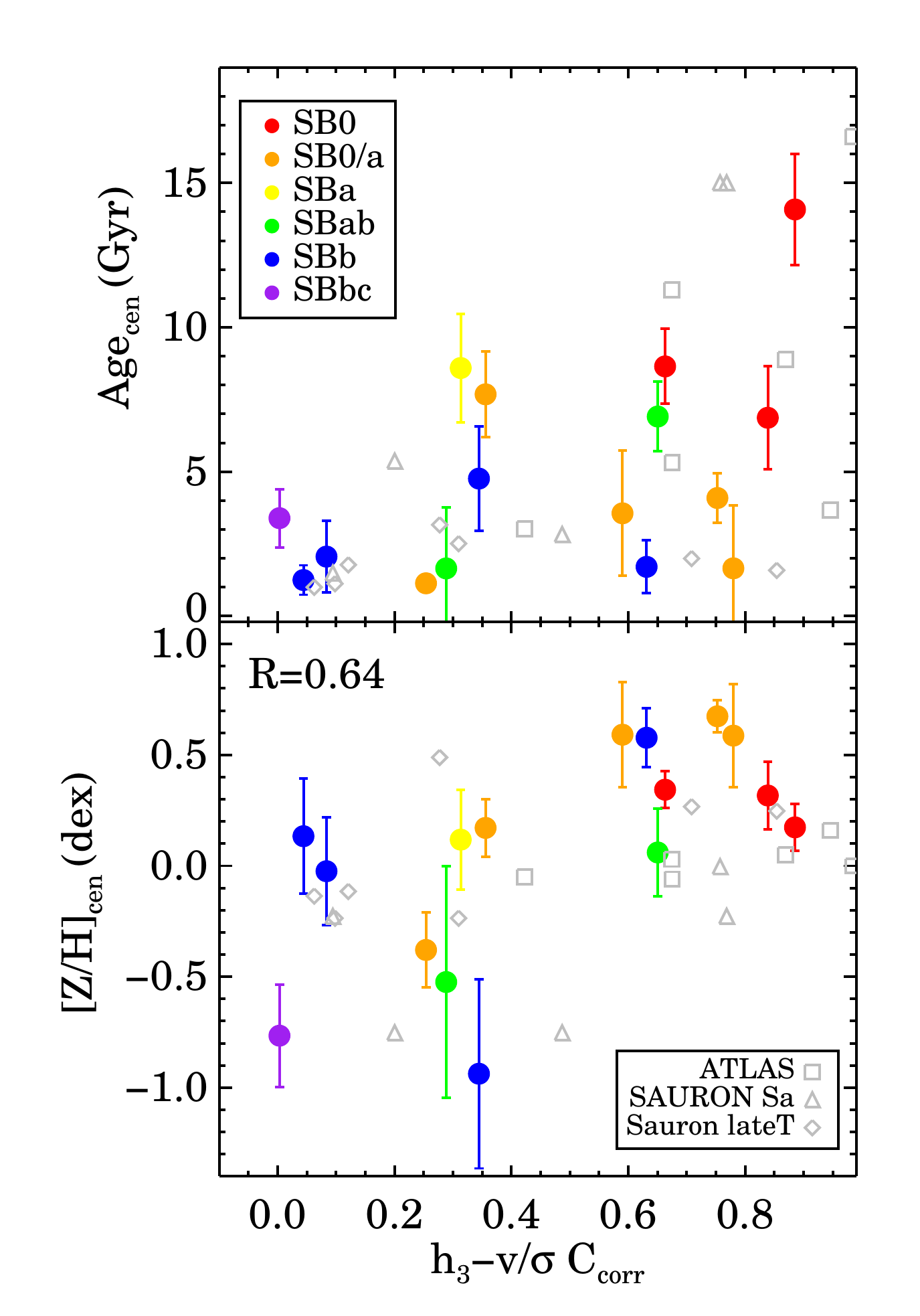}
\caption{Central ages, metallicities, measured within the central aperture of  R$_{\rm eff}$/8, as a function of h$_3$-v/$\sigma$ anti-correlation (analyzed in PaperI). We show the linear Pearson correlation coefficient of the h$_3$-v/$\sigma$ anti-correlation on the x-axis and the central SSP parameters on the y-axis. Hubble types are indicated in colours. And the linear Pearson correlation coefficient is given for the correlation with metallicity in the lower right panel. We also computed the same quantities for our unbarred control sample indicating the Es and S0s from ATLAS and the Sa and Sb-Scd galaxies from SAURON.}
\label{fig:SSPh3}
\end{figure}
%-----------------------------------------------------------------------------

%-----------------------------------------------------------------------------
\begin{figure}
\includegraphics[width=\linewidth]{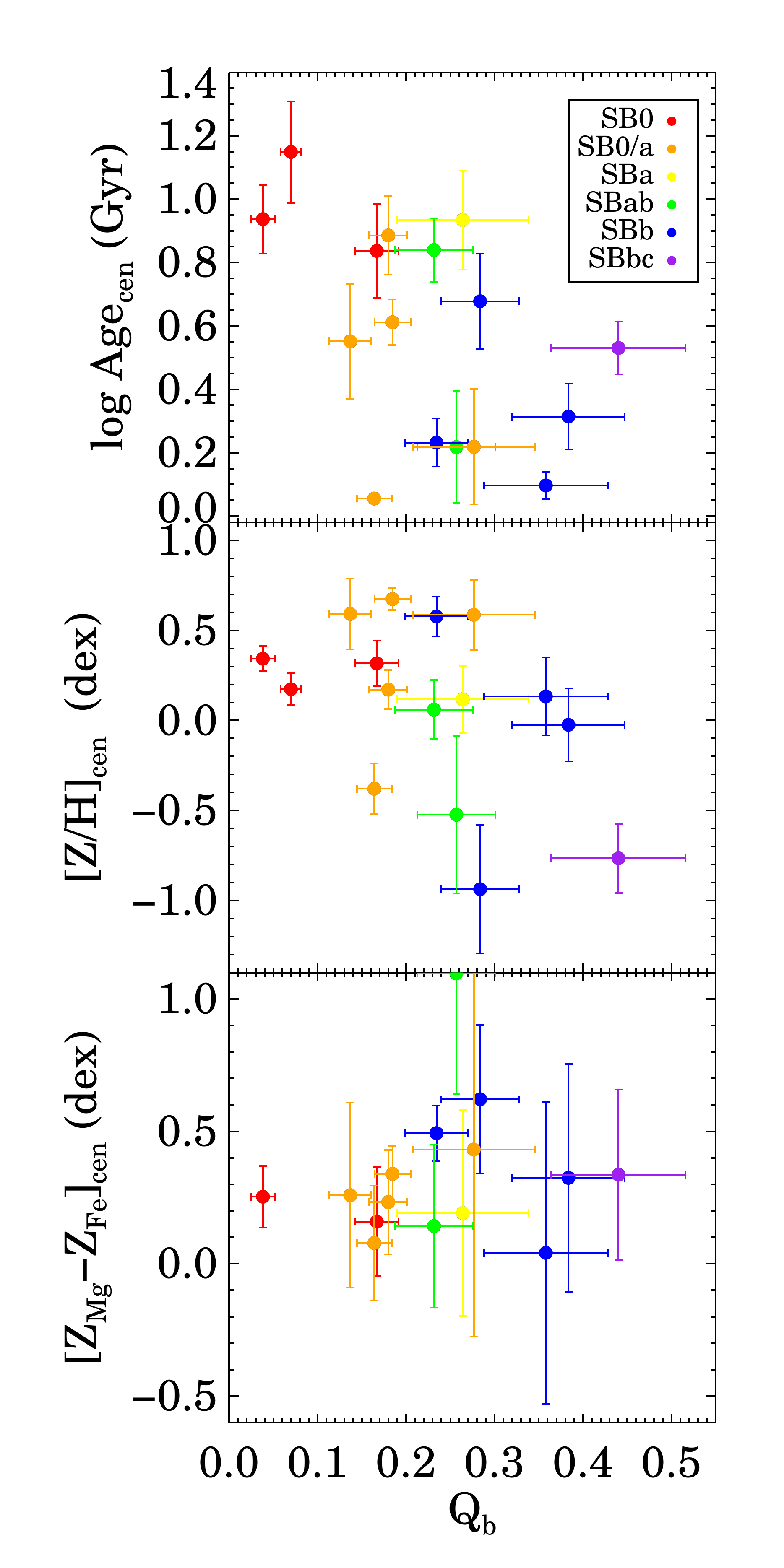} 
\caption{Central ages, metallicities and $\alpha$ abundances, measured within the central aperture of  R$_{\rm eff}$/8, as a function of bar strength and coulor-coded by Hubble type. }
\label{fig:central}
\end{figure}
%-----------------------------------------------------------------------------

Index-index diagrams intuitively relate line-strength indices with single stellar population parameters, using a grid of model predictions. 
Figure~\ref{fig:grid} presents the measurements of absorption line
strengths in an index-index diagram for the galaxies of our sample, showing isophotal integrated radial profiles indicating the centers as large dots (the individual grids per galaxy can be found in the appendix). The values are the average of the index measurements along a given isophote, moving from the center towards the outskirts of the galaxies. We plot H${\beta}$ as an age indicator against the combined index of magnesium and iron, [MgFe50]' (using Mg$b$ and Fe5015, see \S~\ref{sec:methods}) as a proxy for metallicity, independent of [Mg/Fe]. A grid of MILES single stellar population models for Kroupa IMF is overplotted.

Overall, we observe a large variety of distributions within the full grid amongst our sample. Clear differences can be detected between early and late-types: most early types seem to be older and more metal-rich overall with a more prominent gradient in metallicity and exhibit a more confined distribution for all points while late types show a much larger spread among themselves, but also radially in the diagram. This behaviour is of course optically enhanced or even driven by the age difference due to a larger area of the grid covering young ages which makes the range more obvious and spread out than for older populations. Furthermore, we find that for a few galaxies of our sample, the central spectra tend to fall slightly outside the grid (at the high metallicity end). As these are mainly the central spectra, the signal-to-noise is usually well over 100, hence the quality of the spectra should not be the scapegoat. In the appendix of \citet{2015Seidela}, we already analysed this behaviour and found that it is likely due to the complex mixture of populations present in those regions. This would be a very plausible explanation given the peculiarities already observed. In most cases, the central value is slightly younger and particularly metal-rich which hints towards a younger population formed from an already metal-enriched medium.

Within this context, we compared the inner h$_3$-v/$\sigma$ anti-correlation of Paper I (see \citealt{2015Seidelb}, Fig.~8)  with the grids and found that those galaxies that show strong h$_3$-v/$\sigma$ anti-correlations within the central 0.1 bar lengths are the same whose central index measurements are found outside the grid (mainly shifted to the right at the high metallicity end). These are: NGC~2859, NGC~2962, NGC~4245, NGC~4394, NGC~4262 and NGC~4643. Their individual point measurements can be seen in Appendix A. Most of these galaxies exhibit substantial inner components or are even double-barred \citep[e.g.,][]{2013DeLorenzoC}. In her analysis, \citet{2013DeLorenzoC} find that inner bars are younger and more metal-rich than their surrounding bulges and outer bars. 

Those galaxies with straight horizontal or circular accumulated h$_3$-v/$\sigma$ relation (as seen in Fig.~8 in  \citealt{2015Seidelb}) are corresponding to those that seem to have less spread and fall (almost) completely within the grid: NGC~1015, NGC~2893, NGC~3485 and NGC~5701. Those with very large variations in age and metallicity as seen in the grids, also show a very extensive h$_3$-v/$\sigma$ distribution for all radial apertures: NGC~2712, NGC3504 and NGC~5350. Hence, the h$_3$-v/$\sigma$ relation can already give qualitative hints on the present stellar population properties. 

%comparison sample

Trying to quantify this observation, we show the central age and metallicity against the linear Pearson correlation coefficient of the line (h$_3$-v/$\sigma$ anti-correlation) in Fig.~\ref{fig:SSPh3}, top and bottom panels respectively. In colour, we show the BaLROG sample and in open gray symbols unbarred and overall undisturbed galaxies from our control sample. We measured the quantities in the same way and Appendix~\ref{app:controlsampleh3} shows examples of the h3-v/$\sigma$ anti-correlation for the three surveys. For the BaLROG data, we find that a tentative relation exists with metallicity (linear Pearson correlation coefficient R=0.64 and p-value=0.008) and not with age (linear Pearson correlation coefficient R=0.48 and p-value=0.06, hence not significant). Furthermore, the correlation seems to be stronger for earlier types, but is not entirely driven by Hubble type (colour-coded in the plot). This behaviour is however reflected by the control sample where the early type ATLAS galaxies (open squares) seem to lie on a tighter correlation than the other galaxies. It is interesting to note that the unbarred sample seems to show on average lower metallicities, in particular at high h$_3$-v/$\sigma$ C$_{\rm corr}$ values. This might indicate that the relation between higher metallicity and a stronger anticorrelation is independent of a bar, as expected, but that the presence of a bar can increase the central metallicity. Therefore, an inner rotating disc (as suggested by a strong anticorrelation) might be more metal-rich due to more and more continuous gas infall through the bar and stars continously forming from a more enriched medium.

 We also analysed the slope of the h$_3$-v/$\sigma$ anti-correlation (not shown), which seems to be steeper for higher central metallicities. This could be another hint of a connection between the dynamical parameters and their stellar populations.
%-----------------------------------------------------------------------------
\begin{figure*}
\includegraphics[width=1\linewidth]{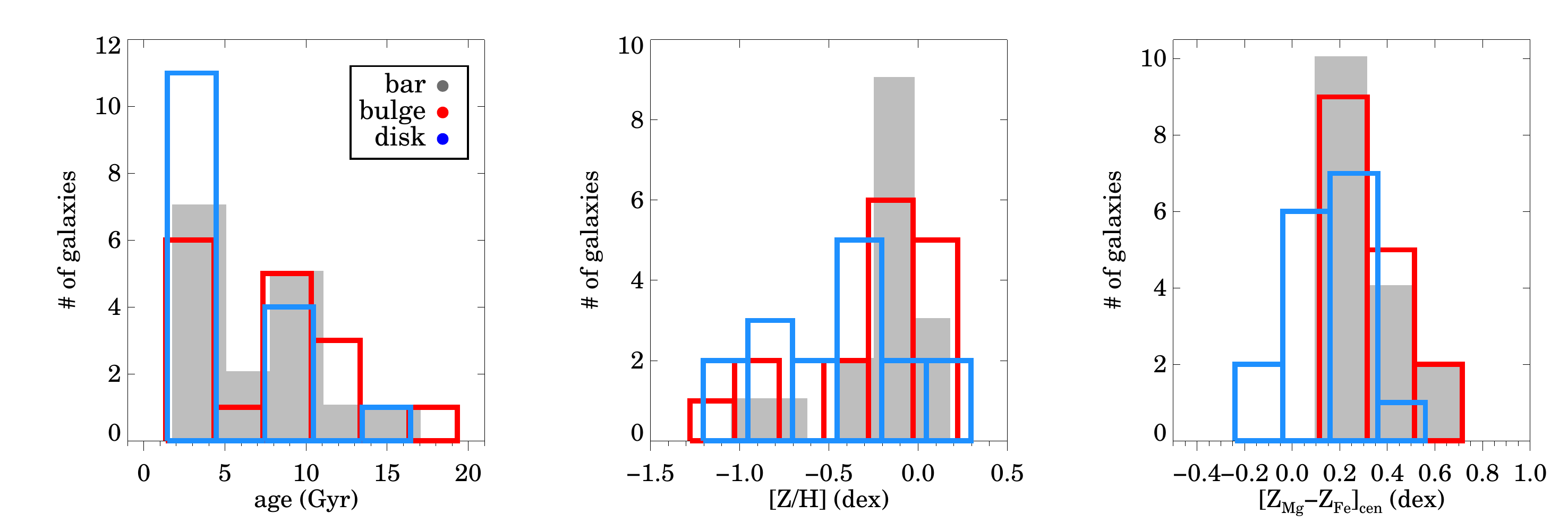} 
\includegraphics[width=1\linewidth]{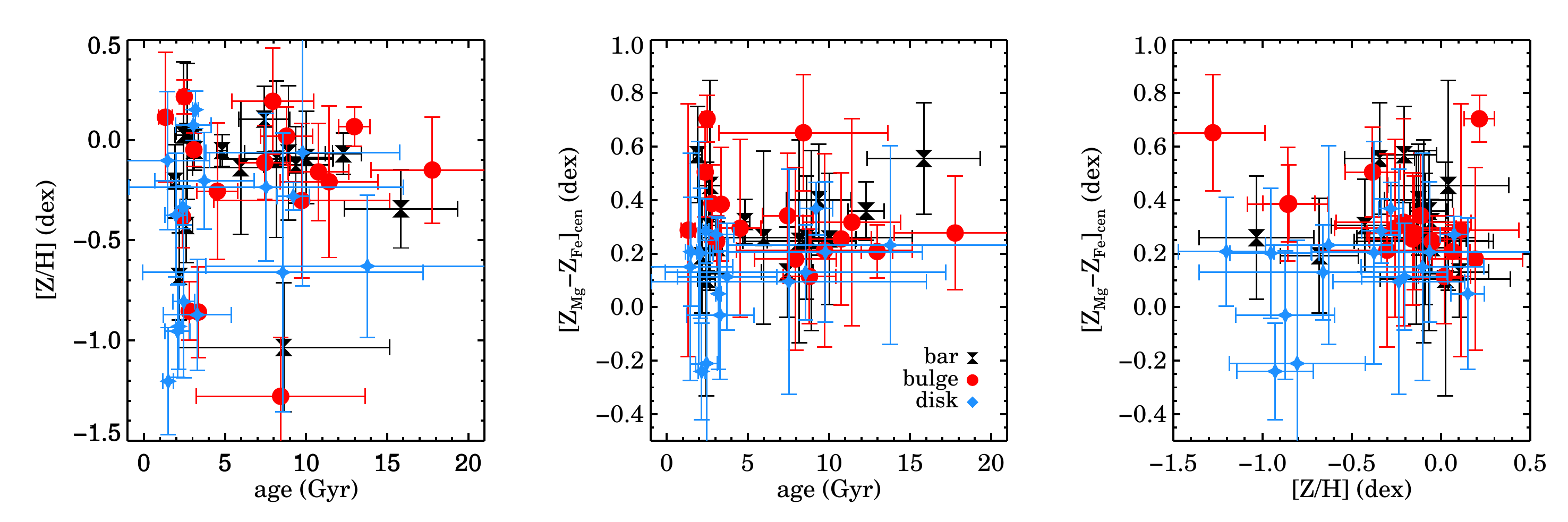} 
\caption{\textit{Top row:} Mean ages, metallicities and $\alpha$ abundances of the bar (gray, filled) in comparison with those of the bulge (red line) and disk (blue line) for the BaLROG sample. \textit{Bottom row:} Correlations between ages, metallicities and $\alpha$ abundances for the different regions.}
\label{fig:barssp}
\end{figure*}
%-----------------------------------------------------------------------------

The overall properties of the isophotal SSP profiles as seen in the index-index diagram are still dominated by their Hubble type and no crucial global influence of the bar can be detected, apart from tentatively on the central values. Those might be influenced by substructures, such as central star formation resulting in inner rings or discs, which are enhanced by the bar's ability to provide the necessary fuel for their creation. In the following section we will further investigate central regions.

\subsection{Central and bar values}

Our previous analysis coupled with hints towards a potential influence of bars in the central regions, predicted by simulations and already seen in observations \citep[e.g.,][]{2004Chung,2003Fathi}, leads us to systematically test the central parameters against the strength of the bar. We use the average found within the aperture of R$_{\rm eff}$/8. 

Figure~\ref{fig:central} shows a weak correlation of age and Q$_{\rm b}$, such that stronger bars have younger central stellar populations (linear Pearson correlation coefficient R=0.57 and p-value=0.02). This is consistent with the proposed scenarios of bars funneling material to the center triggering or nourishing star formation \citep[e.g.,][]{2005Knapen, Allard2006, 2013A&A...551A..81V}. Very recent studies in fact found a central enhancement of the nitrogen-to-oxygen abundance and a larger star formation rate per unit area \citep{2015Florido} for barred galaxies.

At the same time we also observe slightly lower metallicities with increasing bar strength. We notice however, that the galaxies in our sample seem to have metallicity offsets with respect to each other such that most late types have generally lower metallicities, a known Hubble type effect. Overall, the central metallicities are found to not correlate strongly with bar strength (linear Pearson correlation coefficient R=0.45 and p-value=0.08). The bottom pannel of the figure displays the trend with $\alpha$ abundance. We observe again a large scatter similar to the metallicity. Our sample in this case is definitely too small to draw any firm conclusion.

The trends discussed above could also result as a consequence of different Hubble types. Apart from the trends already revealed through the colour-code in the figure, we explored the correlation with Hubble type for our barred sample and other comparison samples, only selecting unbarred galaxies. The age and metallicity trends are both found for Hubble type, too, but bars seem to lower them slightly. In particular, they seem to lower the metallicity in later types. Nevertheless, the statistics are very low and the scatter is large, probably due to yet other mechanisms that come into play in the central parts of galaxies altering these parameters.

%-----------------------------------------------------------------------------
\begin{figure*}
\includegraphics[width=\linewidth]{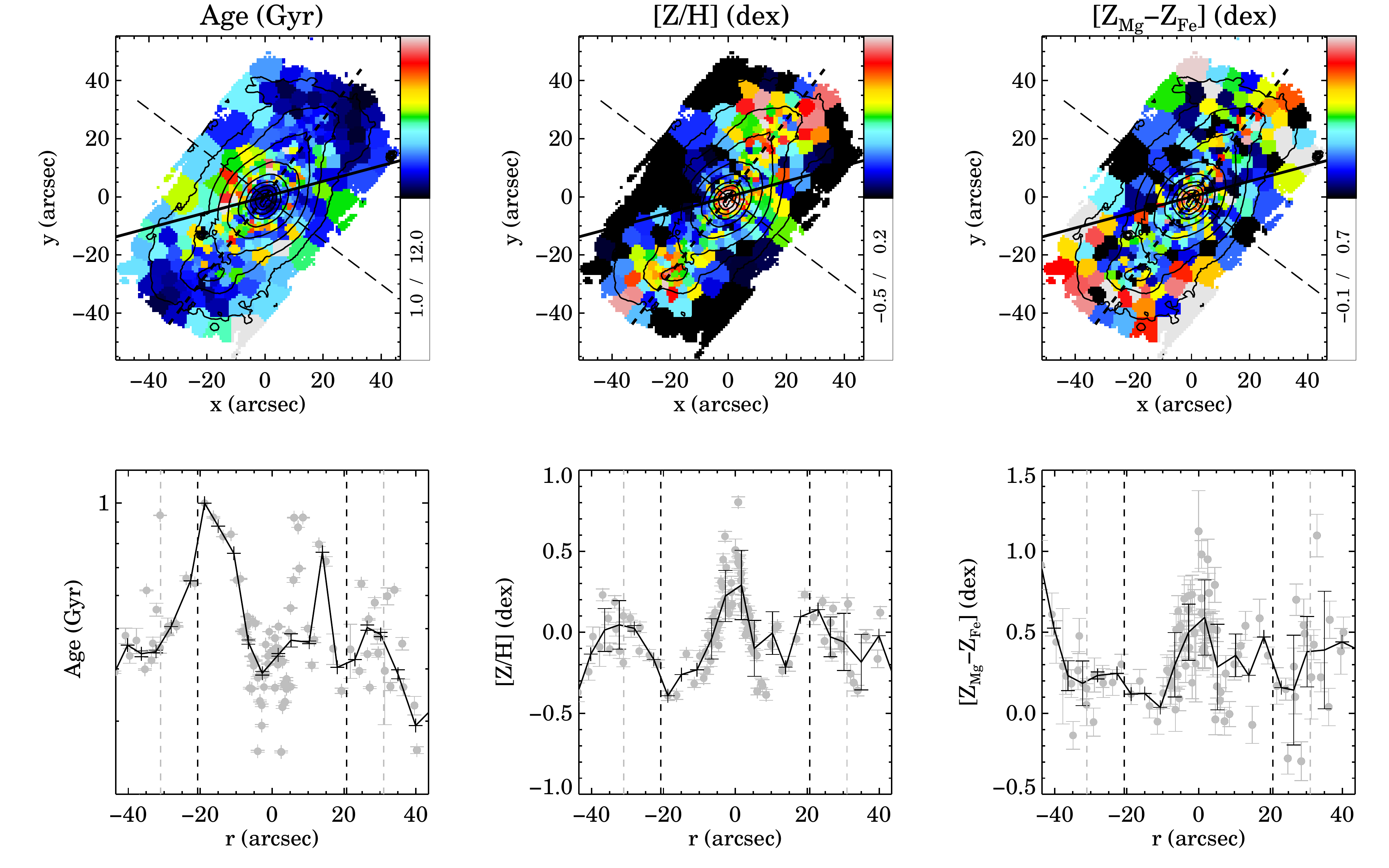} 
\caption{SSP profiles for NGC~4394. On top we present the SSP maps of age, metallicity and $\alpha$ abundance with the major axis indicated by a straight black line, the bar major axis by a thick dashed black line and the bar minor axis by a thin dashed line. The panels below show the profile along the bar major axis directly extracted from the map with a slit of width of 3 arcsec in order to trace along the bins. We show individual measurements in gray and averaged values overplotted in black. Half and three quarters of the bar length are indicated by dashed black and gray lines respectively.}
\label{fig:4394SSP}
\end{figure*}
%-----------------------------------------------------------------------------

Apart from the central values, we also calculated values at specific regions, namely for the bar (filled, gray), bulge (red lines) and disk region (blue lines), shown in Fig.~\ref{fig:barssp} for our sample. The top panels show histograms of the regions and the bottom panels the correlation of the different parameters. The regions are approximated based on the photometric decompositions provided by the S$^4$G Pipeline 4 \citep{2015Salo}. We do not take the exact decompositions because there are large uncertainties and we aim to clearly separate regions. Therefore, we chose the bar region to contain bins well outside the inner regions (from 0.35 to 0.95 barlength, only sampling spaxels that fall in the bar structure). The bulge contains bins between 0.05 up to 0.2 barlength (which corresponds to a clear area of the bulge in the decompositions) and the disk everything further away than one full barlength. We disregard the spaxels in between these areas as those are transition regions that could blur out clear differences.

%-----------------------------------------------------------------------------
\begin{table*}
\centering
\caption{{\textsc Stellar popoulation parameters for the BaLROG sample}.~-~(1) Galaxy name, (2) - (4) central line-strength values and (5), (6) central age and metallicity values, (7), (8) age and metallicity of the bar region.
 Objects forming part of the Virgo cluster are marked with a small $v$ next to their name. {\textsc Notes.~-~}
Line strength indices, ages and metallicities are determined in this work and values are given within R$_{\mathrm{eff}/8}$ as a central measurement. }
\begin{tabular}{lcccccccccc}
\hline
Galaxy         &  H$_{\beta, R_\mathrm{eff/8}}$  &  Fe5015$_{R_\mathrm{eff/8}}$   &  Mg$b$$_{R_\mathrm{eff/8}}$   &  Age$_{R_\mathrm{eff/8}}$  &  [Z/H]$_{R_\mathrm{eff/8}}$ &  Age$_{R_\mathrm{bar}}$  &  [Z/H]$_{R_\mathrm{bar}}$          \\
~              &   (\AA)                &   (\AA)               &   (\AA)              &   (Gyr)           &   (dex)   &   (Gyr)           &   (dex)                         \\
(1)            &   (2)                  &   (3)                 &   (4)                &   (5)             &   (6)    &   (7)             &   (8)                           \\
\hline           
NGC\,1015      &   1.77    & 4.53 & 3.60  & 8.59 &  0.11 &   8.17 &  -0.096 \\
NGC\,2543      &   3.18    & 3.58 & 1.85  & 1.24 &  0.37 &   2.44 &   0.024 \\
NGC\,2712      &   2.49    & 1.77 & 1.83  & 4.76 & -0.93 &   8.62 &  -1.033 \\
NGC\,2859      &   1.78    & 4.99 & 3.88  & 6.86 &  0.31 &   10.0 &  -0.088 \\
NGC\,2893      &   3.75    & 2.23 & 1.12  & 1.13 & -0.37 &   2.17 &  -0.680 \\
NGC\,2962      &   1.82    & 5.54 & 4.33  & 3.55 &  0.59 &   8.91 &  -0.065 \\
NGC\,3485      &   2.86    & 3.33 & 2.10  & 2.06 & -0.02 &   2.59 &  -0.426 \\
NGC\,3504      &   4.20    & 1.88 & 1.41  & 1.65 & -0.52 &   1.91 &  -0.205 \\
NGC\,4245      &   2.18    & 5.24 & 3.72  & 1.02 &  0.67 &   4.81 &  -0.053 \\
NGC\,4262$^v$  &   1.19    & 4.72 & 4.72  & 13.0 &  0.17 &   15.8 &  -0.343 \\
NGC\,4267$^v$  &   1.61    & 5.28 & 4.35  & 8.64 &  0.34 &   12.2 &  -0.069  \\
NGC\,4394$^v$  &   2.58    & 4.35 & 2.92  & 1.70 &  0.57 &   3.10 &  -0.049  \\
NGC\,4643      &   1.97    & 5.55 & 4.01  & 1.65 &  0.58 &   7.42 &   0.103  \\
NGC\,5350      &   2.23    & 3.98 & 2.85  & 3.39 &  0.06 &   2.67 &   0.041  \\
NGC\,5375      &   1.91    & 4.30 & 3.23  & 6.91 &  0.06 &   5.97 &  -0.138   \\
NGC\,5701      &   1.80    & 4.54 & 3.63  & 7.68 &  0.17 &   9.37 &  -0.125  \\ 
\hline
\end{tabular}
\label{tab:ssp}
\end{table*}
%-----------------------------------------------------------------------------

The histograms show a similarity between bulge and bar in all three panels. The disks are also following these overall trends but are more different from the former two components, as expected \citep[e.g.,][]{2007Perez}. In particular in the $\alpha$ abundance (right panel), they show a significant offset towards lower values compared with bulges and bars and in general a broader range of $\alpha$ abundances. 

In the age distribution, we distinguish two peaks, one old and one at a younger age. As there remains a large number of young disks however, some of the older bars and bulges must be residing in these younger disks. Investigating the obtained maps, we find that many early-types show bars of $\sim$10~Gyr in a younger disk. Late-types are overall younger, i.e. the bar, bulge and the disk. The metallicity shows that a large fraction of bars have a common metallicity just below solar. They are slightly less metal-rich than the bulge but significantly more than the disk. Bars and bulges also seem to be very similar in their [Mg/Fe] abundance while disks show significantly lower values. Table~\ref{tab:ssp} summarizes the central and bar measurements for indices and SSP parameters.

In the bottom row of Fig.~\ref{fig:barssp}, we show the correlations between the SSP parameters of the different regions. There seems to be a tentative trend among bars and bulges. Despite being rather more metal-rich overall, they seem to get more metal-rich towards younger ages. In particular at the youngest ages, a large spread of metallicities, especially in disks, is found. The trends between $\alpha$ abundance vs. age are as expected: older ages show higher $\alpha$ abundances. Although some bars and bulges, despite being young, show high $\alpha$ abundances. This can be an effect of the luminosity weighted age, which is biased towards younger populations. Furthermore, we cannot fully exclude an influence of the bulge light contributing to the bar (or vice versa). The offset between bars and bulges compared with disks is best seen in the last panel: while the former show higher metallicities coupled with higher $\alpha$ abundance values, disks show the opposite.

\subsection{SSP Profiles}

Apart from central and regional values, we have the entire two-dimensional maps for all galaxies available. These maps and profiles for all galaxies can be found in the appendix and are displayed for NGC~4394 as an example in Fig.~\ref{fig:4394SSP}. As expected, the SSP maps and radial results are closely related to the indices. The age map shows younger ages corresponding to high H$_\beta$ values (in the presence of spiral arms in this case). The very center seems to be young, but sitting in an older structure, as seen due to the very steep rise at both sides of it. The age of the center in fact compares well to that of the disk, suggesting that their star formation events might have occurred at a similar epoch. However, this could also be an influence of the bar which could produce younger stellar populations at the center and at the end of the bar as predicted by \citet{2007Wozniak}. We also observe that the average age of bars in mainly our early-type galaxies is of the order of the bulge and rather old, usually up to 10~Gyr, no matter if they reside in a young, star forming disk (see Fig.~\ref{fig:barssp} and individual maps in Appendix~A). The later types host younger populations in their bars however as they are overall younger. 

The metallicity reflects best the structure seen in the Fe5015 profiles. The central peak, as well as the peaks around half to three quarters R$_{\rm bar}$ are clearly visible. This is the case for several of the other galaxies, too, although some of the profiles suffer from larger scatter than the index profiles. This dip in between the center and the bar's edges can however also be a contrast effect of the bulge which we cannot fully exclude. About 40\% of our sample shows higher metallicities along the major than the minor axis (in agreement with Mg$b$ and Fe5015 indices). 

The $\alpha$ abundance is very similar and agrees within error bars along both axes. The $\alpha$ abundance map of NGC~4394 in the example figure shows indeed a clear difference between the center, the barred region and the outskirts where the spiral arms start. Within the barred region it remains practically flat and lower than for the inner and outer regions. This is not typically the case. We find a large variety showing central enhancements as well as drops amongst the galaxies of our sample.

%=============================================================================
%=============================================================================

\section{Stellar Population Gradient analysis}
\label{sec:discuss}

One of the most studied aspects of bar-driven secular evolution is the effect on the stellar population gradients. Early theoretical work \citep[e.g.,][]{1994Friedli} suggests a significant flattening of the abundance (metallicity) gradients in barred galaxies. This result is further strengthened by more recent numerical simulations also finding weaker metallicity gradients in the presence of a strong bar \citep[e.g.,][]{2010Minchev}.

\subsection{Break position}

%-----------------------------------------------------------------------------
\begin{figure}
\includegraphics[width=\linewidth]{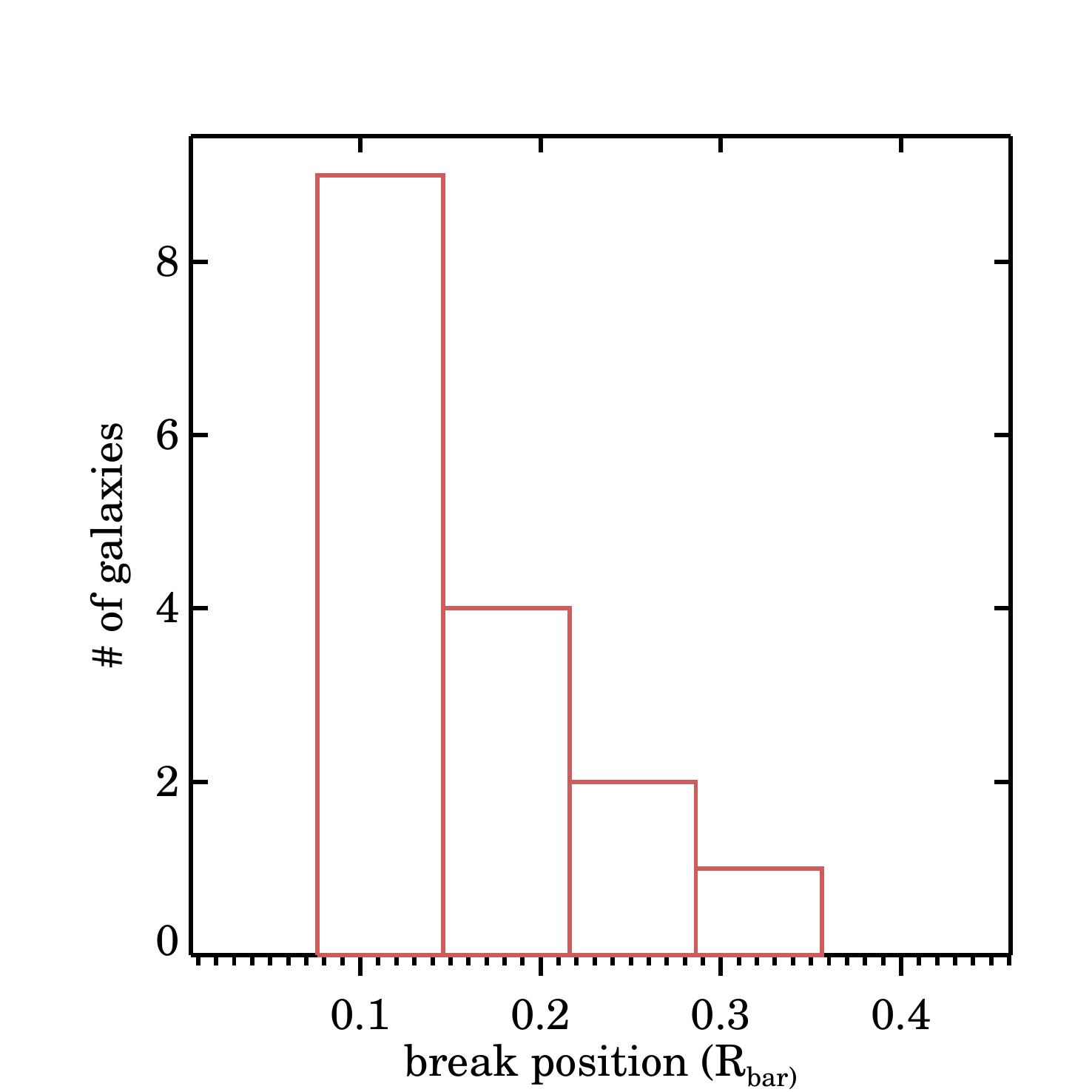}
\caption{Location of the break position for the SSP profiles (almost identical with that for the index profiles). }
\label{fig:histo-pos}
\end{figure}
%-----------------------------------------------------------------------------

%-------------------------------------------------------------
\begin{figure*}
\includegraphics[width=0.88\linewidth]{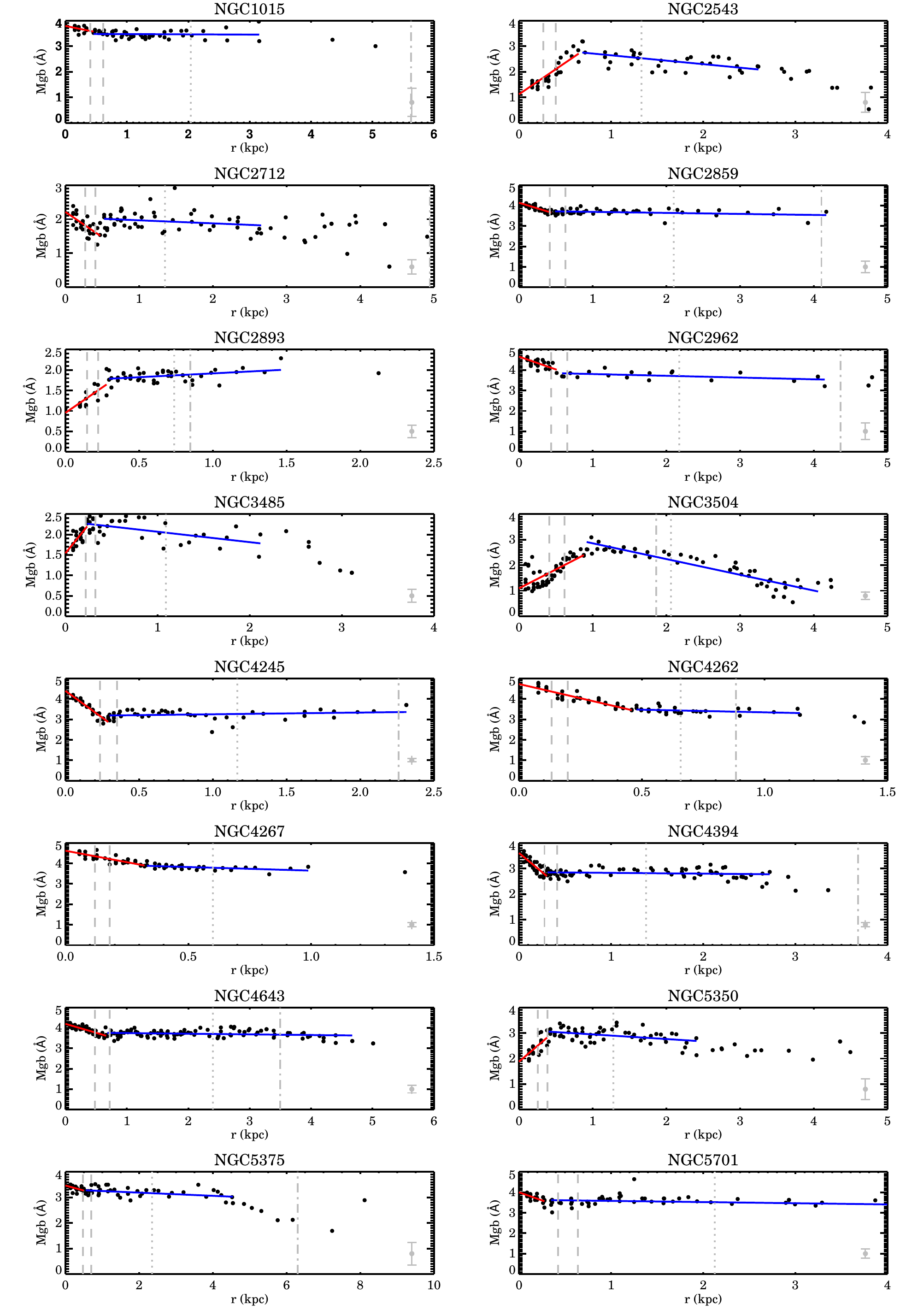}
\caption{Bar major axis profiles and linear inner (red) and outer (blue) gradients of the Mg$b$ index as an example of the index measurements for the entire BaLROG sample. The mean uncertainty is indicated in each panel in the lower right corner. Dashed lines indicate the region between 0.1 and 0.15 R$_{\rm bar}$, dotted lines the position of 0.5 R$_{\rm bar}$ and dashed-dotted lines the position of 1.5 R$_{\rm eff}$.  }
\label{fig:Mggrad}
\end{figure*}
%----------------------------------------------------------------------------

While observations and theory have primarily considered the overall gradient, we attempt to separate gradients and investigate their behaviour in dependence of the host galaxy's bar properties, i.e. their strength. This step is further motivated in the light of our results suggesting a noticeable, although very small, bar influence in the central regions. In fact, changes in the slope of the gas-phase metallicity gradient have been distinguished and studied in numerous works \citep[e.g.,][]{1992Zaritsky, 1992VilaC, 1995MartinRoy, 2011Zahid}. Their existence is usually attributed to variations of the gas density as a result of bar-driven large-scale mixing \citep{1995FriedliB}. However, \citet{2012Sanchez} have found no variation in the gas phase metallicity gradient between barred and unbarred galaxies using the CALIFA sample.

%-----------------------------------------------------------------------------
\begin{figure*}
\includegraphics[width=0.35\linewidth]{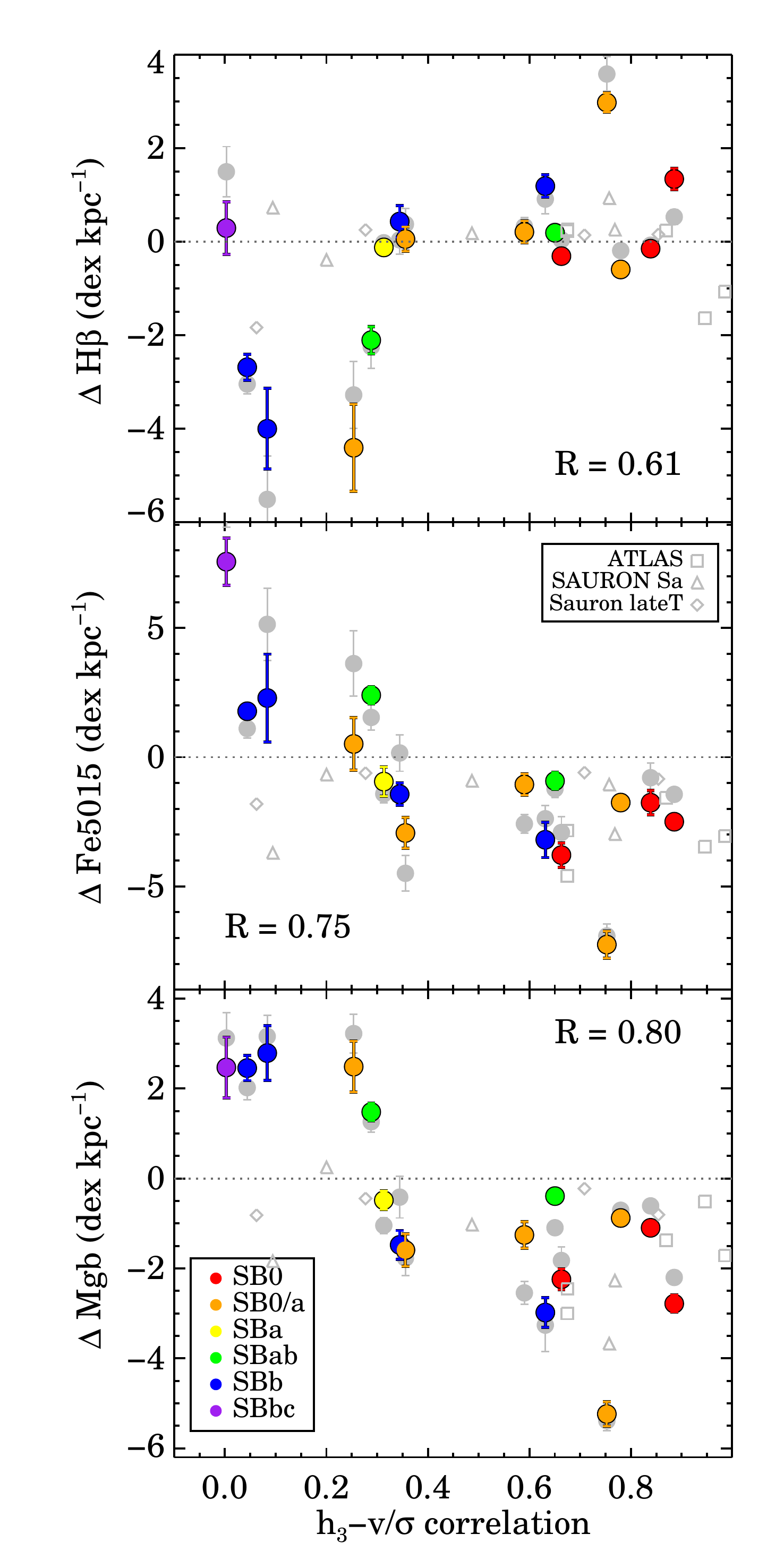}
\includegraphics[width=0.35\linewidth]{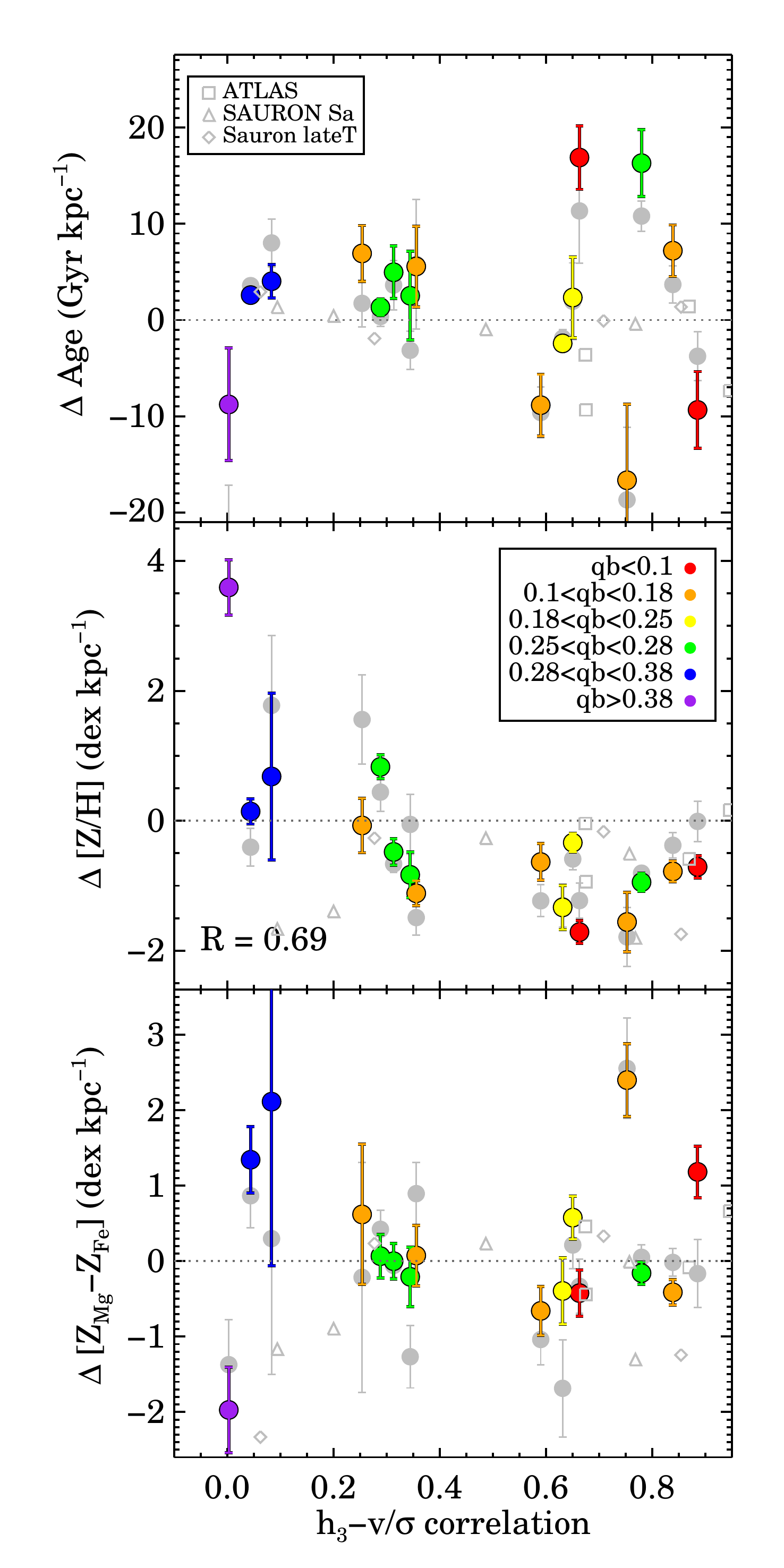}
\caption{Inner gradients as a function of h$_3$-v/$\sigma$ anti-correlation (the latter is analyzed in Paper I and Fig.~\ref{fig:SSPh3} of this paper). Hubble types (and bar strength for the right panel) are indicated in colours for the bar major axis gradients while bar minor axis gradients are shown as dots in gray. Values from our unbarred comparison sample are shown as open symbols distinguishing between the three datasets. The linear Pearson correlation coefficient (calculated for the bar major axis points), if significant, is given in all cases as the value R in the respective panels. }
\label{fig:grad-h3}
\end{figure*}
%-----------------------------------------------------------------------------

Earlier works on stellar metallicity gradients such as \citet{macarthur2009} or \citet{Perez2011} have also separated gradients based on a bulge-disk decomposition. In some of their profiles, this transition coincides with the innate breaks of the profiles. In other cases however, the structure of the profile itself is not taken into account and the gradients are fitted within the predetermined bulge region (bar, disc regions). 
Our analysis is based only on the shape of the index and SSP profiles in order to avoid biases coming from the photometry or applied bulge-disk decompositions. The work by \citet{2014SB} uses this same method to differentiate regions of different metallicity and metallicity gradients. This is crucial, in particular as former studies have not distinguished these breaks \citep[e.g.,][]{2011Sanchez}. The comparison between line-strength indices and full spectral fitting results are in a good agreement, which gives particular confidence in our results. We conclude that it is important to differentiate between areas of distinct slope either by a prior decomposition or by directly observing the profiles. Both attempts bear advantages and disadvantages of course and are simply based on different assumptions.

We investigated the position of the break in the index and SSP profiles (see Fig.~\ref{fig:Mggrad}) which determined two regimes of different gradients. Figure~\ref{fig:histo-pos} illustrates this analysis, showing the locations for the SSP profile breaks (averaged values from all profiles). They are very similar to the positions of the index profile breaks. In both cases more than 50\% of the galaxies have their break at around 0.1 R$_{\rm bar}$, a few up to 0.2 -0.3 R$_{\rm bar}$. Summarizing our kinematic as well as population analysis, it seems that this location, around 0.13$\pm$0.06 R$_{\rm bar}$, is a key location in barred galaxies.

\subsection{Connection between inner gradients and distinct inner dynamics}

%%%%%
Using the inner gradients, we tried to further quantify the impact of bars on creating inner substructures. Figure~\ref{fig:grad-h3} (left panel) shows strong correlations between inner index gradients and h$_3$-v/$\sigma$ anti-correlation, which was already shown in Fig.~\ref{fig:SSPh3} in connection with central SSP values. This correlation is found for all inner index gradients, but mainly for the inner gradients of Mg$b$ and Fe5015. This behaviour is seen along both axes, along the major and minor bar axes, which is expected as the anti-correlation of h$_3$-v/$\sigma$ was obtained using an aperture of 0.13 bar lengths. This roughly corresponds to the area in which the index (and SSP) gradients have been measured, also corresponding to the inner structure, mostly axisymmetric. We also show values for our comparison sample for all three datasets. We find that in particular for strong anti-correlations, the gradients fall within the same area as for the barred galaxies, while lower anti-correlations deviate from the barred sample. The right panel of  Fig.~\ref{fig:grad-h3} shows the same measurement for the SSP gradients. Here, a significant correlation is only found for the metallicity gradients. Again, unbarred galaxies agree with the barred ones for high values of h$_3$-v/$\sigma$ anti-correlation.

%-----------------------------------------------------------------------------
\begin{figure*}
\includegraphics[width=0.551\linewidth]{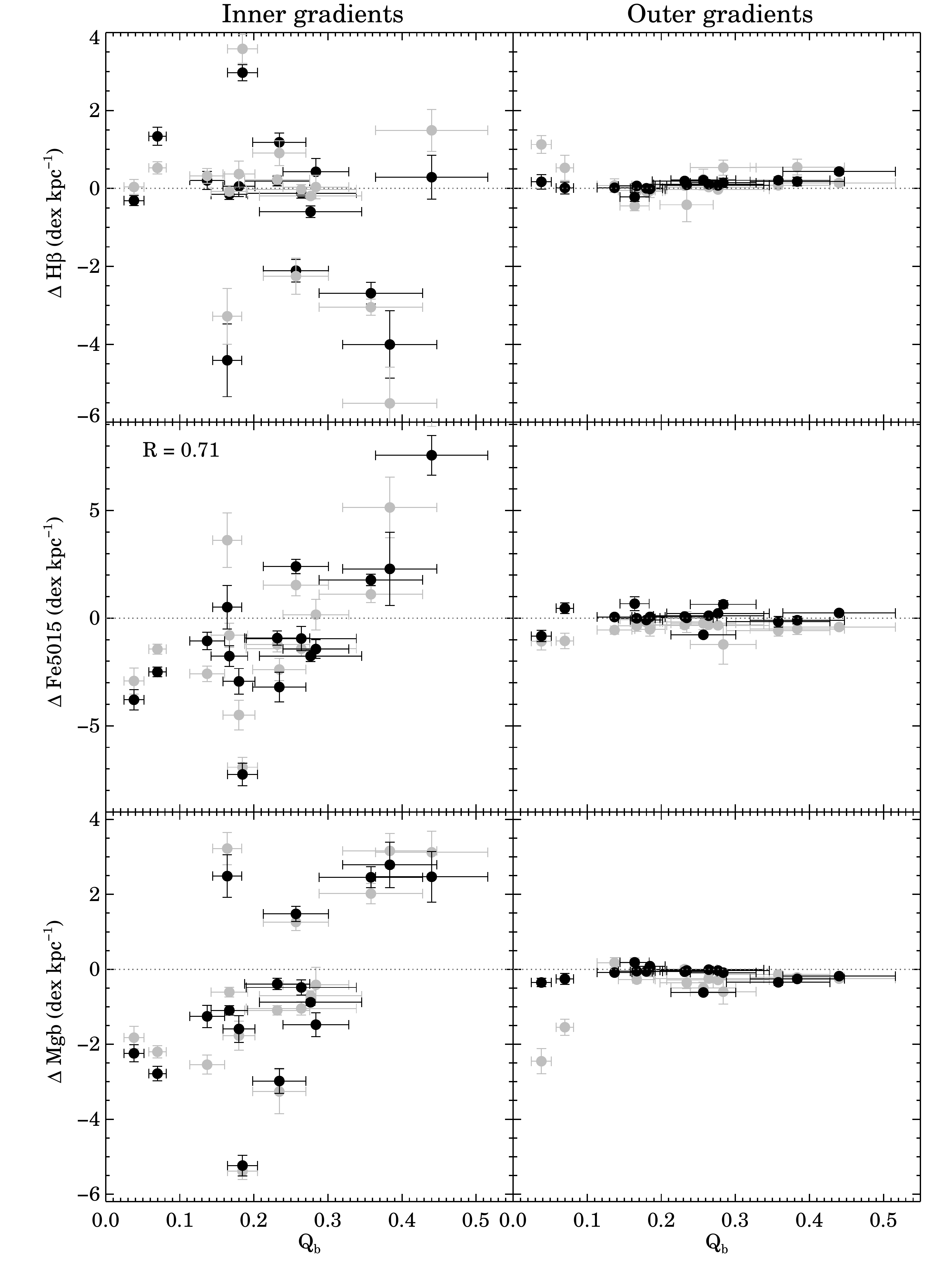}
\includegraphics[width=0.312\linewidth]{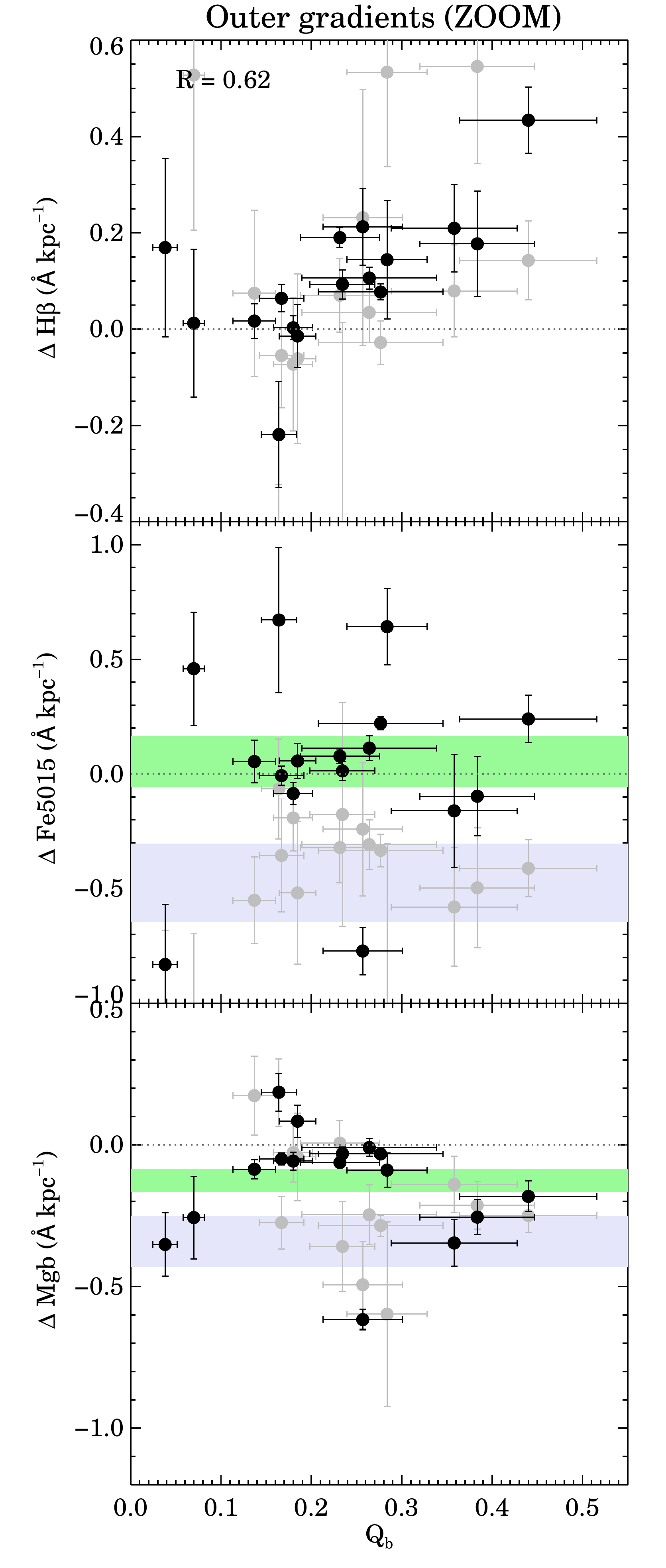}
\caption{Inner and outer index gradients for H$_\beta$, Fe5015 and Mg$b$. We depict the outer gradient twice to 1) highlight its difference in magnitude to the inner gradient (middle panel) and 2) illustrate its values in greater detail (right). Black points represent bar major axis measurements while gray points can be attributed to the corresponding minor axis results. The linear Pearson correlation coefficient is indicated in two cases as R, for the rest it was not significant. In the right panels we furthermore indicate the averaged values (by weight) for the bar major axis measurements (green - upper bar) and for a combination of bar minor axis measurements and measurements on unbarred galaxies from the SAURON and ATLAS3D surveys (gray - lower bar).}
\label{fig:index-q}
\end{figure*}
%-----------------------------------------------------------------------------

The shown correlations illustrate mainly, that the presence of an inner dynamically differently rotating structure seems to be closely linked to a steeper inner negative metallicity gradient. The Hubble type seems to influence this relation but does not appear to be the only driver. Early-type galaxies show in general steeper relations and a stronger anti-correlation, but we lack enough data points for the later types in order to fortify this result. Instead, on the right panel, which is colour-coded with the bar strength values, the relation with this colour-code appears to be slightly stronger. However, it demonstrates that weaker bars show steeper gradients and stronger anti-correlations. So if these structures were due to an influence by the bar, the impact could have occurred at an earlier stage of its evolution and took time to develop. Former works, in particular simulations, suggest that the strongest influence of the bar happens during its buckling phase \citep[e.g.,][]{2006MartinezV}, which is an early phase of the bar formation. 

In any case, these correlations clearly illustrate a relation between distinct inner dynamics and the steepness of inner gradients, possibly under the influence of a barred structure.

\subsection{Line-strength gradients as a function of bar strength}

As the SSP values are based on the line strength values which are directly measured on the data, we first determine the gradients directly from the index values. In Fig.~\ref{fig:Mggrad}, we show the profiles with the gradients indicated for Mg$b$ as an example. Doubtlessly, the break in most profiles stands out clearly for our data. At the same time we wish to point out the necessity for sufficient spatial resolution in order to resolve these profiles and thus be able to distinguish between the two gradients. This distinction is particularly important in barred galaxies as they have rich inner substructures such as inner disks or rings that could alter the overall gradient significantly \citep[e.g.,][]{2002Erwin, macarthur2004, macarthur2009, 2013DeLorenzoC}. 

Figure~\ref{fig:index-q} reveals a clear difference between the inner and outer index gradients, each consistent for major (black points) and minor axis (gray points) measurements. Apart from the major difference in magnitude, much stronger correlations with bar strength are observed for the inner gradients. In particular for the Fe5015 gradients, the correlation is found to be strong, resulting in a linear Pearson correlation coefficient of R=0.71 (p-value=0.002). For the other two indices, tentative correlations with bar strength can be observed, but among our small sample, the scatter is large and it is difficult to deduce a definite conclusion. The linear Pearson correlation coefficients are R=0.31 (p-value=0.24) for the H$_\beta$ gradients and R=0.61 (p-value=0.01) for Mg$b$.  Overall, the H$_\beta$ gradient (as a proxy for age) seems to decrease with bar strength while the two gradients of the metallicity proxies, Fe5015 and Mg$b$ seem to increase consistently.

The outer gradients are all much lower in magnitude. While the H$_\beta$ outer gradients seem to correlate with bar strength, the two metallicity indicators show a scatter around zero or below. 
The fact that the H$_\beta$ gradients seem to increase with bar strength could imply an effect of bars or a relation of them to the disk. The linear Pearson correlation coefficient in this case is R=0.62. Its p-value of 0.01 only gives a tentative significance. As often, this could also be an effect of the galaxies' Hubble type. In our sample (as found in general, see, e.g. \citealt{2007Laurikainen}), later types exhibit stronger bars. Numerous earlier studies using colours \citep[e.g.,][]{deJong1996, 1996PeletierB, macarthur2004, munozm2007, 2009aMunoz} already found bluer outer parts in disk galaxies. This was confirmed by spectroscopic studies measuring elevated H$_{\beta}$ (corresponding to younger ages) in the external regions \citep[e.g.,][]{2008Yoachim, macarthur2009,2011MNRAS.415..709S, 2012Yoachim, 2013hsa7.conf..367R, 2014SB}. Therefore, we expect to find positive values when measuring the gradient towards the edge of the bar. The value of these gradients is further expected to increase when going to later types as the disks are found to be even bluer. This effect could be enhanced by the presence of a bar due to bar-spiral coupling  \citep[e.g.,][]{2010Minchev}. On the one hand, the process of bar-spiral coupling is supposed to make mixing mechanisms more efficient and therefore possibly erase strong gradients, although this effect is much more pronounced at larger radii. On the other hand, through the enhanced perturbations, gas could be funneled more efficiently to these regions and star formation might be triggered easier. However, the flattening of the gradient occurs on a large time scale. The effect of recent young populations could be therefore enough to raise the luminosity weighted line-strength gradient of H$_{\beta}$.

For a comparison with galaxies without bars, we return to our sample of unbarred galaxies from the ATLAS3D and SAURON studies. We performed our analysis analogously to the BaLROG sample on these galaxies and also obtained slopes of inner and outer gradients.  When trying to detect breaks in the index gradients of the unbarred sample, we found that in some cases, they were not as obvious as in the barred galaxies, in particular for the later types. We illustrate examples of these gradients in Fig.~\ref{fig:CSindgrad} in the Appendix, showing Mg$b$ and H$_\beta$ profiles. In certain cases, the profile did not exhibit a strong break, but rather showed a continuity, leading to similar inner and outer gradients. This confirms former results, e.g., by \citet{morelli2008} who only found very weak age gradients in a sample of unbarred galaxies. Hence, we suggest that this lack of breaks in the index profiles is not due to the outer gradients, as their magnitude remains comparable to the ones measured in BaLROG, but to the lack of inner substructures such as nuclear disks or rings or an influence of the bulge. The fact that we find a strong difference of inner and outer gradients in our sample of barred galaxies indicates a possible influence of bars in creating substructures, particularly by providing the fuel for subsequent star formation \citep[e.g.,][]{2005Knapen, 2006Knapen, peletier2007,2010Comeron}. It might however also be due to the influence of a bulge that starts to dominate in the inner regions, as we do detect difference in the profiles also in the unbarred sample. 

Comparing the value of the observed gradients in H$_\beta$, we find a stronger dependence on Hubble type than the presence of bars: our average major and minor bar gradients are practically the same, 0.10$\pm$0.07~\AA/kpc and 0.11$\pm$0.16~\AA/kpc respectively. The ATLAS3D sample of unbarred galaxies yields a value of 0.13$\pm$0.06~\AA/kpc, the SAURON (S0 and Sa galaxies) a value of 0.14$\pm$0.11~\AA/kpc and the late-type SAURON sample 0.24$\pm$0.09~\AA/kpc. In context with former studies finding bluer (younger) outer regions of disk galaxies \citep[e.g.,][]{deJong1996, 1996PeletierB, macarthur2004}, we conclude that the measured gradients are dominated by this effect: for later types the outer disk regions become increasingly bluer and hence steepen the overall H$_\beta$ gradient, with or without a bar.

The gradients determined from  Fe5015 as well as Mg$b$ profiles however are systematically steeper along the minor bar axes compared to the major bar axis. The latter are closer to zero, in particular for Fe5015. 
We indicate the mean values of the bar major axis measurements for iron and magnesium in green (upper bar) in both plots. The lower gray bar is a representative value for minor axis measurements as well as unbarred galaxies where we measured iron and magnesium gradients for the same samples as for H$_\beta$. For these metallicity indicating gradients, we observe a much larger similarity of the unbarred galaxies with the bar minor axis measurements. Quantitatively, we find a mean value for the outer gradients in Fe5015 along the bar major axis in our BaLROG sample of 0.05$\pm$0.1~\AA/kpc while the minor axis shows -0.47$\pm$0.24~\AA/kpc. The effect of the contrast between bulge and bar components could add to the steepening, but is not fully responsible for it. The unbarred SAURON S0-a and Sa galaxies (also within our inclination limits) show -0.55$\pm$0.23~\AA/kpc and the ATLAS3D (mainly S0 and S0-a galaxies) show -0.34$\pm$0.12~\AA/kpc and the late-type galaxies  -0.58 $\pm$0.12~\AA/kpc . The weighted mean value of the comparison samples is -0.49~\AA/kpc in comparison with 0.05~\AA/kpc for the bar major axis measurements. For the Mg$b$ index, we find a mean value of -0.34~\AA/kpc amongst the comparison samples and a value of -0.13$\pm$0.05~\AA/kpc along the bar major axis. 
%-----------------------------------------------------------------------------
\begin{figure}
\includegraphics[width=\linewidth]{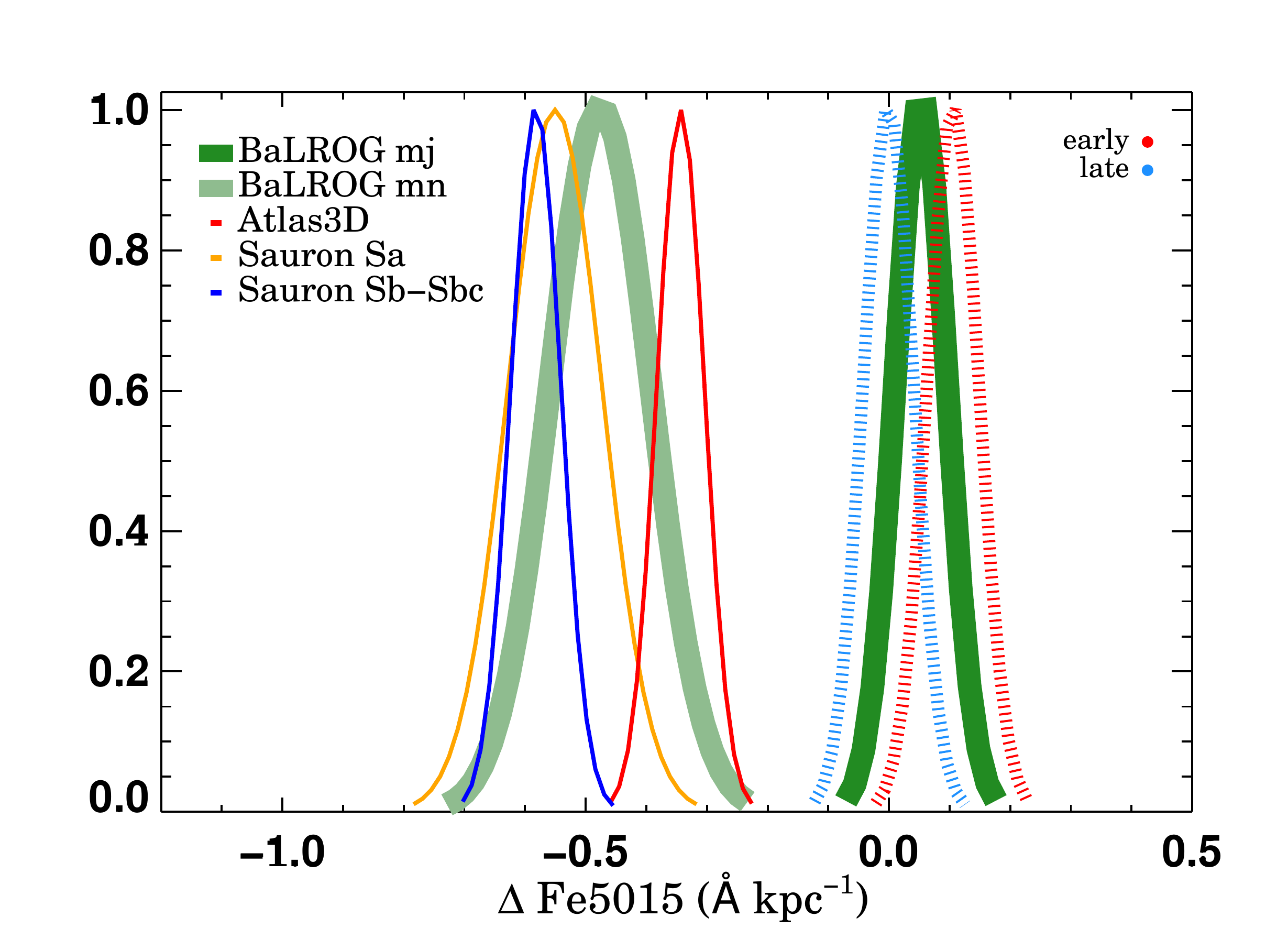}
\caption{Mean values and their uncertainties shown as gaussians for the Fe5015 outer gradients for the BaLROG bar major and minor axis in comparison with the ATLAS3D, SAURON and late-type SAURON samples. Additionally, we separate early and late-types for the BaLROG galaxies . }
\label{fig:fe-hist}
\end{figure}
%-----------------------------------------------------------------------------

%-----------------------------------------------------------------------------
\begin{figure*}
\includegraphics[width=0.551\linewidth]{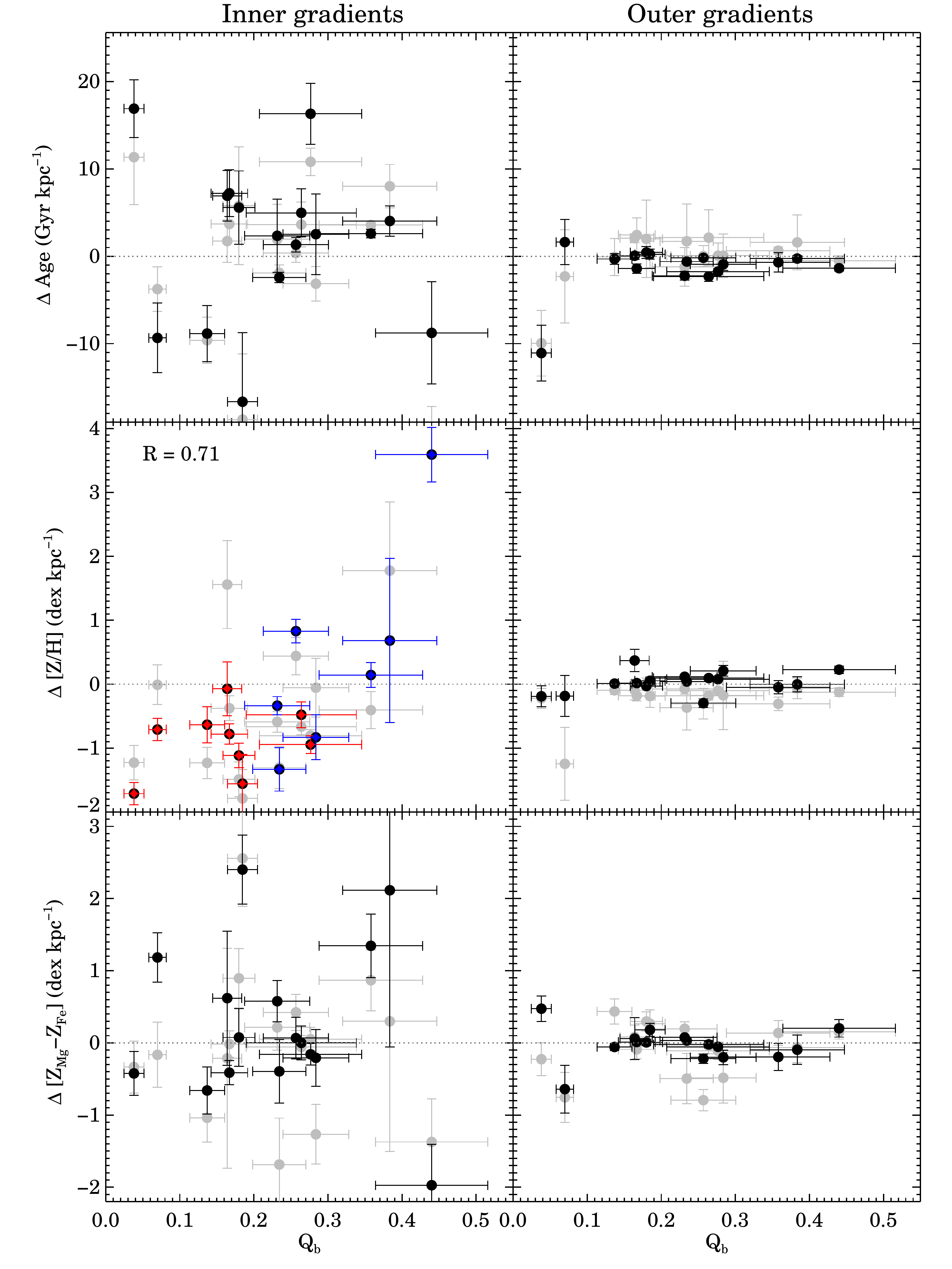}
\includegraphics[width=0.312\linewidth]{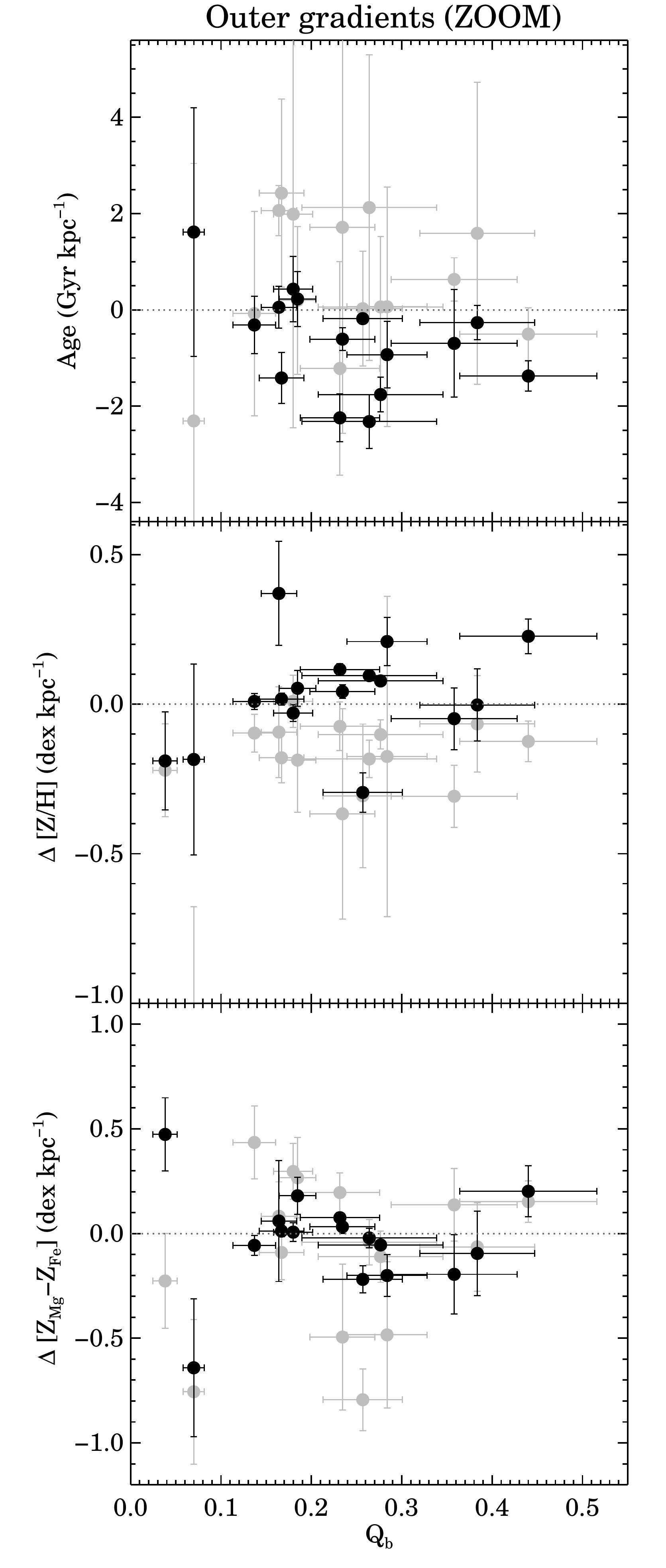}
%\vspace{0.5cm}
\caption{Inner and outer index gradients for age, metallicity and $\alpha$ abundance. We depict the outer gradient twice to 1) highlight its difference in magnitude to the inner gradient (middle panel) and 2) illustrate its values in greater detail (right).  Black and gray points represent the same as in Fig.\ref{fig:index-q}. The linear Pearson correlation coefficient is indicated in the case of the inner metallicity gradient as R=0.71, for the rest it was not significant. In this same panel we also show the difference between early (red) and late (blue) types within our sample.}
%\vspace{0.6cm}
\label{fig:ssp-q}
\end{figure*}
%-----------------------------------------------------------------------------

The results for the Fe5015 gradients are summarized in Fig.~\ref{fig:fe-hist} where each curve represents the mean value and its error. The difference can clearly be appreciated. We also observe a trend that earlier types (ATLAS3D) seem to show shallower gradients than late-types. Therefore, we furthermore separated early- and late-type galaxies for the bar major axis measurements and within this sample, we actually find that the earlier types are not necessarily shallower but actually more positive. Nevertheless, the separation is small and hence we conclude that in this case, the bar is in fact the driver and not the Hubble type.

This analysis confirms our hypothesis that bars display flatter gradients, but only along the bar major axis. A similar result was found in a pilot study by \citet{2011MNRAS.415..709S}, but only using two galaxies. Other studies have not compared profiles along these different axes nor used the distinction of inner and outer gradients based on the observed break in the index profiles. Instead the inner and outer gradients have been separated using the bulge radius which was determined from the photometry. We believe that this different methodology can lead to different results, as the differences between the gradients are not large and can be washed out easily. In any case, the flattening of the Fe5015 and Mg$b$ gradients could be a first indicator of a flattening along the bar. We will look into more details when analysing the SSP gradients in the following section.

\subsection{SSP gradients as a function of bar strength}

Similarly to the index gradients, we determined the slopes along the SSP profiles (age, metallicity and $\alpha$ abundance), again considering two distinct regions within the bar region. An example for the metallicity is shown in Fig.~\ref{fig:Metgrad} in appendix~\ref{app:grad}. The absolute values found for the outer gradients are comparable in magnitude to the ones found in previous studies, such as the recent study of CALIFA galaxies \citep[e.g.,][]{2014SB2}. As found for the index profiles, there is a clear difference between two gradients in our SSP profiles. To make this distinction and to identify the break in the profiles is crucial for unravelling the influence of bars.

As already seen for the indices, the magnitude of the inner gradients is much higher (see Fig.~\ref{fig:ssp-q}). This time, we only find a correlation of the inner metallicity gradients with bar strength. Weaker bars show a stronger negative metallicity gradient, however this seems to be influenced by the Hubble type, too. Early types do not show a strong correlation whereas the correlation is maintained for the late types. Supposing an initial negative metallicity gradient \citep[e.g.,][]{2013DiMatteo}, this finding implies that weak bars allow it to survive - or that there is a significant difference between the initial metallicity gradient between early and late type galaxies. Towards stronger bars, this gradient seems to flatten and even turns into a positive gradient in fact, such that metallicities get higher in the bar regions compared to the center. In addition, we observe once more an agreement of the results along major and minor axes. This could also suggest that the influence of the bar, if it reaches these regions, is rather quickly distributed radially around the center or that bars have only a limited influence in these galaxies.

While the gradients agree for the inner regions between major and minor axes, the outer gradients along the different axes are slightly different. In the amplification of the panels on the right side of Fig.~\ref{fig:ssp-q}, it can be seen that the age gradients along the major axis are slightly negative. The opposite is the case for the age gradients along the bar minor axis. As our gradients stop at the maximum bar length, this could be the result of the entering spiral arms along the major axis which leads the gradients to be slightly negative, hence going towards younger populations towards the outer parts. Simulations by \citet{2007Wozniak} suggest that young populations are indeed found at the edges of bars. Furthermore, the disc region with the spiral arms might not have been reached along the minor axis.Therefore, the gradient in those cases can result to be slightly positive, i.e. the population is younger in the inner parts.

\begin{figure}
\includegraphics[width=\linewidth]{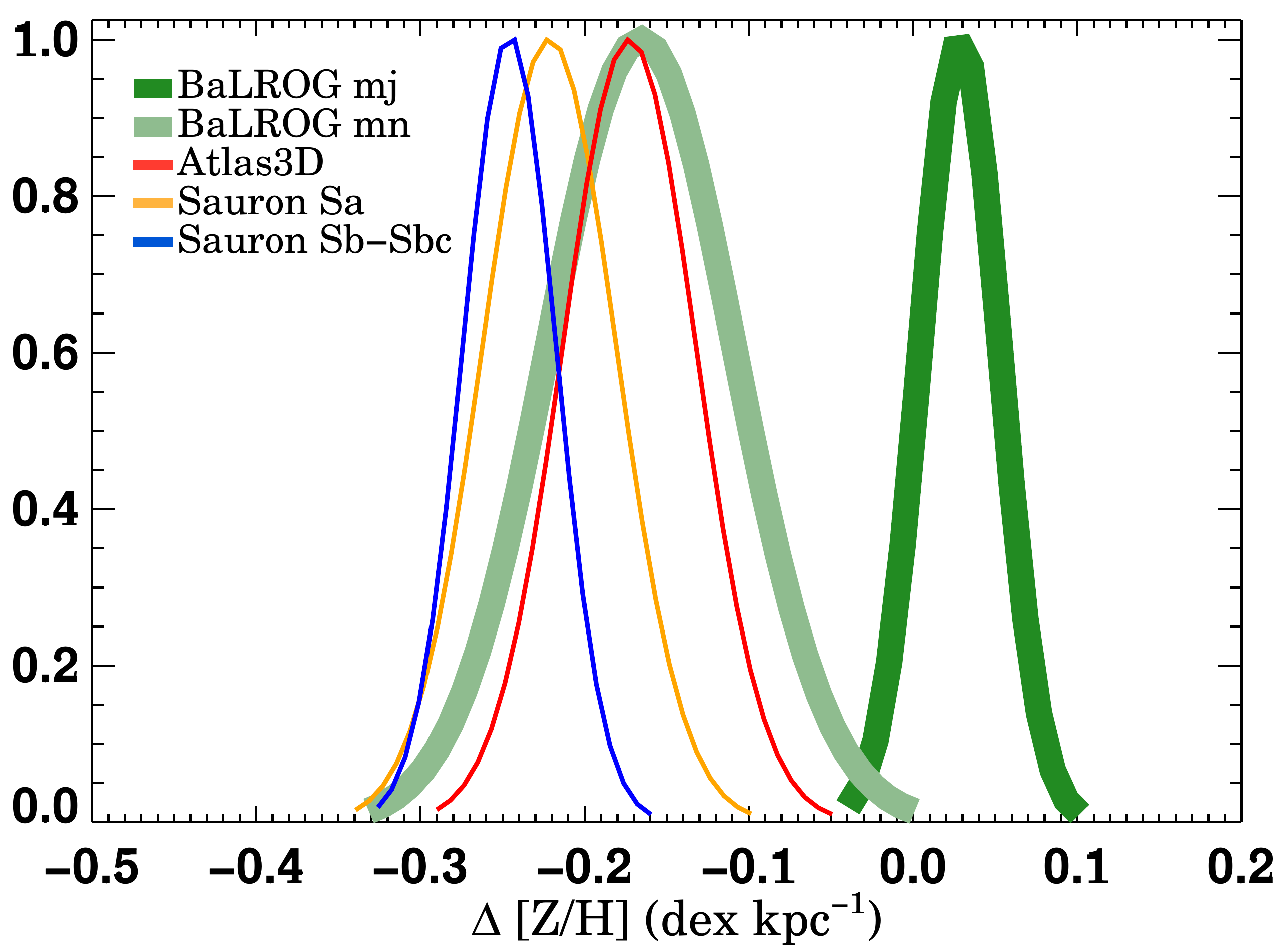}
\caption{Mean values and their uncertainties shown as gaussians for the metallicity gradients for the BaLROG bar major and minor axis in comparison with the ATLAS3D, SAURON and late-type SAURON samples.}
\label{fig:met-hist}
\end{figure}

The outer metallicity gradient exhibits a similar behaviour as the Fe5015 (and Mg$b$) index: bar major axis gradients are found to be closer to zero, while bar minor axis gradients are found to be slightly more negative. This supports the notion of increased flattening of the metallicity gradient along the bar major axis, also found for two galaxies in \citet{2011MNRAS.415..709S}. We furthermore compute metallicity gradients of three unbarred control samples already mentioned above, namely  ATLAS3D \citep{2011Cappellari}, SAURON S0 and Sa galaxies \citet{peletier2007} and late-type SAURON galaxies \citet{ganda2007}. We use the same method as for our sample: we first derive the metallicities with {\tt rmodel} from the line strength maps provided by the published work; then we determine the profile and identify its break to measure the two slopes. Figure~\ref{fig:met-hist} illustrates the clear difference between the outer metallicity gradient along the bar major axis and the significantly steeper metallicity gradients found along the minor axis and in all unbarred samples. Along the bar major axis we find the metallicity gradient to be 0.03$\pm$0.07~dex/kpc whereas the mean value of the minor axis and unbarred control sample is -0.20$\pm$0.04~dex/kpc. We mention once more that the contrast effect between the bulge and bar component (see Sec.~\ref{sec:profind}) can alter the gradient, but cannot account for the general flattening along the bar major axis. It might lead however to some of the positive gradients that we measure, which are not necessarily expected according to bar-driven secular evolution. We further note a very mild trend with Hubble types among the unbarred control samples in the sense that steeper negative gradients are found in later type galaxies. Consistent with our former results on Mg$b$ and Fe5015 gradients, this trend is well within the uncertainties and the difference of barred (\textit{along the bar major axis}) vs. unbarred (or barred, but not along the bar axis) is much higher. We therefore suggest the bar as the responsible agent for this flattening.

The $\alpha$ abundance profiles do not reveal any tendency. We note however, that those values need to be taken with care due to the mixture of populations already revealed and observed in the index-index diagrams. 
%-----------------------------------------------------------------------------
%-----------------------------------------------------------------------------

%We note however, that the enhancement can also be an effect of contrast between the bulge and the bar components. Considering the high steep gradient of the inner structure and the light contribution of the bulge, compared with the flatter gradient of the bar and its light, the combination of both can make the gradient in the inner bar to be less steep, flat or even positive.

\section{Discussion}
\label{sec:discuss2}

\subsection{Discrepancies in the literature}
\label{lite}

As already illustrated in the introduction, numerous works have been studying the influence of bars on the metallicity and $\alpha$ abundance gradients. Simulations and theoretical studies mainly propose a flattening of these gradients, which is even stronger for older stellar populations \citep[e.g.,][]{1994Friedli, 2010Minchev}. 
These theoretical results have only been partly corroborated by observations. While some studies \citep[e.g.,][]{1994MartinR, 2011MNRAS.415..709S, 2012Williams} confirm the flattening, others observe a large variety \citep[e.g.,][]{2009PerezSBZ} and recent results do not find any difference in the gradients of barred or unbarred galaxies for stellar and gas-phase metallicity \citep[e.g.,][]{2014Sanchez, 2014SB2, 2015bCheung, 2015Ho}. There are however significant differences between the studies that could lead to this disagreement. In particular, it is crucial \textit{where} the gradient is measured and also \textit{along which axis} and at which possible \textit{spatial resolution} - in order to distinguish different types of gradients which are related to different drivers.

To start with, simulations disagree on the initial overall gradient and its temporal evolution. While most chemical evolution models predict an initially negative metallicity gradient that flattens in time \citep[e.g.,][]{1997Molla,2005MollaD,2009SchoenrichB,2009Fu} others start with flat or even initially inverted gradients that steepen with time \citep[e.g.,][]{ 1997Samland, 2001Chiappitini}. These differences in the models are attributed to i) the degree of enrichment of infalling material and ii) distinct efficiencies between enrichment processes of inner and outer disc regions. The works of \citet{2012Pilkington} and \citet{2013Gibson} further analyse metallicity gradients and their evolution in fully cosmological hydrodynamical simulations. They attribute the discrepancies in the resulting gradients to the influence of the sub-grid physics employed in the hydrodynamical codes used in cosmological simulations, where galaxy formation crucially depends on the implemented feedback schemes. Observationally, an inside-out formation scenario, leading to final overall negative metallicity gradients is commonly confirmed \citep[e.g.,][]{deJong1996, bell2000, macarthur2004,munozm2007}. Our results on the outer metallicity gradients confirm this finding. We typically observe negative metallicity gradients, in particular when extending to the outer disk regions. Only when measuring the gradient along the bar major axis, we observe differences.

So, which influence can bars have? Do they or not produce a flattening and if so where and why? A flattening of the overall gradient is supposed to be most obvious in the outer parts, beyond bar corotation due to a bar-spiral coupling \citep[e.g.,][]{2013Minchev, 2013DiMatteo}. Studies such as done by \citet{2014SB2} examined exactly this metallicity gradient and found no hint for a flattening introduced by the bar. This result is supported by other works also studying the global (or integrated) metallicity and metallicity gradients and which also do not detect any difference between barred or unbarred galaxies \citep[e.g.,][]{2014Cacho, 2015bCheung}.

However, one has to distinguish between this outer (and global) and yet another inner gradient: due to the bar providing the necessary fuel to trigger central star formation in the nuclear region, abundance gradients can steepen significantly in these regions \citep[e.g.,][]{1994Friedli}. In fact, even in unbarred galaxies, a clear break in the stellar metallicity profile has been detected and inner and outer gradients have been distinguished \citep[e.g.,][]{2014SB}. Furthermore, the axis along which the gradient is measured is important. Scarce former studies have compared the metallicity gradients along the bar with that perpendicular to it or along the disk major axis and detected flatter gradients along the bar major axis \citep{2009PerezSBZ,2011MNRAS.415..709S,2012Williams}. This difference is predicted by simulations \citep[e.g.,][]{2013DiMatteo, 2013MVG}, although the absolute value of the gradient is strongly dependent on the initial metallicity gradient in the disk. In simulations by  \citet{2013MVG} the initial gradient along the bar of -0.4~dex/kpc evolves to a final value of -0.26~dex/kpc, while \citep{1998Friedli} starts with an inital gradient of -0.1~dex/kpc and hence also obtains a lower final value.

Our work clearly distinguishes the regions of different gradients. This allows us to differentiate between the influence of inner substructures and overall properties along the bar compared to the overall galaxy. We do observe a flattening of the metallicity gradient, but only \textit{along the bar major axis}. This flattening is a small effect as expected from former studies that did not reach a consensus on a difference in stellar metallicity gradients in barred or unbarred galaxies, and confirms the flattening seen along the bar major axis \citep{2007Perez,2009PerezSBZ,2011MNRAS.415..709S,2012Williams,2014SB2,2015bCheung}. The implication of our finding of a flattening along the bar compared with steeper gradients observed along other axes or in unbarred galaxies is discussed in the following Sec..

\subsection{The role of bars in galaxy evolution}

The results obtained in our work clearly indicate only a very weak influence of bars on the stellar populations of their host galaxies, definitely less than suggested by numerical simulations. As explained above, this minor yet existing bar influence does not contradict former studies that claim no difference between stellar population parameters of inner regions (bulge) or metallicity gradients between barred and unbarred galaxies. Most of the differences and discrepancies between former studies arise due to different analysis techniques and/or spatial resolution effects (see above, Sec.~\ref{lite}).

Our results reveal a subtle but noticeable influence of bars in the central regions. They are able to trigger or at least favour the build-up of substructures leading to younger ages, higher metallicities and different orbital configurations. This confirms numerous former studies on central star formation, nuclear-rings and inner disks \citep[e.g.,][]{1994Heller, 1995Knapen,Allard2006,2010Knapen, 2013vdLaan, 2015Florido}.

Furthermore, we demonstrate for the first time with a representative sample of barred galaxies, compared with a large number of unbarred galaxies, that the metallicity gradients along the bar are significantly shallower than along another axis or in the unbarred control sample. We stress here that we do not sample these gradients in the outer discs, but within the bar regions. Simulations of chemical evolution models investigating bar-driven secular evolution predict a flattening of the metallicity gradients of barred versus unbarred galaxies in their discs \citep[e.g.,][]{1994Friedli, 2013DiMatteo}. This has yet to be observed, in particular in these outer parts, beyond corotation, where the effect is supposed to be strongest \citep[e.g.,][]{2011Brunetti}. So far, studies have found no difference (see above and \citealt{2014SB2}). However, \citet{2013DiMatteo} also predicts a flattening in the inner regions of barred systems and as such the behaviour observed in \citet{2011MNRAS.415..709S} and our study. In other words, their predictions fit our findings of bars showing higher metallicities and flatter gradients than disk stars in the same region. Simulations of the Milky Way confirm this behaviour \citep[e.g.,][]{2013MVG}.

Our interpretation of the flattened metallicity gradients along the bar (and no other axis) coupled with the absence of a difference of gradients of barred and unbarred galaxies beyond corotation \citep{2014SB2} or integrated \citep{2014Cacho, 2015bCheung} is that bars seem to be confined structures, rotating as cylinders within the galaxy. Under the assumption that bars are long-lived, consistent with the latest numerical simulations \citep[e.g.,][]{2013Atha}  (but proposed since a long time) and observations detecting bars out to redshift z$\sim$2 \citep[e.g.,][]{2014Simmons}, we would expect to find a mixing effect if present. However, it seems that bars mostly affect themselves and remain rather confined structures within the galactic disks without too much interaction with it. As gas is more susceptible to non-axisymmetric components, such as bars, they seem to be able to funnel it to the center and trigger the growth of circumcentral substructures. The properties along the bar however resemble the bulge properties largely (see Fig.~\ref{fig:barssp}) and additionally exhibit higher metallicities (as predicted by  \citealt[e.g.,][]{2013DiMatteo}). The flatter metallicity gradient along the bars could hence be the result of orbital mixing - but only within the bar, as a confined structure, because it is only observed along it. 

Another possibility is that the flat gradient is already produced at higher redshifts due to strong radial mixing which can be achieved by strong galaxy interactions \citep[e.g.,][]{2010Rupke}. Then, in some galaxies, bars formed from this material which already showed enhanced metallicity and flatter metallicity gradients, while in others, only the bulge remains and consists of this material. In both cases, the disk with different properties settles around the system. The similarity of bars and bulges concerning their age, metallicity and also $\alpha$ abundance supports this idea. There could be an effect of overlap between the two components, but its effect would be minor. The clear offset in $\alpha$ abundance to lower values for disks further underscores their distinct formation scenario and the different timescale of it. While higher $\alpha$ abundances, such as seen in the bars and bulges of our sample, suggest short formation timescales, lower values indicate more extended formation scenarios. The fact that bars are observed up to high redshifts also enables the possibility of this scenario. Further, recent simulations also find that bars form very fast (within a few rotation timescales) and early on \citep[e.g.,][]{2015Saha} as soon as a cool stellar disk is assembled around the bulge spheroid, which in turn is then altered by bar-driven secular processes. As our sample is slightly biased towards early-types, the strong connection between bars and their bulges might be even more explicit amongst this sample and could be valid in particular for bars in early-type galaxies. In particular, bars also empty the region around them, so anything that happens in the disc, does not reach the bar such that the bar is a proxy of the state of the disc when the bar formed.

In fact, the average age of bars depends on the Hubble type. However, in the early-type galaxies of our sample, it is of the order of the bulge and rather old, usually up to 10~Gyr. At the same time they often reside in a younger, star forming disk. The fact that the age of the bar in those systems is old, despite the presence of younger populations in other parts of the galaxy, argues against a recent bar formation from disk stars, but for an early formation of a bar that survived a long time, already put forward by \citet{2006GadottideSouza, 2007Perez, 2011MNRAS.415..709S} and supported by recent studies of the bar fraction at high redshift finding bars out to z$\sim$2, using the HST CANDELS data \citep{2014Simmons}.

Nonetheless, for both early- and late-type galaxies, we conclude that bars produce a noticeable \textit{local} effect on their host galaxies. This is to be expected as barred and unbarred galaxies share the same large-scale dynamical properties, i.e. they both belong to the same Tully-Fisher relations \citep{2003Courteau}. This should indicate that they share a similar baryon/dark matter global distribution, which implies that secular evolution should be a local phenomenon - as we observe it. We summarize the most important results and conclusions in the following section.

%=============================================================================
\section{Summary and conclusions}
\label{sec:summ}

In this work, we explored the BaLROG (Bars in Low Redshift Optical Galaxies) sample further, focusing on the stellar populations. The sample comprises 16 large mosaics of nearby barred galaxies observed with the integral field unit SAURON. Similar to our kinematic analysis (\citealt{2015Seidelb}, Paper I), we note once again the necessity of our spatial resolution (typically 100~pc) in order to be able to detect features induced by bars. 

Using the classical method of line strength index measurements, we determine SSP ages, metallicities and $\alpha$ abundances and their gradients which lead us to the following results and conclusions: 

\begin{itemize}

\item Bars among our sample exhibit a large variety of index measurements, and hence ages and metallicities. Global values, such as integrated quantities within the effective radius, do not seem to be influenced by the bar as already observed in larger samples  \citep[e.g.,][]{2015McDermid}, but follow general galaxy trends according to Hubble type and central velocity dispersions.

\item Elevated Fe5015 values are found at the edges of 7 of the bars of our sample, compared to their discs, reflected by an elevation in the metallicity profile around 0.5 to 0.75 bar lengths. We cannot fully exclude that this is due to a contrast effect of the bulge. 

\item Galaxies with their central spectra falling outside the SSP grid (towards the high metallicity end) overlap with those exhibiting a strong h$_3$-v/$\sigma$ anti-correlation within 0.1 R$_{\rm eff}$ (found in our Paper I), suggesting a strong connection between this dynamical feature and the presence of a complex mixture of populations.

\item Ages, metallicities and $\alpha$ abundances are found to be similar in bars and bulges, while disks show offsets. They exhibit in particular younger ages and lower $\alpha$ abundances. Although some of the early-type bars are rather old, up to 10~Gyr, they reside in a young, star forming disk. This supports theoretical predictions that bars are long-lived structures and might have formed a long time ago and survived until the present day. 

\item We identify breaks among all index as well as SSP profiles and therefore measure two different gradients. The location of the break between the two different slopes is mainly located at 0.13$\pm$0.06 bar lengths, confirming this place as a particular region already identified in Paper I, and possibly linked to an inner Lindblad resonance \citep{1994Elmegreen}. 

\item Inner gradients are found to be much steeper than outer gradients, for both index and SSP profiles in bar major and minor axes. Inner gradients commonly agree for bar major and minor axes and might be a result of the bulge dominated region here. They become larger when there is a central rotating component. This implies a close connection between gradients with dynamics and orbits.

\item Outer gradients show offsets between major and minor bar axes. In particular for Fe5015, Mg$b$ and metallicity, we find that major axis gradients scatter around zero while minor axis gradients are steeper and negative towards the outside. This might be a hint of flattening of these parameters along the bar major axis. This is confirmed by analyzing unbarred galaxies from the SAURON and ATLAS3D surveys that also show steeper outer gradients comparable to the minor axis gradients found in the barred galaxies. Along the bar major axis the metallicitiy gradient is found to be 0.03$\pm$0.07~dex/kpc  whereas the mean value of the minor axis and the unbarred control sample is -0.20$\pm$0.04~dex/kpc. This trend was seen for 2 galaxies analysed in \citet{2011MNRAS.415..709S} and is predicted by models \citep[e.g.,][]{2013DiMatteo, 2013MVG}. We note however, that the exact values of the mean gradients are still based on a small sample and could be slightly altered due to the contrast effect of the overlying structures of bulge, disc and bar. 

\item We further observe a good correlation between inner Fe5015 and metallicity gradients versus bar strength, such that stronger bars show positive gradients, while weak bars show negative gradients. This might be an influence of the bar providing fuel for nuclear star formation. This result is supported by the increasing central H$_\beta$ values and decreasing younger central ages with bar strength. Hence, the bar would alter the composition of the bulge. We note that all results on individual galaxies are shown in the appendix.

\end{itemize}

In conclusion, we reiterate our statement from Paper I which is that bars do not seem to alter global galaxy properties but do influence their host galaxies in a consistent way, \textit{localized}, in different regions. In particular the region around 0.13$\pm$0.06 bar lengths seems to be a common point where this influence can be detected, but sufficient spatial resolution is needed. Furthermore, in agreement with former pilot studies \citep[e.g.,][]{2011MNRAS.415..709S}, we find a flattening of the metallicity gradient - only along the major axis of the bar - and when taking the gradient separately from the inner slope which is clearly different. The spatial resolution coupled with the method to obtain gradients and the area where they are taken is very likely the reason for the lack of difference found in recent large statistical studies \citep{2014SB2, 2015bCheung}.  In upcoming work we will investigate in detail the resonance points in order to identify reasons for the behaviour found in kinematics and stellar populations. Furthermore, observations by even better (higher spatial and spectral resolution, larger FoV) IFUs such as MUSE will enable us to improve our understanding of substructures in galaxies, their origin and relation to large-scale structures such as bars and bulges.

%###############################################################################
\section*{Acknowledgments}

We would like to thank Lorenzo Morelli and Francesco La Barbera for stimulating 
comments and discussion. MKS and JFB wish to express their gratitude
to the Roque de los Muchachos Observatory on La Palma and the different
operators of the William Herschel Telescope during our numerous runs, as well as
observing support by Agnieszka Rys, Carolin Wittman and Thorsten Lisker. We also
acknowledge support from grant AYA2013-48226-C3-1-P from the Spanish Ministry of
Economy and Competitiveness (MINECO). IMV acknowledges support from grant
AYA2009-11137. MKS acknowledges the support of the Instituto de Astrof\'isica
de Canarias via an Astrophysicist Resident fellowship. We also acknowledge
support from the FP7 Marie Curie Actions of the European Commission, via the
Initial Training Network DAGAL under REA grant agreement number 289313. 
This research made use of Montage, funded by the National Aeronautics and Space Administration's Earth Science Technology Office, Computational Technnologies Project, under Cooperative Agreement Number NCC5-626 between NASA and the California Institute of Technology. The code is maintained by the NASA/IPAC Infrared Science Archive. This research has made use of the NASA/IPAC Extragalactic Database (NED) which is operated by the Jet Propulsion Laboratory, California Institute of Technology, under contract with the National Aeronautics and Space Administration. The paper is based on observations obtained at the William Herschel Telescope, operated by the Isaac Newton Group in the Spanish Observatorio del Roque de los Muchachos of the Instituto de Astrof\'isica de Canarias. 
%\mks{put everywhere:  [See the electronic edition of the Journal for a color
%version of this figure.]}

%\label{lastpage}

\bibliographystyle{mn2e_fixed.bst}
\bibliography{refs_jabref_v5}

\newpage

\appendix{}

\section{Complete set of maps of line strength indices, SSP-grids and SSP parameters for the BaLROG sample}
\label{sec:allmaps}

\begin{figure*}
\includegraphics[width=0.79\linewidth, angle=90]{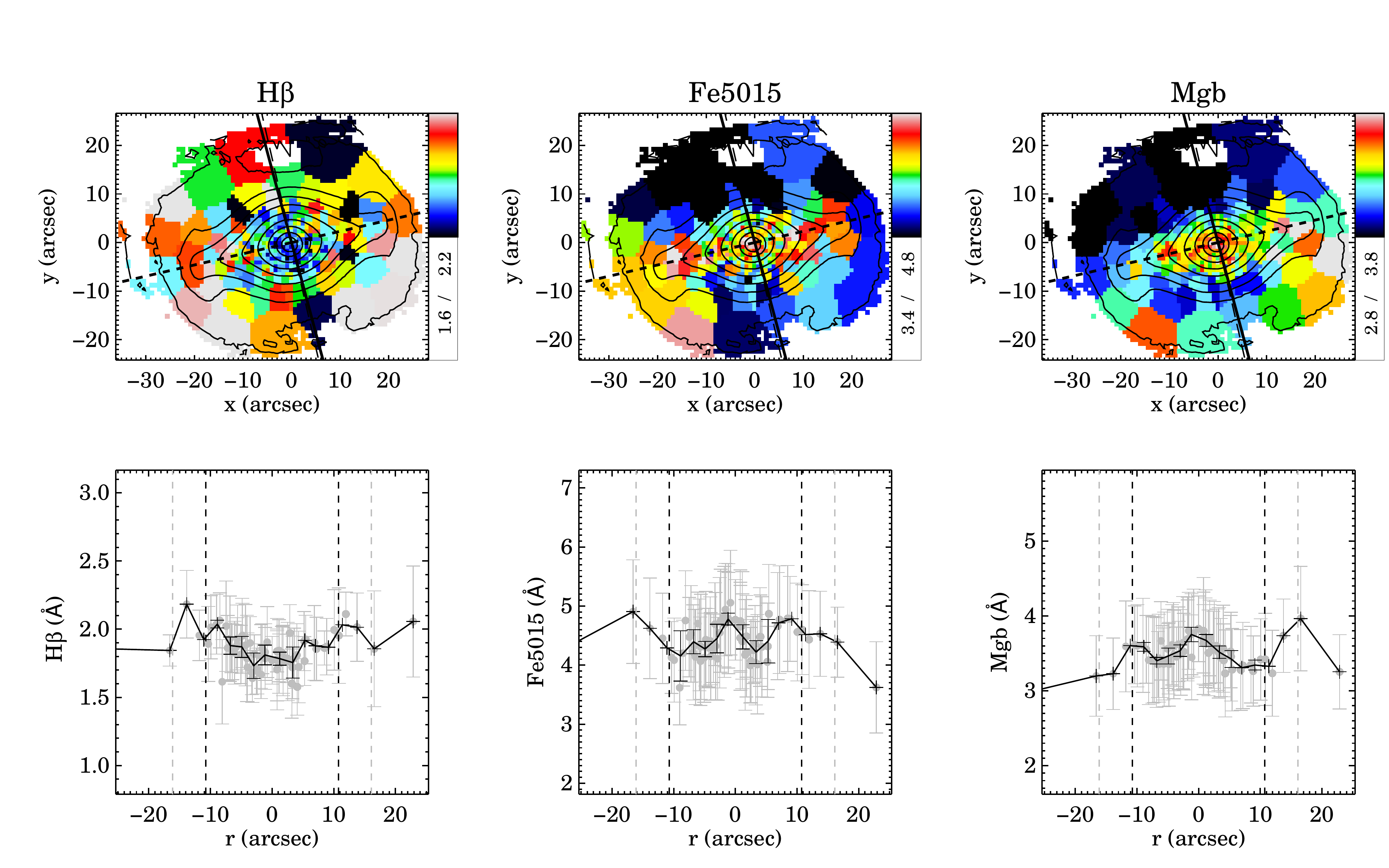}
\includegraphics[width=0.79\linewidth, angle=90]{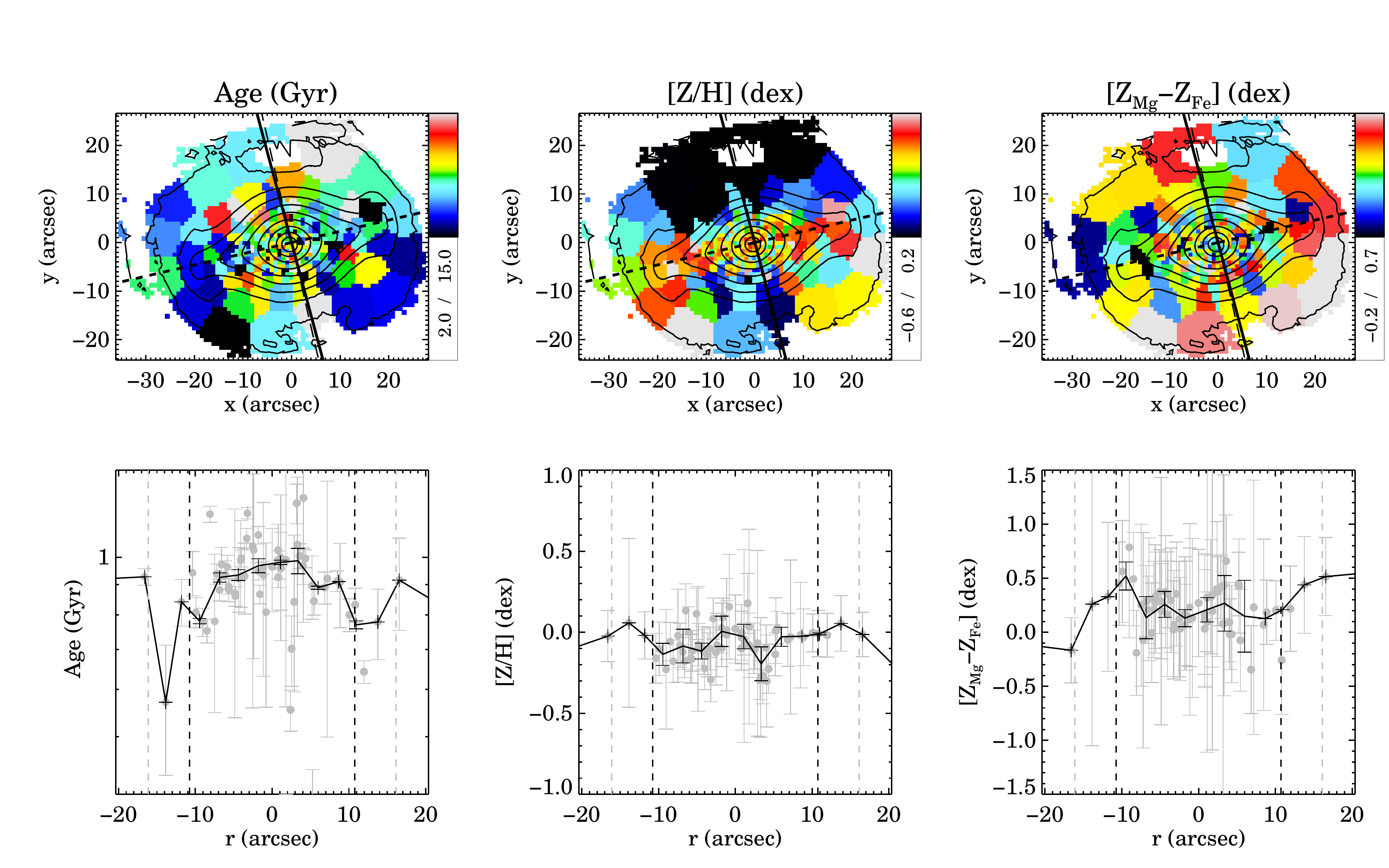}
\includegraphics[width=0.33\linewidth, angle=90,bb = -100 -150 600 600]{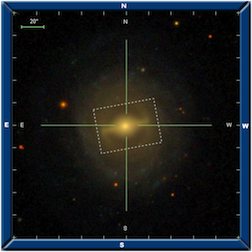}
\includegraphics[width=0.33\linewidth, angle=90,bb = -400 -40 50 -300]{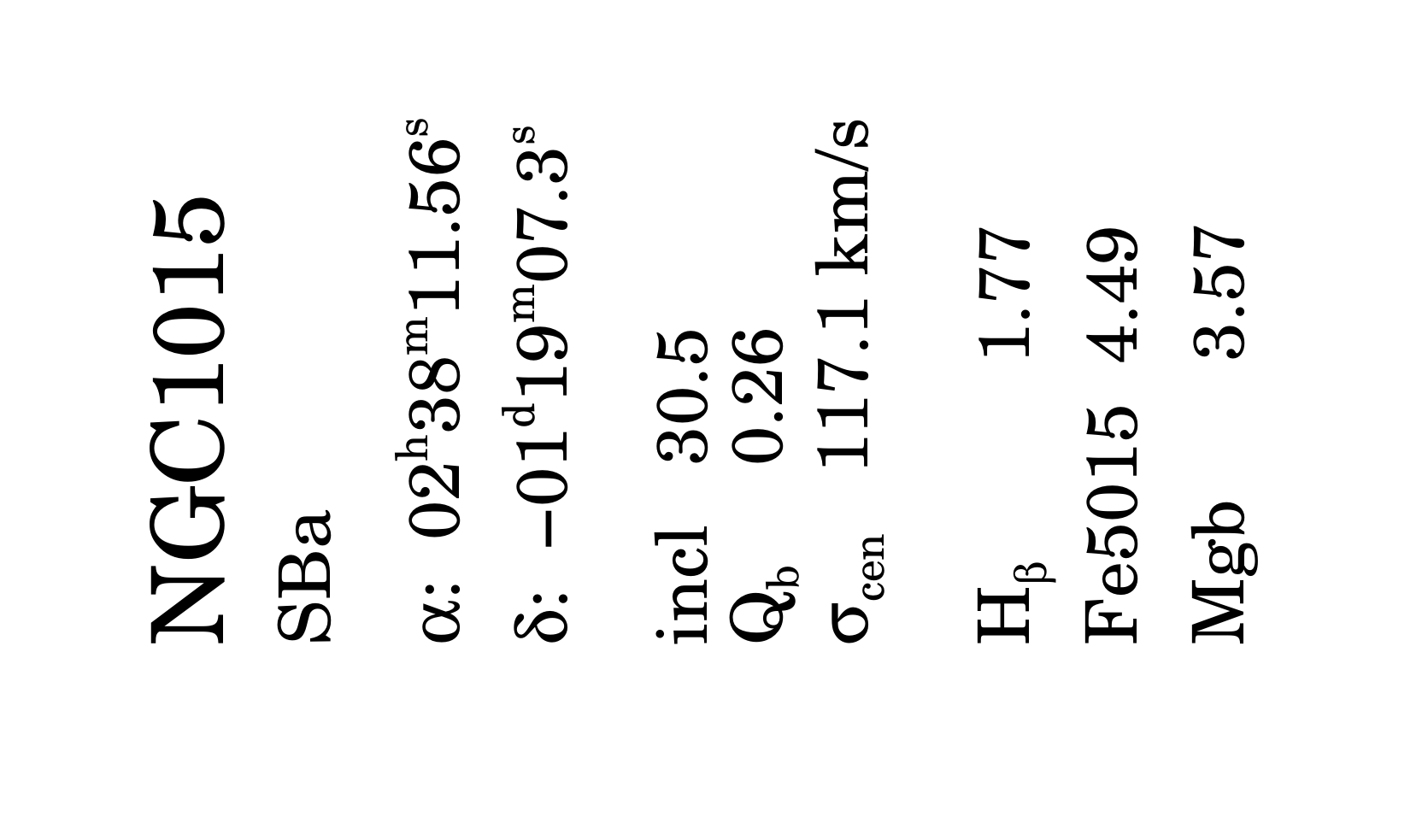}
\includegraphics[width=0.5\linewidth, angle=90,bb = 50 150 800 900]{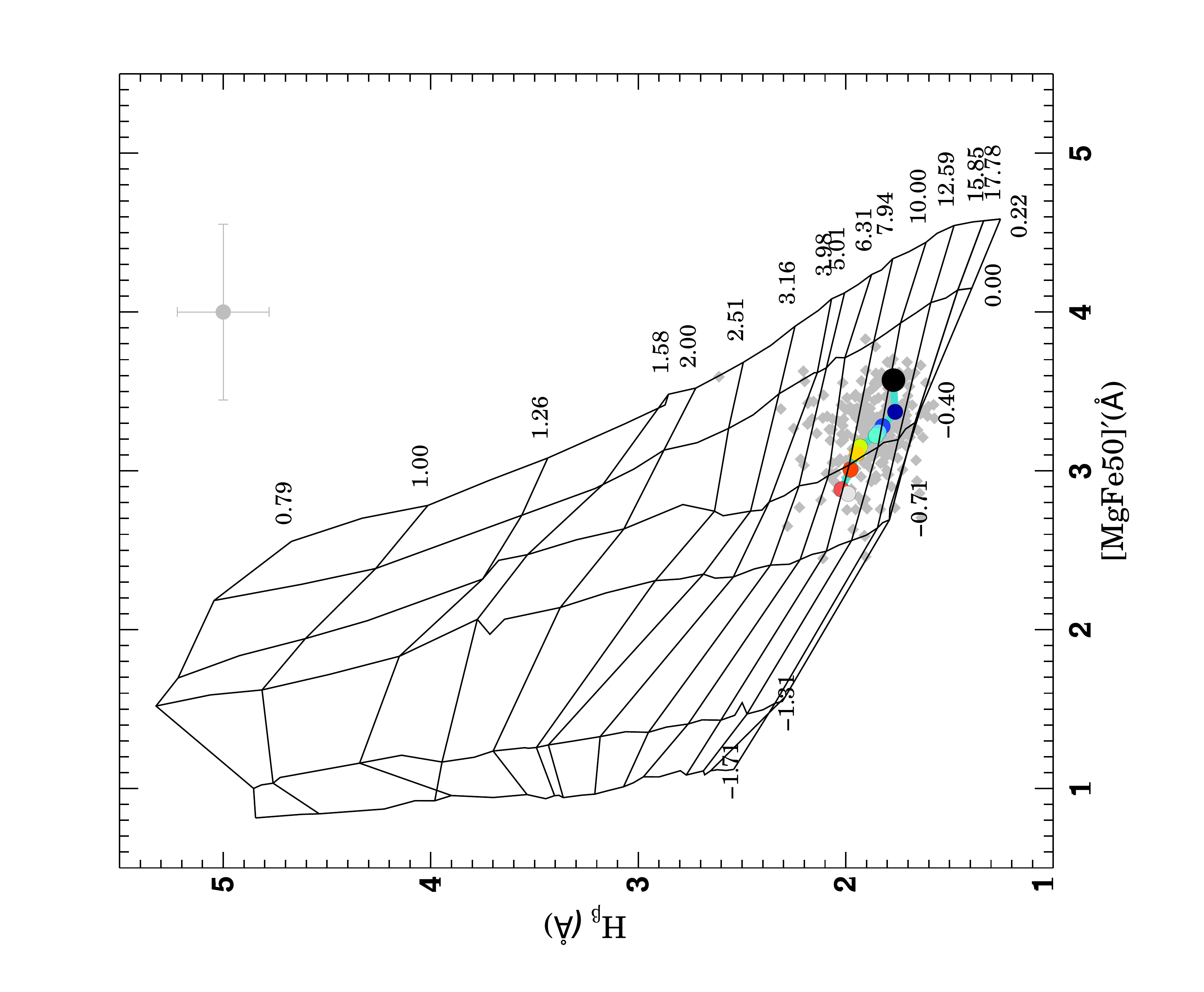}
\caption{Summary of the stellar population maps and important parameters for each galaxy. On the top left we show a colour SDSS image with an indication of the IFU mosaic in white dashed lines, below an index-index diagram and to the right index and SSP profiles. For more details please refer to the text.}
\label{fig:summ1015}
\end{figure*}

\begin{figure*}
\includegraphics[width=0.79\linewidth, angle=90]{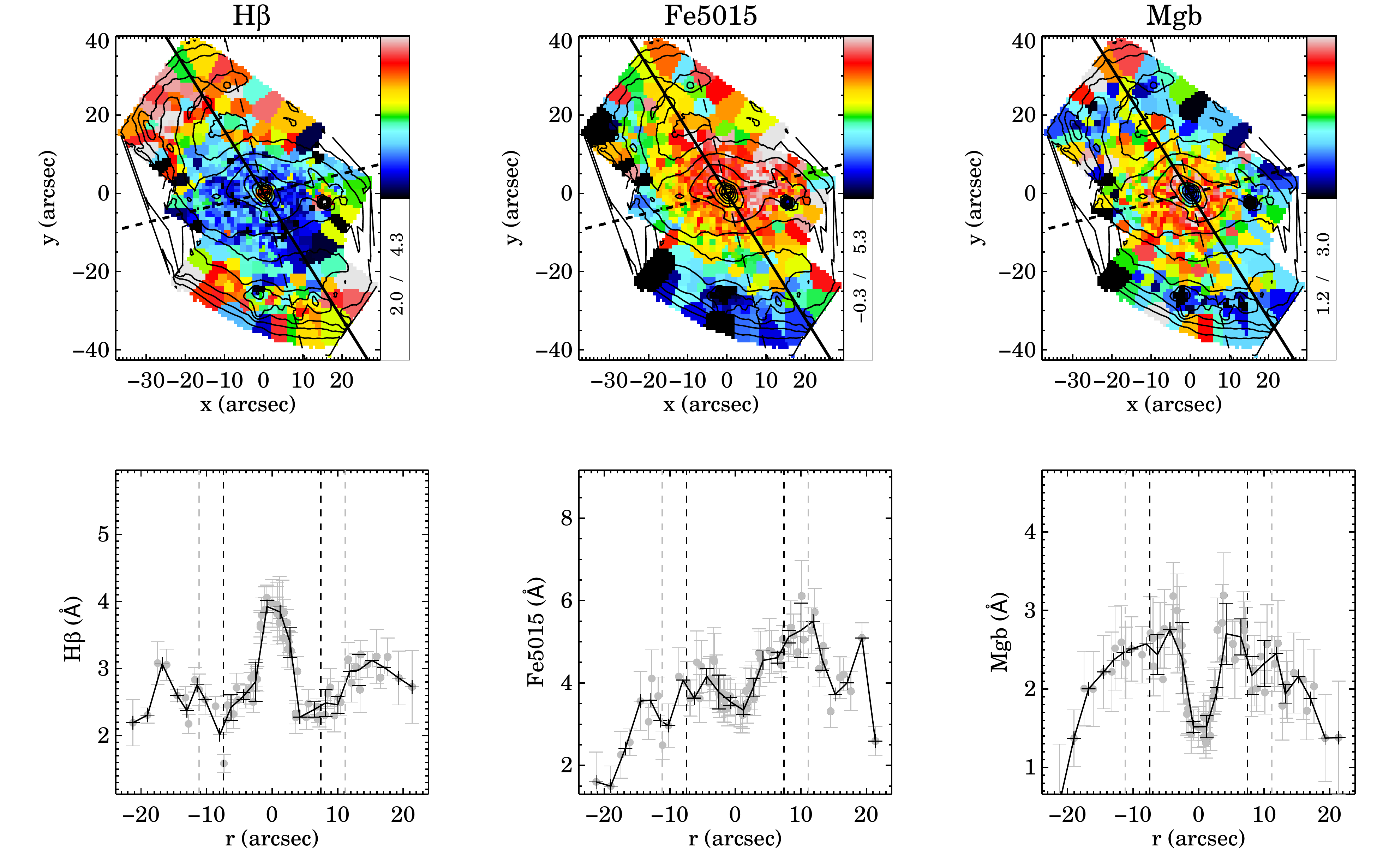}
\includegraphics[width=0.79\linewidth, angle=90]{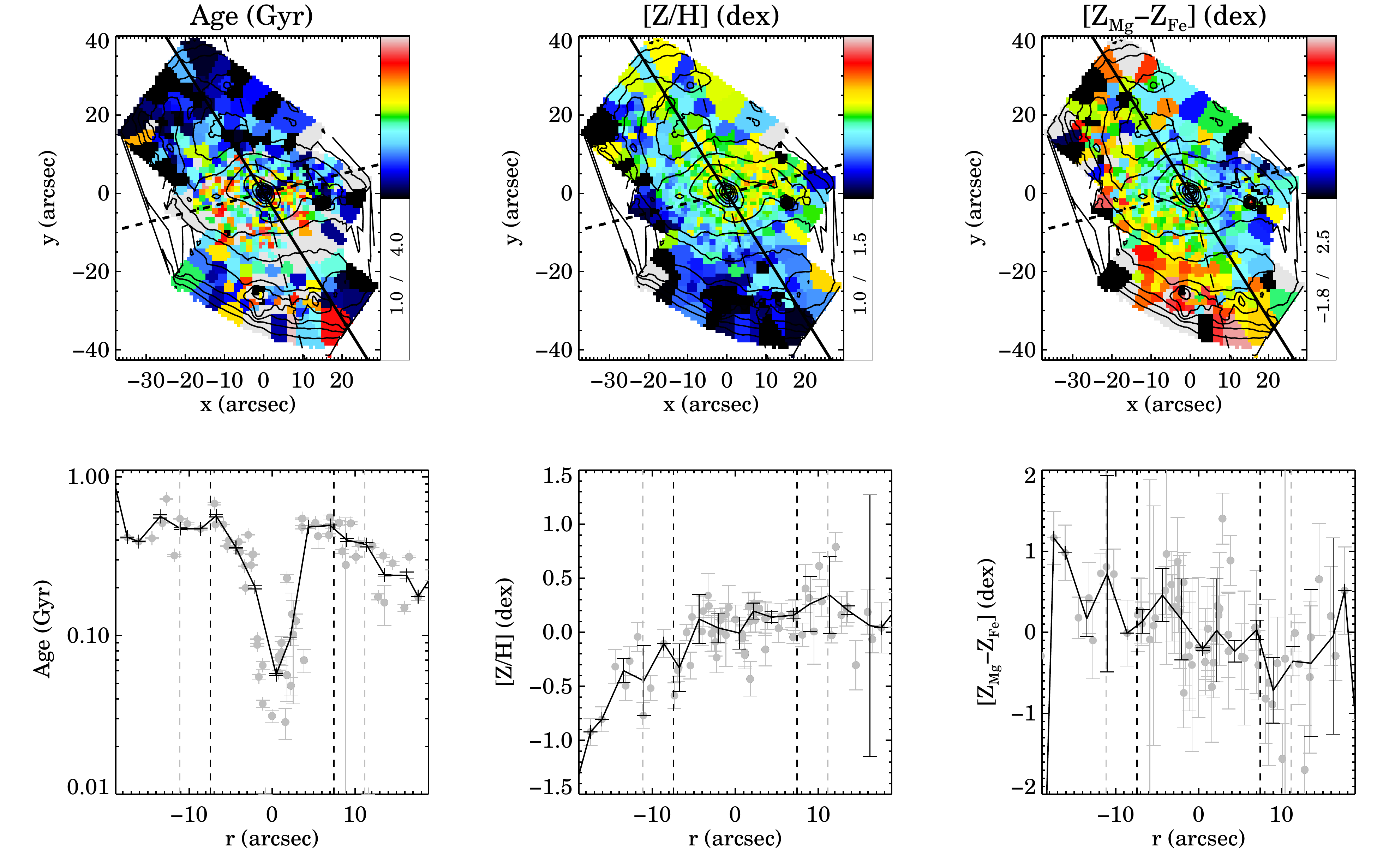}
\includegraphics[width=0.33\linewidth, angle=90,bb = -100 -150 600 600]{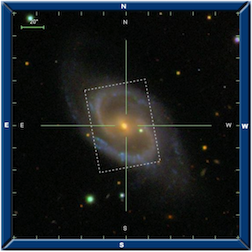}
\includegraphics[width=0.33\linewidth, angle=90,bb = -400 -40 50 -300]{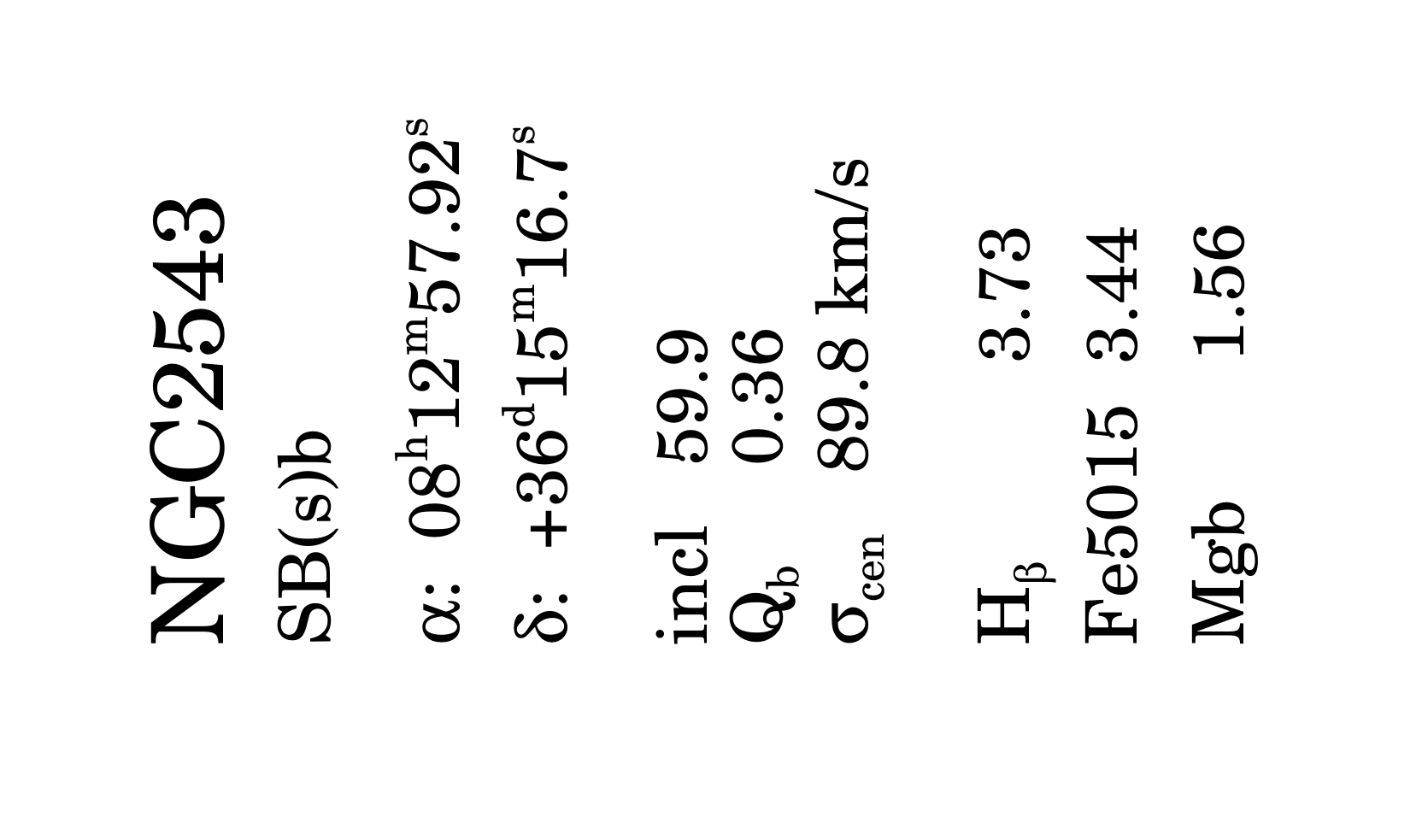}
\includegraphics[width=0.5\linewidth, angle=90,bb = 50 150 800 900]{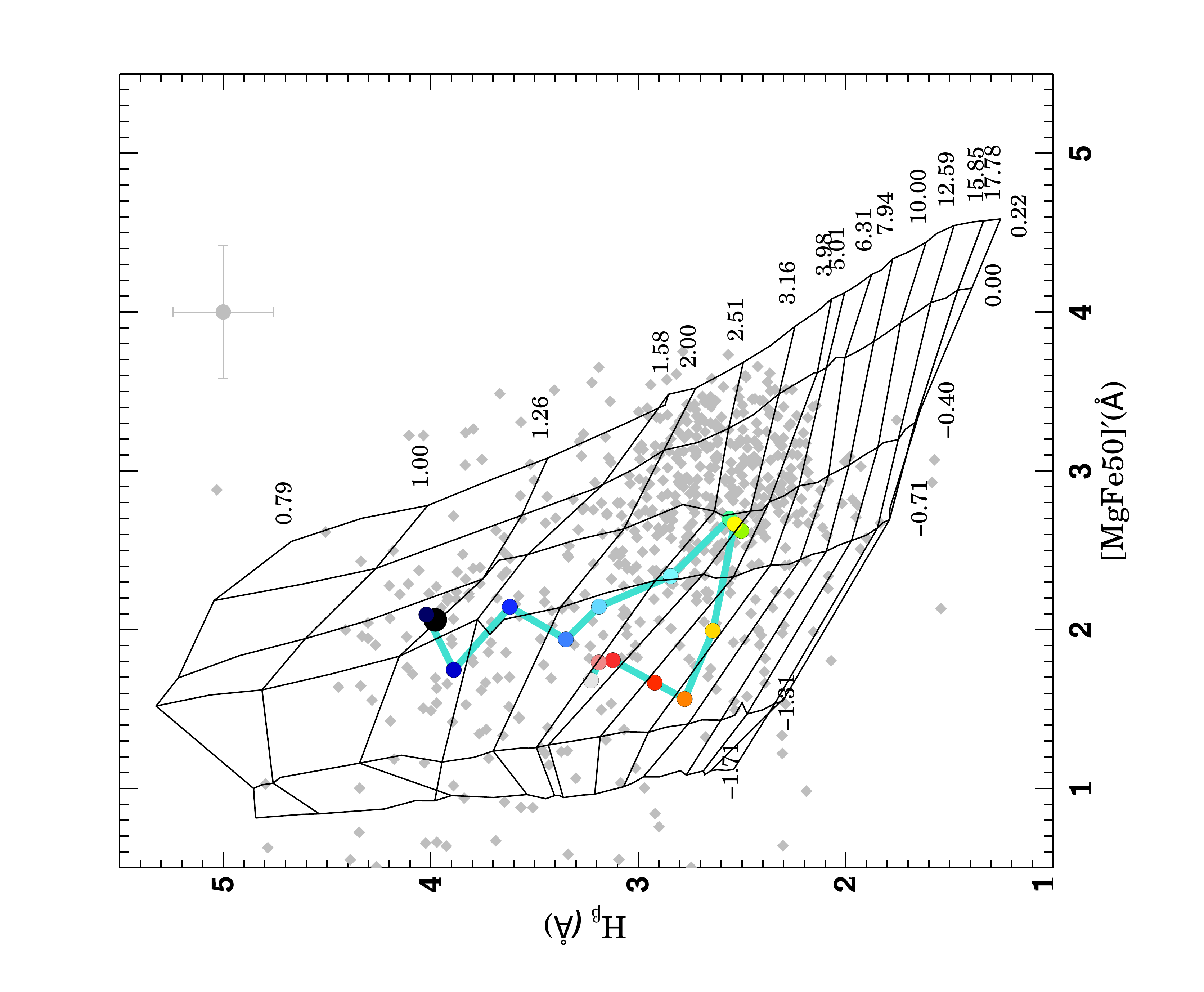}
\caption{Fig~\ref{fig:summ1015} continued.}
\label{fig:summ2543}
\end{figure*}

\begin{figure*}
\includegraphics[width=0.79\linewidth, angle=90]{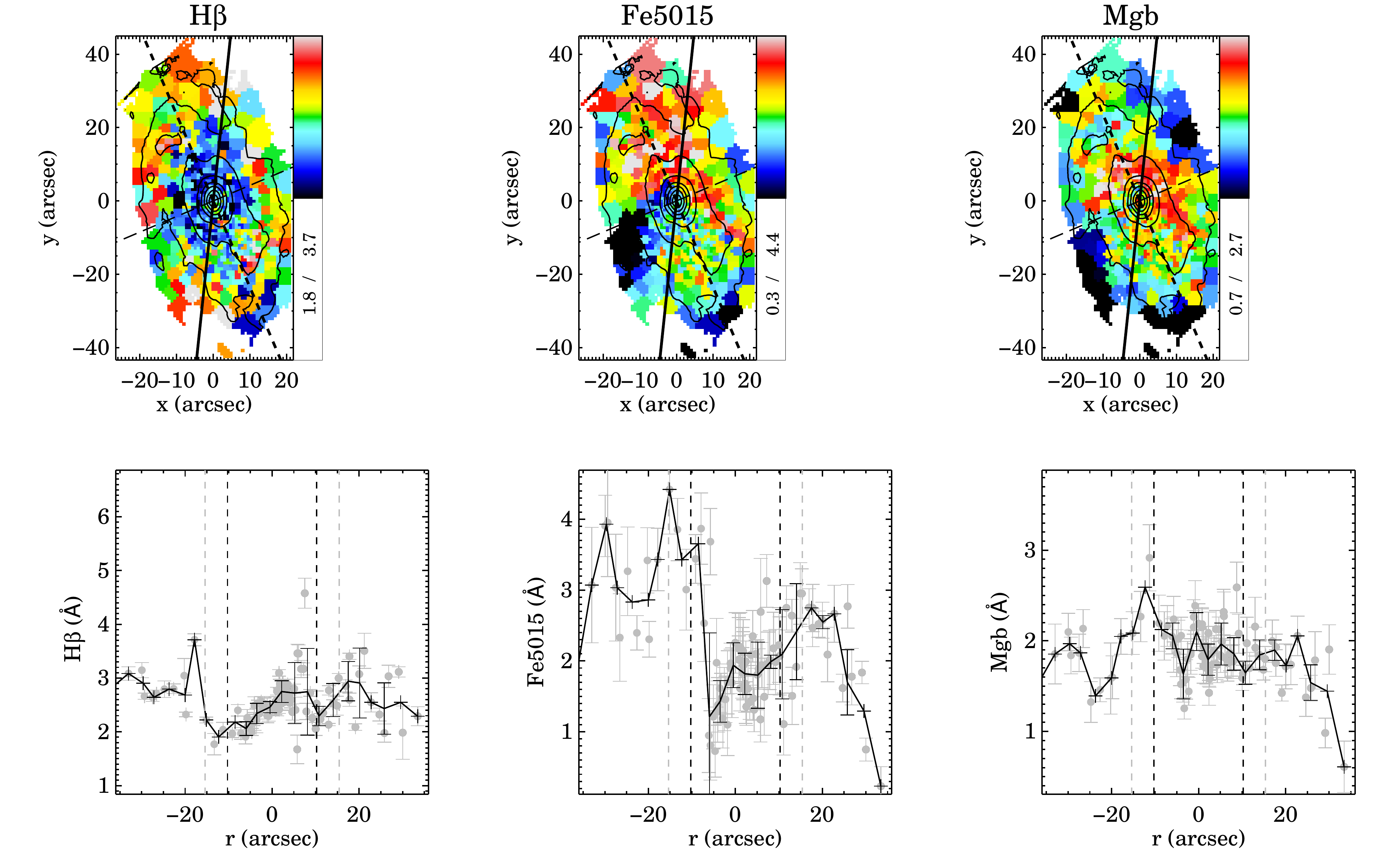}
\includegraphics[width=0.79\linewidth, angle=90]{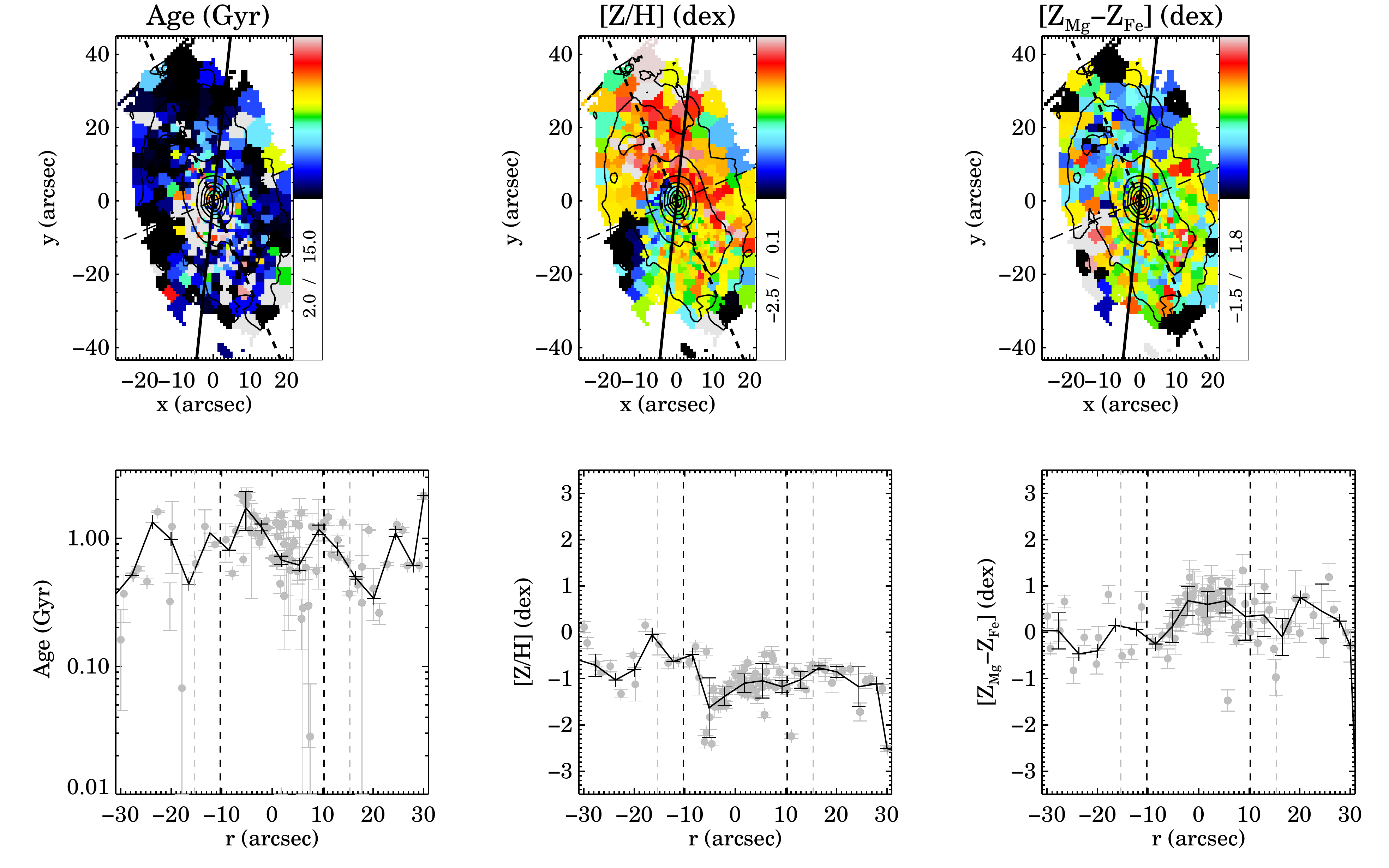}
\includegraphics[width=0.33\linewidth, angle=90,bb = -100 -150 600 600]{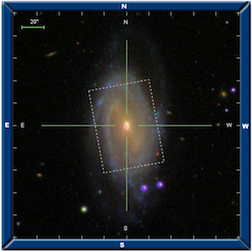}
\includegraphics[width=0.33\linewidth, angle=90,bb = -400 -40 50 -300]{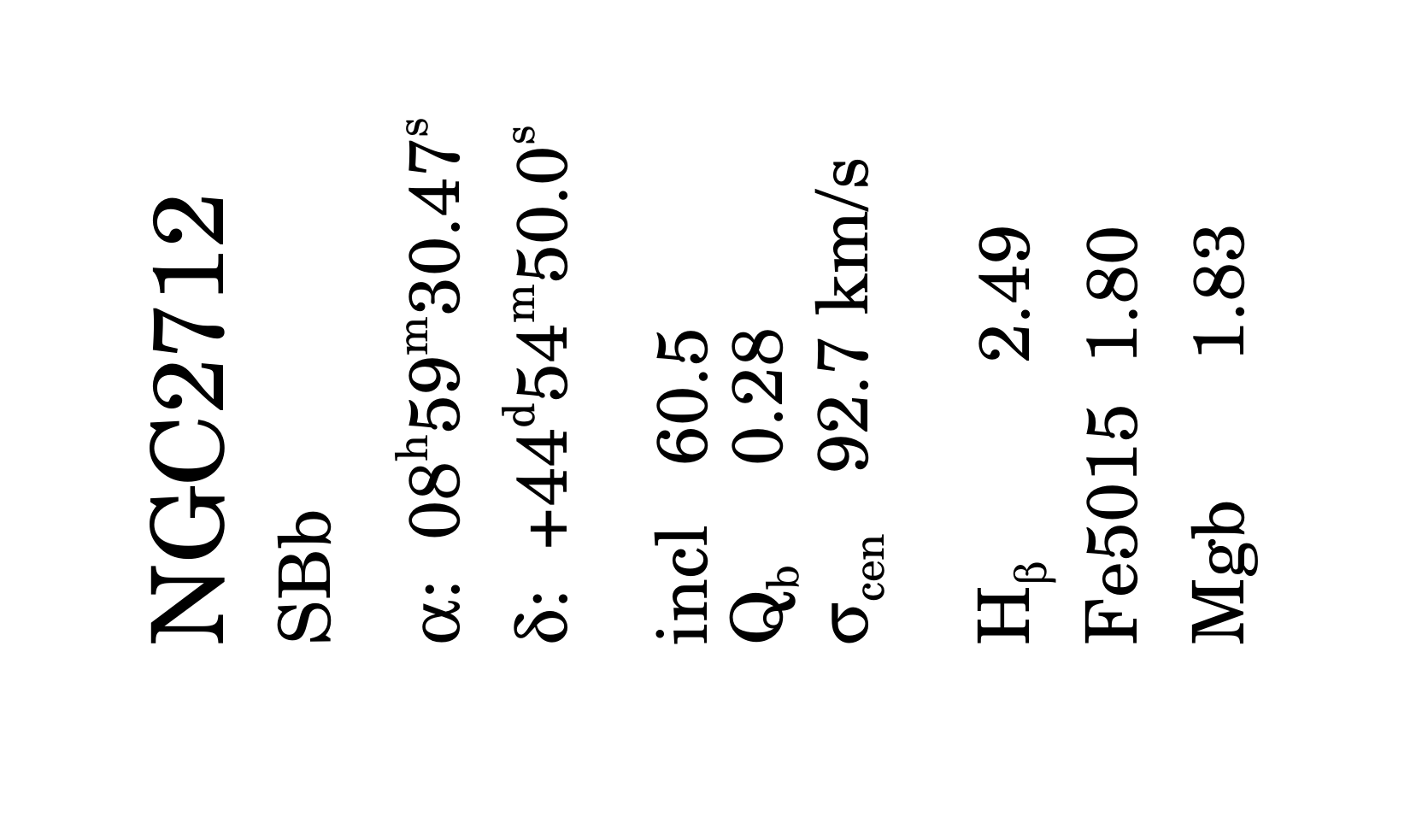}
\includegraphics[width=0.5\linewidth, angle=90,bb = 50 150 800 900]{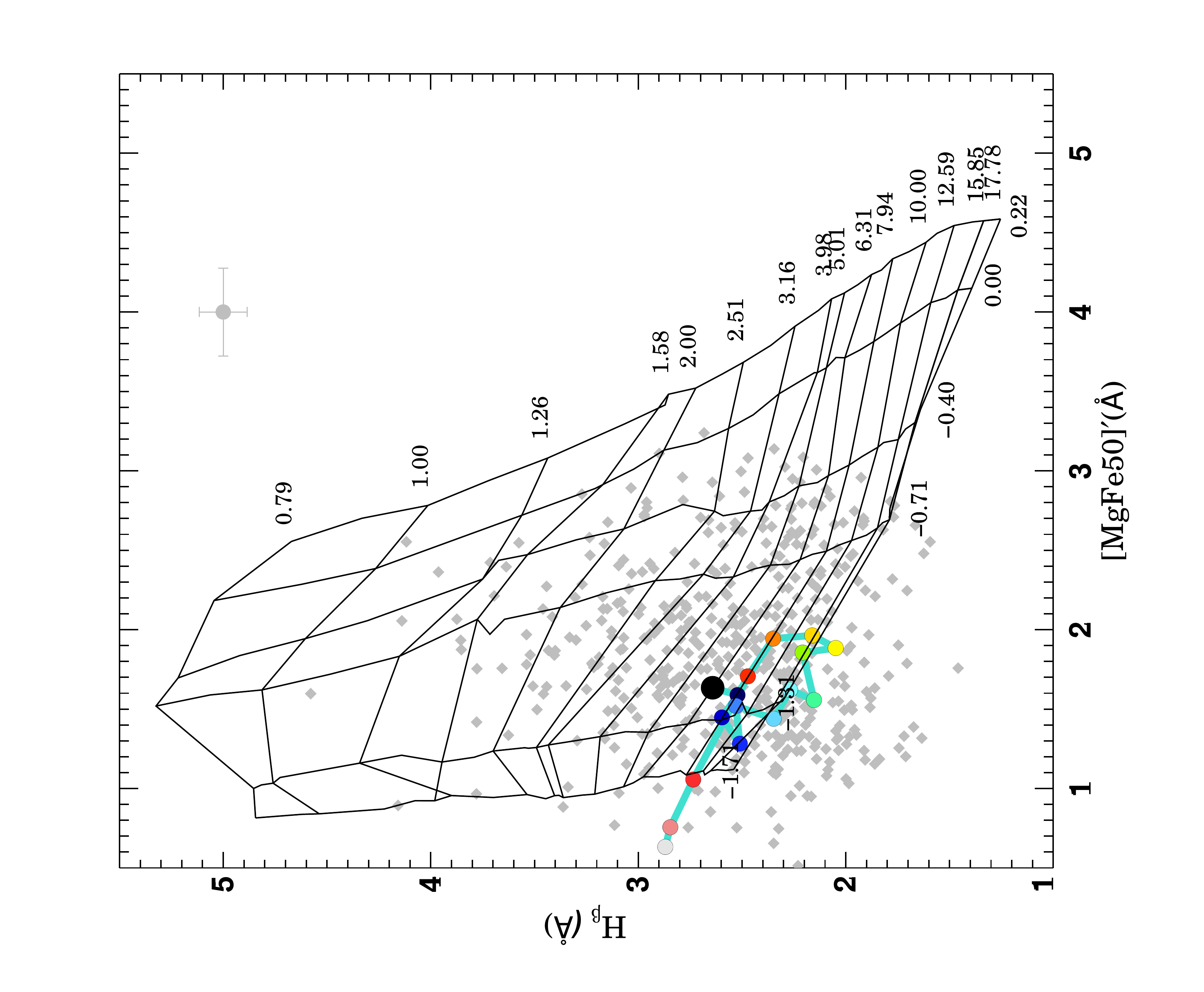}
\caption{Fig~\ref{fig:summ1015} continued.}
\label{fig:summ2712}
\end{figure*}

\begin{figure*}
\includegraphics[width=0.79\linewidth, angle=90]{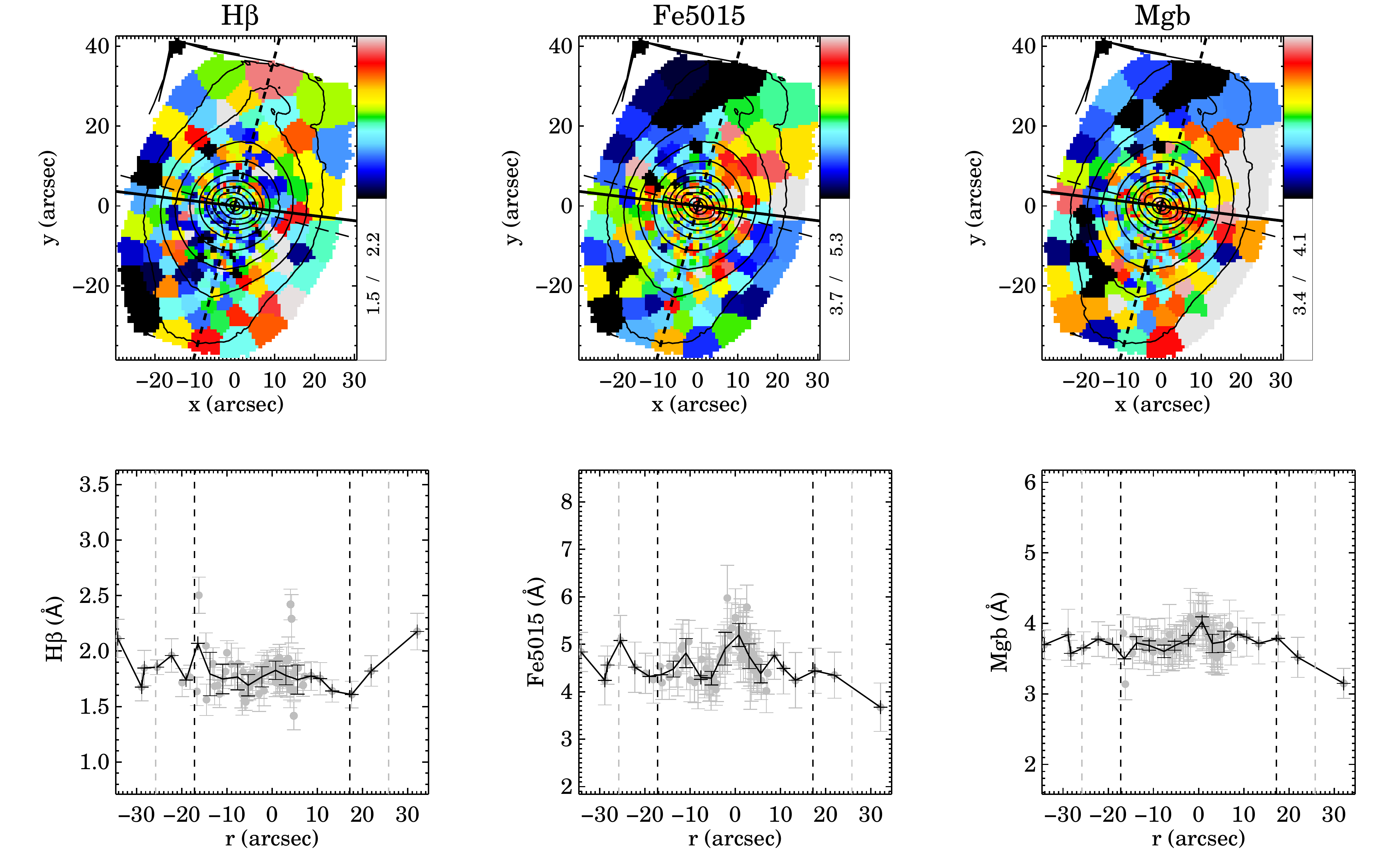}
\includegraphics[width=0.79\linewidth, angle=90]{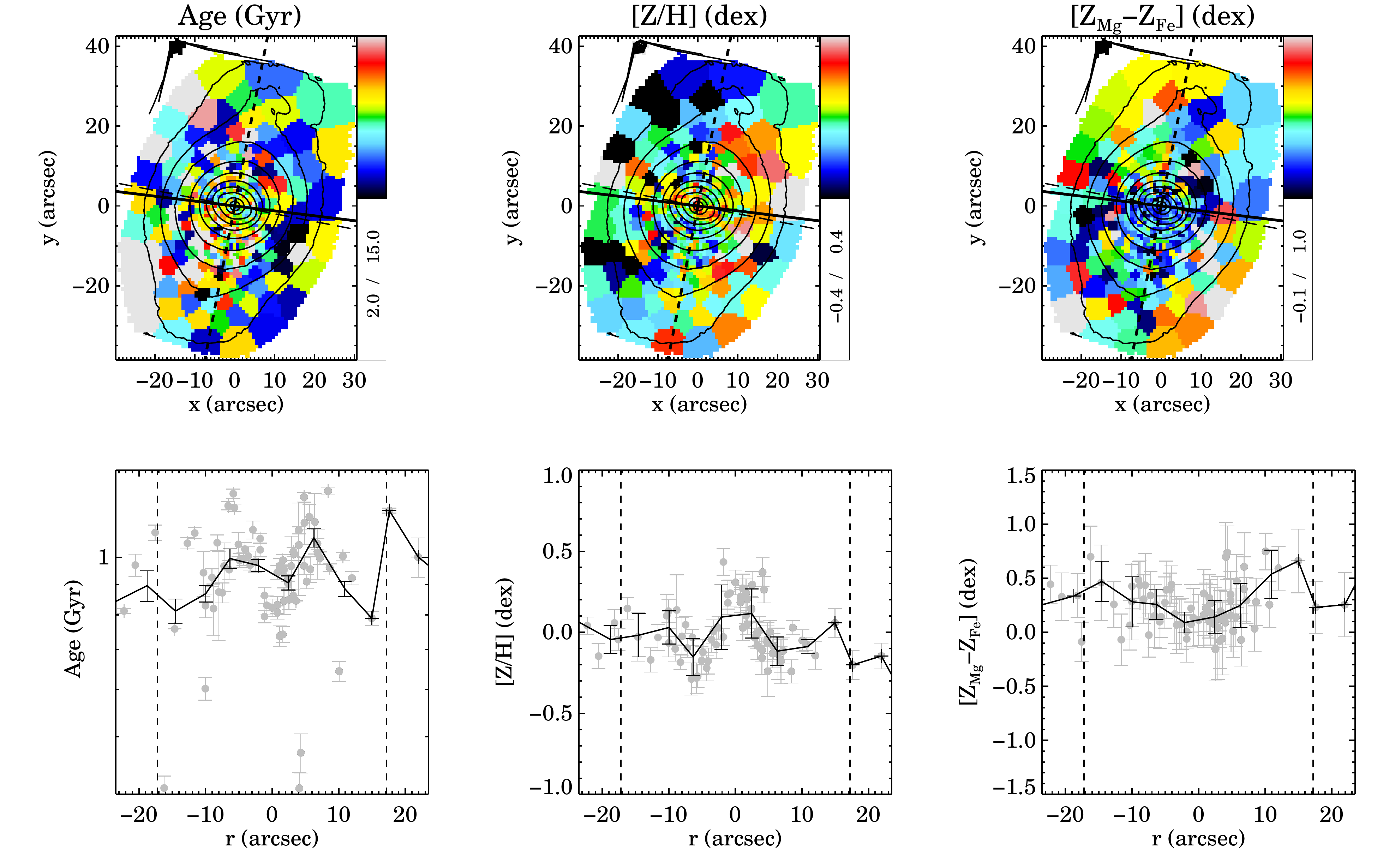}
\includegraphics[width=0.33\linewidth, angle=90,bb = -100 -150 600 600]{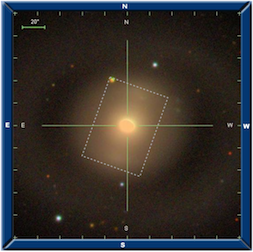}
\includegraphics[width=0.33\linewidth, angle=90,bb = -400 -40 50 -300]{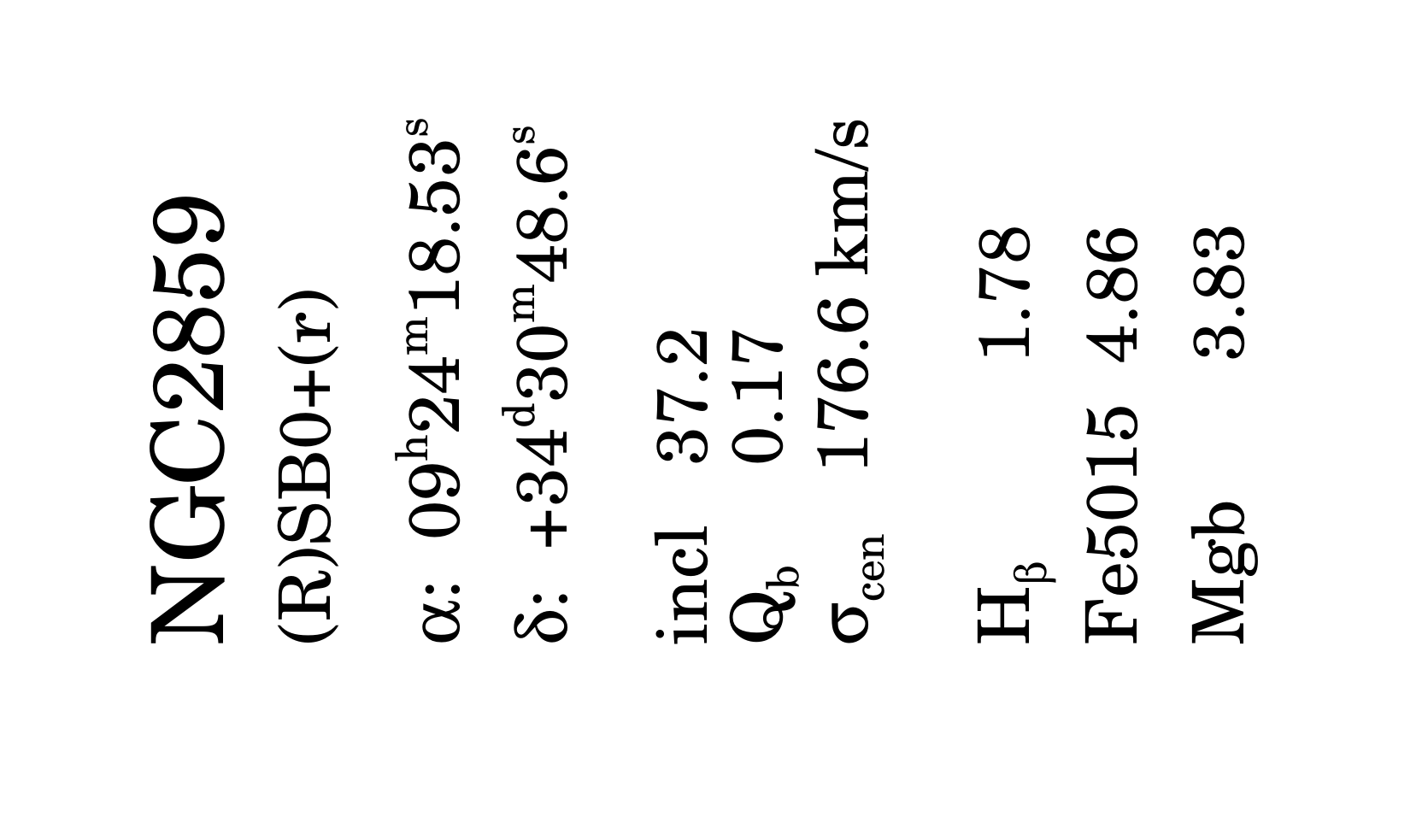}
\includegraphics[width=0.5\linewidth, angle=90,bb = 50 150 800 900]{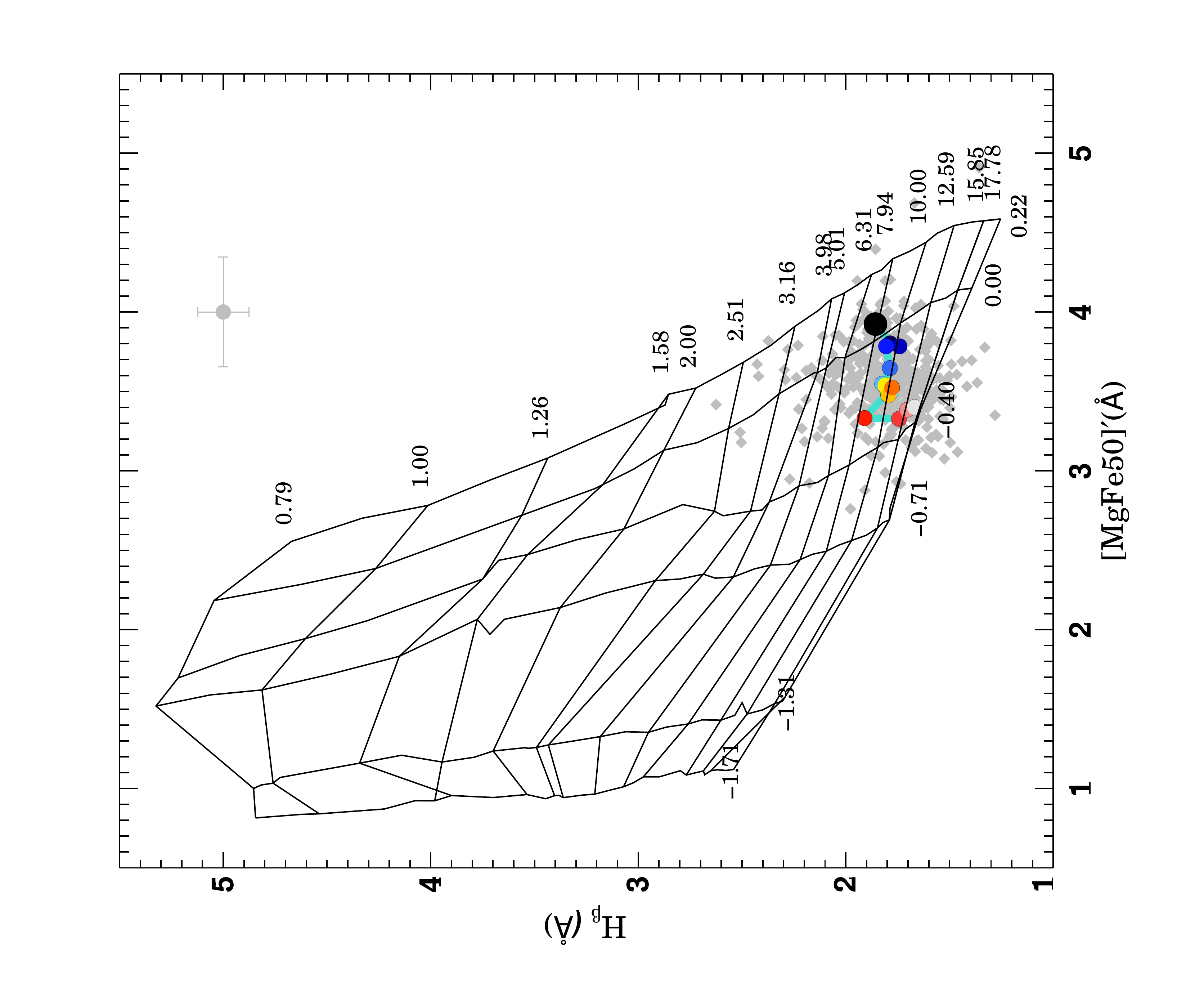}
\caption{Fig~\ref{fig:summ1015} continued.}
\label{fig:summ2859}
\end{figure*}

\begin{figure*}
\includegraphics[width=0.79\linewidth, angle=90]{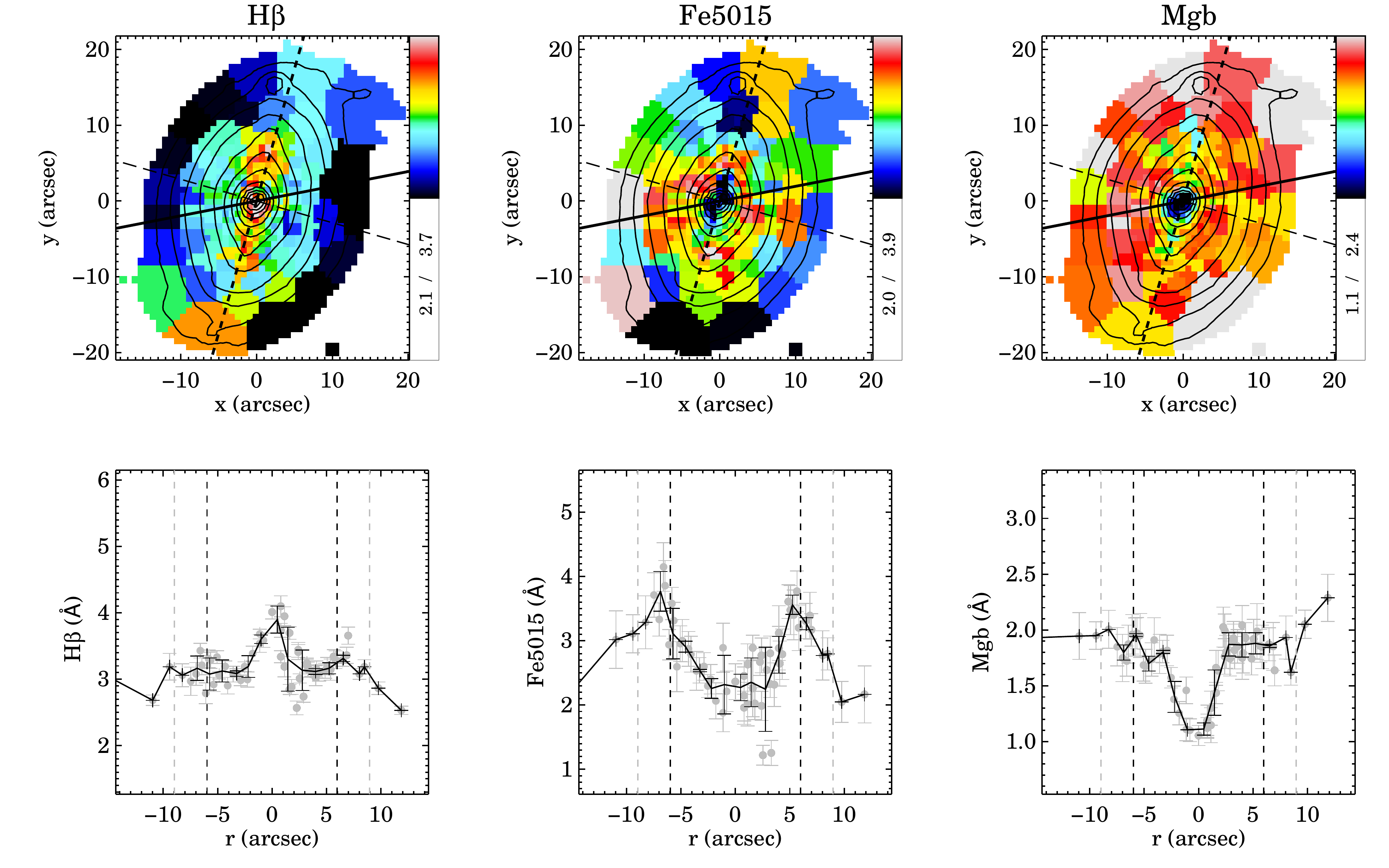}
\includegraphics[width=0.79\linewidth, angle=90]{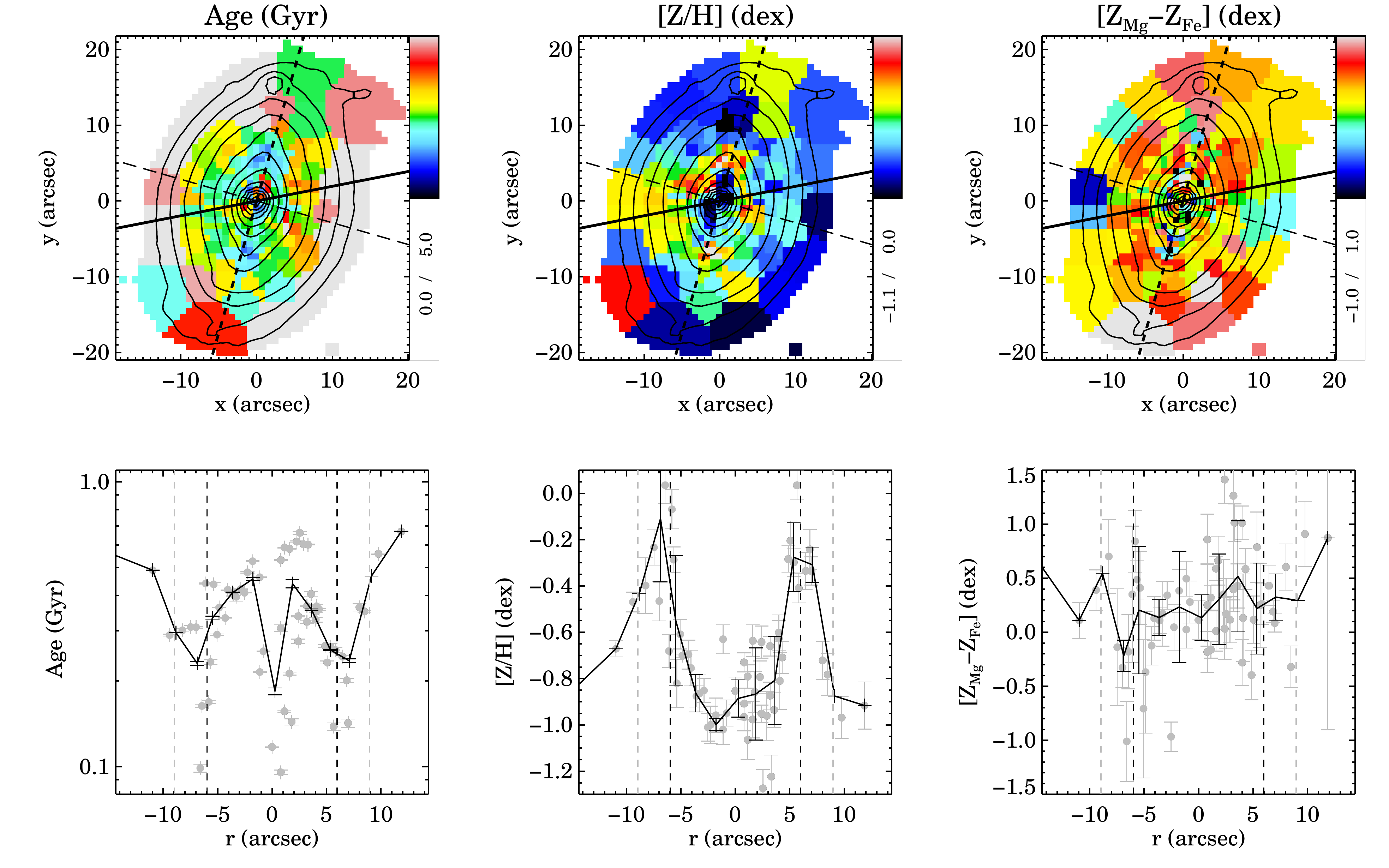}
\includegraphics[width=0.33\linewidth, angle=90,bb = -100 -150 600 600]{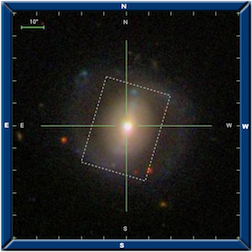}
\includegraphics[width=0.33\linewidth, angle=90,bb = -400 -40 50 -300]{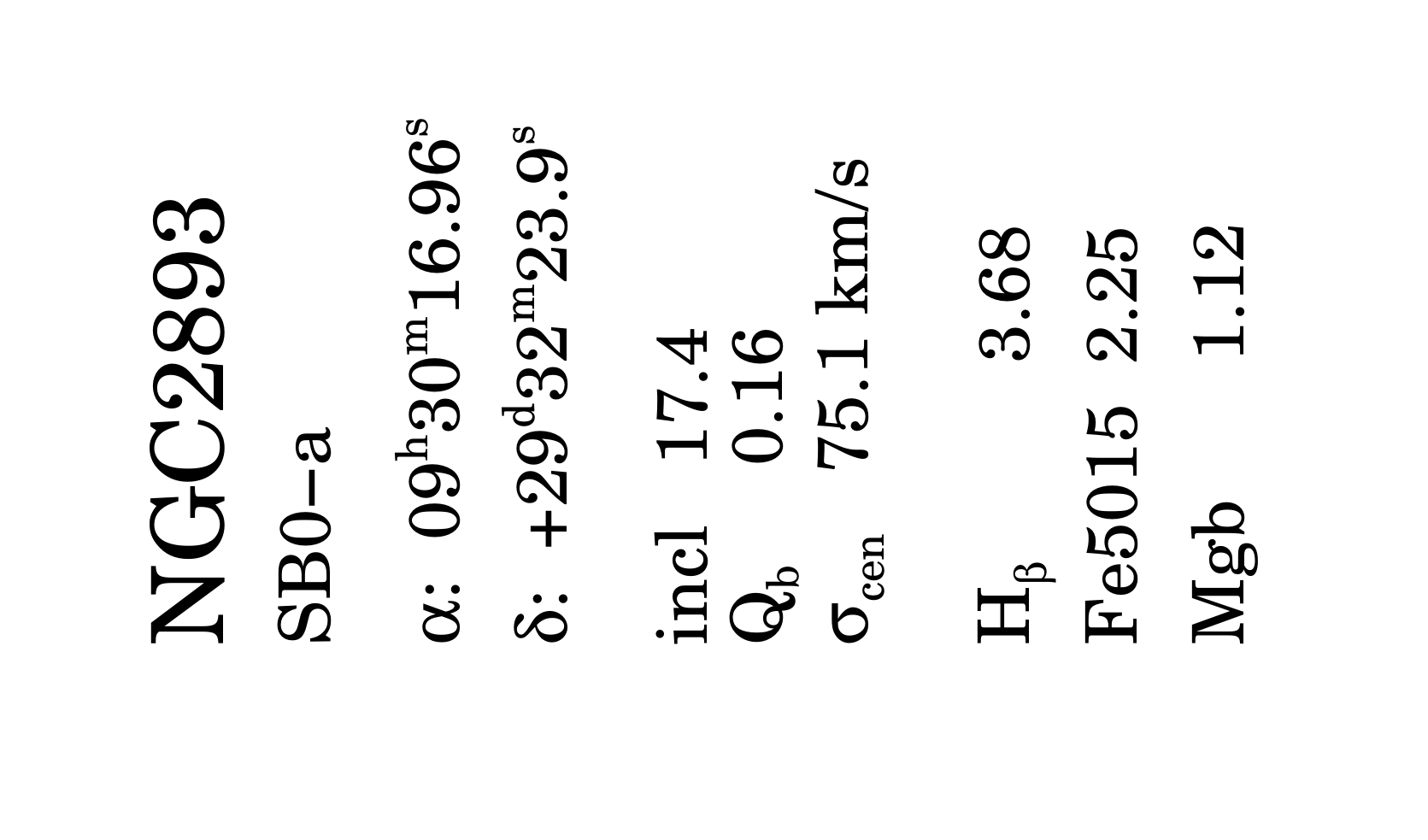}
\includegraphics[width=0.5\linewidth, angle=90,bb = 50 150 800 900]{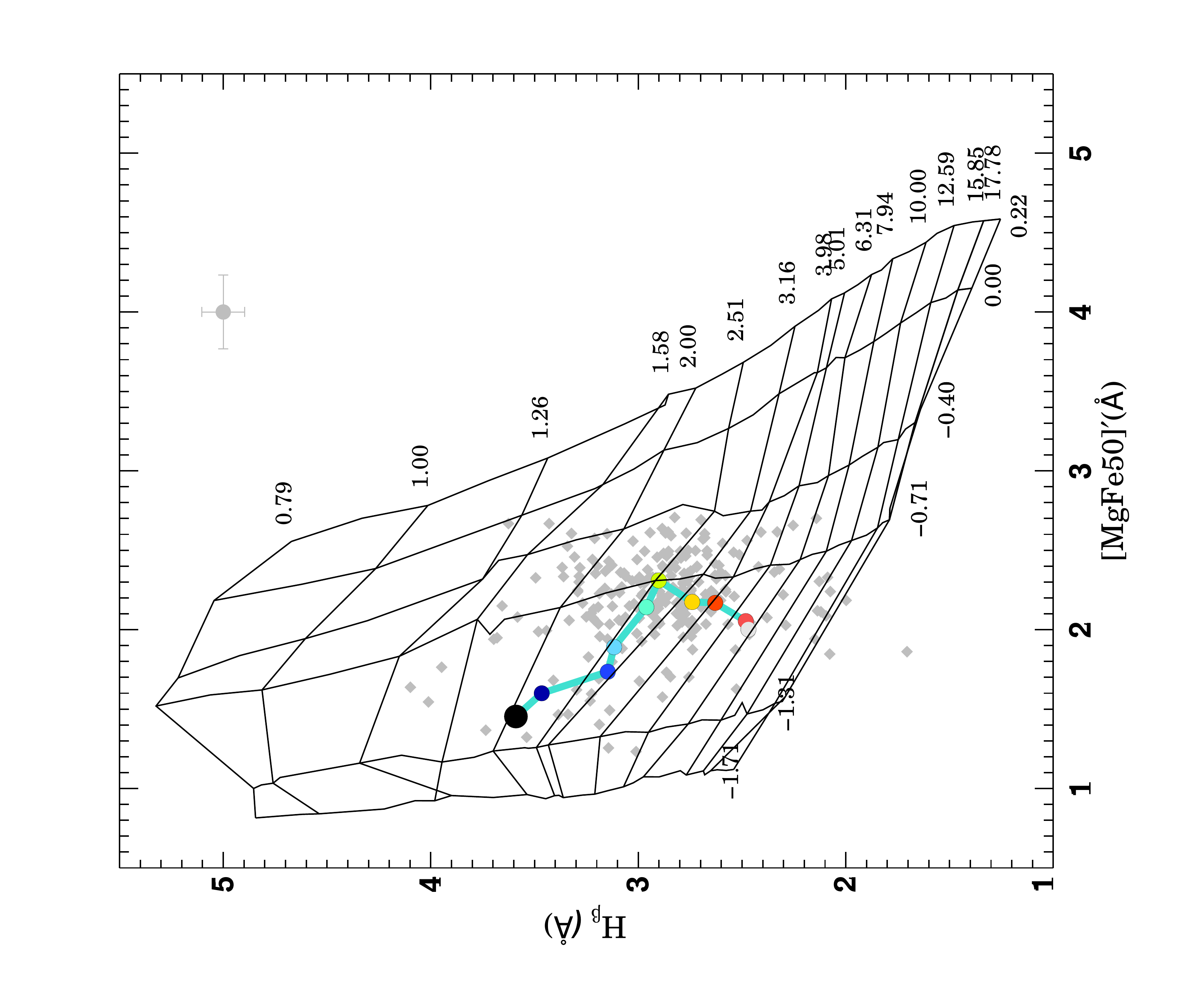}
\caption{Fig~\ref{fig:summ1015} continued.}
\label{fig:summ2893}
\end{figure*}

\begin{figure*}
\includegraphics[width=0.79\linewidth, angle=90]{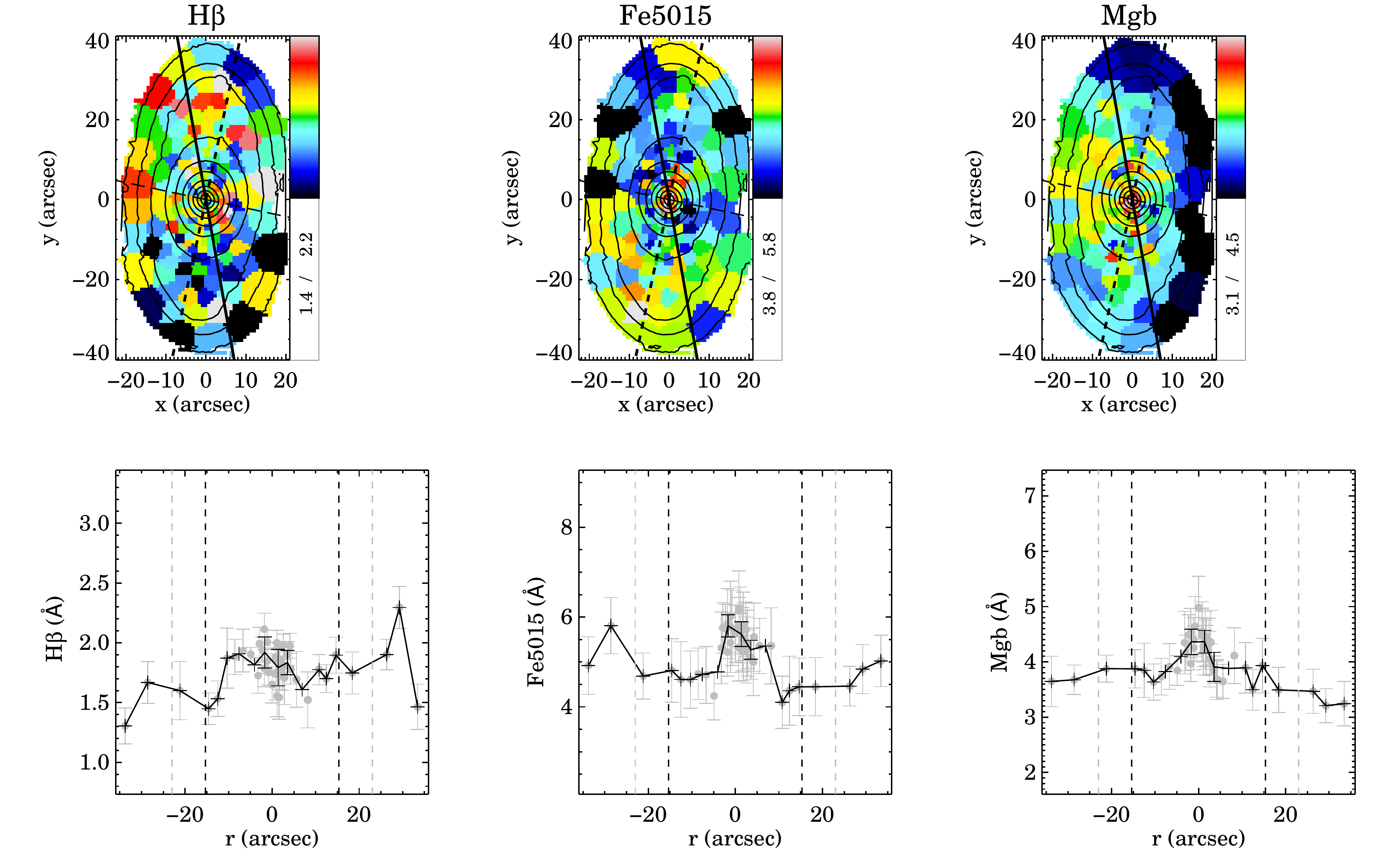}
\includegraphics[width=0.79\linewidth, angle=90]{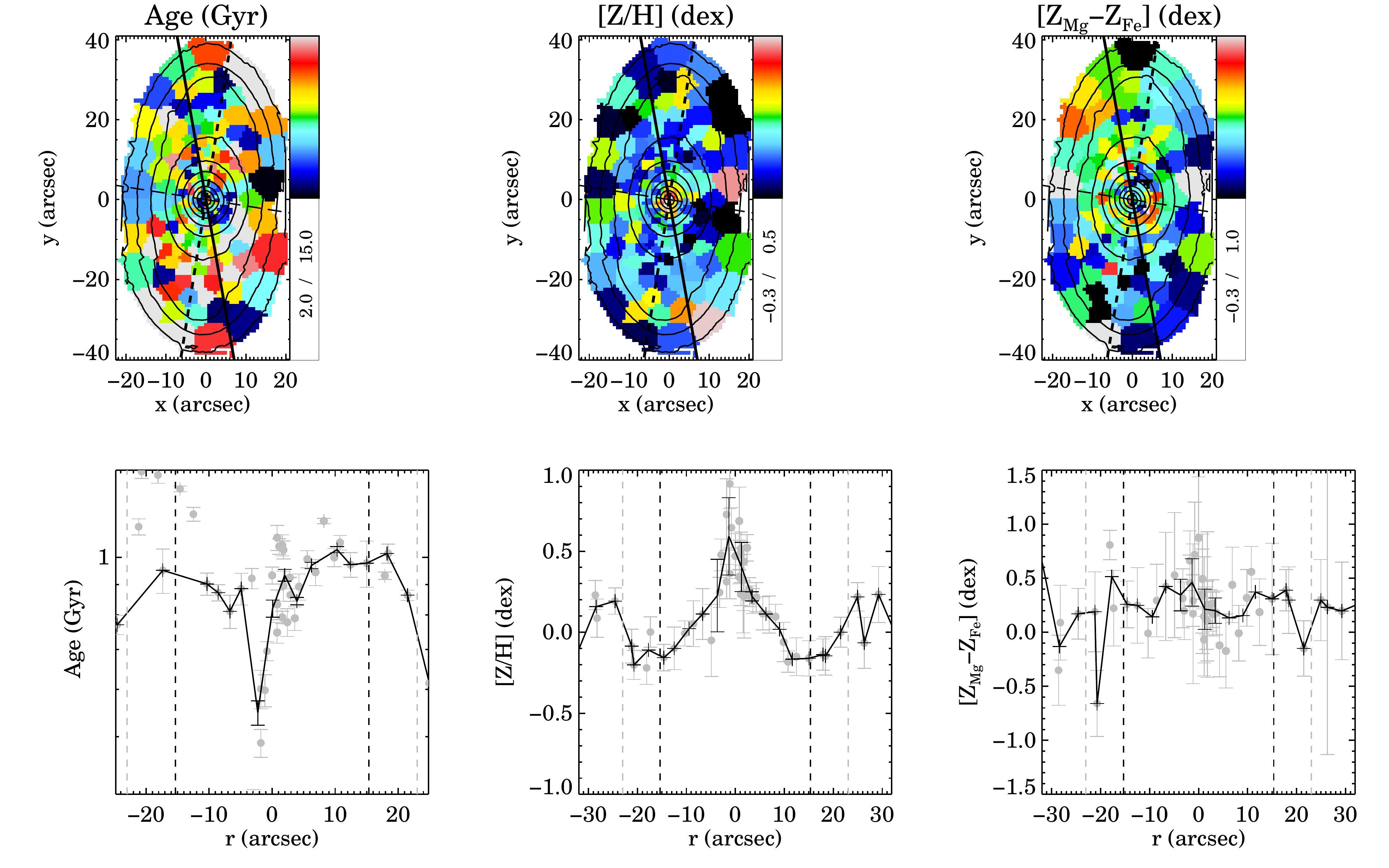}
\includegraphics[width=0.33\linewidth, angle=90,bb = -100 -150 600 600]{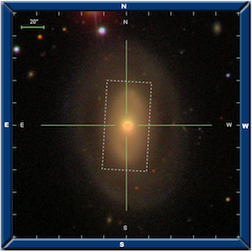}
\includegraphics[width=0.33\linewidth, angle=90,bb = -400 -40 50 -300]{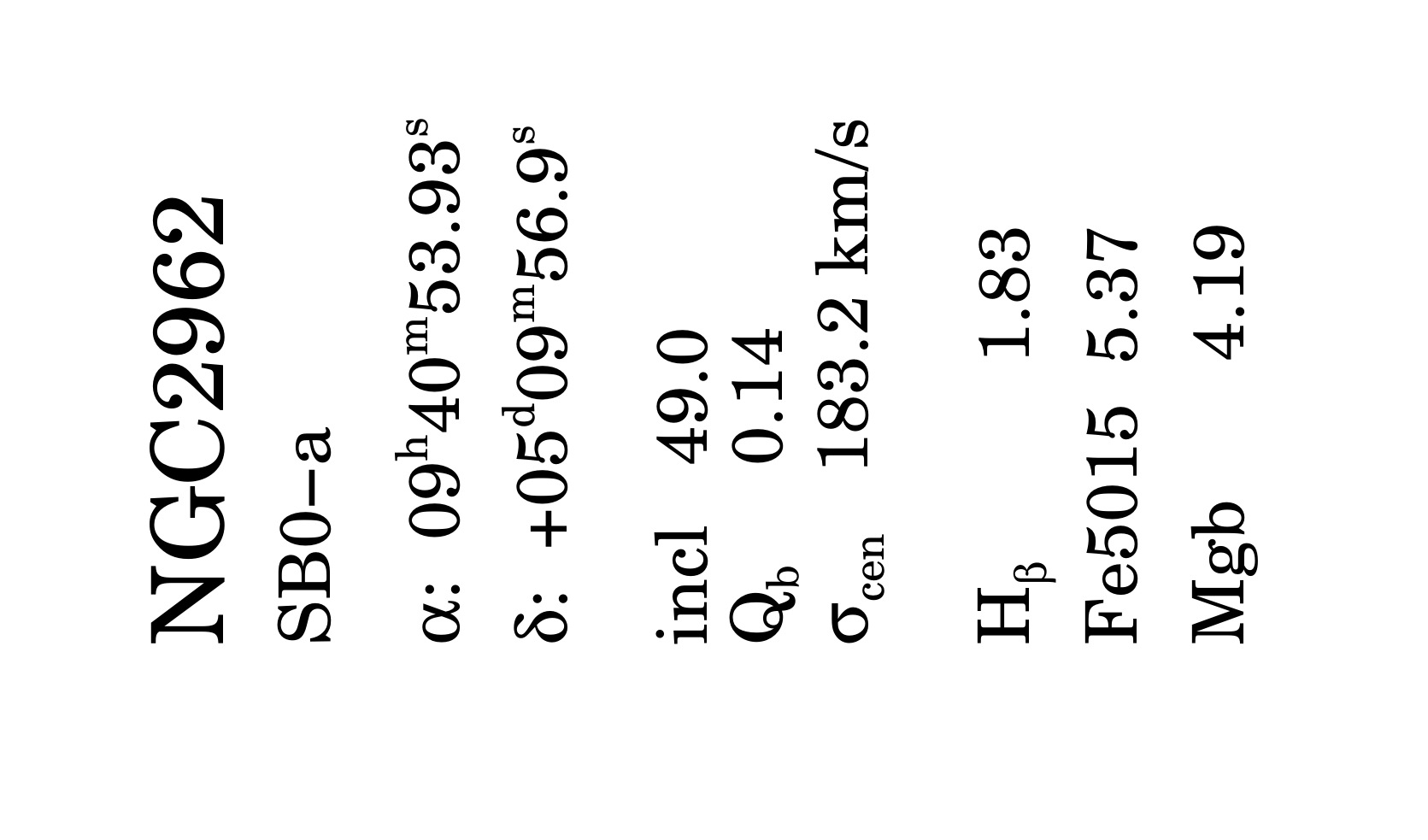}
\includegraphics[width=0.5\linewidth, angle=90,bb = 50 150 800 900]{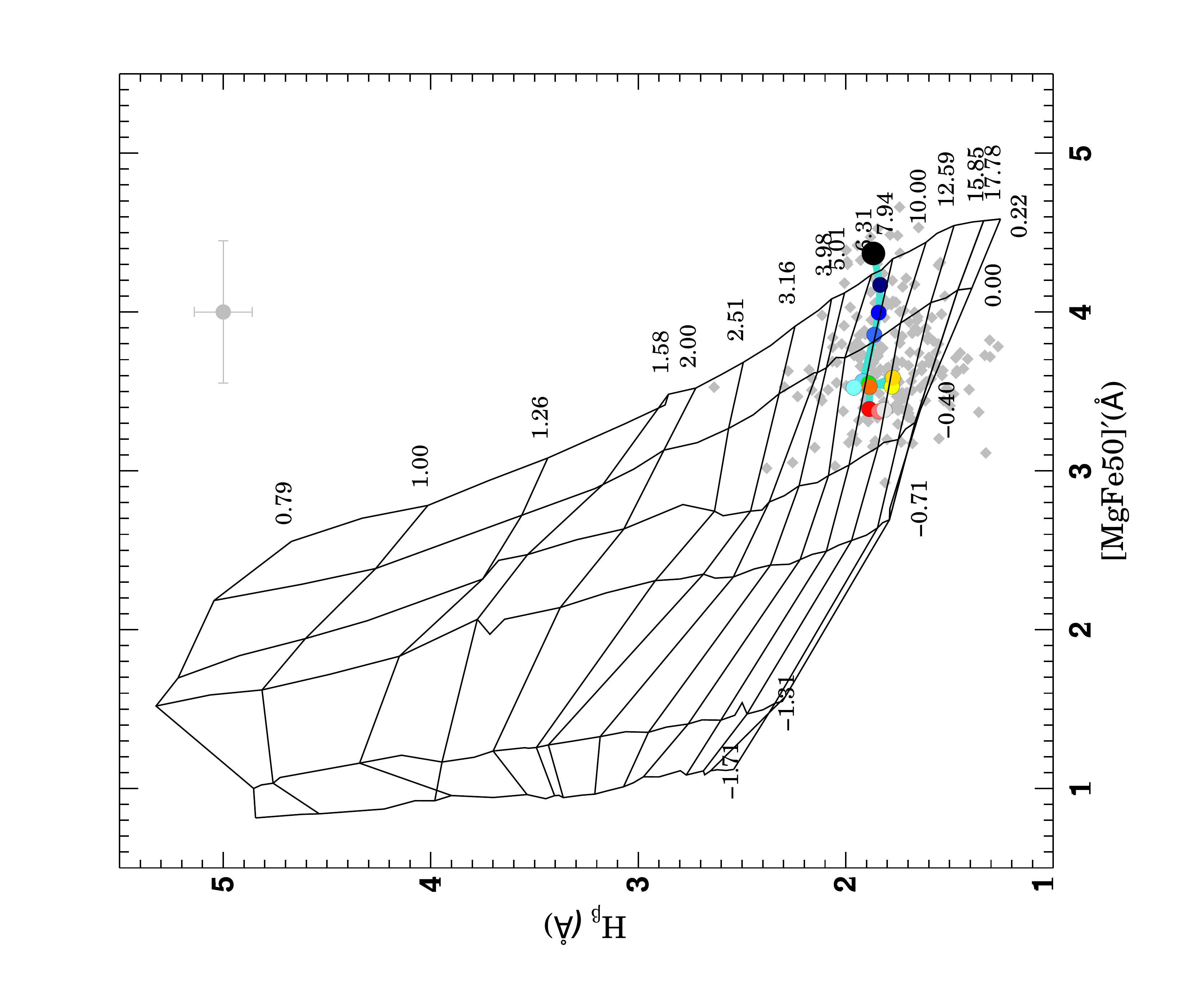}
\caption{Fig~\ref{fig:summ1015} continued.}
\label{fig:summ2962}
\end{figure*}

\begin{figure*}
\includegraphics[width=0.79\linewidth, angle=90]{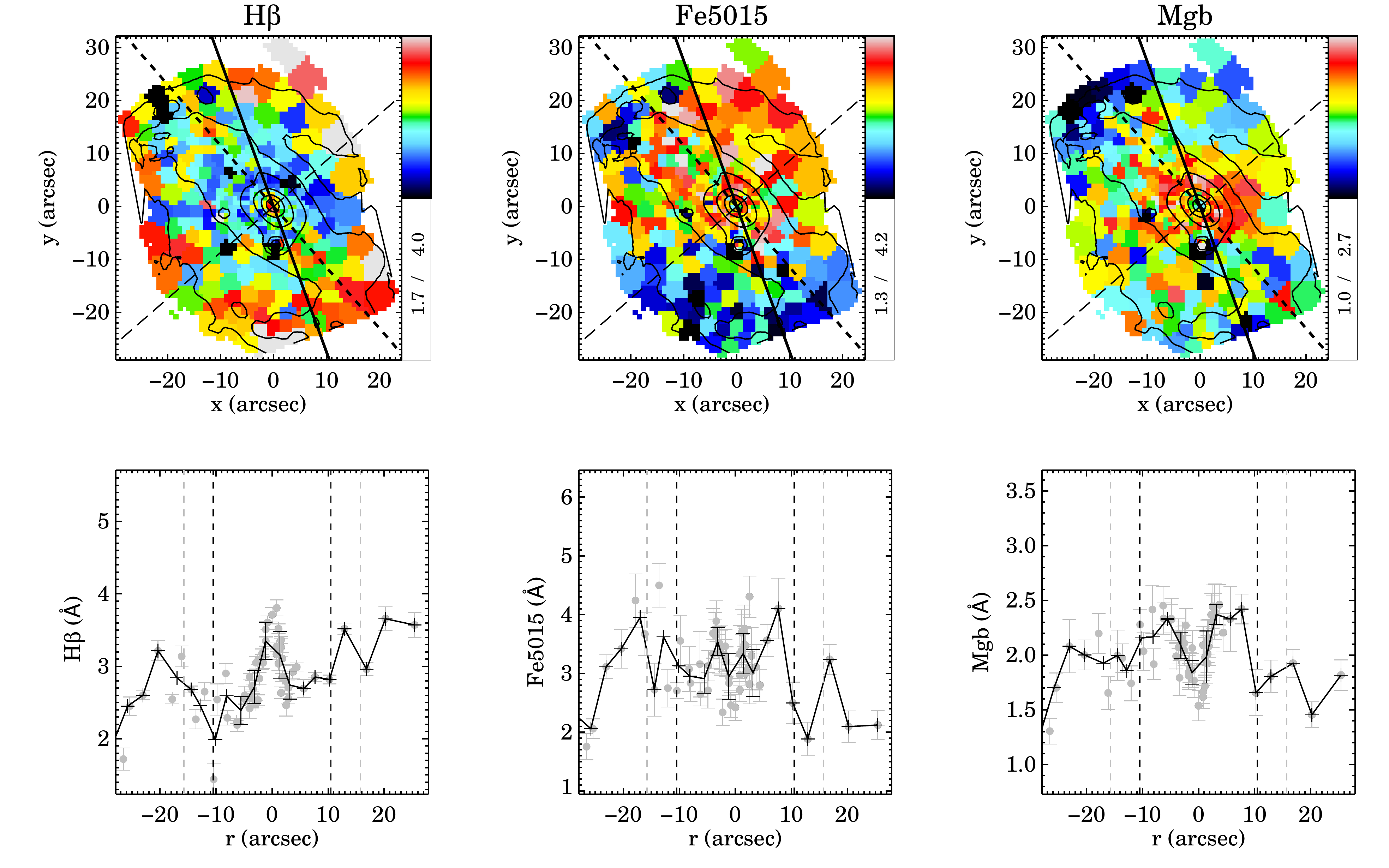}
\includegraphics[width=0.79\linewidth, angle=90]{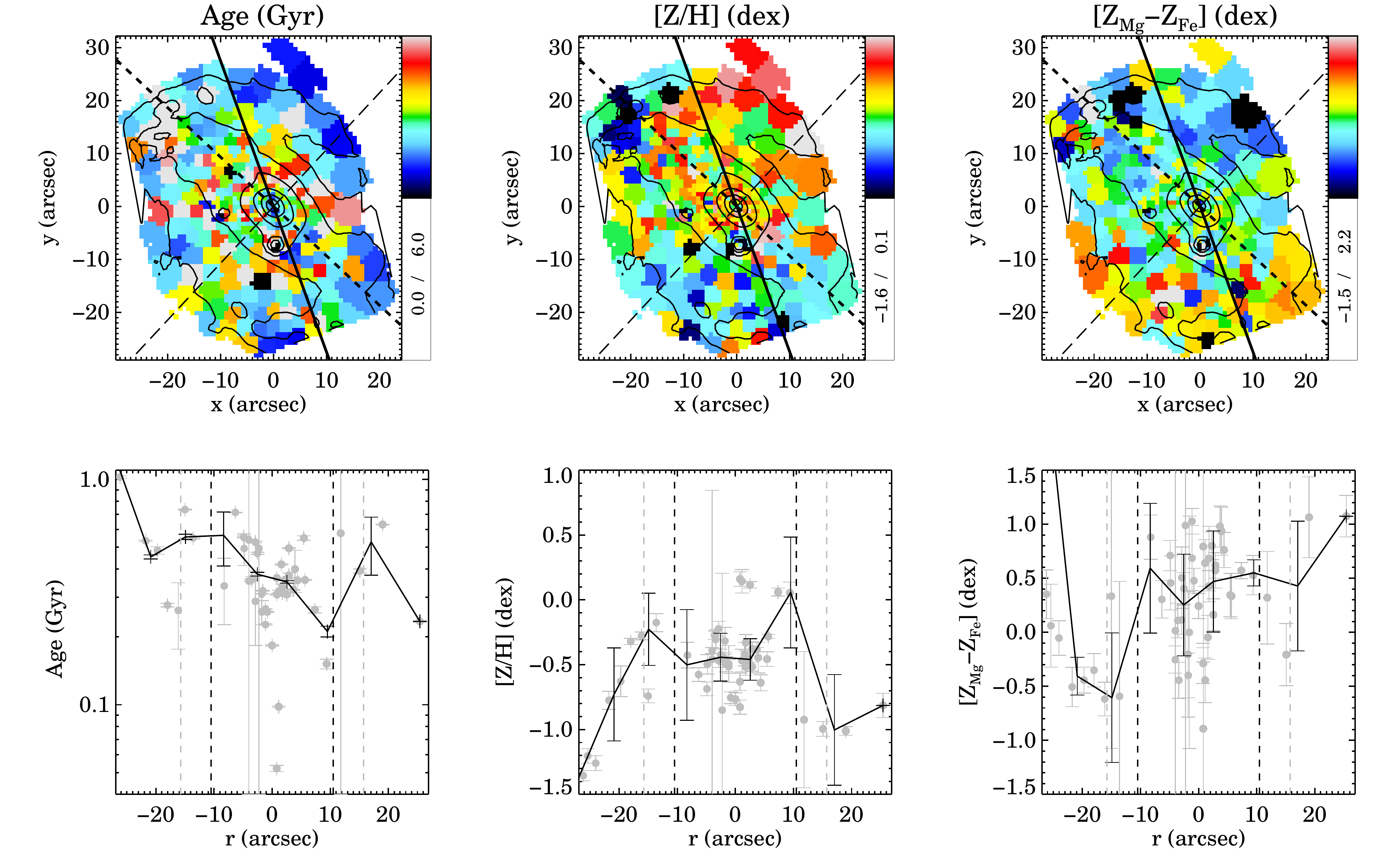}
\includegraphics[width=0.33\linewidth, angle=90,bb = -100 -150 600 600]{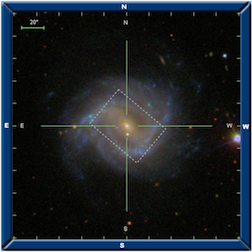}
\includegraphics[width=0.33\linewidth, angle=90,bb = -400 -40 50 -300]{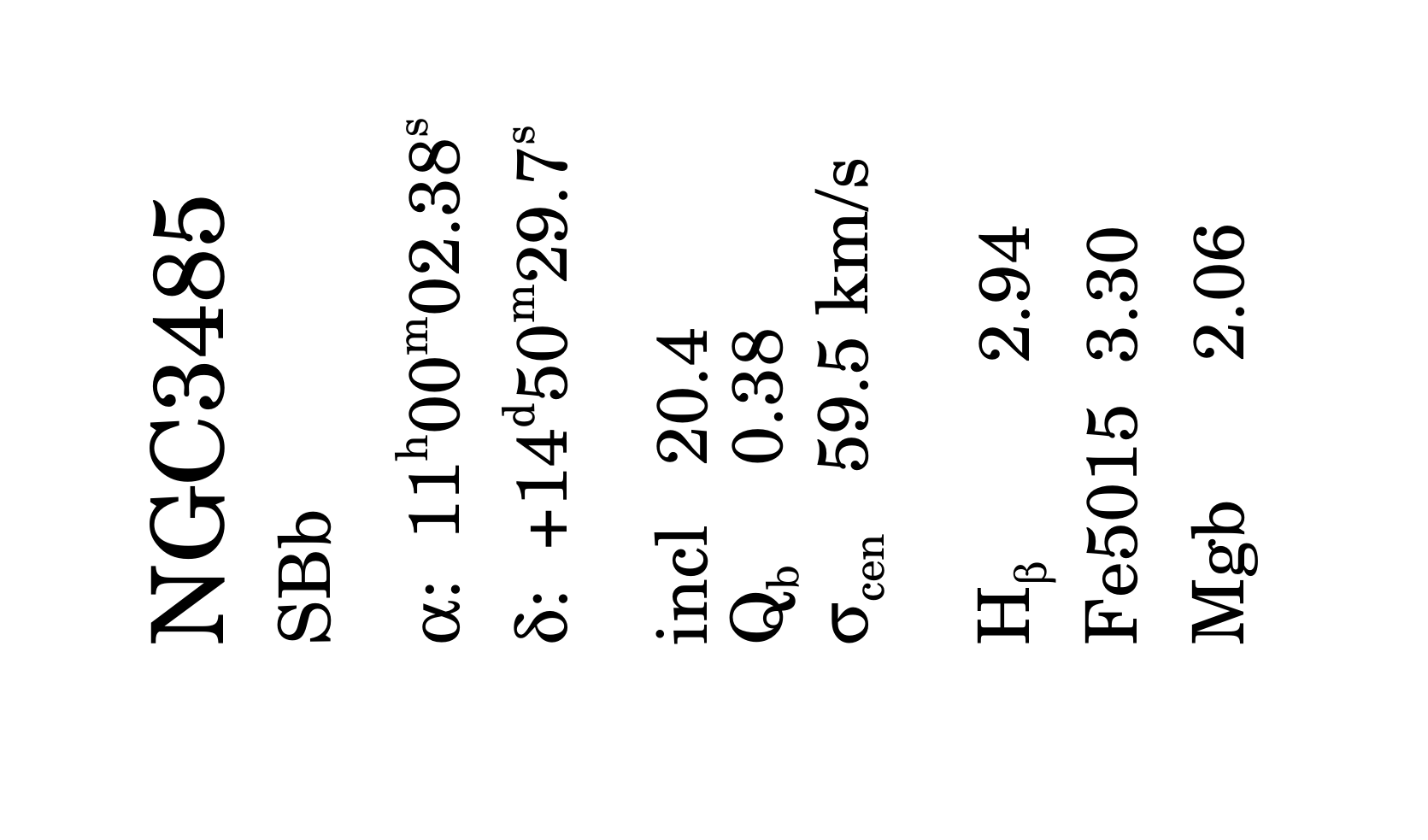}
\includegraphics[width=0.5\linewidth, angle=90,bb = 50 150 800 900]{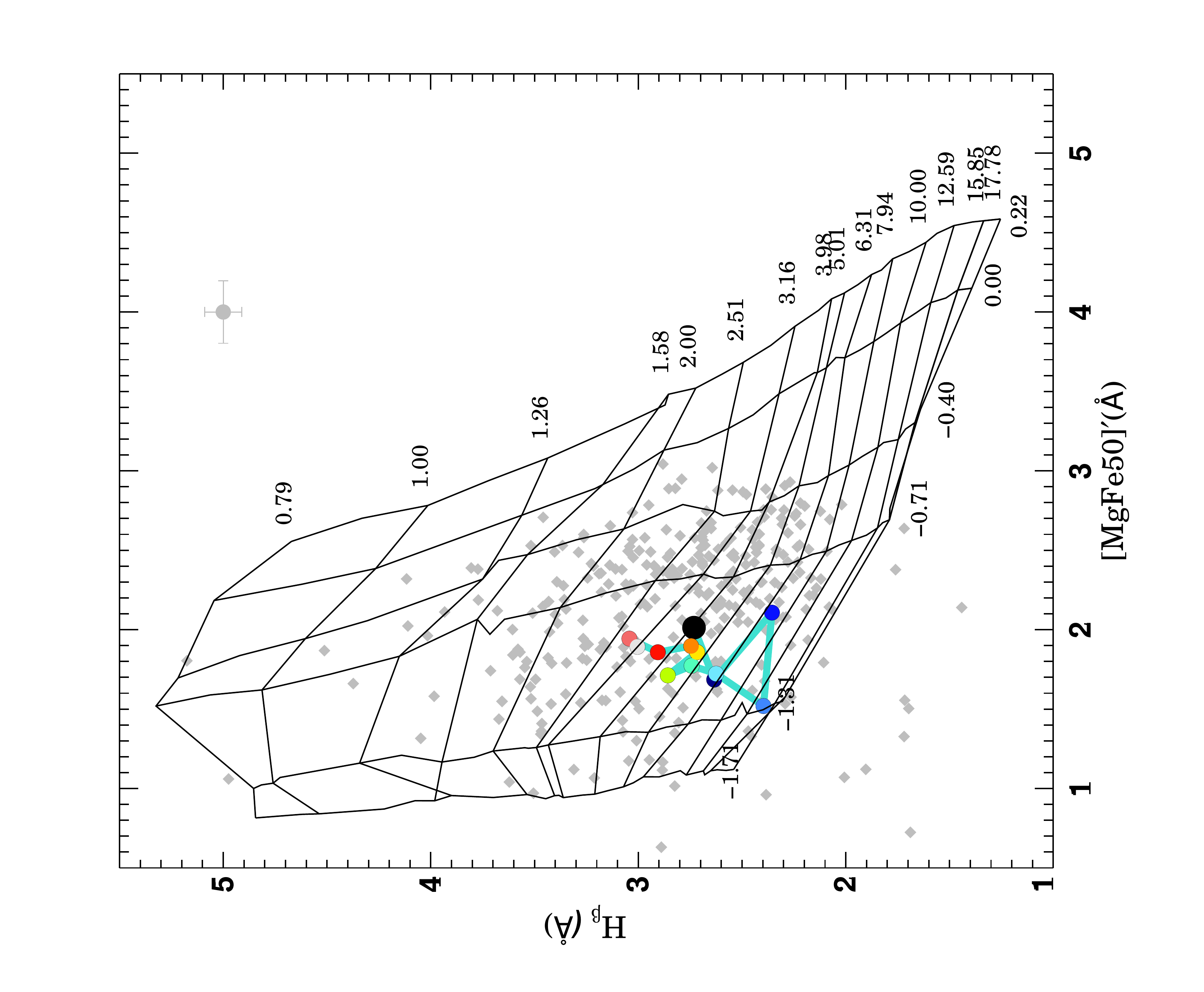}
\caption{Fig~\ref{fig:summ1015} continued.}
\label{fig:summ3485}
\end{figure*}

\begin{figure*}
\includegraphics[width=0.79\linewidth, angle=90]{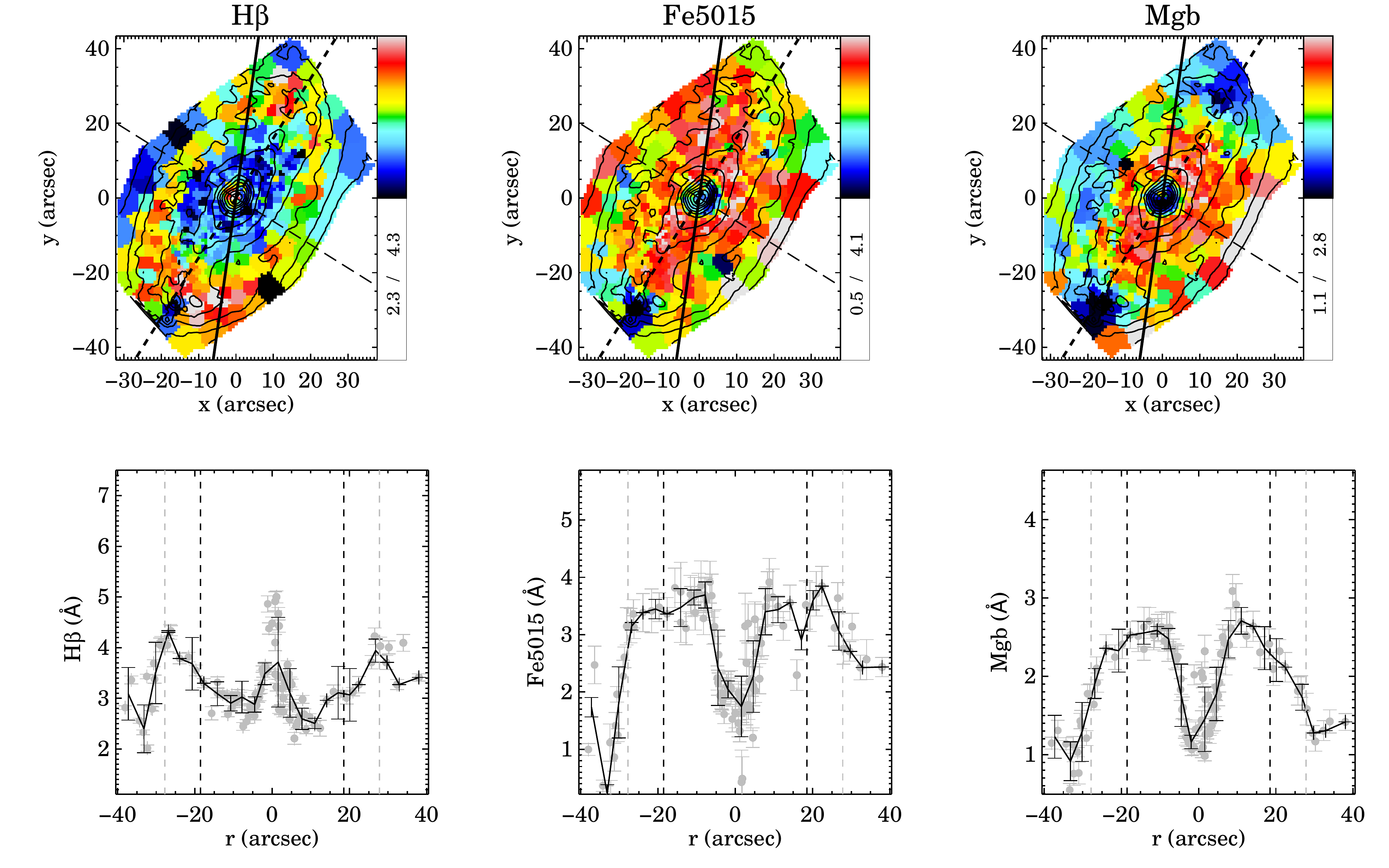}
\includegraphics[width=0.79\linewidth, angle=90]{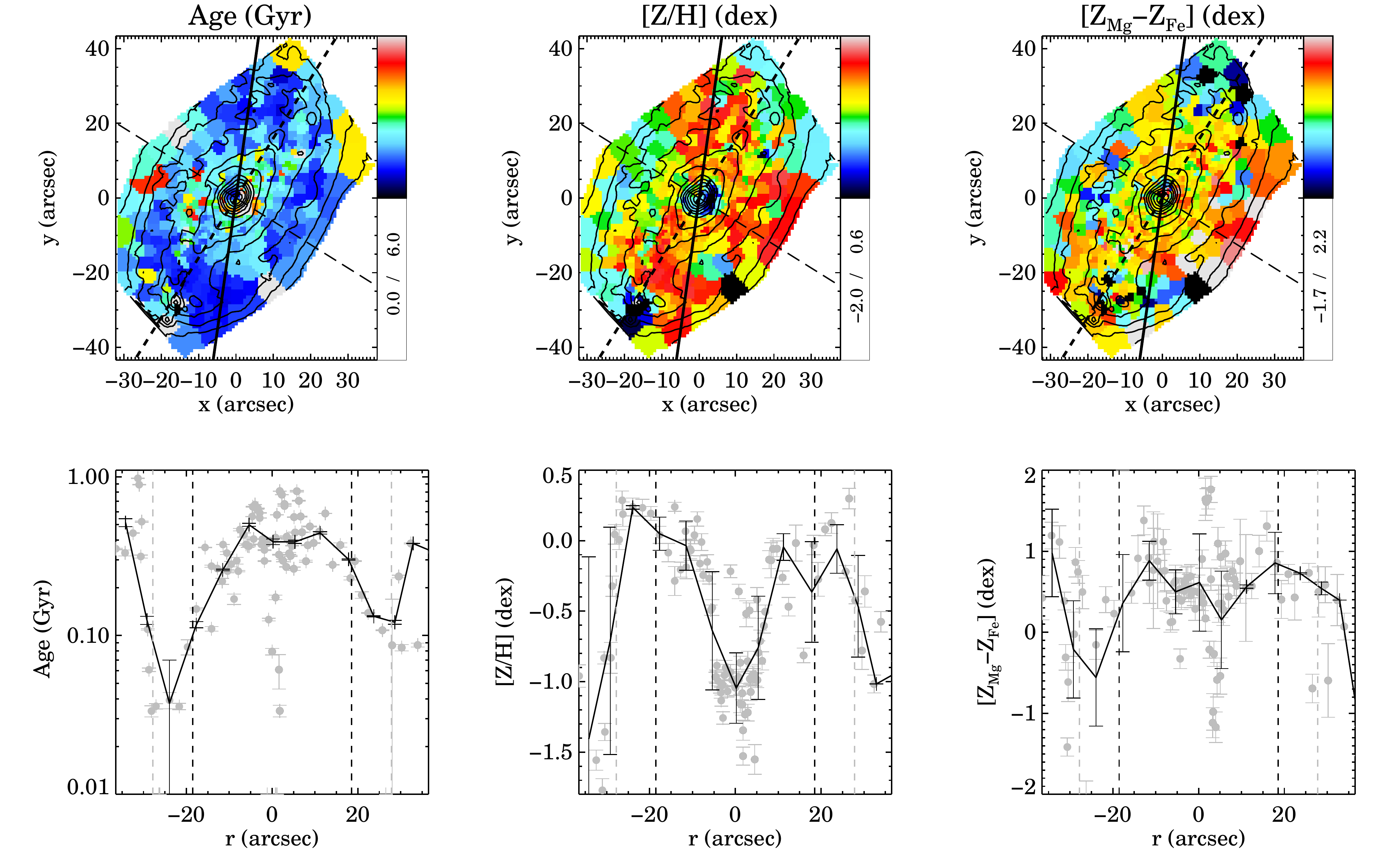}
\includegraphics[width=0.33\linewidth, angle=90,bb = -100 -150 600 600]{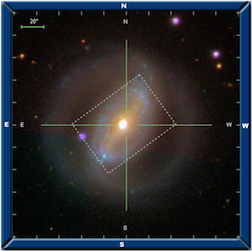}
\includegraphics[width=0.33\linewidth, angle=90,bb = -400 -40 50 -300]{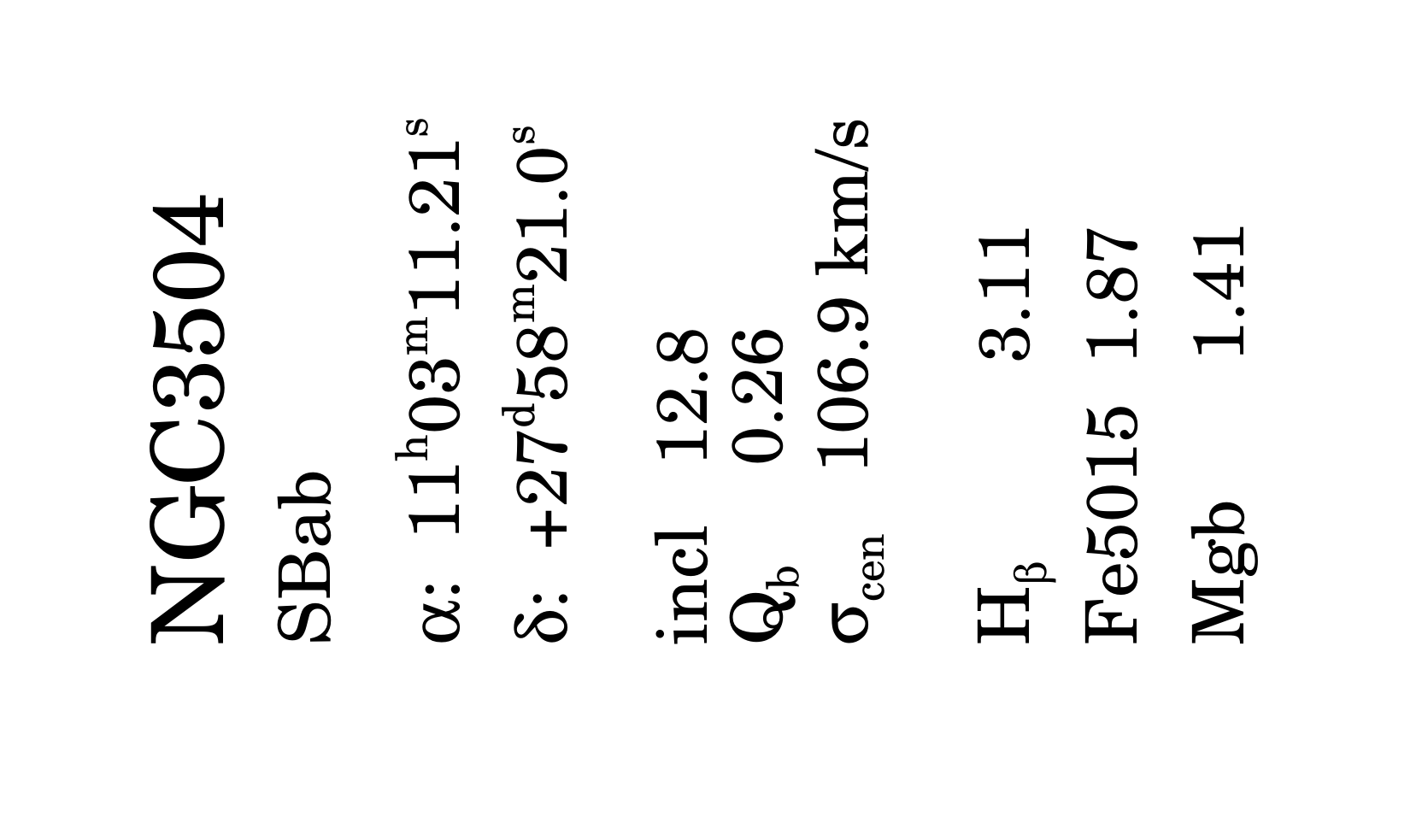}
\includegraphics[width=0.5\linewidth, angle=90,bb = 50 150 800 900]{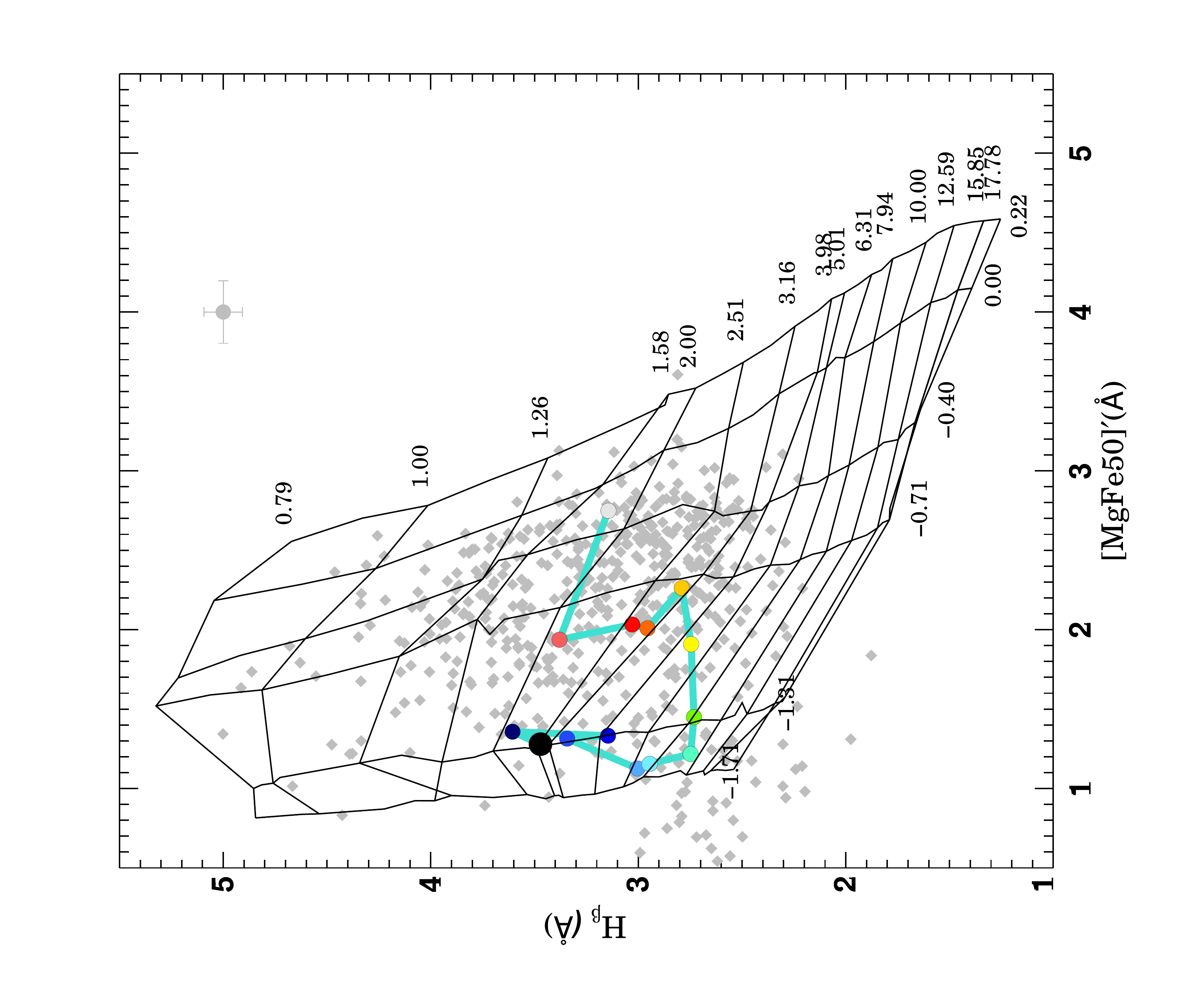}
\caption{Fig~\ref{fig:summ1015} continued.}
\label{fig:summ3504}
\end{figure*}

\begin{figure*}
\includegraphics[width=0.79\linewidth, angle=90]{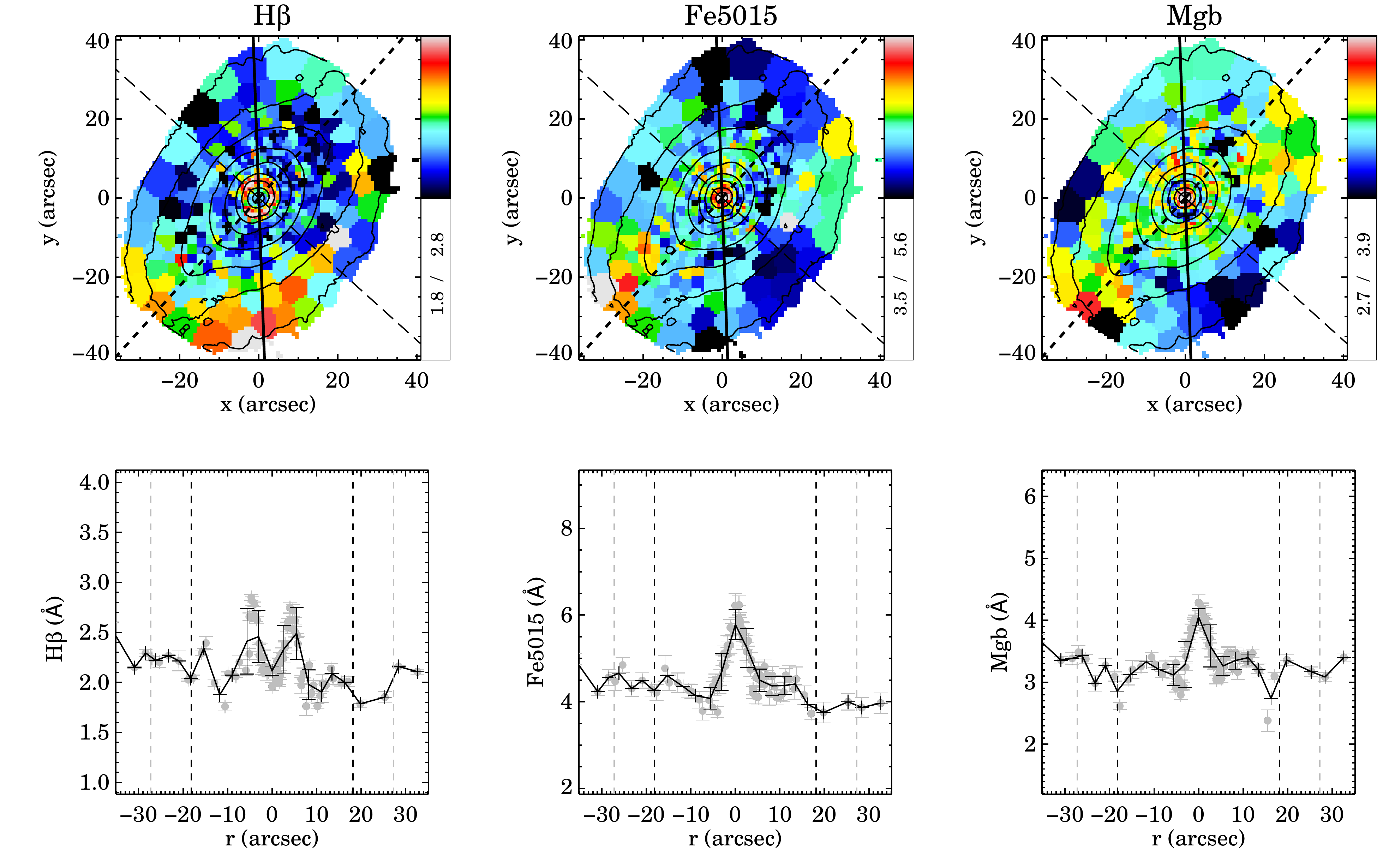}
\includegraphics[width=0.79\linewidth, angle=90]{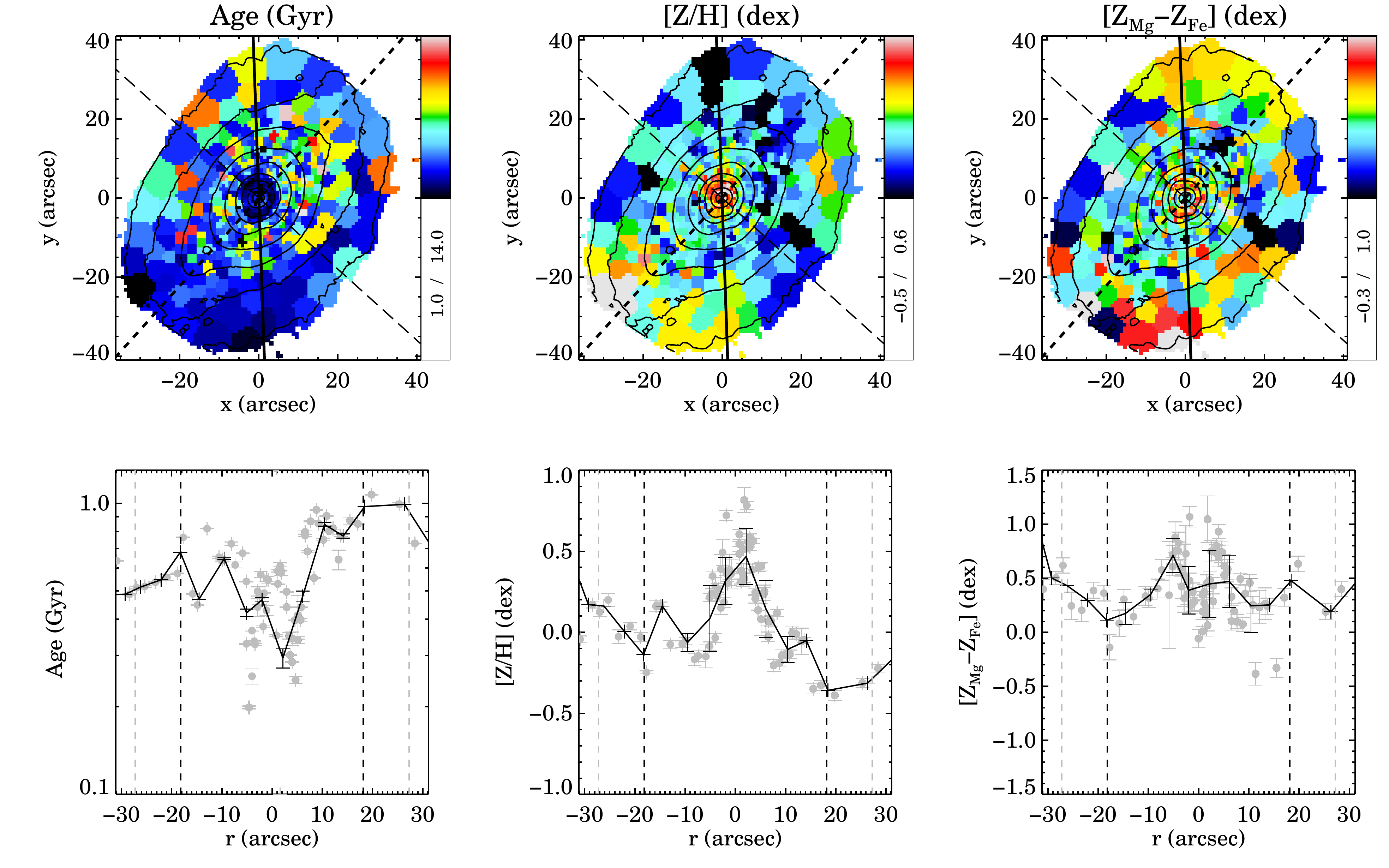}
\includegraphics[width=0.33\linewidth, angle=90,bb = -100 -150 600 600]{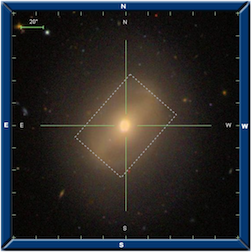}
\includegraphics[width=0.33\linewidth, angle=90,bb = -400 -40 50 -300]{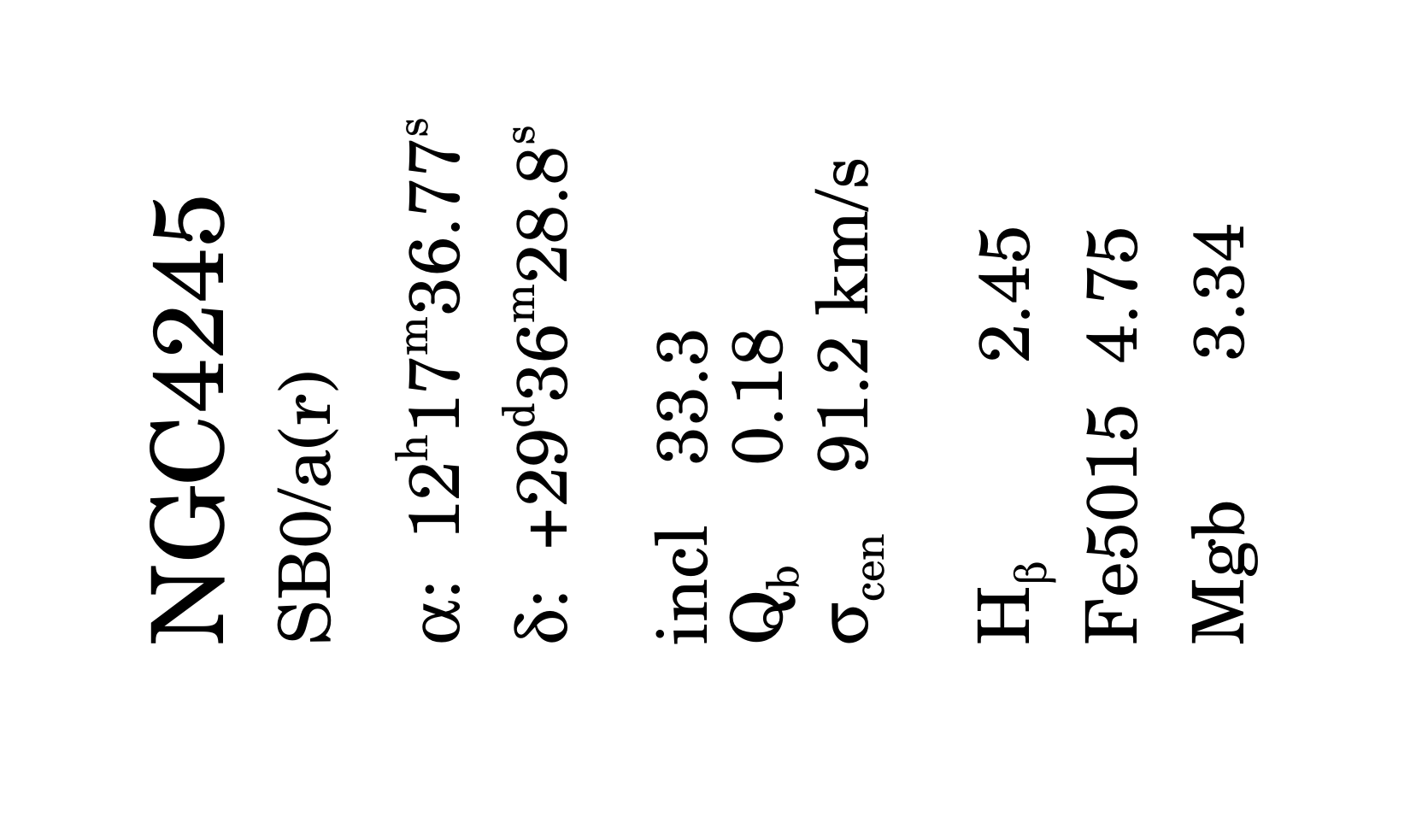}
\includegraphics[width=0.5\linewidth, angle=90,bb = 50 150 800 900]{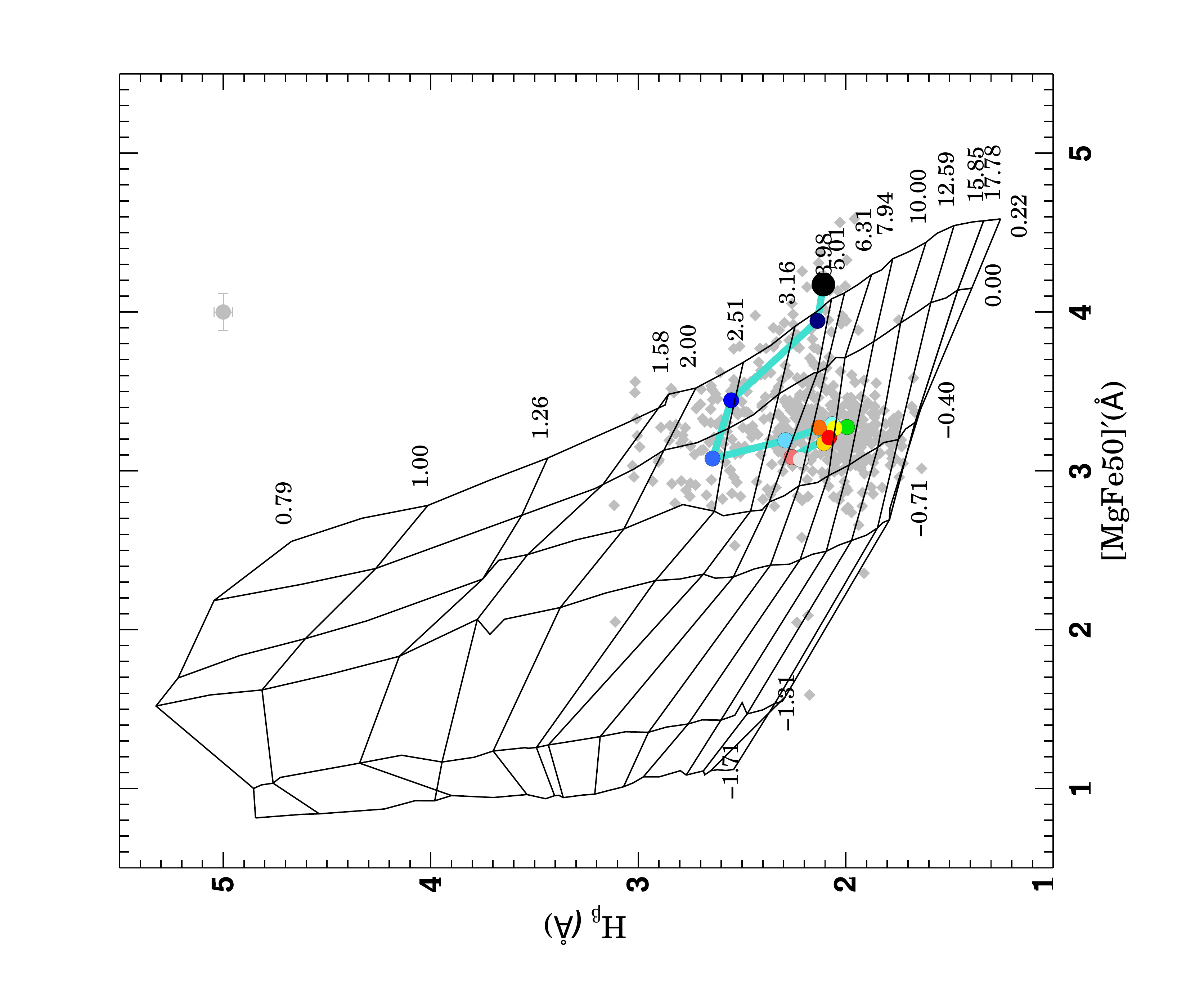}
\caption{Fig~\ref{fig:summ1015} continued.}
\label{fig:summ4245}
\end{figure*}

\begin{figure*}
\includegraphics[width=0.79\linewidth, angle=90]{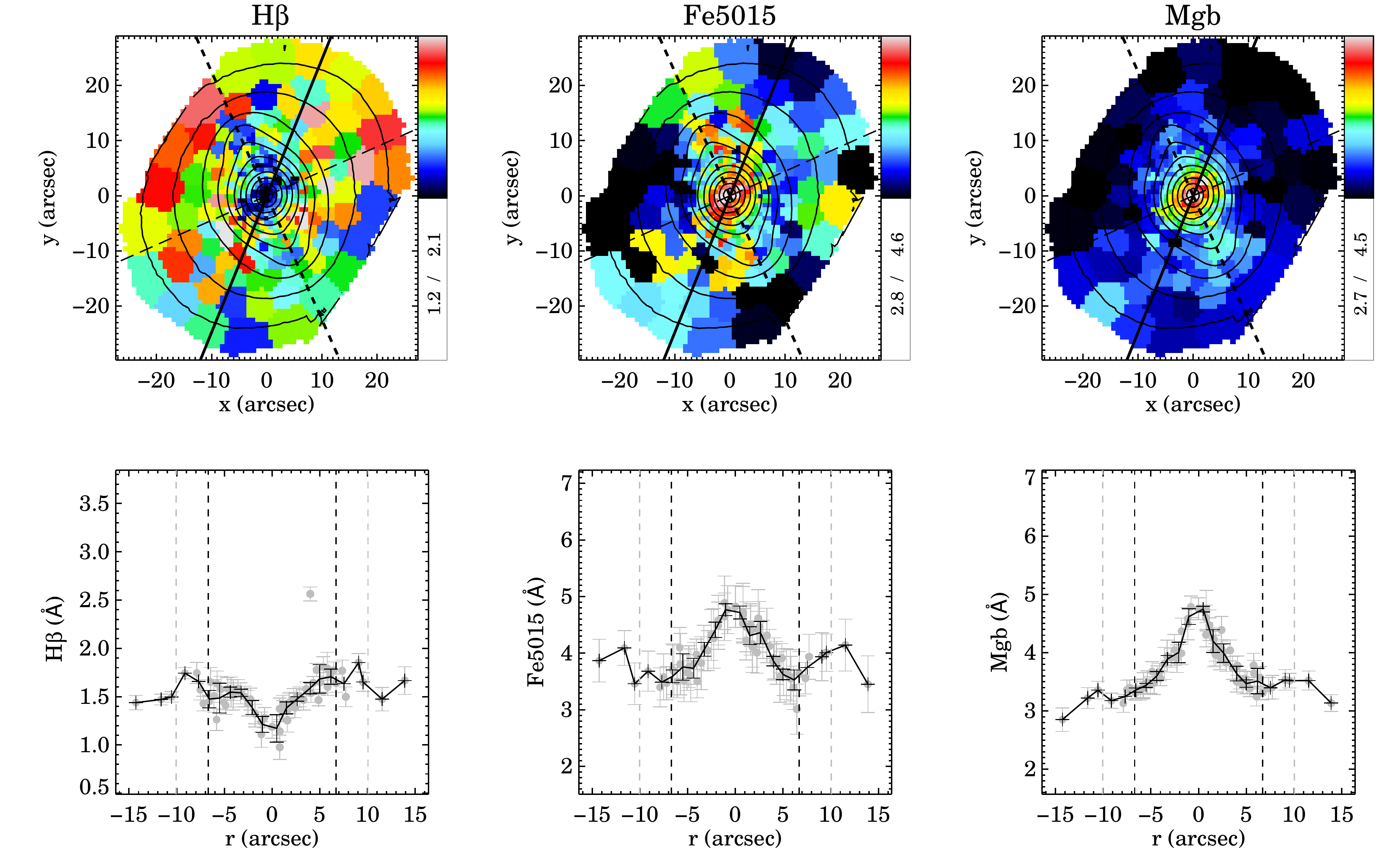}
\includegraphics[width=0.79\linewidth, angle=90]{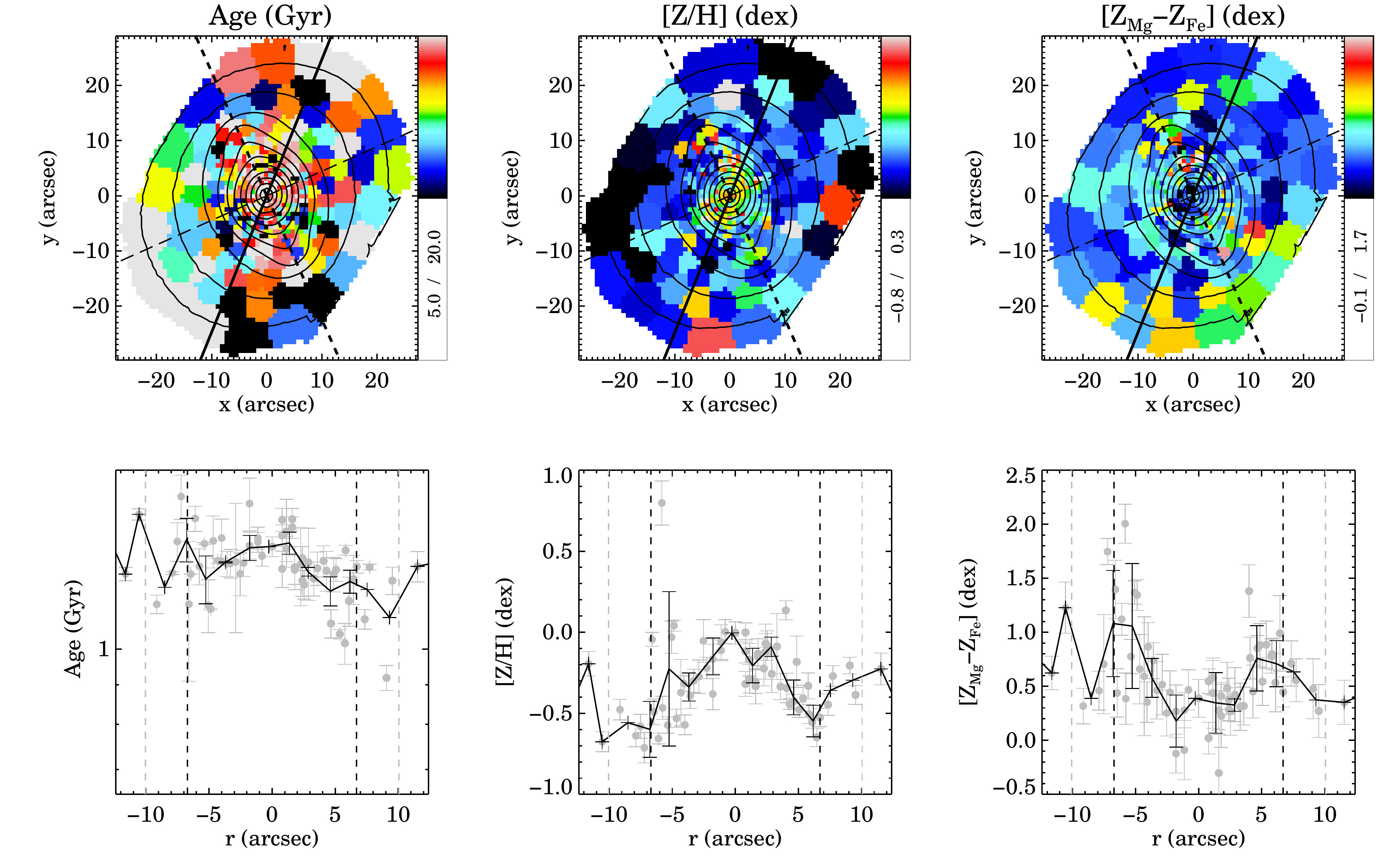}
\includegraphics[width=0.33\linewidth, angle=90,bb = -100 -150 600 600]{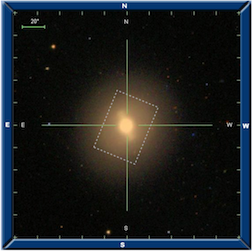}
\includegraphics[width=0.33\linewidth, angle=90,bb = -400 -40 50 -300]{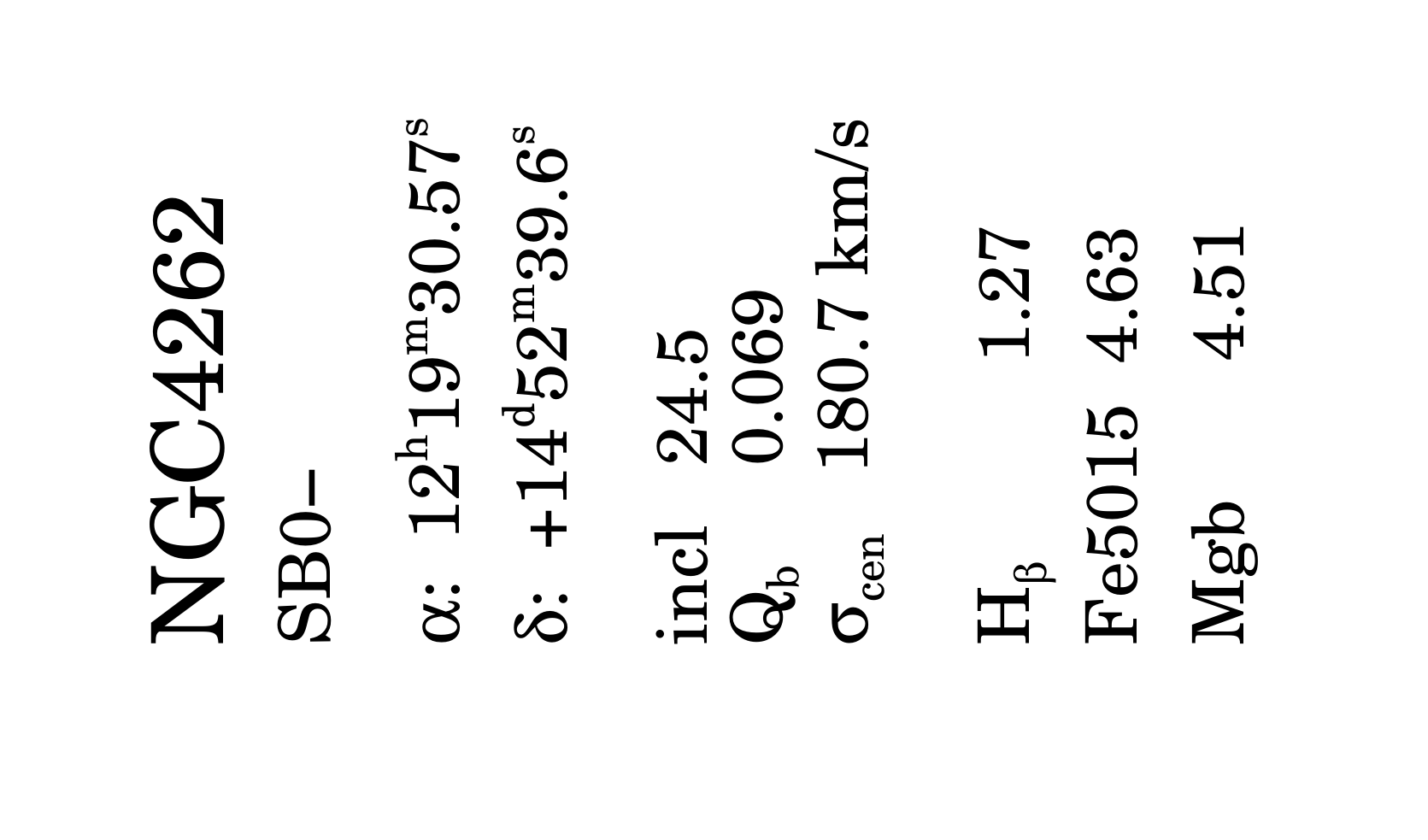}
\includegraphics[width=0.5\linewidth, angle=90,bb = 50 150 800 900]{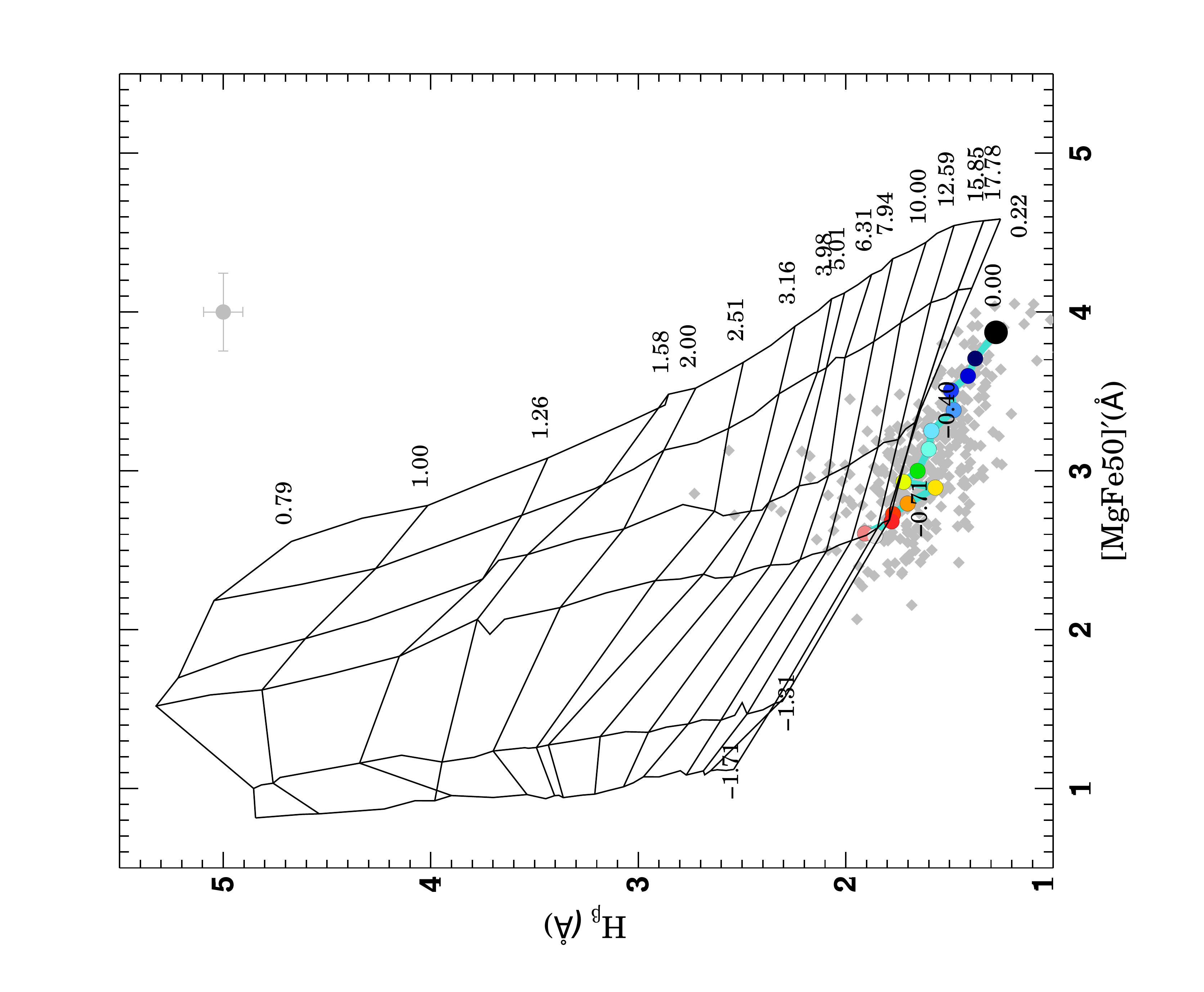}
\caption{Fig~\ref{fig:summ1015} continued.}
\label{fig:summ4262}
\end{figure*}

\begin{figure*}
\includegraphics[width=0.79\linewidth, angle=90]{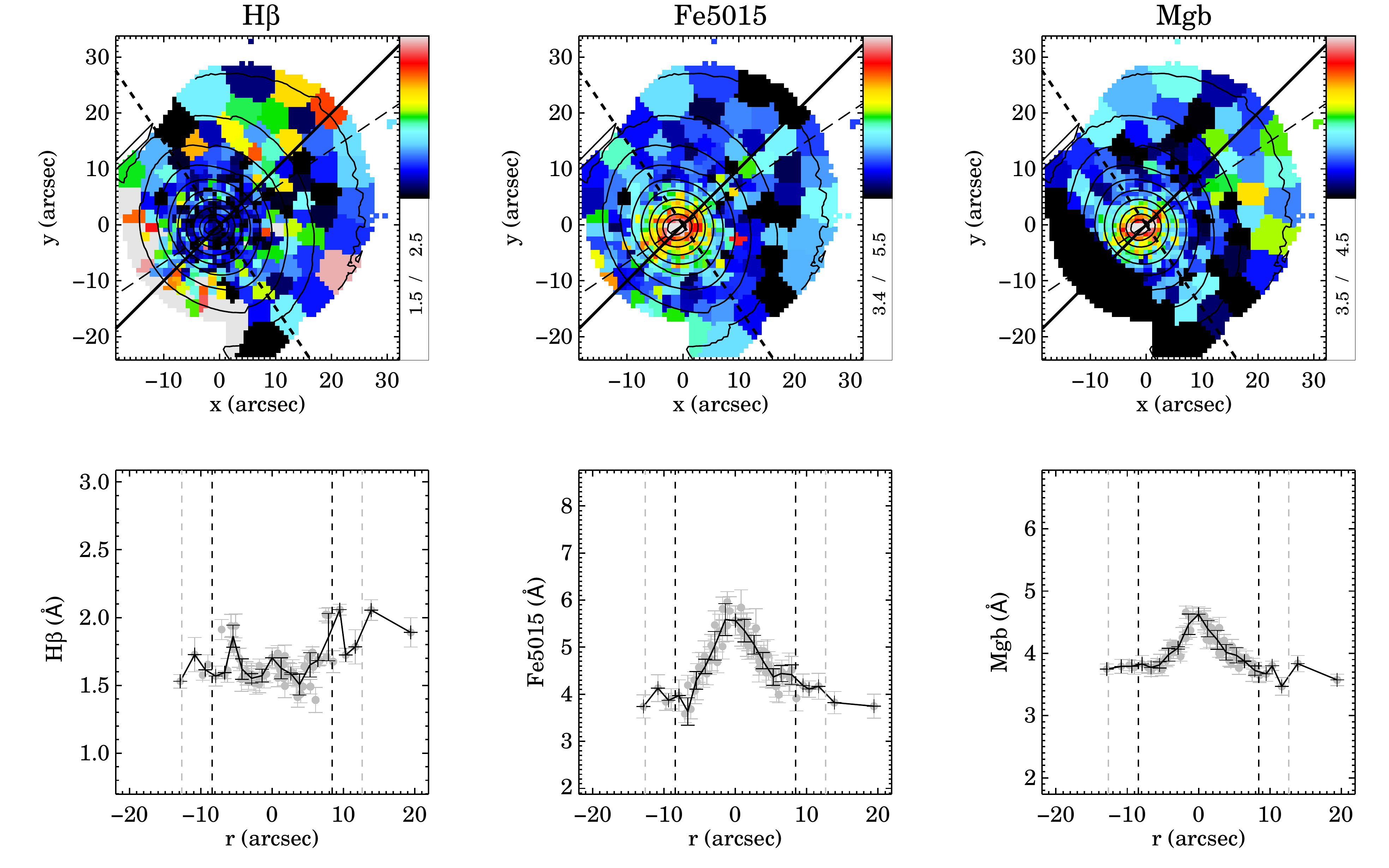}
\includegraphics[width=0.79\linewidth, angle=90]{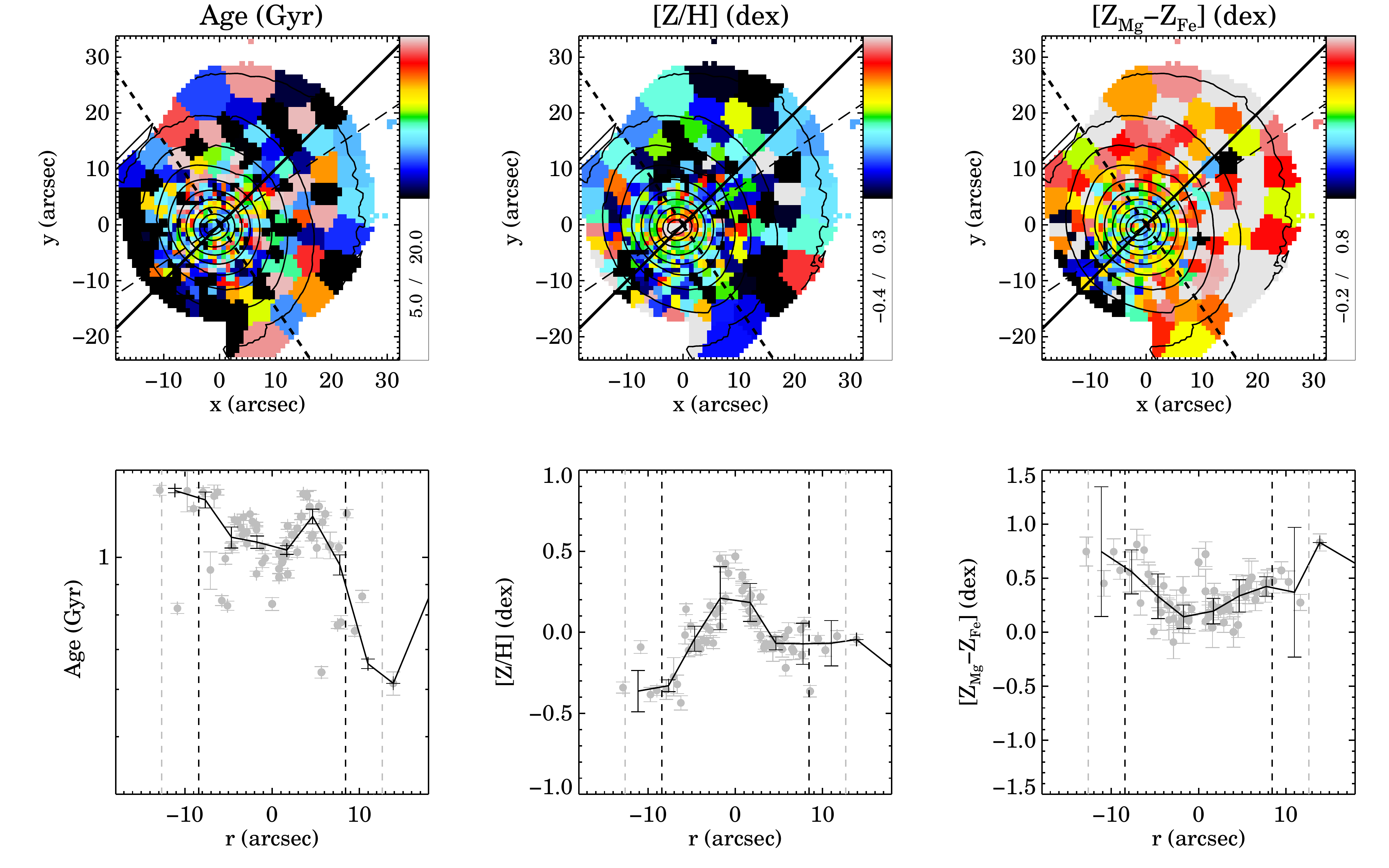}
\includegraphics[width=0.33\linewidth, angle=90,bb = -100 -150 600 600]{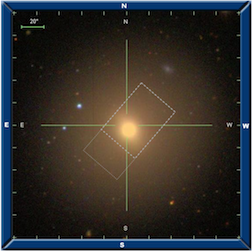}
\includegraphics[width=0.33\linewidth, angle=90,bb = -400 -40 50 -300]{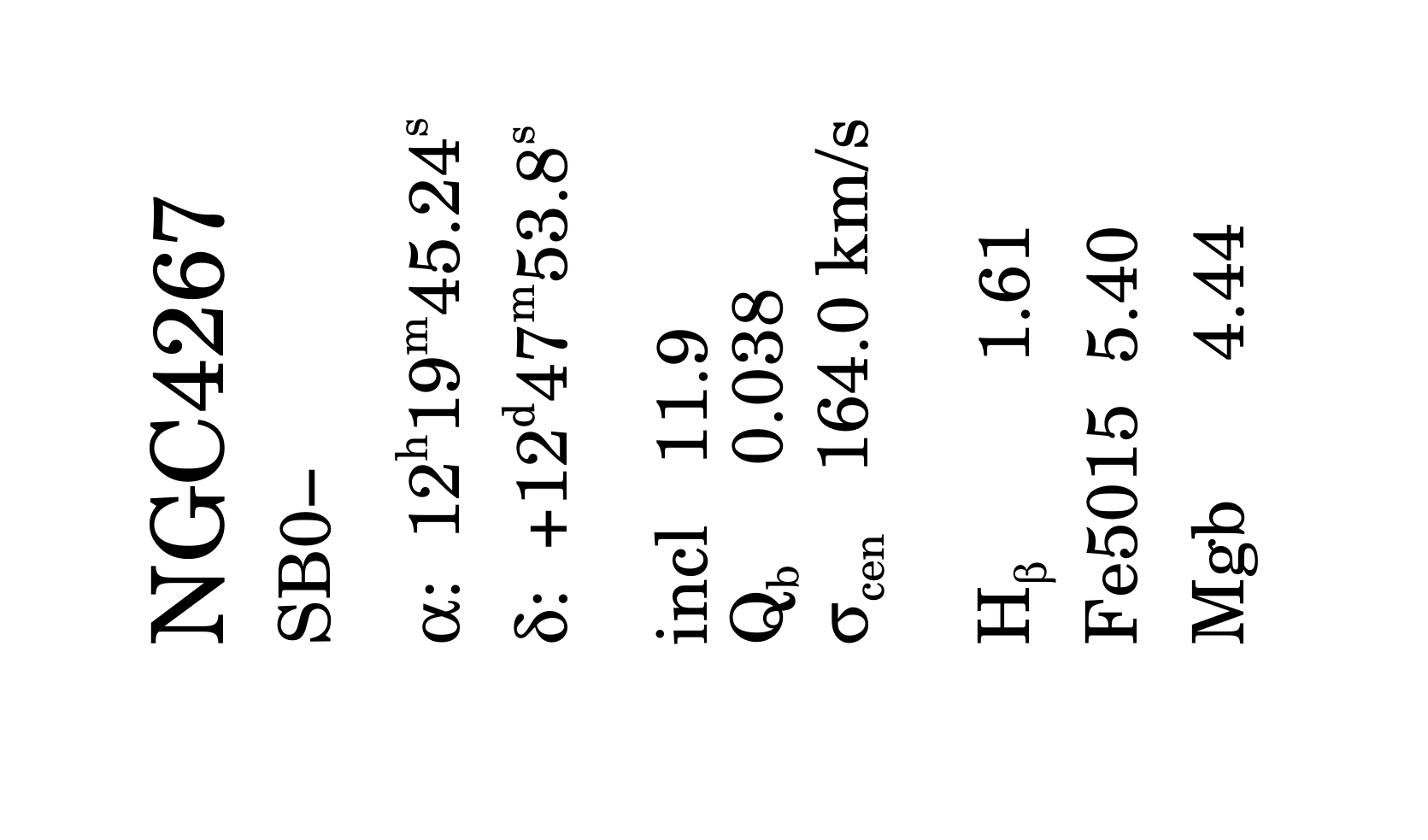}
\includegraphics[width=0.5\linewidth, angle=90,bb = 50 150 800 900]{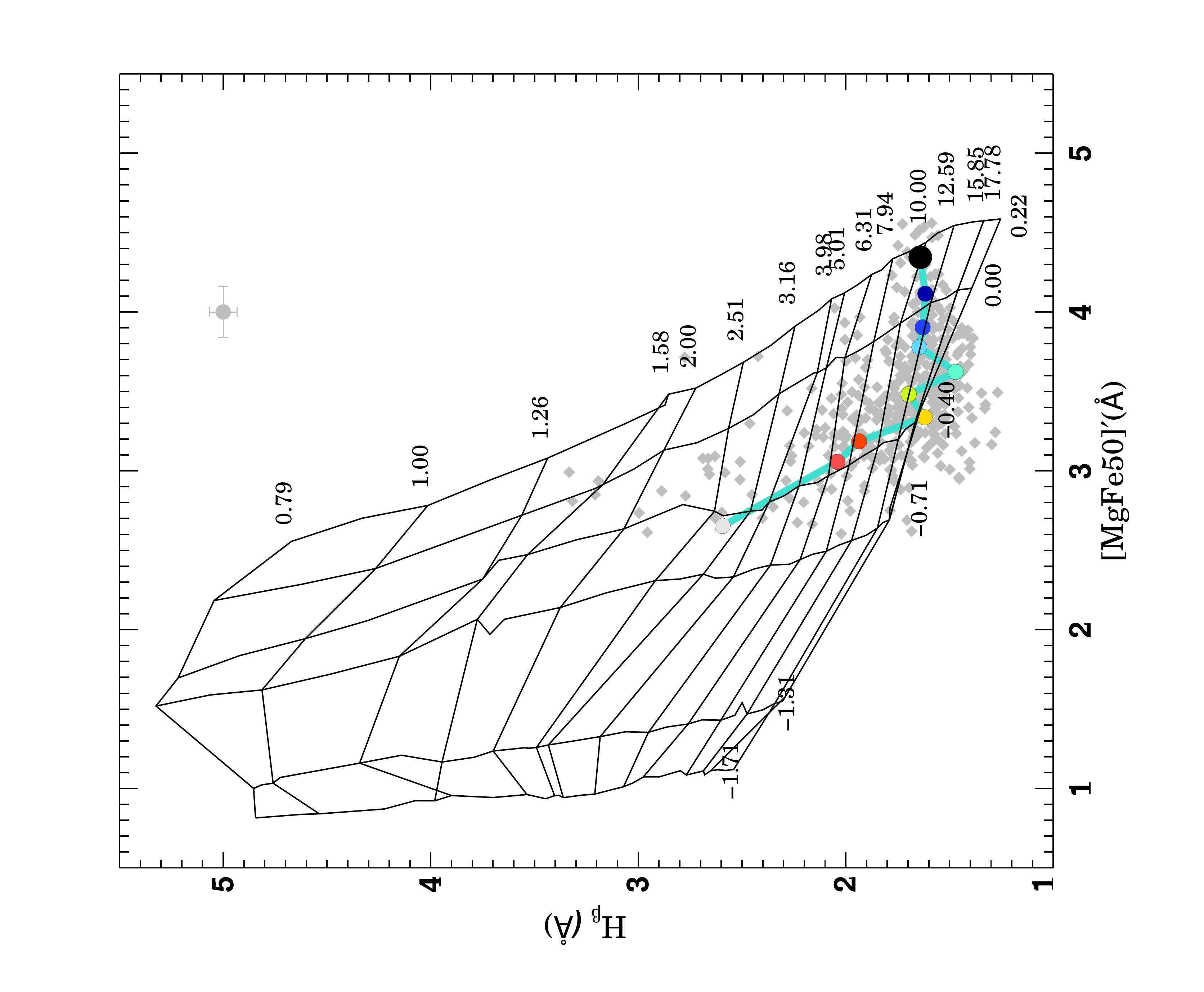}
\caption{Fig~\ref{fig:summ1015} continued.}
\label{fig:summ4267}
\end{figure*}

\begin{figure*}
\includegraphics[width=0.79\linewidth, angle=90]{NGC4394_indices_v7-eps-converted-to.pdf}
\includegraphics[width=0.79\linewidth, angle=90]{NGC4394_ssp_multilog_v6-eps-converted-to.pdf}
\includegraphics[width=0.33\linewidth, angle=90,bb = -100 -150 600 600]{dr7_IFU_prints012.png}
\includegraphics[width=0.33\linewidth, angle=90,bb = -400 -40 50 -300]{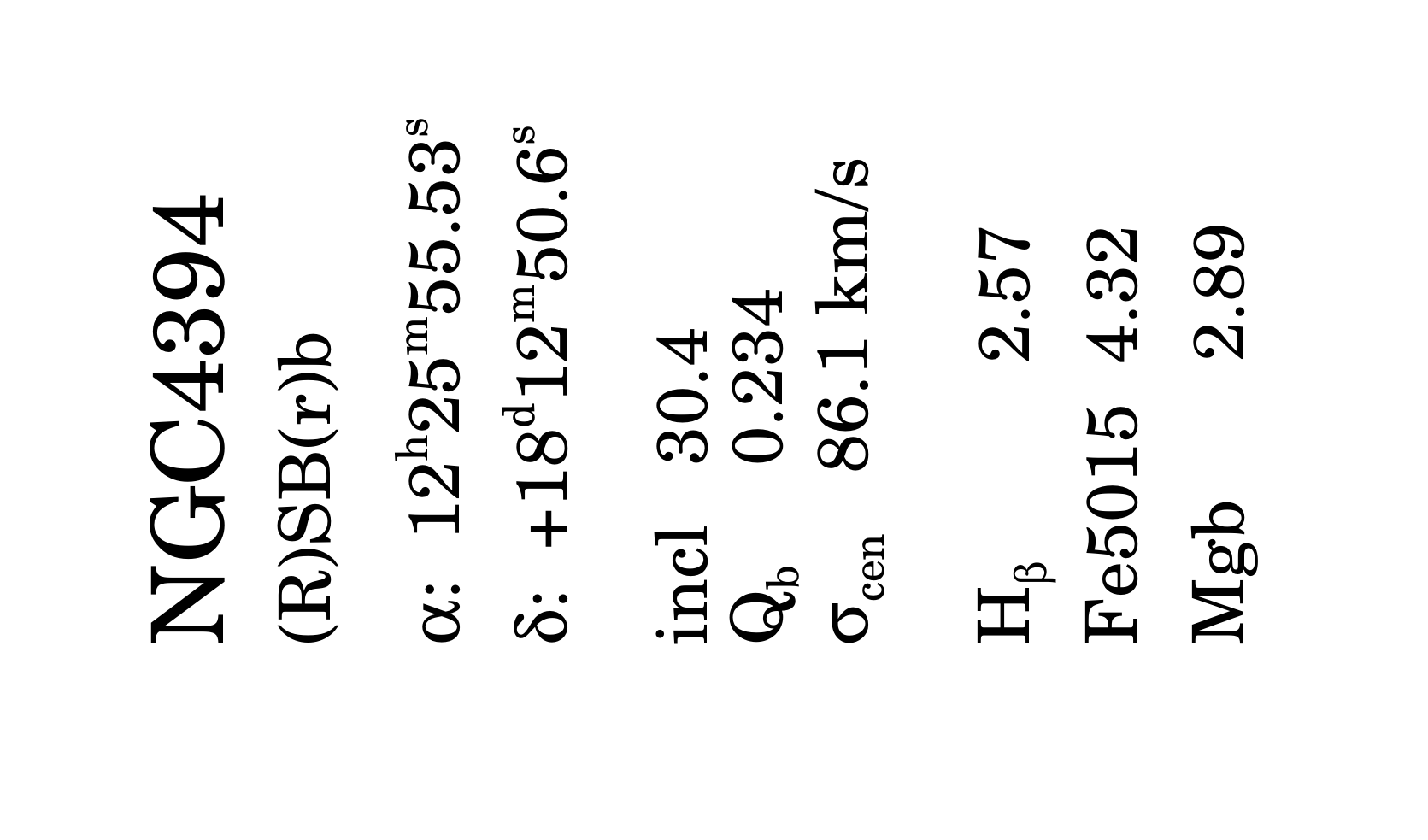}
\includegraphics[width=0.5\linewidth, angle=90,bb = 50 150 800 900]{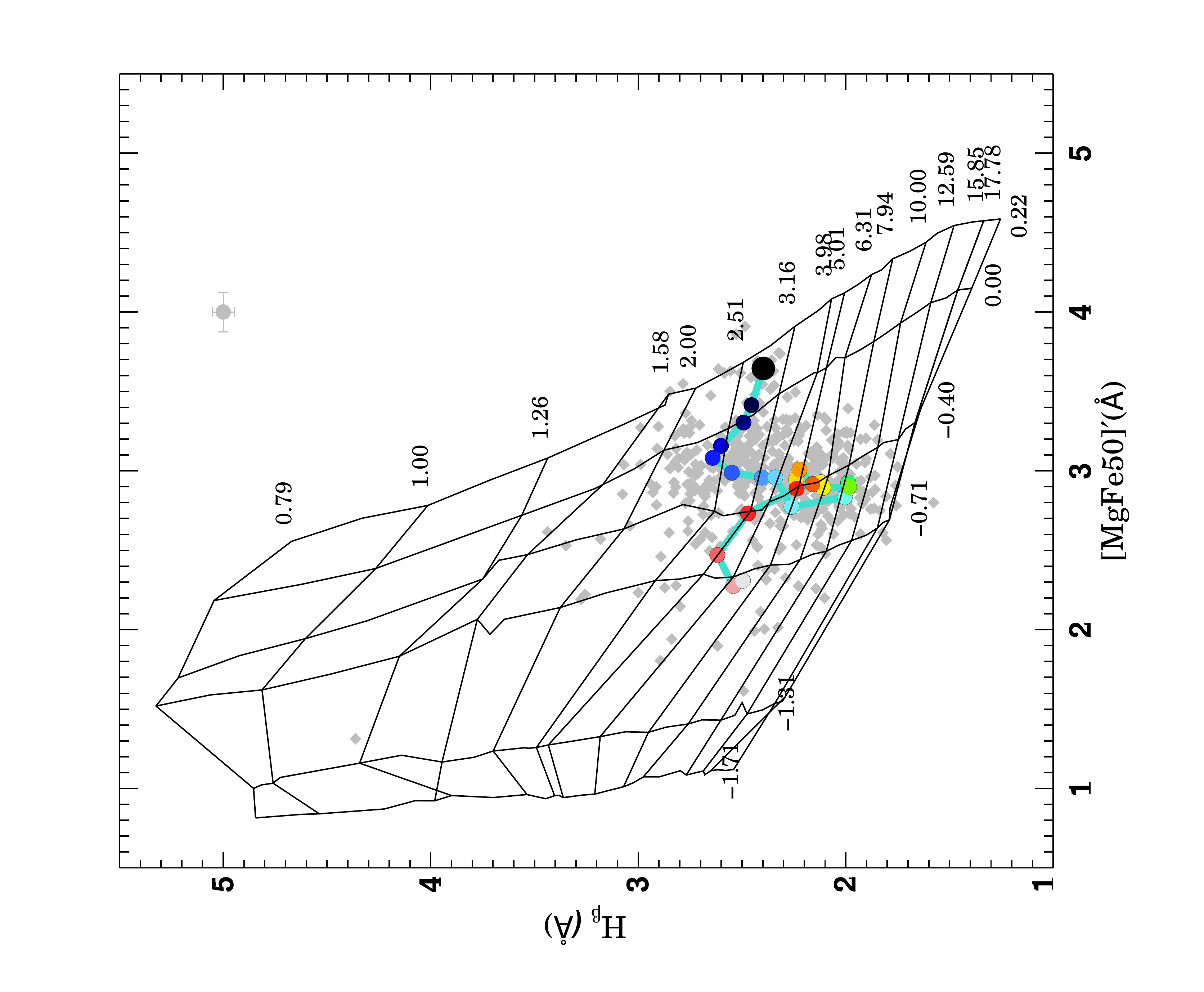}
\caption{Fig~\ref{fig:summ1015} continued.}
\label{fig:summ4394}
\end{figure*}

\begin{figure*}
\includegraphics[width=0.79\linewidth, angle=90]{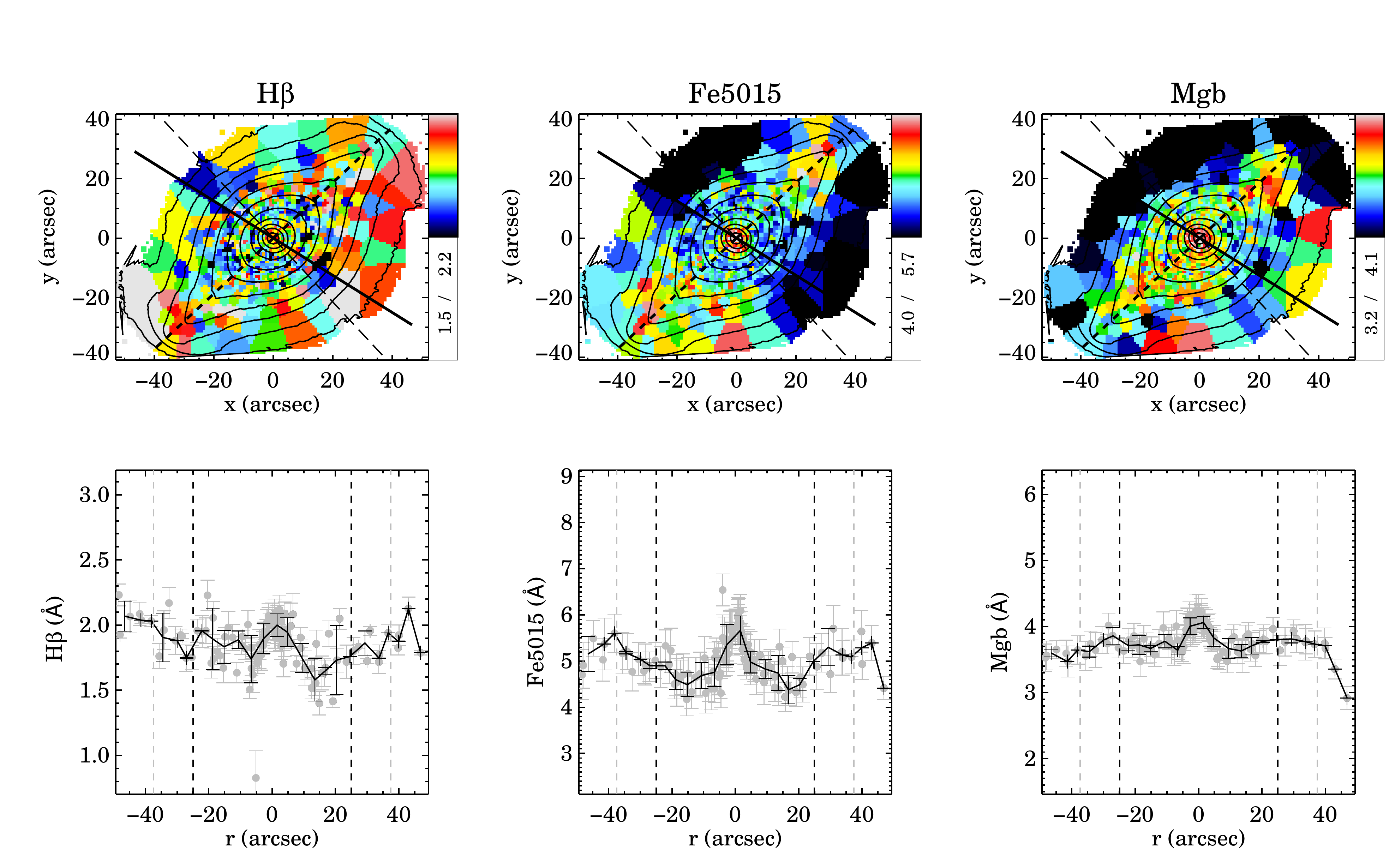}
\includegraphics[width=0.79\linewidth, angle=90]{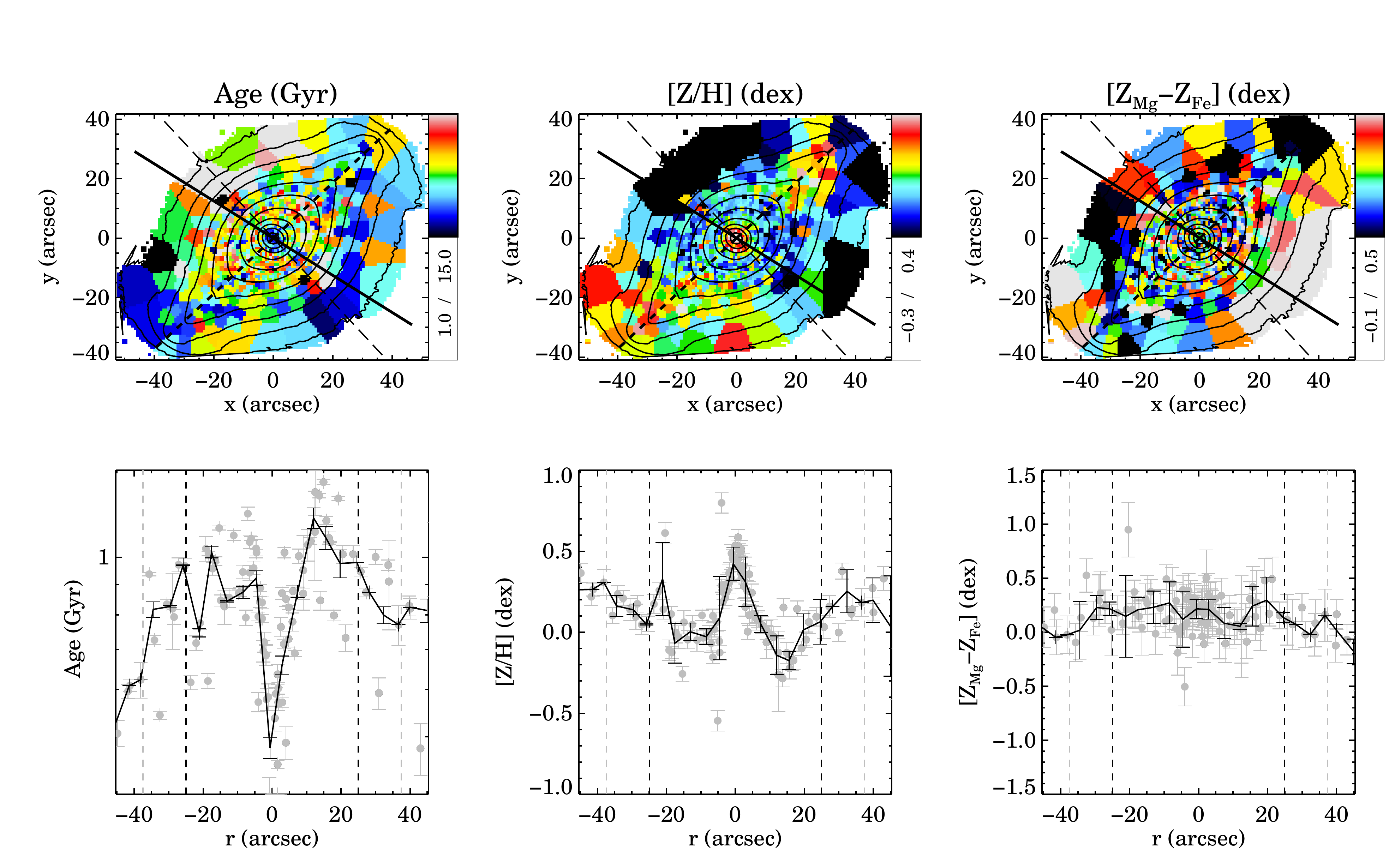}
\includegraphics[width=0.33\linewidth, angle=90,bb = -100 -150 600 600]{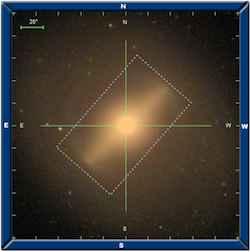}
\includegraphics[width=0.33\linewidth, angle=90,bb = -400 -40 50 -300]{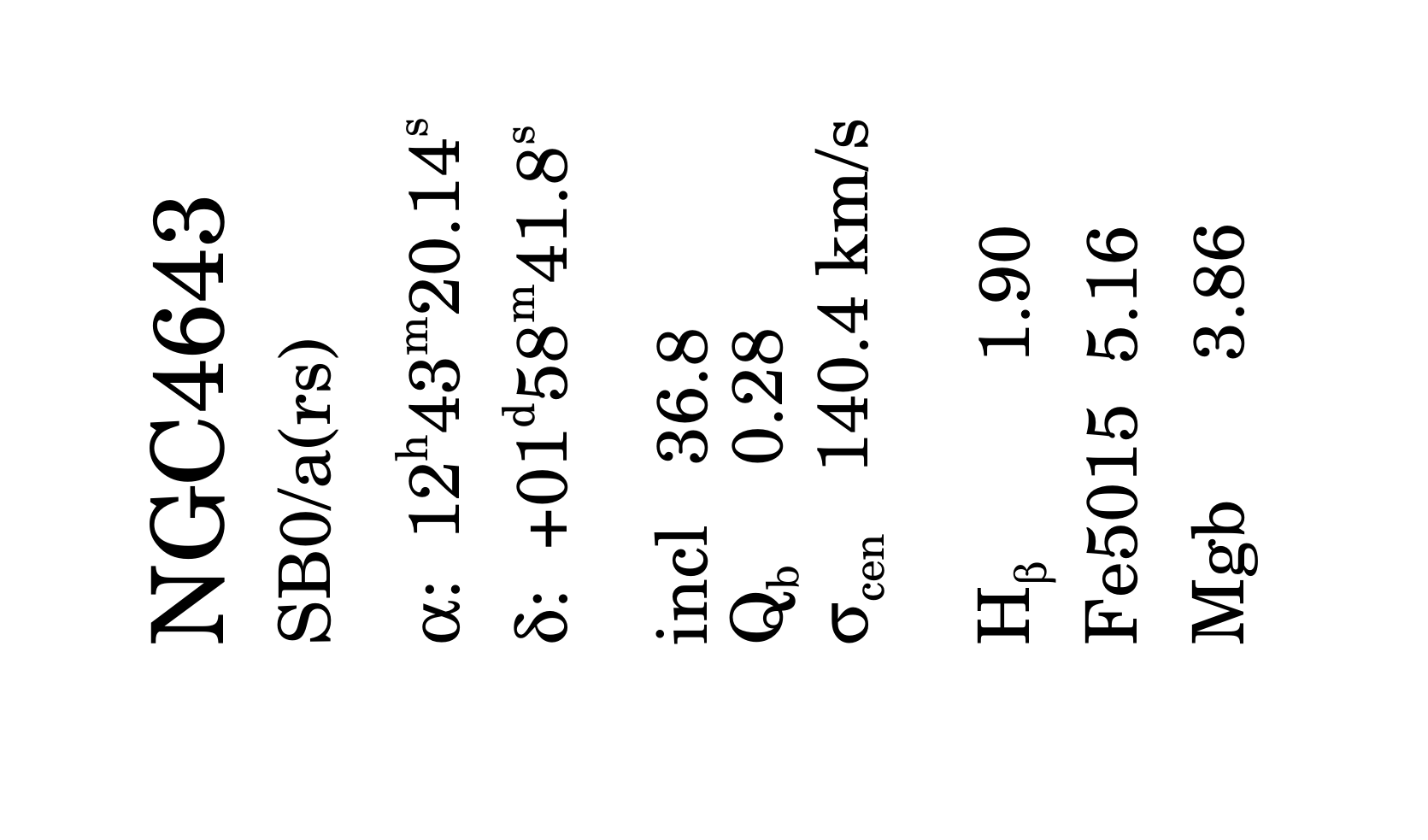}
\includegraphics[width=0.5\linewidth, angle=90,bb = 50 150 800 900]{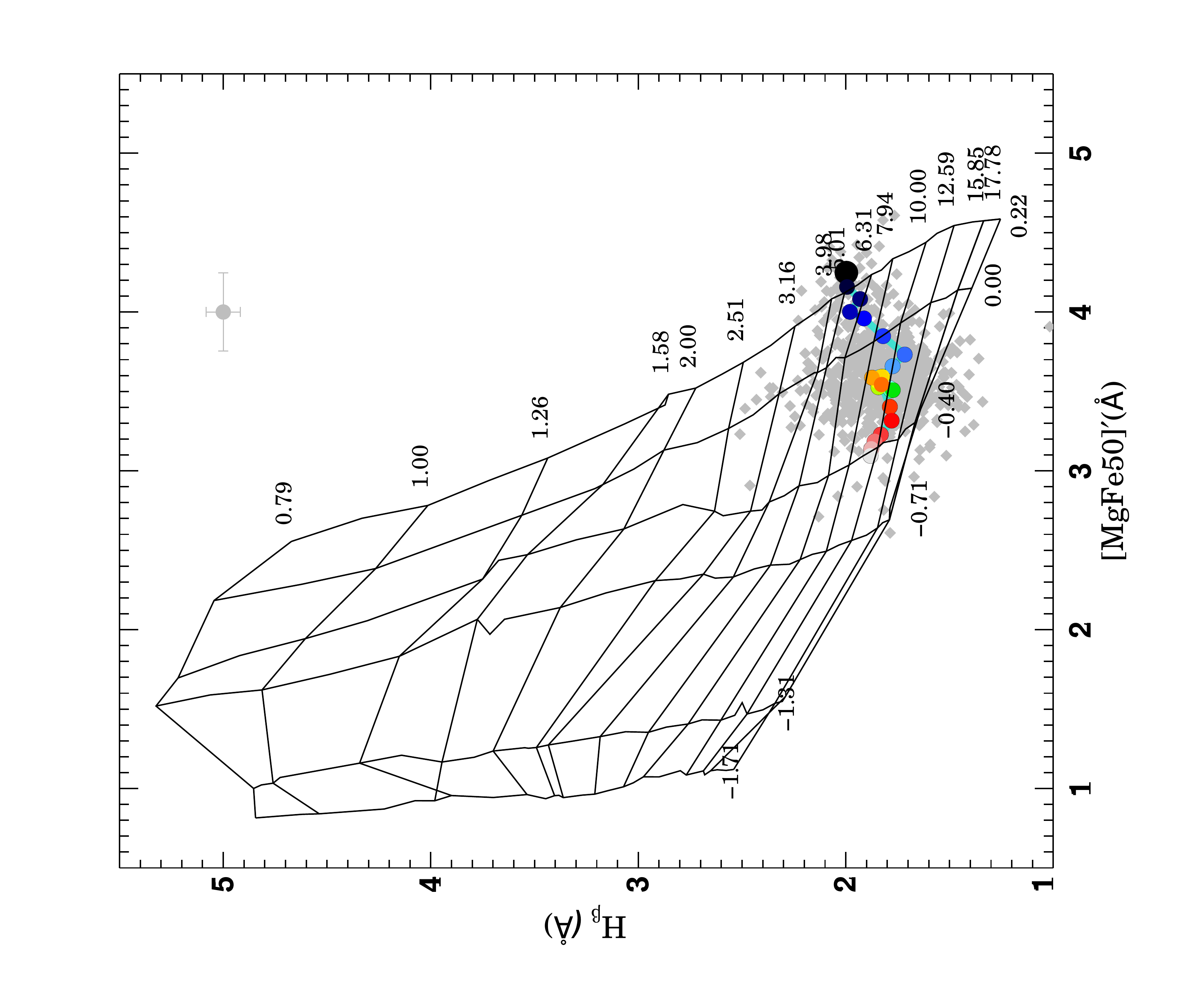}
\caption{Fig~\ref{fig:summ1015} continued.}
\label{fig:summ4643}
\end{figure*}

\begin{figure*}
\includegraphics[width=0.79\linewidth, angle=90]{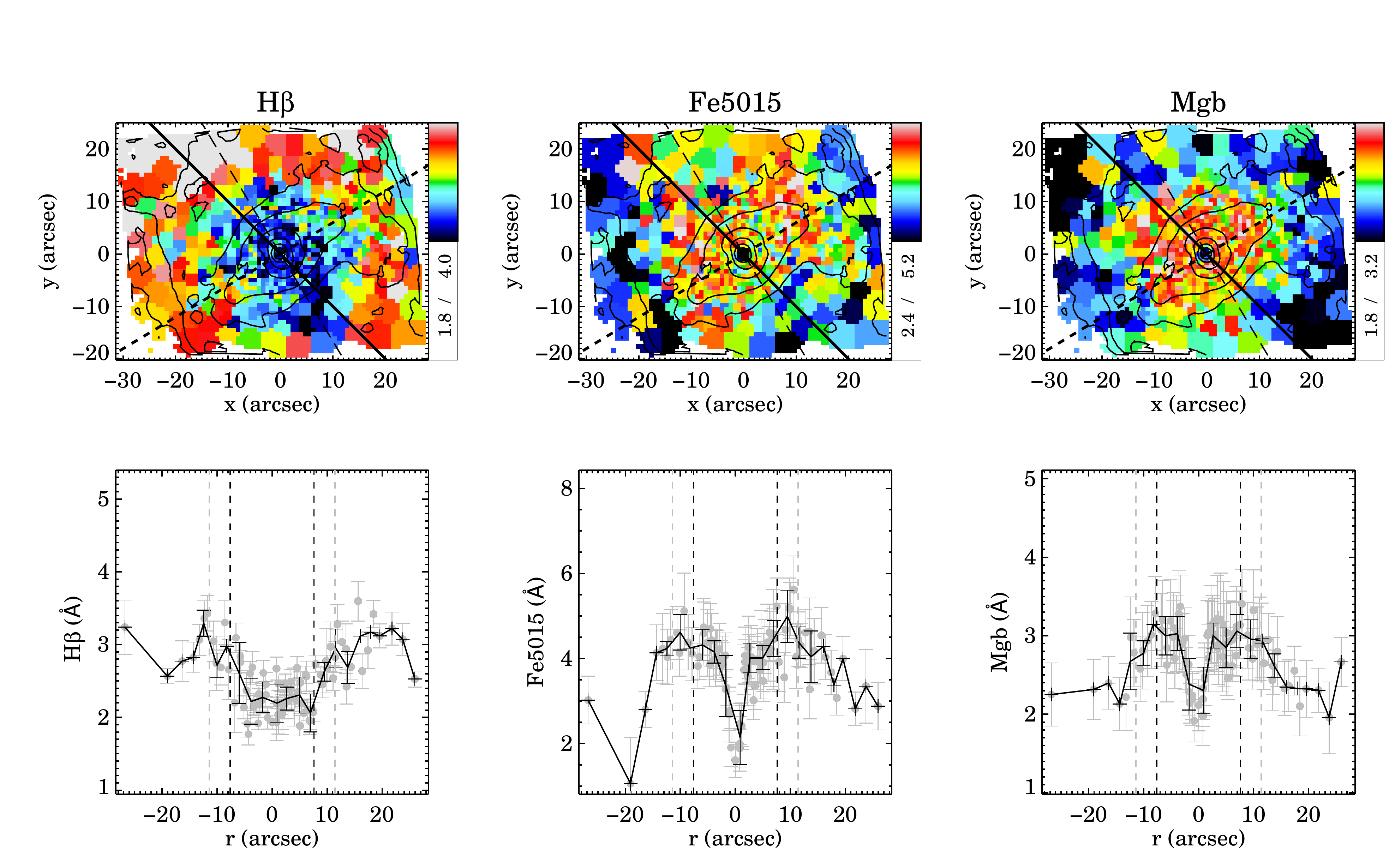}
\includegraphics[width=0.79\linewidth, angle=90]{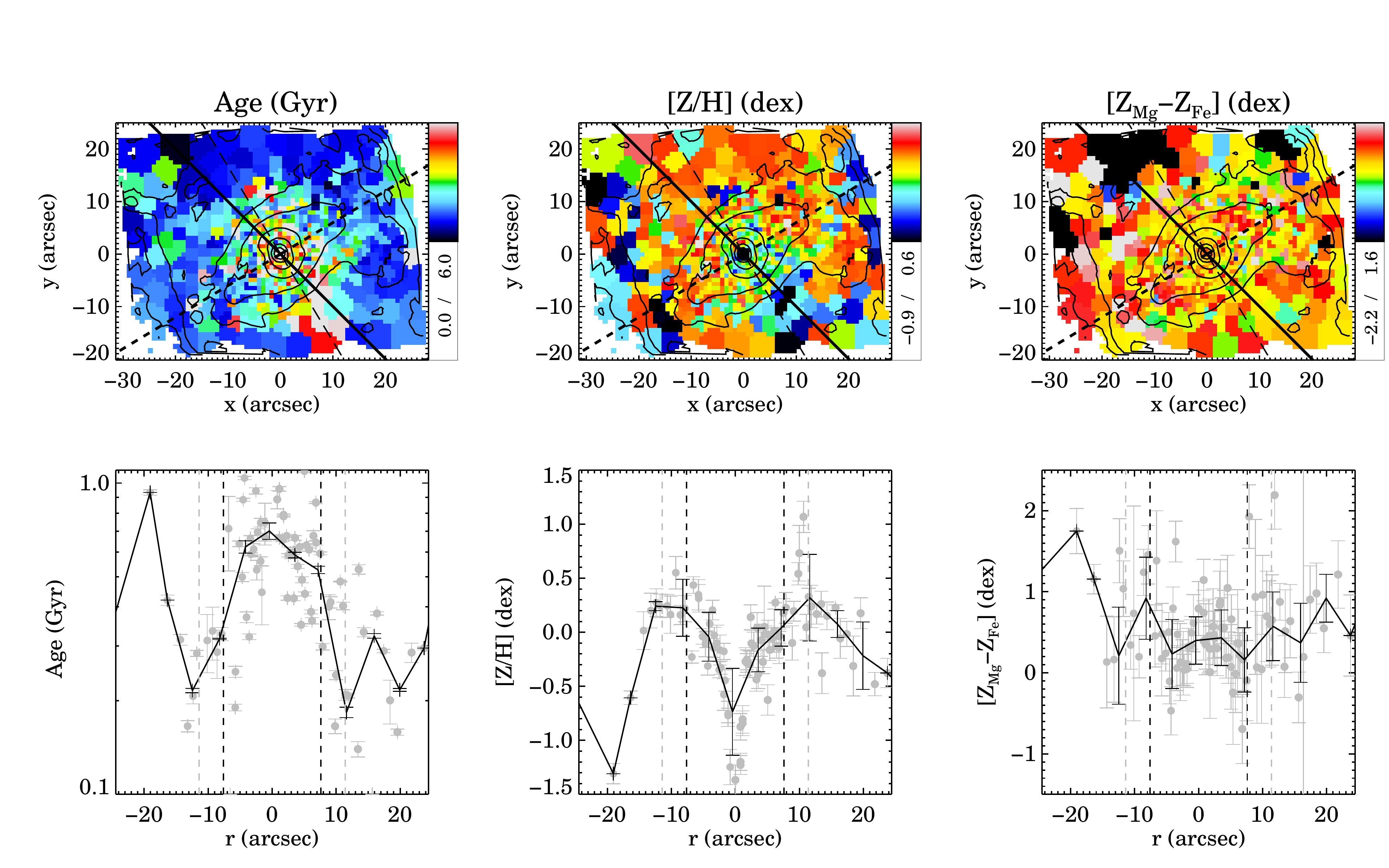}
\includegraphics[width=0.33\linewidth, angle=90,bb = -100 -150 600 600]{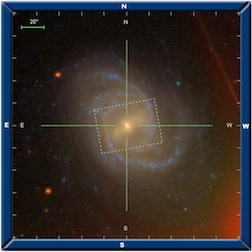}
\includegraphics[width=0.33\linewidth, angle=90,bb = -400 -40 50 -300]{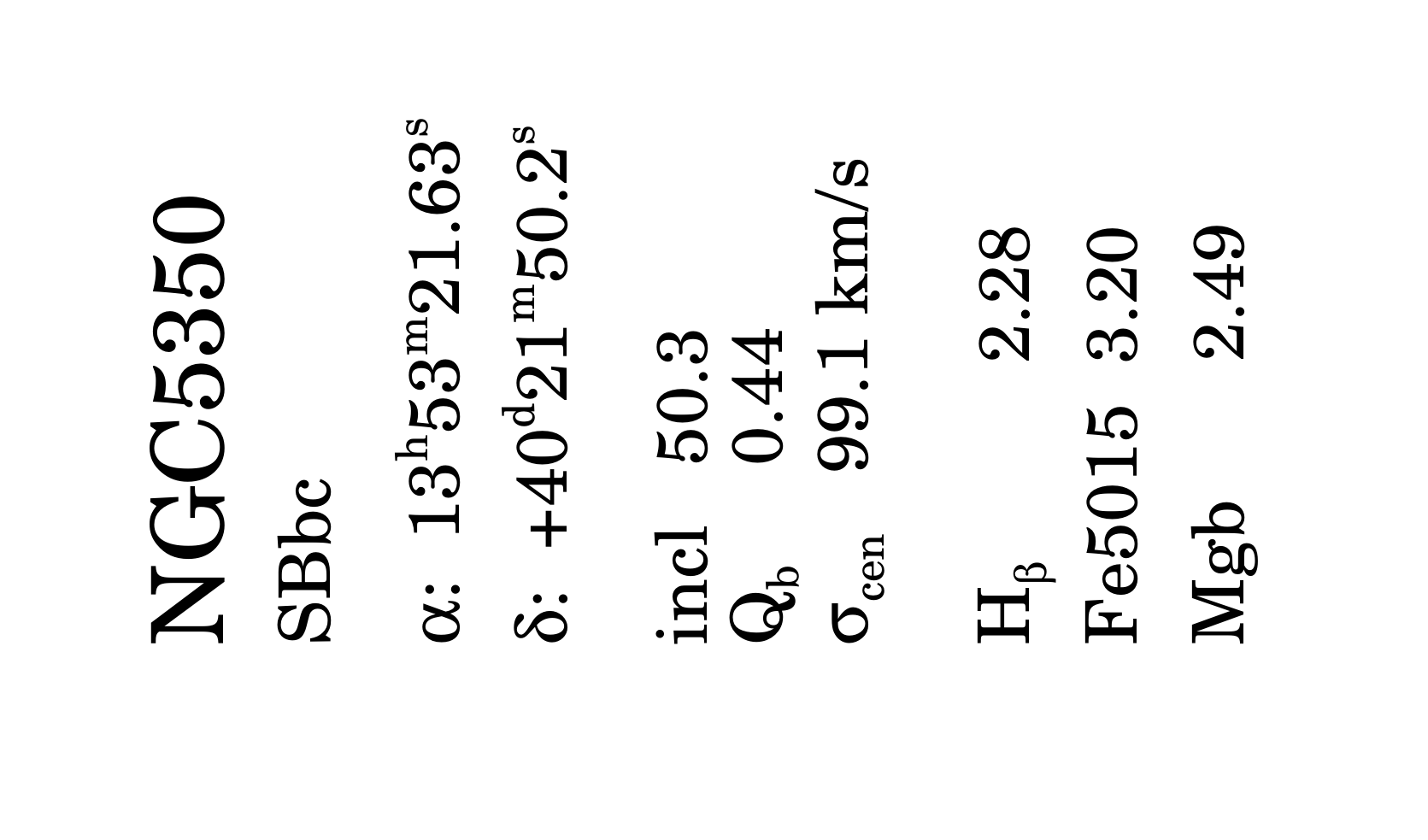}
\includegraphics[width=0.5\linewidth, angle=90,bb = 50 150 800 900]{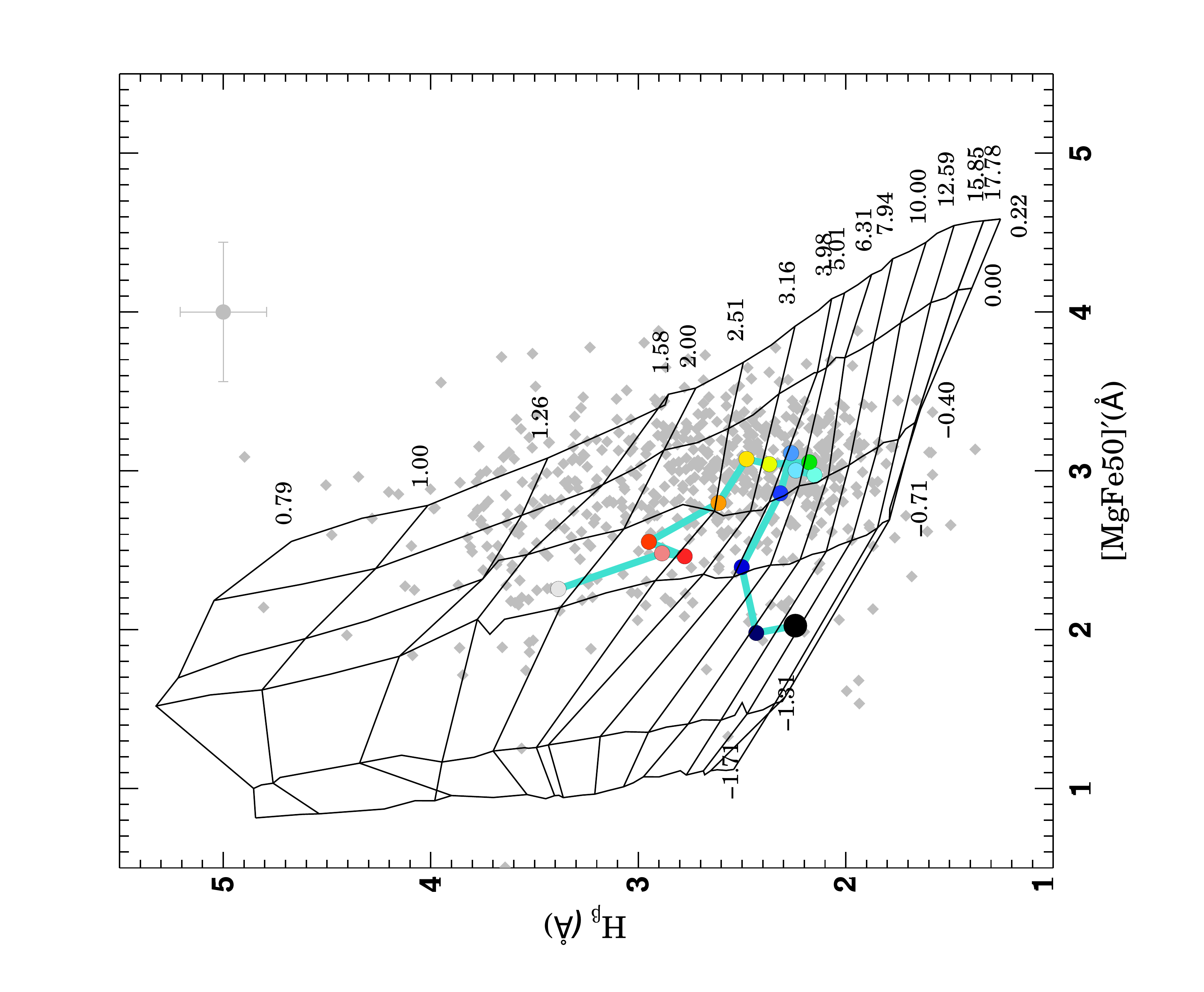}
\caption{Fig~\ref{fig:summ1015} continued.}
\label{fig:summ5350}
\end{figure*}

\begin{figure*}
\includegraphics[width=0.79\linewidth, angle=90]{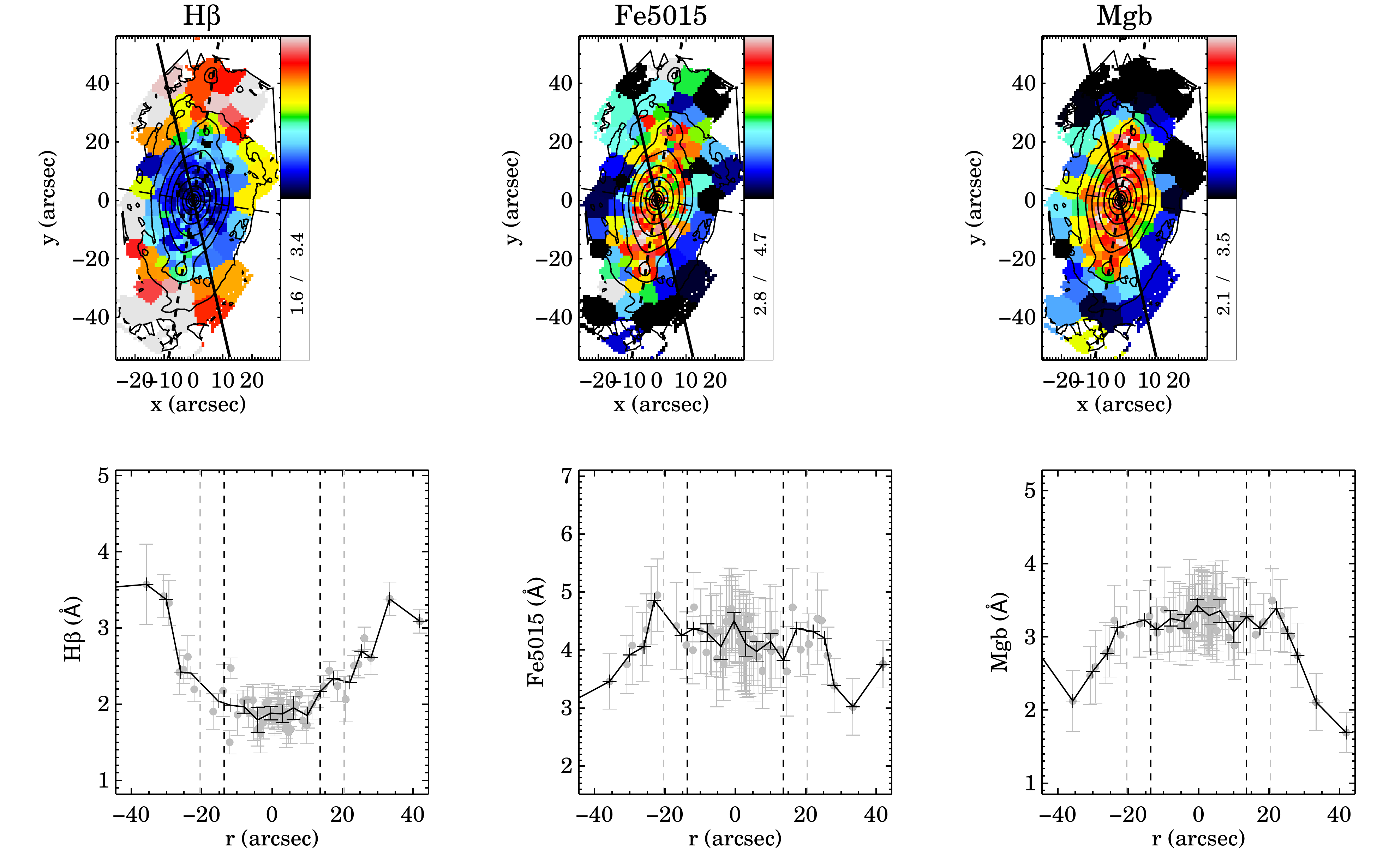}
\includegraphics[width=0.79\linewidth, angle=90]{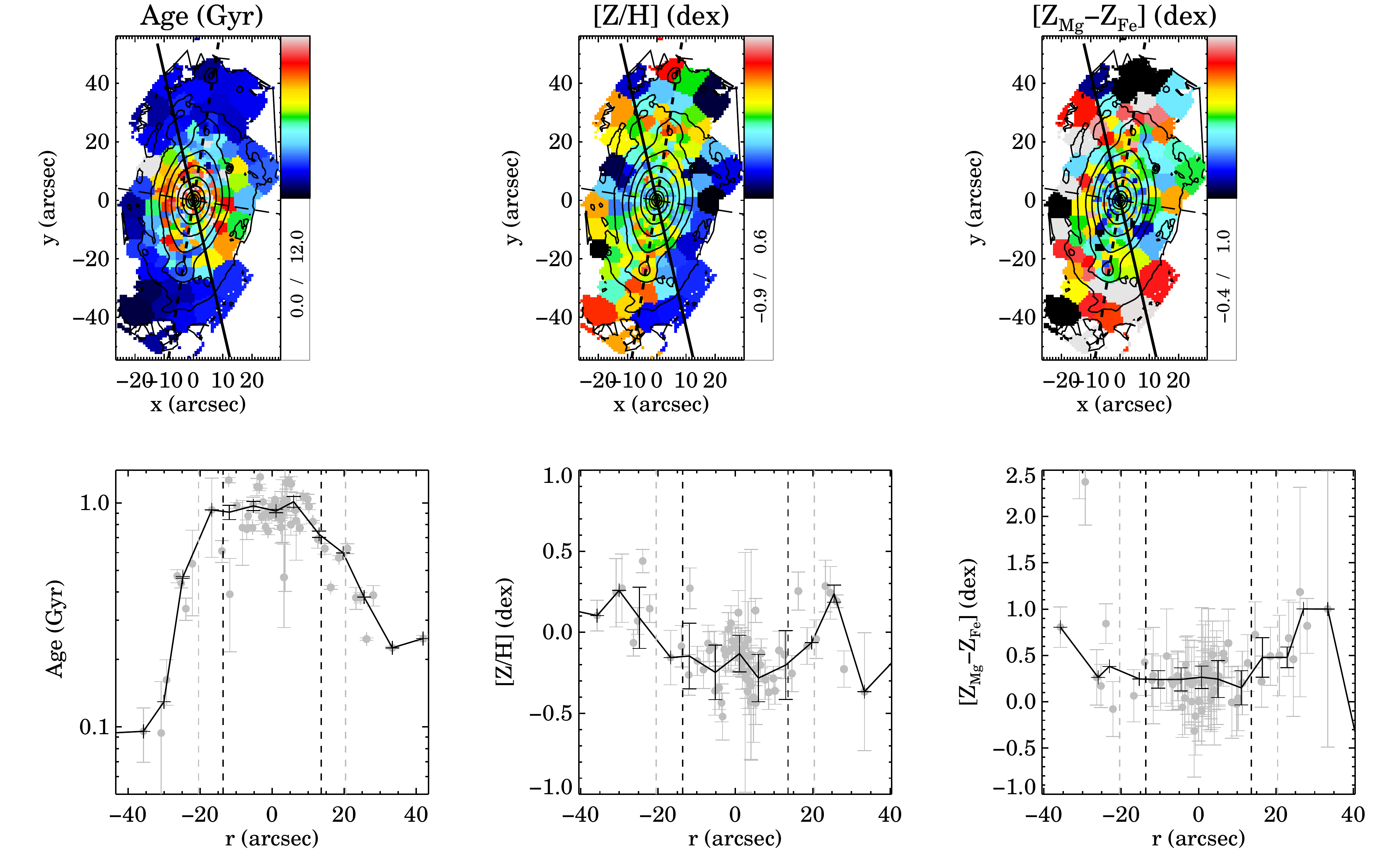}
\includegraphics[width=0.33\linewidth, angle=90,bb = -100 -150 600 600]{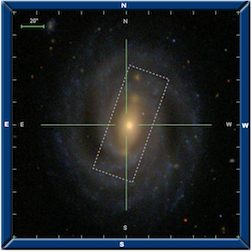}
\includegraphics[width=0.33\linewidth, angle=90,bb = -400 -40 50 -300]{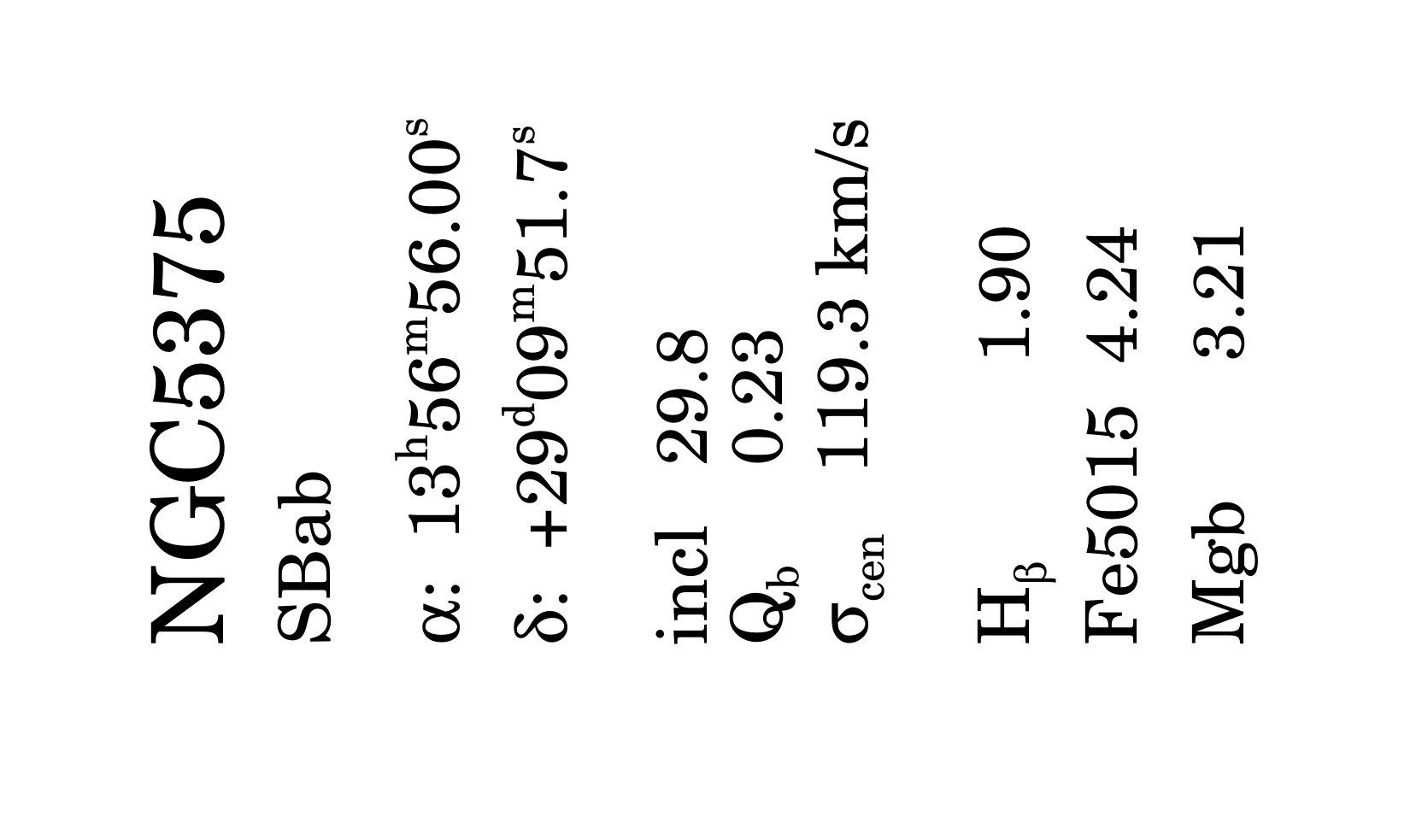}
\includegraphics[width=0.5\linewidth, angle=90,bb = 50 150 800 900]{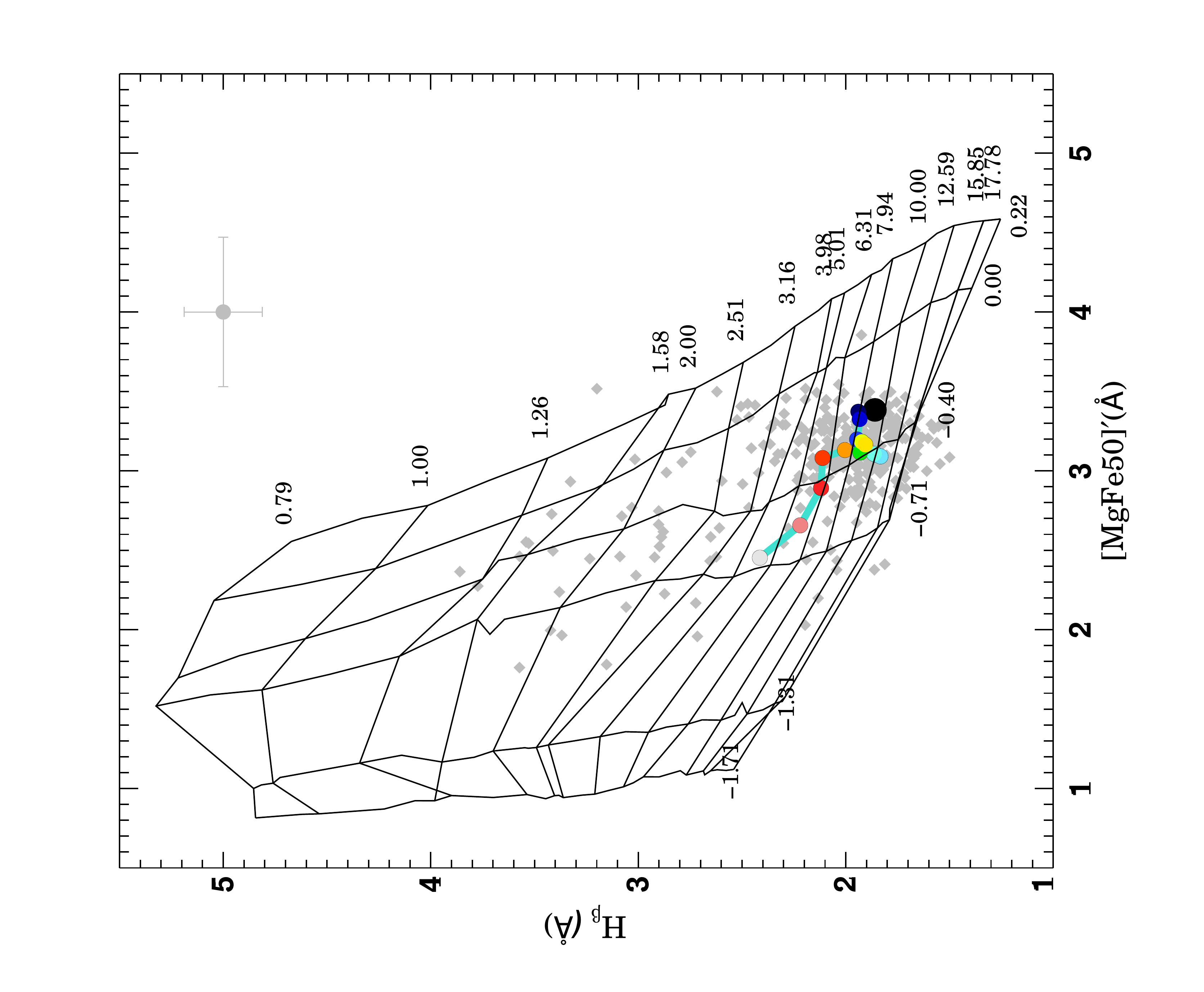}
\caption{Fig~\ref{fig:summ1015} continued.}
\label{fig:summ5375}
\end{figure*}

\begin{figure*}
\includegraphics[width=0.79\linewidth, angle=90]{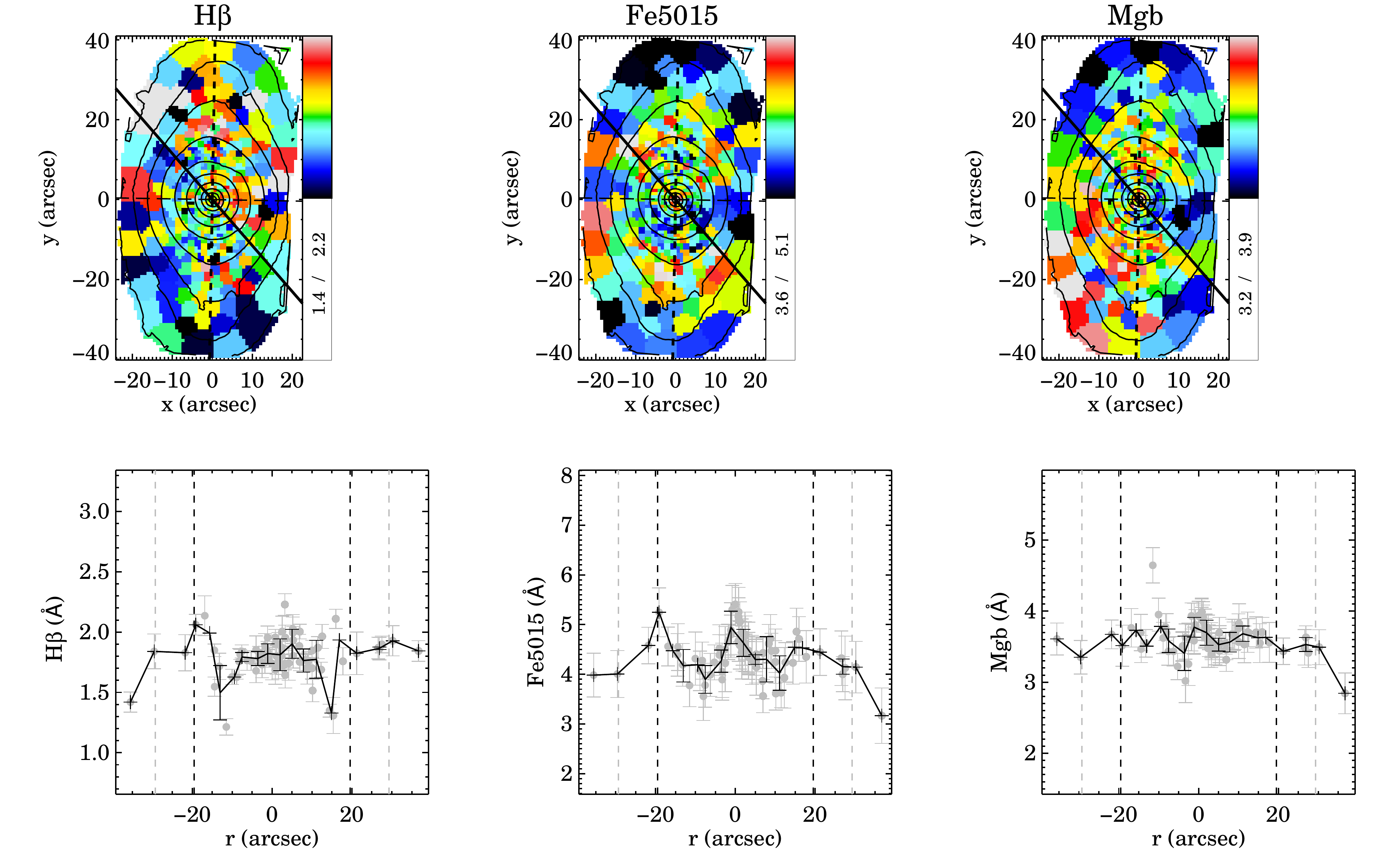}
\includegraphics[width=0.79\linewidth, angle=90]{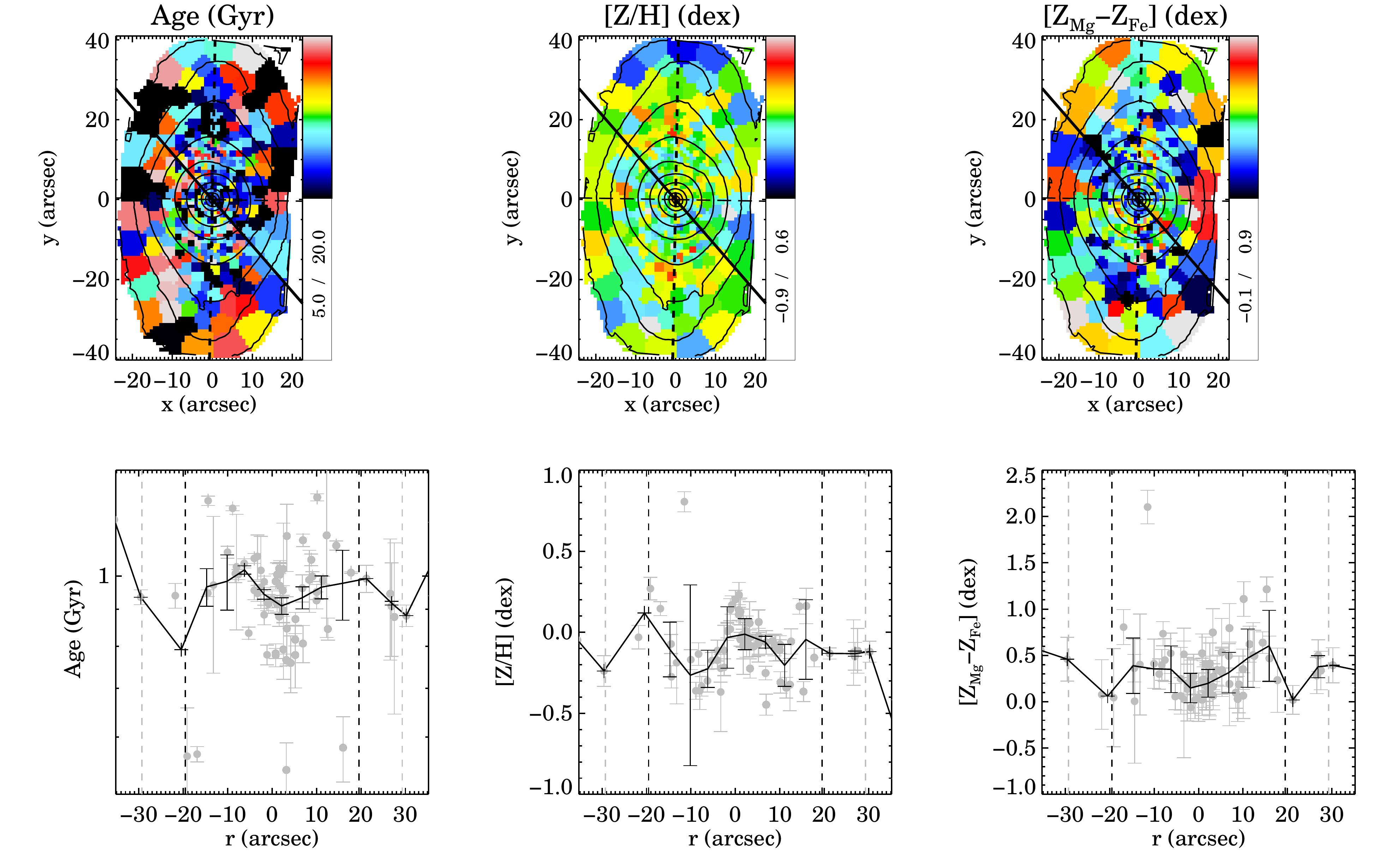}
\includegraphics[width=0.33\linewidth, angle=90,bb = -100 -150 600 600]{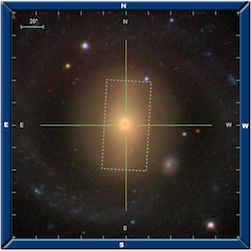}
\includegraphics[width=0.33\linewidth, angle=90,bb = -400 -40 50 -300]{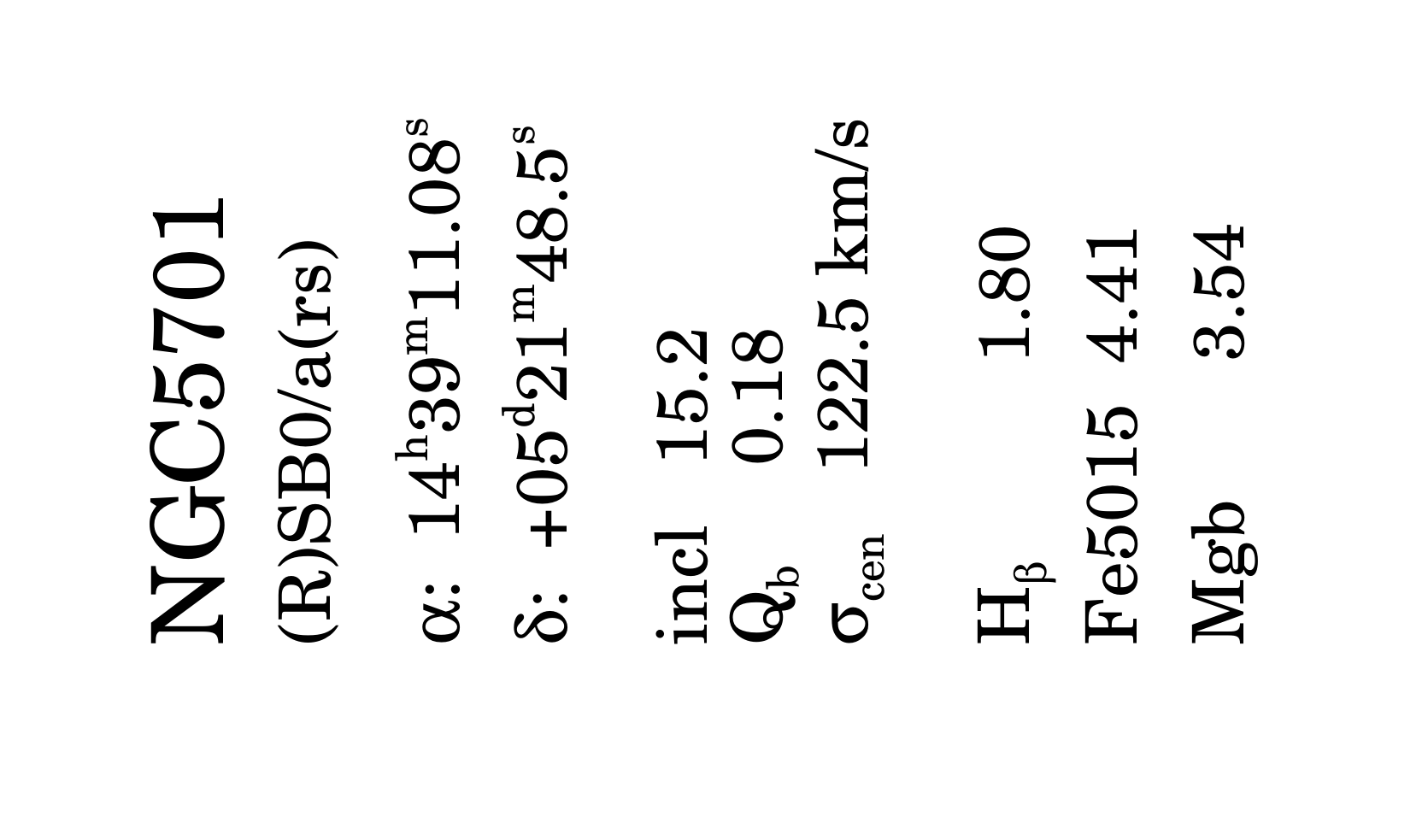}
\includegraphics[width=0.5\linewidth, angle=90,bb = 50 150 800 900]{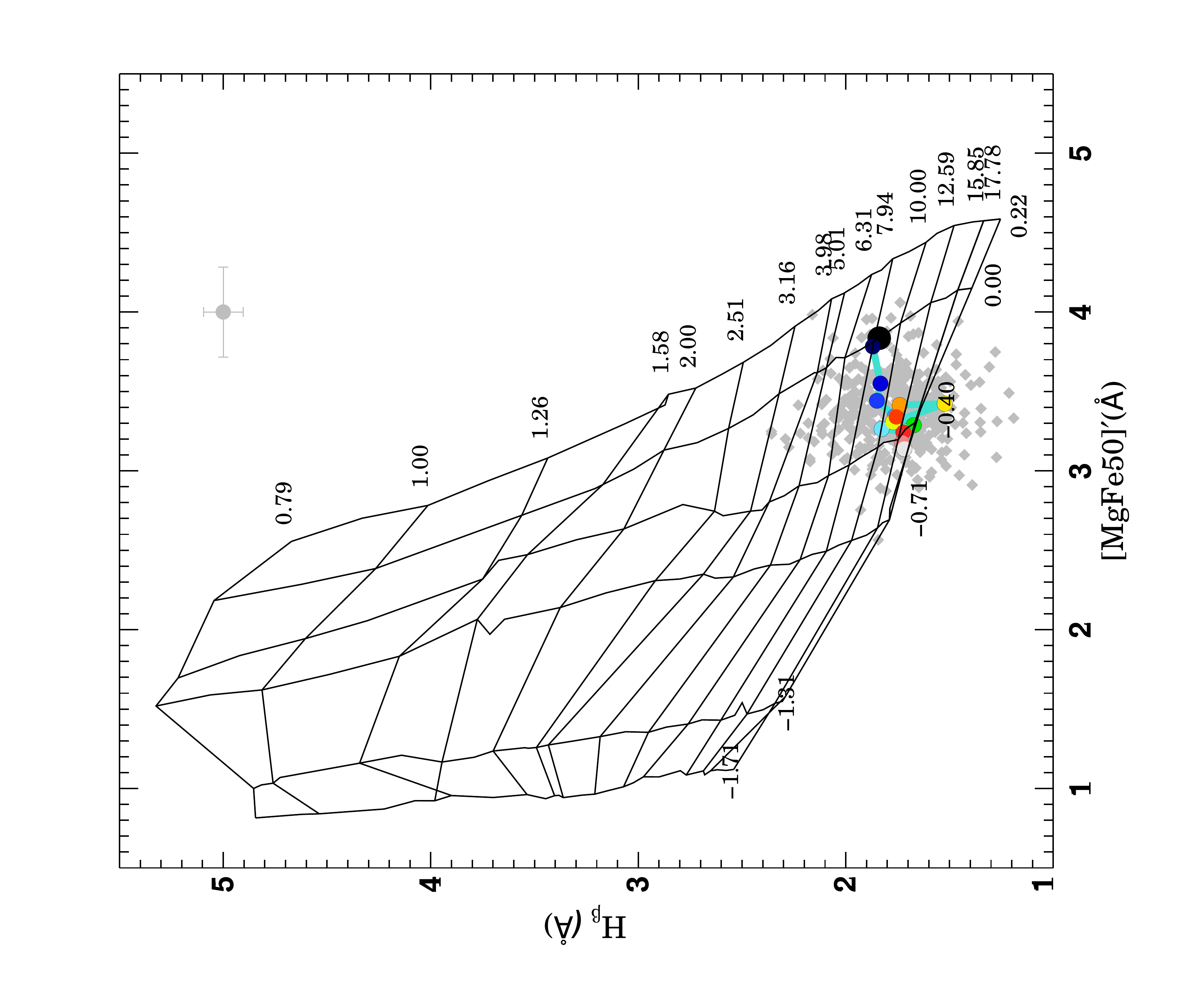}
\caption{Fig~\ref{fig:summ1015} continued.}
\label{fig:summ5701}
\end{figure*}

We show maps of the obtained line strength indices and derived SSP parameters for the entire BaLROG 
sample of galaxies in figures~\ref{fig:summ1015} to \ref{fig:summ5701}. In each figure
 we show different maps of each galaxy in landscape format, top to bottom and left to right:  top left: (i) SDSS ugri-colour image of the galaxy , (ii) parameters of
 the galaxy along with the inclination, the bar strengths measured, the central stellar velocity dispersion and central line strength indices for H$_\beta$, Fe5015 and Mg$b$. Bottom left: (i) index-index diagram with age-sensitive index H$_\beta$ versus metallicity-sensitive combined index of MgFe50', individual measurements from each bin are shown in gray and the isophotal profile with dark blue (and larger) dots showing the central measurement going from yellow to red towards the outer parts, representative uncertainties are indicated in the right top corner for individual measurements. On the left, top to bottom we show index maps and SSP maps and their corresponding profiles along the bar major, after the maps we show the (i) major axis profile: gray values show individual Voronoi measurements while black ones indicate averaged quantities, gray dashed vertical lines indicate 3/4 of the bar length and black dashed lines half the bar length. From left to right we show (i) H$_\beta$ index, (ii) Fe5015 index, (iii) Mg$b$ index; below the SSP parameters: (i) age (in logarithmic units and in Gyr), (ii) metallicity (in dex), (iii) $\alpha$ abundance (as a measure of the difference between metallicities of Mg and Fe). In all maps, isophotes are separated by half a magnitude.

% JUST COMMENTED FOR NOW
%
%%-----------------------------------------------------------------------------
%----------------------------------------------------------------------------
%%-----------------------------------------------------------------------------
%----------------------------------------------------------------------------
%
%-----------------------------------------------------------------------------
%----------------------------------------------------------------------------
%-----------------------------------------------------------------------------
%----------------------------------------------------------------------------
%-----------------------------------------------------------------------------
%----------------------------------------------------------------------------
%-----------------------------------------------------------------------------
%----------------------------------------------------------------------------
%-----------------------------------------------------------------------------
%----------------------------------------------------------------------------
%-----------------------------------------------------------------------------
%----------------------------------------------------------------------------
%-----------------------------------------------------------------------------
%----------------------------------------------------------------------------
%-----------------------------------------------------------------------------
%----------------------------------------------------------------------------
%-----------------------------------------------------------------------------
%----------------------------------------------------------------------------
%-----------------------------------------------------------------------------
%----------------------------------------------------------------------------
%-----------------------------------------------------------------------------
%----------------------------------------------------------------------------
%-----------------------------------------------------------------------------
%----------------------------------------------------------------------------
%-----------------------------------------------------------------------------
%----------------------------------------------------------------------------
%-----------------------------------------------------------------------------
%----------------------------------------------------------------------------
%

\section{Measurement of the h$_3$-v/$\sigma$ anti-correlation on the control sample}
\label{app:controlsampleh3}

\begin{figure}
\includegraphics[width=\linewidth]{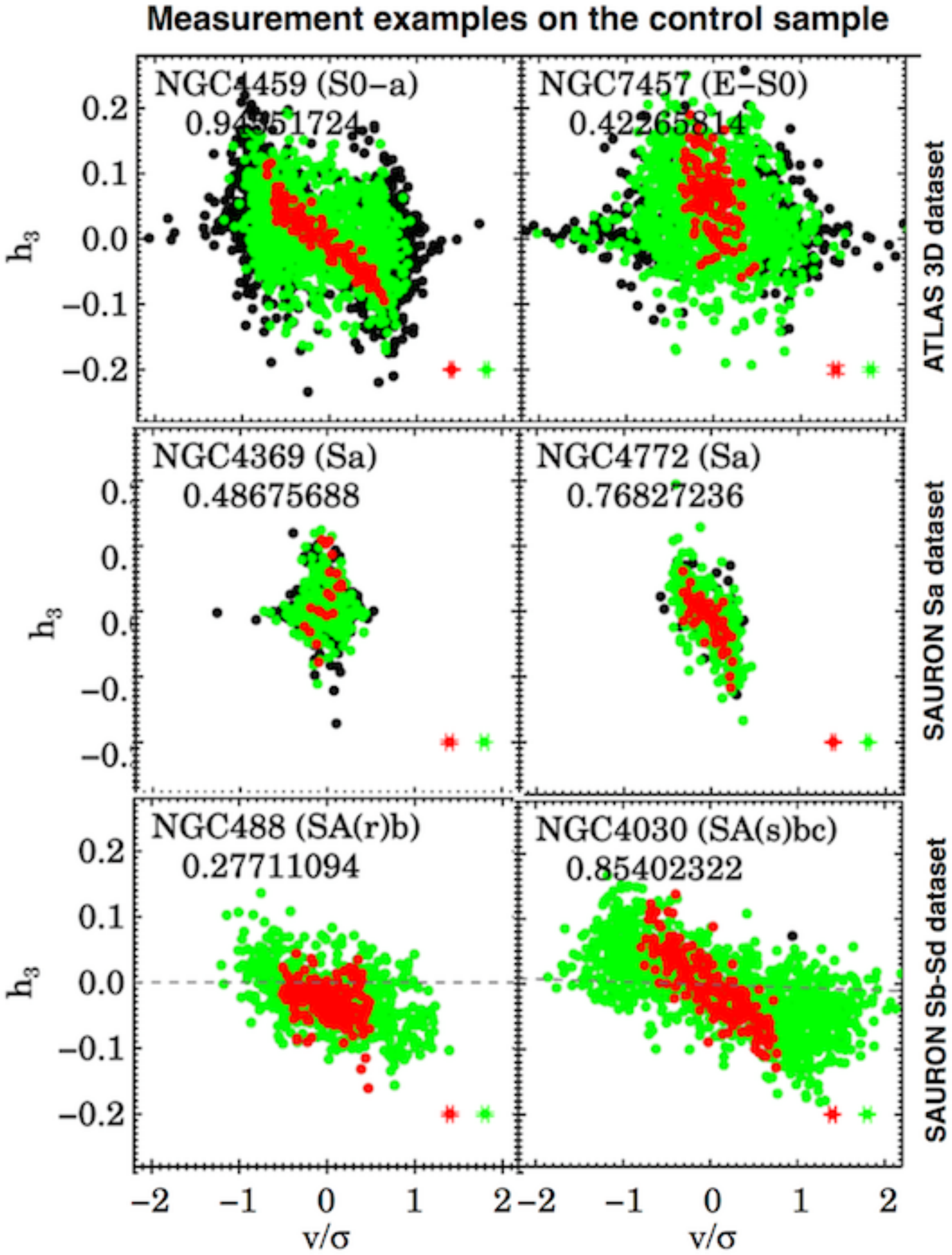}
\caption{Example of 6 galaxies (2 per survey) of the h$_3$-v/$\sigma$ anti-correlation measurement on the unbarred control sample. Black points indicate the extend of one effective radius, green points within half the effective radius and red points within 0.1 of the effective radius. }
\label{fig:CSh3}
\end{figure}

Just for illustration, we show the h$_3$-v/$\sigma$ anti-correlation measurement done on the control sample analogue to Paper I. Figure~\ref{fig:CSh3} shows this measurement for all the three surveys, two galaxies as examples. The top left panel shows a strong anti-correlation within 0.1 effective radii (red points) for example while the right panel does not show this. In every dataset we find both scenarios. 

%-------------------------------------------------------------
%----------------------------------------------------------------------------

%
%
\section{Index Gradients}
\label{app:indexgrad}

\begin{figure*}
\includegraphics[width=0.88\linewidth]{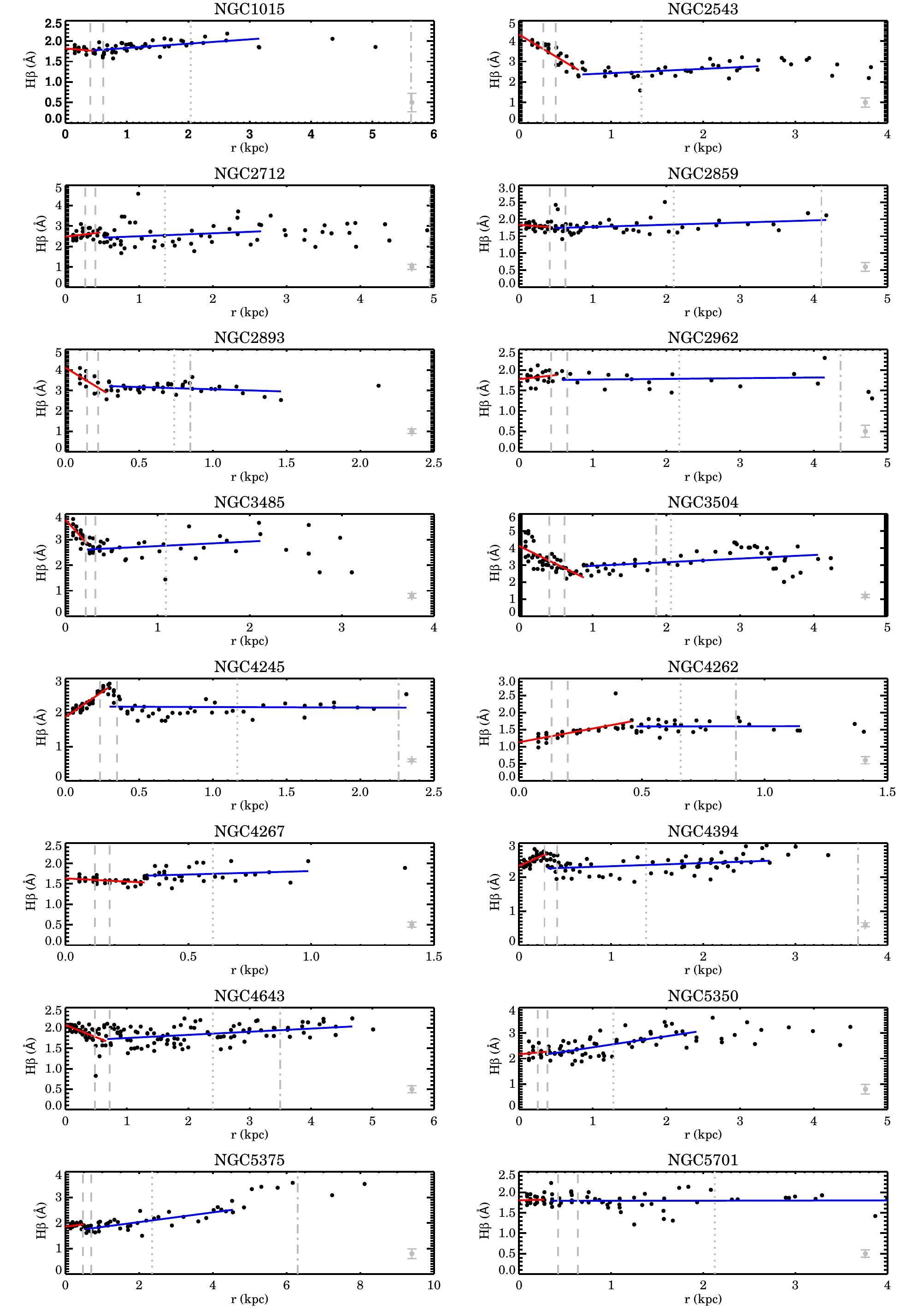}
\caption{Bar major axis profiles and linear inner (red) and outer (blue) gradients of the H$_\beta$ index. The mean uncertainty is indicated in each panel in the lower right corner. Dashed lines indicate the region between 0.1 and 0.15 R$_{\rm bar}$, dotted lines the position of 0.5 R$_{\rm bar}$ and dashed-dotted lines the position of 1.5 R$_{\rm eff}$.  }
\label{fig:hbgrad}
\end{figure*}

\begin{figure*}
\includegraphics[width=0.88\linewidth]{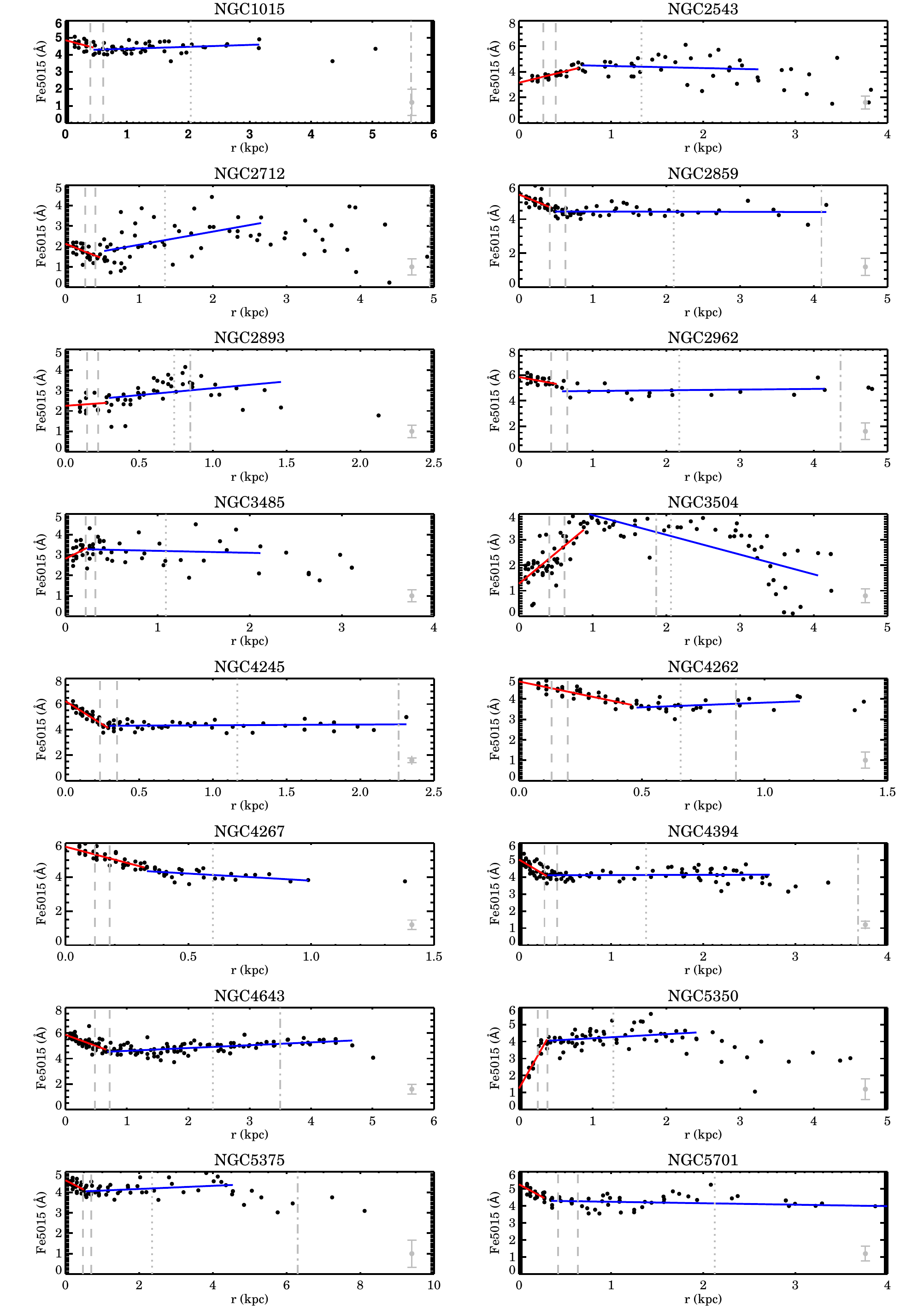}
\caption{Bar major axis profiles and linear inner (red) and outer (blue) gradients of the Fe5015 index. The mean uncertainty is indicated in each panel in the lower right corner. Dashed lines indicate the region between 0.1 and 0.15 R$_{\rm bar}$, dotted lines the position of 0.5 R$_{\rm bar}$ and dashed-dotted lines the position of 1.5 R$_{\rm eff}$.  }
\label{fig:fegrad}
\end{figure*}

%We determine two gradients along the index and SSP profiles for major and minor axes of our galaxies. As examples we show in the text the major axis gradients for Mg$b$ and here the metallicity in Figs.~\ref{fig:Mggrad} and \ref{fig:Metgrad} respectively. 
We determine two gradients along the index profiles for major and minor axes of our galaxies. As examples we show in the text the major axis gradients for Mg$b$ in Fig.~\ref{fig:Mggrad}. Here, we show the other gradient measurements: i) H$_\beta$ gradients in Fig.~\ref{fig:hbgrad} and ii) Fe5015 gradients in Fig.~\ref{fig:fegrad}. 

%-------------------------------------------------------------
%----------------------------------------------------------------------------
\clearpage
%-------------------------------------------------------------
%\begin{figure*}
%\includegraphics[width=0.88\linewidth]{Bar_mj_Mg-gradients_paper_v1-eps-converted-to.pdf}
%\caption{Bar major axis profiles and linear inner (red) and outer (blue) gradients of the Mg$b$ index as an example of the index measurements for the entire BaLROG sample. The mean uncertainty is indicated in each panel in the lower right corner. Dashed lines indicate the region between 0.1 and 0.15 R$_{\rm bar}$, dotted lines the position of 0.5 R$_{\rm bar}$ and dashed-dotted lines the position of 1.5 R$_{\rm eff}$.  }
%\label{fig:mgbgrad}
%\end{figure*}
%----------------------------------------------------------------------------
\clearpage
%-------------------------------------------------------------
%----------------------------------------------------------------------------
\clearpage

\section{Control Sample Gradients}
\label{app:CSgrad}

\begin{figure*}
\includegraphics[width=0.88\linewidth]{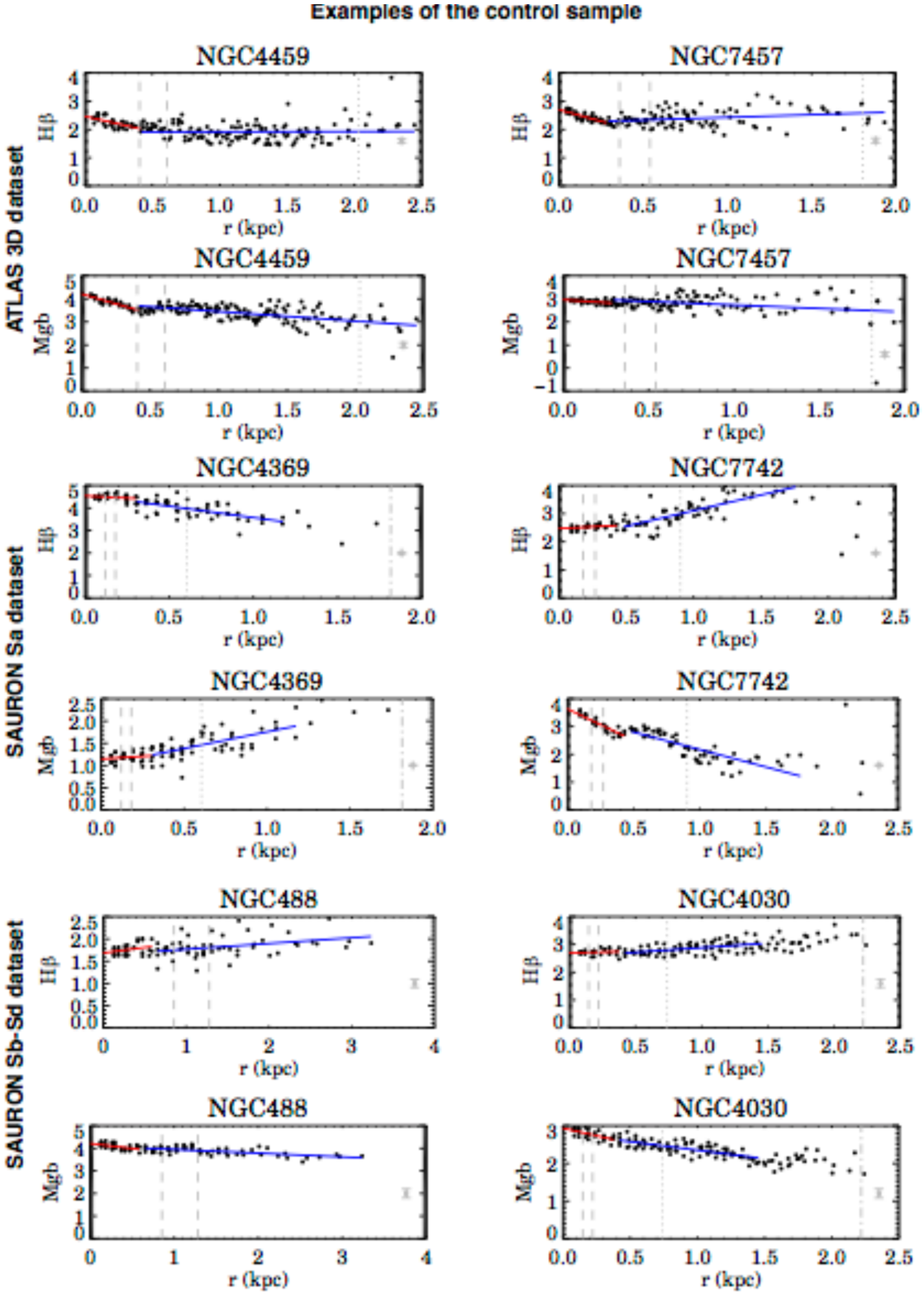}
\caption{Radial profiles and linear inner (red) and outer (blue) gradients of the H$_\beta$ and Mg$b$ index for examples of the three control datasets. The mean uncertainty is indicated in each panel on the right. Dashed lines indicate the region between 0.1 and 0.15 R$_{\rm eff}$, dotted lines the position of 0.5 R$_{\rm eff}$ and dashed-dotted lines the position of 1.5 R$_{\rm eff}$.  }
\label{fig:CSindgrad}
\end{figure*}

%We determine two gradients along the index and SSP profiles for major and minor axes of our galaxies. As examples we show in the text the major axis gradients for Mg$b$ and here the metallicity in Figs.~\ref{fig:Mggrad} and \ref{fig:Metgrad} respectively. 
Figure~\ref{fig:CSindgrad} shows the profiles of examples of the unbarred control datasets. The breaks are clearly less visible in all of the profiles compared to the barred sample. We indicate fractions of the effective radius, but it does not seem to be strongly related to the break position.

%-------------------------------------------------------------
%----------------------------------------------------------------------------

\label{lastpage}

\section{SSP Gradients}
\label{app:grad}

\begin{figure*}
\includegraphics[width=0.88\linewidth]{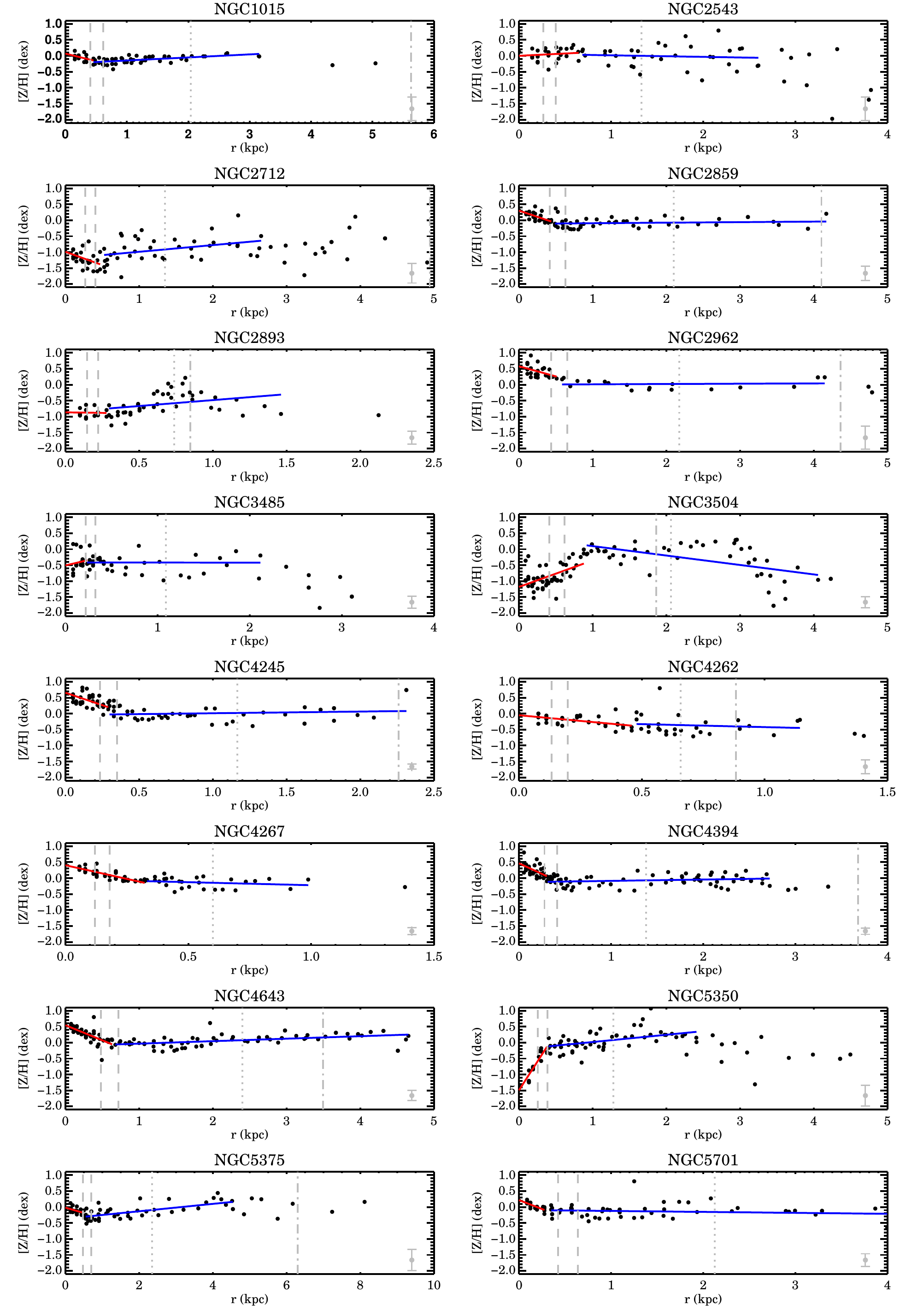}
\caption{Bar major axis profiles and linear inner (red) and outer (blue) gradients of the metallicity. The mean uncertainty is indicated in each panel in the lower right corner. Dashed lines indicate the region between 0.1 and 0.15 R$_{\rm bar}$, dotted lines the position of 0.5 R$_{\rm bar}$ and dashed-dotted lines the position of 1.5 R$_{\rm eff}$. }
\label{fig:Metgrad}
\end{figure*}

\begin{figure*}
\includegraphics[width=0.88\linewidth]{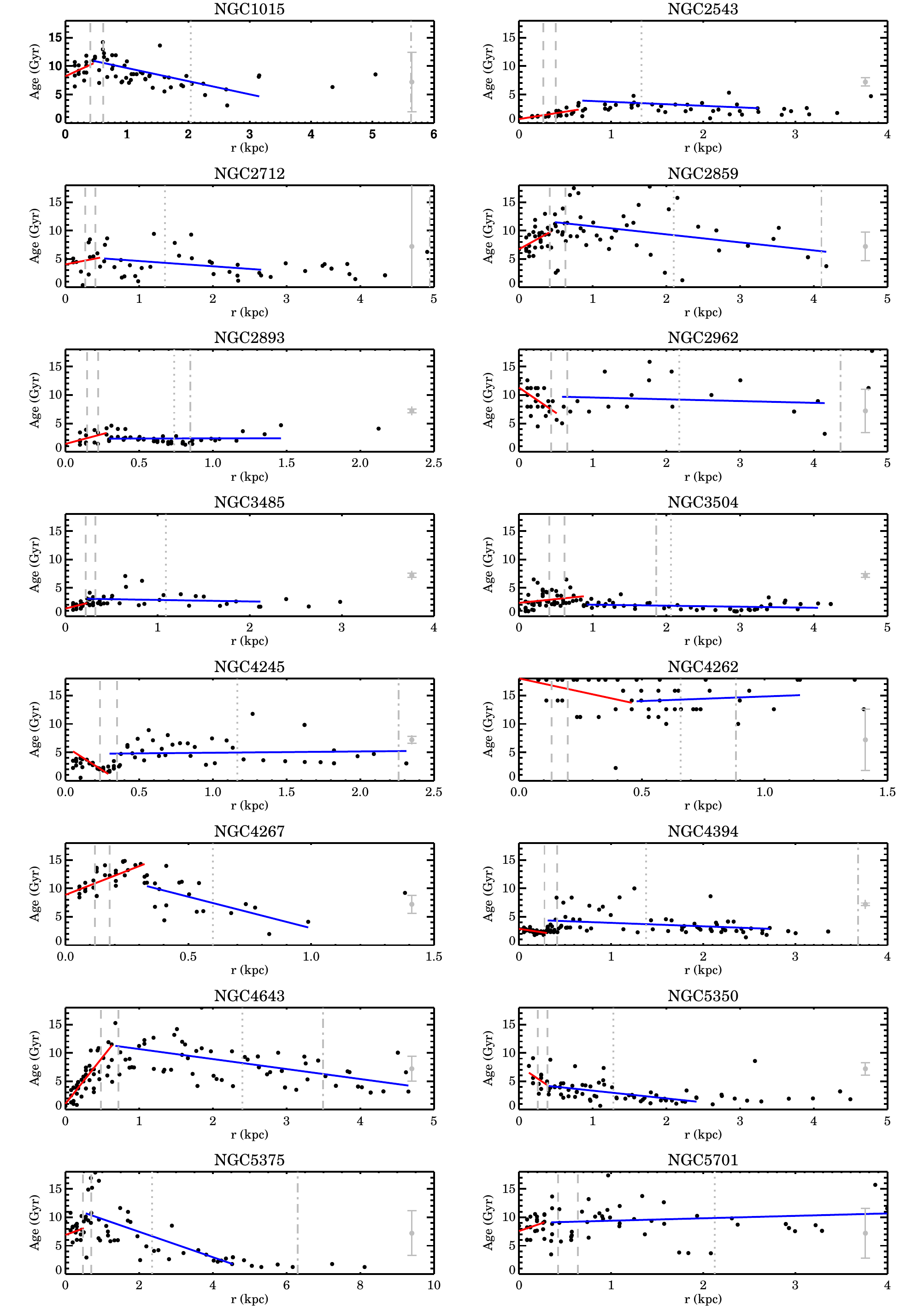}
\caption{Bar major axis profiles and linear inner (red) and outer (blue) gradients of the age. The mean uncertainty is indicated in each panel in the lower right corner. Dashed lines indicate the region between 0.1 and 0.15 R$_{\rm bar}$, dotted lines the position of 0.5 R$_{\rm bar}$ and dashed-dotted lines the position of 1.5 R$_{\rm eff}$. }
\label{fig:agegrad}
\end{figure*}

\begin{figure*}
\includegraphics[width=0.88\linewidth]{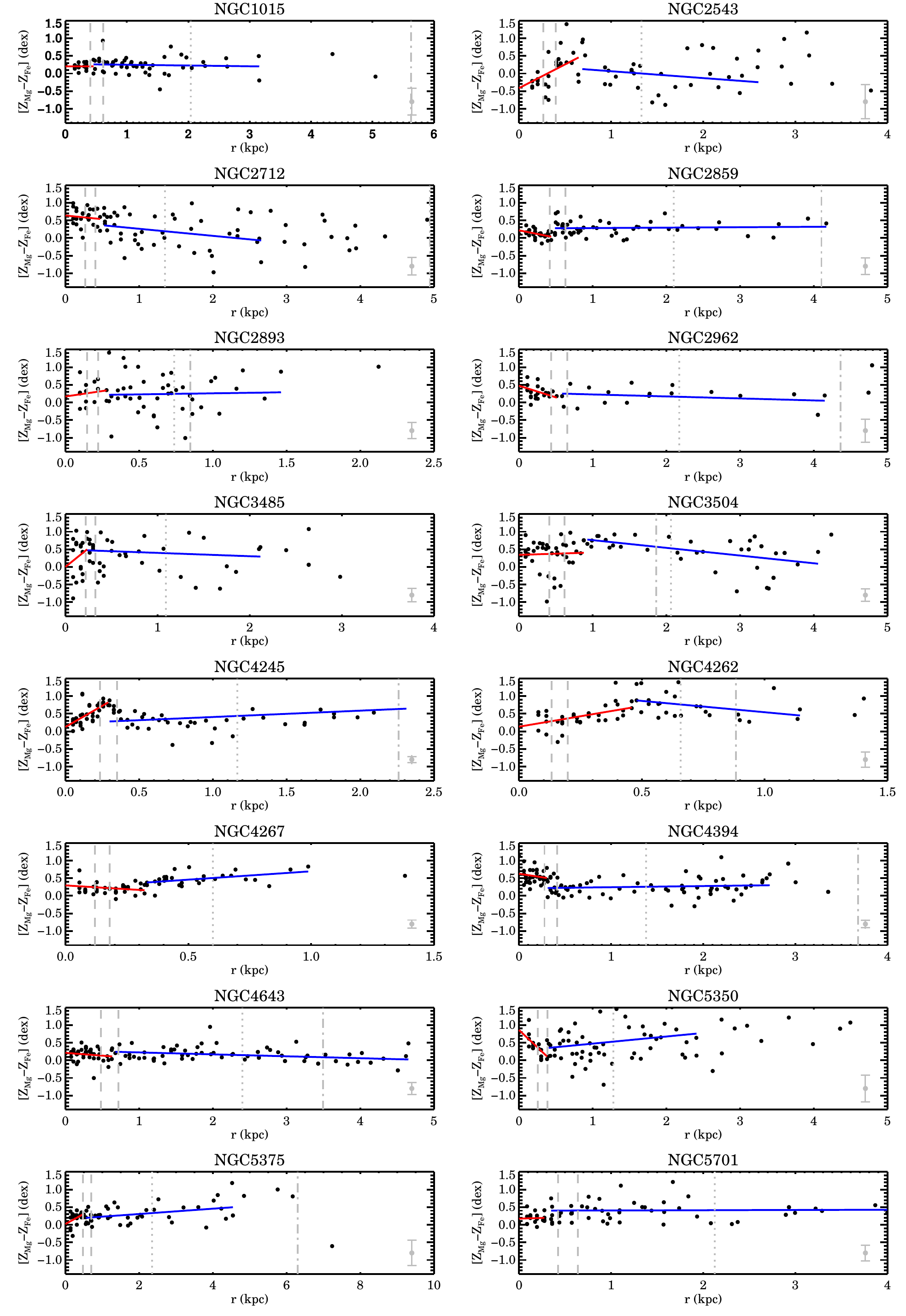}
\caption{Bar major axis profiles and linear inner (red) and outer (blue) gradients of the $\alpha$ abundance. The mean uncertainty is indicated in each panel in the lower right corner. Dashed lines indicate the region between 0.1 and 0.15 R$_{\rm bar}$, dotted lines the position of 0.5 R$_{\rm bar}$ and dashed-dotted lines the position of 1.5 R$_{\rm eff}$. }
\label{fig:abungrad}
\end{figure*}

 We also determined the two gradients along the SSP profiles, shown here for the metallicity gradients in Fig.~\ref{fig:Metgrad}, age gradients in Fig.~\ref{fig:agegrad} and $\alpha$ abundance gradients in Fig.~\ref{fig:abungrad}.

%-----------------------------------------------------------------------------
%----------------------------------------------------------------------------
\clearpage
%-----------------------------------------------------------------------------

%----------------------------------------------------------------------------
\clearpage
%-----------------------------------------------------------------------------

%----------------------------------------------------------------------------

%example_plots.pdf

%\begin{landscape}[total={17.8cm,24.0cm}]
%\noindent
%\rule{\linewidth}{4pt}
%\rule{\geometry}{total={24.0cm,17.8cm}}
%\thispagestyle{empty}
%\vspace{1cm}
%\advance\leftskip-3cm
%\advance\rightskip-3cm
%\includegraphics[width=0.3\linewidth, angle=90]{4394_dr7.jpg}
%\vspace{5cm}
%\includegraphics[width=0.01\linewidth, angle=0]{white.pdf}
%\includegraphics[width=0.13\linewidth, angle=90]{4394_dr7a.pdf}
%\end{landscape}

%\label{lastpage}

\end{document}